\documentclass[a4paper,12pt,twoside]{ThesisStyle}

\usepackage{amsmath,amssymb}             
\usepackage[latin1]{inputenc}
\usepackage[T1]{fontenc}
\usepackage[left=1.in,right=.9in,top=1.1in,bottom=1.1in,includefoot,includehead,headheight=13.6pt]{geometry}

\usepackage{cancel}
\usepackage{tabularx}

\usepackage{aecompl}

\usepackage{ifpdf}

\ifpdf
  \usepackage[pdftex]{graphicx}
  \DeclareGraphicsExtensions{.jpg}
  \usepackage[a4paper,pagebackref,hyperindex=true]{hyperref}
\else
  \usepackage{graphicx}
  \DeclareGraphicsExtensions{.ps,.eps}
  \usepackage[a4paper,dvipdfm,hyperindex=true]{hyperref}
\fi

\graphicspath{{.}{images/}}

\usepackage{color}
\definecolor{linkcol}{rgb}{0,0,0.4}
\definecolor{citecol}{rgb}{0.5,0,0}

\hypersetup
{
bookmarksopen=true,
pdftitle="Design and Use of Anatomical Atlases for Radiotherapy",
pdfauthor="Olivier COMMOWICK",
pdfsubject="Creation of atlases and atlas based segmentation", 
pdfmenubar=true, 
pdfhighlight=/O, 
colorlinks=true, 
pdfpagemode=None, 
pdfpagelayout=SinglePage, 
pdffitwindow=true, 
linkcolor=linkcol, 
citecolor=citecol, 
urlcolor=linkcol 
}

\usepackage{rotating}                    
\usepackage{fancyhdr}                    

\let\headruleORIG\headrule
\renewcommand{\headrule}{\color{black} \headruleORIG}

\fancypagestyle{plain}{
  \fancyhead{}
  \fancyfoot{}
  
}

\usepackage{algorithm}
\usepackage[noend]{algorithmic}

\makeatletter

\def\cleardoublepage{\clearpage\if@twoside \ifodd\c@page\else%
  \hbox{}%
  \thispagestyle{empty}
  \newpage%
  \if@twocolumn\hbox{}\newpage\fi\fi\fi}

\makeatother

{%

\hrulefill
\vspace*{0.5cm}%
\end{minipage}
}

\usepackage{multirow}
{ \begin{list}%
    {$\bullet$}%
    {\setlength{\labelwidth}{25pt}%
     \setlength{\leftmargin}{30pt}%
     \setlength{\itemsep}{\parsep}}}%
{ \end{list} }

\renewcommand{\epsilon}{\varepsilon}

\usepackage{graphicx,color}

\usepackage{amssymb}
\usepackage{hyperref} 

\usepackage{amsfonts}
\usepackage[all]{hypcap} 
\usepackage{verbatim} 
\usepackage{amsmath}
\usepackage{color}
\usepackage{siunitx}
\usepackage{lmodern}

\newcommand{\abbrev}[1]{\uppercase{#1}}

\newcommand{\kBoltzmann}{k_\text{\abbrev{b}}}

\newcommand{\y}{}

\usepackage{dcolumn}
\usepackage{bm}
\usepackage[normalem]{ulem} 
\usepackage{psfrag}

\usepackage{dcolumn}
\usepackage{bm}
\usepackage{subfigure}

\newcommand{\ph}{\textit{p}-H$_2$}

\newcommand{\bs}{\boldsymbol}
\newcommand{\mz}{\mathbb{Z}}
\newcommand{\mr}{\mathbb{R}}
\newcommand{\ii}{{\rm i}}
\newcommand{\di}{\,{\rm d}}
\newcommand{\lims}{\stackrel{s \to 0}{\longrightarrow}}
\newcommand{\erfc}{{\rm erfc}}
\newcommand{\tabnote}{\caption}

\newcommand{\E}{{\cal E}}
\newcommand{\F}{{\bf F}}

\newcommand{\dR}{\di {\bf R}}
\newcommand{\intint}{\int\!\!\!\!\int}

\newcommand{\m}{{\bs{m}}}

\renewcommand{\k}{{\bs{k}}}

\newcommand{\R}{{\bf R}}
\newcommand{\Rc}{{\bf{\cal R}}}
\newcommand{\Rm}{R_{m}}
\newcommand{\rve}{{\bf r}}
\newcommand{\rv}{\rve}

\newcommand{\dr}{{\di \rve}}
\newcommand{\ri}{\rve_i}
\newcommand{\rj}{\rve_j}
\newcommand{\rk}{\rve_k}
\newcommand{\rik}{{\ri-\rk}}
\newcommand{\rjk}{{\rj-\rk}}
\newcommand{\rij}{{\ri-\rj}}
\newcommand{\rN}{{\rve_1,...,\rve_N}}

\renewcommand{\kappa}{\varkappa}
\newcommand{\tr}{\mathop{\rm tr}\nolimits}

\newcommand{\arctg}{\mathop{\rm arctan}\nolimits}

\newcommand{\angstrom}{\textup{\AA}}

\addtocounter{secnumdepth}{3}
\addtocounter{tocdepth}{3}
\hypersetup{pdfpagemode=FullScreen}

\begin{document}

\thispagestyle{empty}
{\bf
\Large
\centerline{Universitat Polit\`ecnica de Catalunya}
\vspace{1.0cm}
\centerline{Departament de F\'isica i Enginyeria Nuclear}
\vspace{1.5cm}
\large
\centerline{Tesi Doctoral en F\'isica}
}
\vspace{4.0cm}

\Huge
{\bf
\centerline{MONTE CARLO STUDY}
\vspace{0.5cm}
\centerline{OF QUANTUM PHASE TRANSITIONS}
\vspace{0.5cm}
\centerline{AT ZERO TEMPERATURE}
}

\vspace{3.0cm}

\normalsize
$$
\begin{array}{p{6cm}p{3cm}p{6cm}}
\bf Directors:&&\bf Candidate:\\
{\bf\large Prof.~Dr.~Jordi Boronat,}&&{\bf\large O.~N.~Osychenko} \\
{\bf\large Dr.~G.~E.~Astrakharchik}&&{\, }

\end{array}
$$

\vspace{2cm}

\centerline{\bf Doctorat de Recerca en F\'isica Computacional i Aplicada}
\vspace{0.2cm}
\centerline{\bf 23 de octubre de 2012}

\pagenumbering{roman}
\setcounter{page}{1}

\tableofcontents

\pagenumbering{arabic}
\setcounter{page}{1}

\newpage

\chapter*{Introduction}
\addcontentsline{toc}{chapter}{Introduction}
Phase transition is a common term for a wide range of phenomena generally described as a transition between  different states of matter with the variation of one or more physical parameters of the system. The transition is accompanied by an abrupt change of some of its physical quantities or its derivatives, 
whereas the relevant physical magnitudes change continuously within a given phase.
The simplest example of a phase transition is the melting of a crystal to form a fluid when its temperature is increased, 
which produces a discontinuous behavior of its density and some other physical properties. Historically, the first classification of the phase transitions was given by P. Ehrenfest~\cite{Ehrenfest33} and relies on a definition of a phase as a state with the minimum thermodynamic free energy. The first-order transition in this framework is a transition with an abrupt change of the first derivative of the system's free energy with respect to a certain parameter. 
The second-order transitions are those when the first derivative has a cusp when the parameter is changed, 
that is when a finite discontinuity appears in the second derivative of the free energy. 
Ehrenfest's original proposal was later extended to the cases of infinite discontinuities of physical parameters. 
The higher-order transitions are defined similarly, as the ones possessing a discontinuity in $n$-th ($n>1$) derivative of the free energy with respect to the parameter. 
Landau theory~\cite{LandLif1980} describes the second-order phase transitions as a result of a symmetry breaking,
 with a rapid change of a so-called order parameter, characterizing the symmetry properties of a phase. 
 Well-known examples of second-order phase transitions are the transitions between a normal fluid and a superfluid,
  with the superfluid fraction being the order parameter, or
   the ferromagnetic-paramagnetic transition with the magnetization as order parameter.

Quantum phase transitions are a broad subclass of these phenomena related to quantum matter,
 most generally described as a transition between phases at zero or low enough temperature,
  where quantum effects play an important role. The profound difference of quantum phase transitions
   from the classic phase transitions lay in the absence of entropy due to the Nernst heat theorem~\cite{Nernst07}.
    A classical description of a zero-temperature system can prescribe only one phase
     (an ideal crystal), whilst a quantum system is capable to undergo a transition,
      but only with the change of a certain non-thermal parameter, as for instance its density.
       The role of entropy in classical systems is played in quantum phases by quantum fluctuations.
One of the first experimental evidences of a quantum phase transition was the solidification to hcp solid $^4$He at low temperatures with a growth of pressure, made by Keesom~\cite{Keesom42}. The recent advances in methods of manipulation of ultracold matter, especially in the topics of cooling and trapping of atoms~\cite{Chu98,Cohen98,Phillips98} and Feshbach resonances~\cite{Timmermans99,Chin10}, demonstrated possibilities to 
produce systems with unique and highly tunable interparticle potentials~\cite{Cohen11}. 
 The tunability of the interaction 
in terms of non-thermal parameters, which was achieved in a number of experiments,
plays a key role for quantum phase transitions. 
One of the first theoretical proposals for a quantum phase transition 
was the bosonic superfluid-Mott insulator transition, based on the Bose-Hubbard model~\cite{Fisher1989,Jaksch1998}, 
that was finally experimentally confirmed in the work of Greiner \textit{et al.}~\cite{Greiner2002} 
and a number of subsequent experimental set-ups~\cite{Orzel01,Tuchman06,Trotzky11}. 
Recently, the phase diagram and essential thermodynamics of the three-dimensional Bose-Hubbard model 
was obtained in quantum Monte Carlo simulations by Capogrosso-Sansone \textit{et al.}~\cite{Capogrosso2007}. 

The energy of a quantum system, described by the Schr\"odinger equation in a state of 
a certain symmetry can be obtained with the help of 
Quantum Monte Carlo methods. From the equations of state, corresponding to different phases, one can 
find the pressure as a function of the relevant parameters. 
The double-tangent Maxwell construction, based on the equality of pressures and chemical potential 
along the transition line, allows to obtain the first-order transition point and the width of the transition zone.

Quantum Monte Carlo (QMC) techniques are \textit{ab initio} quantum calculation algorithms that might provide deep insight into the design of quantum matter, with a capability to describe a multitude of relevant properties and phenomena of the system. 
Among them the possibility to locate quantum and temperature phase transitions, 
and to quantify correlations in the system (e.g. pair correction function, structure factors, and even non-local properties, such as superfluid fraction and Bose--Einstein condensate). 
Bose-Einstein condensation (BEC), i.e., a macroscopic occupation of the 
zero-momentum quantum state of a system, despite being proposed by A. Einstein and S.~N.~Bose 
in the mid-twenties of the previous century~\cite{Einstein24},~\cite{Einstein25}, used to be 
considered for many decades more as a mathematical abstraction than an achievable state of matter. 
The superfluid properties of $^4$He at low temperatures, found in the experiments of Kapitza, 
Allen and Misener~\cite{Kapitza38,Allen38} are believed to be related to the presence of a Bose--Einstein condensate. The BEC-like phase transitions have also been 
observed in excitonic systems~\cite{Lin93}.
Long-lasting efforts of numerous experimental groups to actually observe a signature of a 
condensed phase in ultracold gases finally gave a positive result: in 1995 Bose-Einstein condensate 
was found by E.~Cornell and C.~Wieman~\cite{Anderson95} in gaseous $^{87}$Rb 
and later the same year in the other alkali vapours of $^{23}$Na (W.~Ketterle \textit{et al.}~\cite{Davis95}) 
and $^{7}$Li (R.~Hulet \textit{et al.}~\cite{Bradley95}). The experimental set-ups to 
produce a condensate are generally quite complex, partially due to the strict 
temperature requirements ($T\sim 10^{-7}$~K). From the theoretical sight, QMC simulations can yield 
accurate predictions about the properties of Bose--Einstein condensate, provided the interaction in the system is known.


Let us explain in more detail the techniques, challenges and results that we present in this thesis. 
We are usually concerned with the properties of a bulk system in its thermodynamic limit, but its 
QMC description of a bulk is generally 
performed with periodic boundary conditions (p.b.c.) applied to \textit{finite} size system. Therefore 
any Quantum Monte Carlo method yields results with a certain error, 
related to the size of a simulated system and one has to study the limit $N_{p}\rightarrow \infty$, with $N_p$ the number of
 particles. 
The properties of a system in the thermodynamic limit are therefore found out by extrapolating the data for limited system sizes to infinity. 
In the condensed systems that we consider, the convergence in the energy goes as $1/N_{p}$. 
The dependence of a certain physical quantity is then found for a set of different 
numbers of particles, and the result is extrapolated to infinity. 
Nevertheless the lowest number of particles, for which the asymptotic $1/N_{p}$ behavior is reached 
within an acceptable precision is a priori unknown. In practice, a number of probe simulations is 
performed in order to observe the needed linear dependence, but sometimes the accessible 
system sizes are too small to obtain the thermodynamic limit correctly. 
In some cases this problem can be greatly relieved by using the Ewald summation technique. 
In the framework of the Ewald method the potential energy is calculated for the infinite system, 
consisting of periodically replicated copies of the original simulation box. 
Although the method makes the calculation more time-consuming, its use shows that in particular systems, especially
when the long-range interactions are present, the $1/N_{p}$ dependence of the 
correction is substantially reduced, making the extrapolation to the thermodynamic limit possible.

The Ewald technique is a well-known simulation tool, often used with 
some modifications in a number of applications, involving molecular dynamics, 
Monte Carlo and other algorithms. Despite its popularity, the scope of its utilization is 
mostly limited to the summation of the Coulomb interactions, where its use is essential due to the long-range nature of the potential. 
Nonetheless, conceptually the Ewald technique may be applied to a broad variety of various pairwise potentials, for example, of the generic power-law type $1/|\bs{r}|^k$. In this Thesis we present a detailed step-by-step derivation of the Ewald sums for power-law interaction potentials and for all the terms we give explicit formulas, ready to be used straightforwardly in actual simulations. 
The derived expressions have been used in the simulations of systems consisting of Rydberg atoms 
and particles interacting via the Yukawa interaction potential.

The Yukawa potential has been used in the past as the simplest model interaction in atomic nuclei, in dusty plasmas and other systems, but recently this interaction appeared in the field of ultracold gases.
In recent experiments~\cite{exp:LiK1,exp:LiYb2} ultracold systems made up of two kinds of fermions, one heavy and another light, have been realized in actual setups, with 
an effective cooling achieved by means of an additional bosonic component. 
Theoretic treatment of quasi-two-dimensional systems with this kind of fermionic mixtures has been done in Ref.~\cite{YukawaPot2D}. It is argued that the effective interaction potential between light-heavy pairs of fermions is of Yukawa (screened Coulomb) type, with a feasibility of reaching a gas-crystal phase transition in two-dimensional geometry. 
In the present Thesis we extend the study of crystallization to a fully three-dimensional case at zero temperature. 
A similar problem was pursued before~\cite{Ceperley1}, but unfortunately this problem was not solved entirely, and an approximate Lindemann criterion was used instead of full scale quantum simulations. 
We find by means of the diffusion Monte Carlo method the exact phase diagram as a function of the 
relevant parameters, that is density and mass ratio between the two Fermi species. 
The obtained diagram provides valuable information on the minimum requirements 
for the mass ratio to achieve a phase transition in actual experiments.
Thanks to advances in the field of optical 
lattices there arises a possibility to produce particle mixtures with extremely high ratios of effective masses.
In this Thesis we argue that certain existing setups, involving optical lattices, allow to increase the effective mass ratio enough for potentially reaching the crystallization. 

In the last decade, there is a new wave of interest in ultracold systems 
consisting of Rydberg atoms, and a number of interesting experiments has been 
performed~\cite{Phau2007-EvidenceForCoherentCollectiveRydbergExcitationInTheStrongBlockadeRegime, Pfau2009-UniversalScalingInAStronglyInteractingRydbergGas, Pfau2011-ArtificialAtomsCanDoMoreThanAtomsDeterministicSinglePhotonSubtractionFromArbitraryLightFields}. 
A Rydberg atom is a neutral atom with a single electron excited to a high orbital. 
The important properties of this quantum object are its simplicity and similarity to a hydrogen atom. 
Furthermore, its unique properties of possessing very strong and controllable interactions over long distances, 
together with the novel techniques of ultracold atom manipulation, attracted a great deal of 
attention due to a prospectively rich behavior of mixtures of excited and unexcited atoms. 
In a typical experiment a trapped cloud of cold atoms is exposed to a laser field, 
exciting a small fraction of atoms to a particular Rydberg state. These excited states 
interact in a much stronger way among themselves than with the unexcited background.
The possibility of tuning, turning on, and off, the large magnitude of the forces, 
as well as a number of other advantageous properties, suggest its use as a quantum gate, 
that is, a basic element of quantum circuits. Currently, there is a wide range of 
proposals for physical systems to realize quantum information processing units: 
trapped ions~\cite{Blatt08}, linear optics~\cite{Kok07}, superconductors~\cite{Clarke08}, 
quantum dots~\cite{Li03}, and so on. The one, based on systems made up of Rydberg atoms 
is unique in terms of the range and the amplitude of the interaction, 
the working frequency and other advantageous properties~\cite{Saffman10}. 
The basic principles of a trapped Rydberg system as a quantum gate stem from the 
idea of a so-called Rydberg blockade, that is when a single excited atom 
shifts the energy levels of the nearby unexcited atoms out of the 
resonance with the driving laser pulse. Further excitations, injected 
into the cloud, can bring a macroscopic fraction of the cloud to a blockaded mode, 
allowing for a partial or complete saturation. In actual experiments, the 
fraction of unexcited atoms permits over $10^3$ excitations before the suppression of the new ones appear  
~\cite{Phau2007-EvidenceForCoherentCollectiveRydbergExcitationInTheStrongBlockadeRegime,%
Gould2004-LocalBlockadeOfRydbergExcitationInAnUltracoldGas,%
Weidemuller2004-SuppressionOfExcitationAndSpectralBroadeningInducedByInteractionsInAColdGasOfRydbergAtoms}. The actual physical phase of the excited atoms cannot be accessed directly in the reported experiments, although their arrangement is considered as a relevant information, both  as a standalone physical problem and for the implementation of the quantum gate. A direct observation of a quantum phase transition and a presence of 
long-range ordering is argued to be a feasible task in similar systems~\cite{Phau2008-QuantumCriticalBehaviorInStronglyInteractingRydbergGases,%
Pfau2009-UniversalScalingInAStronglyInteractingRydbergGas}. In the field of quantum computations,  Pohl \textit{et al.}~\cite{Pohl2010-DynamicalCrystallizationInTheDipoleBlockadeOfUltracoldAtoms} proposed that the presence of a crystal-like ordering could provide a better control over the quantum states. An insight to the spatial ordering in a cloud of Rydberg atoms may also shed light to the phenomenon of the so-called ``antiblockade''~\cite{Pohl2007-ManyBodyTheoryOfExcitationDynamicsInAnUltracoldRydbergGas}.

As mentioned before, the behavior of an ultracold mixture of excited Rydberg 
atoms and unexcited background is profoundly rich and complex. 
It can also depend substantially on particular experimental conditions, 
like the cloud geometry, laser field properties, etc. We perform a study of a model system, 
in which we neglect the interactions related to the unexcited background, 
and using the  pairwise repulsive van der Waals $C_6/|\bs{r}|^6$ for the excited atoms. 
The general aim of this study is to fully understand the phase diagram of the system. 
A perspective comparison with future experimental results can demonstrate, how 
well the properties of the system can be derived from this simple model. 
Since the number of Rydberg atoms, typically present in the current experimental works, 
is of the order of thousands or greater, in our simulations 
we look for all the relevant results in the thermodynamic limit. 
There is also a variety of possible crystal packings, which might be realized in the solid phase, 
hence we give a discussion on which of them are energetically preferable.

Another physically relevant system, considered in the Thesis, is bulk molecular \textit{para}-hydrogen (\ph). 
This system in the quantum regime (at low temperature) was 
 proposed theoretically as a
possible candidate for superfluidity, but it crystallizes at the temperature substantially
higher than transition temperature, making it impossible to observe a transition to the superfluid phase.
In this work, our Group has studied a
metastable non crystalline phase of bulk \ph  by means of  Quantum 
Monte Carlo methods in order to find out the temperature at which
this system still contains a noticeable superfluid fraction. 
The ultimate goal that our Group pursued, was 
to frustrate the formation of the crystal in the simulated system and
to calculate the temperature dependence of the one-body density matrix and
of the superfluid fraction.
I present the study of the limit of zero temperature
using the diffusion Monte Carlo method. Results for the energy, condensate
fraction, and structure of the  metastable liquid phase  at $T=0$ are
reported and compared with the ones of the stable solid phase. The simulation at zero temperature 
is used by our Group as a starting point for the simulation of the system at 
low temperatures by using Path Integral Monte Carlo  
technique.

The structure of the Thesis is as follows.

In Chapter~\ref{secTools} we discuss the analytical approaches and
approximations used in the subsequent Chapters; also we describe the general concepts of the two-particle 
scattering problem as a tool to construct Jastrow terms in trial wave functions.  Chapter~\ref{secQMC} explains in
details the Quantum Monte Carlo methods employed in our calculations from the theoretical and 
practical points of view.  In Chapter~\ref{secEwald} we explain the Ewald summation technique, 
applied to a power-law $1/|\bs{r}|^k$  interaction potential, and a generic approach to obtain the Ewald terms. 
The obtained expressions of this analytic work are implemented into simulations 
of different physically relevant systems (Rydberg atoms and Yukawa particles). 
Chapter~\ref{secRydberg} is devoted to the modelling of a system, governed 
by the model potential between Rydberg atoms $1/|\bs{r}|^6$. The phase diagram of the system is 
obtained for a relevant range of densities and temperatures, combining quantum simulations at low 
temperature and classical treatment at higher temperature. A special attention is paid to the 
classical description of this system, composed of Rydberg atoms, 
and its comparison to the quantum system. In Chapter~\ref{secYukawa} we  
present the simulation of a system with the Yukawa interaction potential. 
The following Chapter~\ref{secHydrogen} presents the results of the Quantum Monte Carlo 
simulations of molecular para-hydrogen at zero and finite temperatures, performed in our Group. 
Conclusions are drawn in Chapter~\ref{secConclusions}.

\chapter{Tools\label{secTools}}
\section{Introduction}
This Chapter is intended to provide theoretical basis for the following Chapters. 
The quantities characterizing properties of a quantum system (correlation functions, static structure factor, and so on)
are introduced here and are later used in the subsequent chapters. 
We also discuss the two-body scattering problem in a three-dimensional system geometry
that sheds light to the short-range properties of many-body systems. The two-body 
scattering solution can be used in the development of trial wave functions needed 
in the Quantum Monte Carlo algorithms. 

The structure of the Chapter is the following.

In Section~\ref{secIntrCorr} we introduce experimentally relevant magnitudes and functions that are present in a quantum
system. First of all, we consider the
analytic forms of the first and second quantization (Secs.~\ref{secCorr2nd},
\ref{secCorr1st}). Special attention is given to the relation between correlation functions and mean values of 
quantum operators.
 Some correlation functions may be
greatly simplified in case of a homogeneous system, 
which is presented in Section~\ref{secCorr1st}. The definitions and general comments on static structure factor and momentum
distribution are drawn in Section~\ref{secnk}.

Section~\ref{secIntrScat} is devoted to the study of two-body scattering processes in three-dimensional  geometry. 
The solutions of the two-body scattering problem are provided for a number of
physically relevant interaction potentials. The main aim of this last Section is to 
give an efficient tool to construct two-body Jastrow terms of the trial wave function for 
Quantum Monte Carlo simulations (for details on QMC methodology see Chapter~\ref{secQMC}).

\newpage
\section{Correlation functions\label{secIntrCorr}}

\subsection{Second quantization form\label{secCorr2nd}}

A quantum system of identical particles of variable number is generally described with the 
help of annihilation and creation operators. The commonly used notations for the auxiliary field operators are 
$\hat\Psi^\dagger(\rve)$ for an operator creating a particle in the position $\rve$, 
and $\hat\Psi(\rve)$ for an operator destroying a particle in the same position.  
By means of the creation operator $\hat a_m$ for $m$-th state, that puts a particle to an orbital $\varphi_m(\rve)$,
and the annihilation operator, $\hat a_\m^\dagger$, that removes a particle from the orbital $\varphi_m(\rve)$,
these field operators can be easily represented in the following form: 

\begin{eqnarray}
\left\{
\begin{array}{lll}
\hat\Psi^\dagger(\rve) &=& \sum\limits_\m \varphi^*_m(\rve)\,\hat a_m^\dagger\\
\hat\Psi(\rve) &=& \sum\limits_m \varphi_m(\rve)\,\hat a_m
\end{array},
\right.
\label{Psi}
\end{eqnarray}
If we consider a uniform gaseous system within a volume $V$, single particle states
are evidently plain waves $\varphi_m(\rve) = e^{\ii\k_m\rve}/\sqrt{V}$. 
Bosonic operators (\ref{Psi}) obey commutation relation $[\Psi(\rve),\Psi^\dagger(\rve')]=\delta(\rve-\rve')$,
$[\Psi(\rve),\Psi(\rve')]=0$, while fermionic operators obey commutation relations.

First of all, let us discuss the relation between the correlation functions and the mean values of 
one- and two-body quantum mechanical operators. Let us consider the simplest case when the Hamiltonian of the system
is a sum of only one-body and two-body operators
\begin{eqnarray}
\hat H = \hat F^{(1)}+\hat F^{(2)},
\end{eqnarray}
where the one-body operator $\hat F^{(1)}$ stands for a sum of the one-body terms, 
and the two-body operator $\hat F^{(2)}$ is a sum of corresponding two-body terms, depending on $\rve_i,\:\rve_j$:
\begin{eqnarray}
\hat F^{(1)} &=& \sum\limits_{i=1}^N \hat f^{(1)}_{i}\:,\\
\hat F^{(2)} &=& \frac{1}{2}\sum\limits_{i\ne j}^N \hat f^{(2)}_{i,j}\:.
\end{eqnarray}

Obvious examples for one-body operators are an external potential field, depending only on the particles'
coordinates: $f^{(1)}(\rve) = V_{\rm ext}(\rve)$, or the kinetic energy: $f^{(1)}(p)=p^2/2m$. The first operator is diagonal in 
coordinate space, while the second one is diagonal in momentum representation. 
A typical example of a two-body operator is a pairwise interaction potential, 
given in coordinate space: $f^{(2)}(\rve_1,\rve_2)
= V_{\rm int}(\rve_1,\rve_2)$.

The representation of one- and two-body operators $\hat A^{(1)}$ and
$\hat A^{(2)}$ in terms of field operators (see (\ref{Psi}))  is straightforward:
%
\begin{eqnarray}
\hat F^{(1)} &=&
\intint\hat\Psi^\dagger(\rve)f^{(1)}(\rve,\rve')\hat\Psi(\rve')\,\dr\dr'\\
\hat F^{(2)} &=&
\frac{1}{2}\intint\hat\Psi^\dagger(\rve)\hat\Psi^\dagger(\rve')
f^{(2)}(\rve,\rve')\hat\Psi(\rve')\hat\Psi(\rve)\,\dr\dr'
\label{thF2}
\end{eqnarray}
where the factor $1/2$ is introduced to take into account the double summation. 

Up to now, we did not restrict ourselves to local one-body operators (that is those satisfying the relation for 
the quantum averages $\langle\rve|a^{(1)}|\rve'\rangle = a^{(1)}(\rve)\delta(\rve-\rve')$), but we also
consider non-local operators, that is, ones allowed 
to depend on two arguments $a^{(1)}=a^{(1)}(\rve,\rve')$ in the corresponding integral of (\ref{Psi}). 

Correlation functions can be introduced in terms of the field operators in the 
following way:
\begin{eqnarray}
\label{G1}
C_1(\rve,\rve')&=&\langle\hat\Psi^\dagger(\rve)\hat\Psi(\rve')\rangle\:,
\end{eqnarray}
\begin{eqnarray}
\label{G2}
C_2(\rve,\rve')&=&\langle\hat\Psi^\dagger(\rve)\hat\Psi^\dagger(\rve')\hat\Psi(\rve')\hat\Psi(\rve)\rangle\:,
\end{eqnarray}
Note that in one-body correlation function (\ref{G1}) we consider non-local dependence,
and it has two arguments. At the same time we consider only local two-body operators, that is why we keep two arguments instead of four  
 in (\ref{thF2}). 

The quantum averages of the operators $\hat F^{(1)}$ and $\hat F^{(2)}$ may be obtained from
$\hat f^{(1)}$ and $\hat f^{(2)}$, when the correlation functions are known:

%
\begin{eqnarray}
\label{F1}
\langle\hat F^{(1)}\rangle &=&\int f^{(1)}(\rve)C_1(\rve,\rve)\,\dr\\
\langle\hat F^{(2)}\rangle &=&\frac{1}{2}\intint f^{(2)}(\rve,\rve')C_2(\rve,\rve')\,\dr\dr'\:.
\label{F2}
\end{eqnarray}

The correlations of the field operator between 
two distinct points $\rve$ and $\rve'$, are characterized by the one-body correlation function $C_1(\rve,\rve')$. 
The diagonal component of (\ref{G1}) $\rve=\rve'$ yields the density of the particles
$\rho(\rve)=\langle\hat\Psi^\dagger(\rve)\hat\Psi(\rve)\rangle=C_1(\rve,\rve)$, hence the sum over the 
diagonal terms, i.e., the trace of
the matrix $C_1$, is equal to the total number of particles $N=\tr C_1 = \int C_1(\rve,\rve)\,\dr$. 
 The two-body correlation function $C_2(\rve,\rve')$ defines correspondingly the density correlations between the
particles  at positions $\rve$ and $\rve'$, respectively.

It is convenient to introduce dimensionless versions of these functions (\ref{G1}) and
(\ref{G2}):
\begin{eqnarray}
\label{g1}
c_1(\rve,\rve') &=& \frac{C_1(\rve,\rve')}{\sqrt{\rho(\rve)}\sqrt{\rho(\rve')}}\\
c_2(\rve,\rve') &=& \frac{C_2(\rve,\rve')}{\rho(\rve)\rho(\rve')}
\label{g2}
\end{eqnarray}

It can be seen that for bosons the range of any of the functions (\ref{g1}-\ref{g2}) is $[0,1]$, and the function can be interpreted as a
probability to remove a particle from the position $\rve$ and place it to the position $\rve'$. The obvious relation $c_1(\rve,\rve)=1$ reflects the fact, that there is always a possibility to put the particle back to its initial location.
If no Bose-Einstein condensate is present, the non-diagonal terms asymptotically vanish in the long range limit 
$c_1(\rve,\rve')\to 0, |\rve,\rve'|\to\infty$. The function $c_2(\rve,\rve')$ can be understood as a joint probability to find one particle in the point $\rve$ and another one in the point $\rve'$.

Additional information on correlation functions can be found 
in Refs~\cite{Feenberg67,Mahan00,Glauber63,Naraschewski99,Gangardt03b}.

\subsection{First quantization form\label{secCorr1st}}

The  physical meaning of the correlation function written in the form of the second quantization 
has been briefly discussed in Section \ref{secCorr2nd}. 
We will use the Monte Carlo methods in order to evaluate averages over the wave function $\psi(\R)$ 
of the system. For that one needs to represent the averages as integrals of the operators over the wave function $\psi(\R)$. 
We will look for the mean values of the operators in 
forms, similar to that of (\ref{F1}) and (\ref{F2}). In the first quantization the expectation value of 
a one-body operator reads as
\begin{eqnarray}
\nonumber
\langle A^{(1)} \rangle
= \frac{\int \psi^*(\R) A^{(1)}\psi(\R)\,\dR}{\int |\psi(\R)|^2\,\dR}=
\frac{\sum\limits_{i=1}^N\int
\psi^*(\rve_1, ..., \rve_N) a^{(1)}_i \psi(\rve_1, ...., \rve_N)\,\dR}{|\psi(\R)|^2\,\dR}
= \\= N\frac{ \int a^{(1)}(\rve_1)
|\psi(\rve_1, ..., \rve_N)|^2\,\dR}{\int|\psi(\R)|^2\,\dR}
=\int a^{(1)}(\rve) C_1(\rve, \rve)\,\dr,
\label{F1mean}
\end{eqnarray}
with the notation $C_1(\rve,\rve')$ used for the expression
\begin{eqnarray}
C_1(\rve, \rve') =N
\frac{ \int \psi^*(\rve, \rve_2, ..., \rve_N) \psi(\rve', \rve_2, ..., \rve_N)\,\dr_2 ... \dr_N}
{\int \psi^*(\rve_1, ..., \rve_N) \psi(\rve_1, ..., \rve_N)\,\dr_1 ... \dr_N}\label{c1obdm}
\end{eqnarray}

An average of a two-body operator (\ref{F2}) can be expressed in 
terms of the two-body correlation function (\ref{G2}) in the following way:
\begin{eqnarray}
\langle A^{(2)} \rangle
= \frac{\int \psi^*(\R) A^{(2)}(\R) \psi(\R)\,\dR}{\int |\psi(\R)|^2\,\dR}=
\frac{\frac{1}{2}\sum\limits_{i\ne j}^N\int
\psi^*(\rve_1, ..., \rve_N) a^{(2)}(\rve_i, \rve_j) \psi(\rve_1, ...., \rve_N)\,\dR}{|\psi(\R)|^2\,dR}
= \\= \frac{N(N-1)}{2} \frac{\int a^{(2)}(\rve_1, \rve_2)
|\psi(\rve_1, ..., \rve_N)|^2\,\dR}{\int|\psi^*(\R)|^2\,\dR}
=\frac{1}{2}\intint a^{(2)}(\rve_1, \rve_2) C_2(\rve_1, \rve_2)\,\dr_1\dr_2,
\label{secQnt}
\end{eqnarray}
with the following expression for the first quantization form of the two-body correlation function
\begin{eqnarray}
\label{g2fq}
C_2(\rve', \rve'') =
\frac{N(N-1) \int |\psi(\rve', \rve'', \rve_3,..., \rve_N)|^2\,\dr_3 ... \dr_N}
{\int|\psi(\rve_1, ..., \rve_N)|^2\,\dr_1 ... \dr_N}
\end{eqnarray}


The situation can get easier if we stick to the case of a homogeneous system, as it possesses translational symmetry.
In the case of one-body operator, the diagonal $\rve=\rve'$ element of Eq.~(\ref{c1obdm}) is just a constant,
which value is fixed by the density $n=N/V$, $C_1(\rve,\rve) = n$.
 The non-diagonal elements of the normalized 
matrix of one-body operator in the first quantization (or simply {\it one-body density matrix}) read as
%
\begin{eqnarray}
g_1(r) =
\frac{N}{n}\frac{\int \psi^*(\rve, \rve_2, ..., \rve_N) \psi(0, \rve_2, ..., \rve_N)\,\dr_2 ... \dr_N}
{\int |\psi(\rve_1, ..., \rve_N)|^2\,\dr_1 ... \dr_N}
\label{g21}
\label{g1hom}
\end{eqnarray}
where $n=N/V$ is the average density of a homogeneous system. 
The normalized two-body density matrix, also called {\it pair distribution function} is represented by 
\begin{eqnarray}
g_2(r) = \frac{N(N-1)}{n^2}
\frac{ \int |\psi(\rve, 0, \rve_3,..., \rve_N)|^2\,\dr_3 ... \dr_N}
{\int|\psi(\rve_1, ..., \rve_N)|^2\,\dr_1 ... \dr_N}
\label{g2hom}
\end{eqnarray}

The basic properties of the pair distribution function in the zero temperature limit  can be
deduced in the following way. 
In a gas phase, density-density correlations vanish for large interparticle distances, which 
corresponds to $g_2(r)\rightarrow 1-\frac{1}{N}$, hence in the thermodynamic limit 
$g_2(r)$ asymptotically tends to 1. 

One faces the opposite situation at short distances, where the particle correlations are strong, and the
value of $g_2(0)$ can vary depending on the interaction potential. 
For instance, in case of a repulsive potential $g_2(0)<1$, 
on the contrary for a purely attractive potential $g_2(0)>1$, and in the case of a hard-core interaction  
when the particles are not allowed to overlap, $g_2(0)=0$, when $|\rve|<R_{core}$.


Let us consider the expectation value of a two-body operator (\ref{F2}) and see how it can be simplified in a homogeneous system:
%
\begin{eqnarray}
\left\langle A^{(2)}\right\rangle =\frac{n^{2}}{2}\int\!\!\!\!\int
g_2(\rve_1,\rve_2)~a^{(2)}(|\rve_1-\rve_2|)\,\dr_1\dr_2=\frac{nN}{2}\int
g_{2}(r)~a^{(2)}(r)\,dr
\end{eqnarray}

In the integration performed above we made use of the mentioned fact, that the operator depends only on the distance between particles, allowing to integrate one of the coordinates out. 

In the simplest case of a contact delta-potential $V_{\rm int}(r) = g \delta(r)$ 
($g$ stands for a coupling constant, defining 
the interaction strength) the
potential energy is simply proportional to the value of the pair distribution function
at $\rve=0$:
\begin{eqnarray}
\frac{E_{\rm pot}}{N} = \frac{1}{2}g\, n g_2(0)
\label{meanEcontact}
\end{eqnarray}

\subsection{Static structure factor\label{secnk}}

By virtue of the field operator (\ref{Psi}) the momentum distribution $n_\k$ is represented as
\begin{eqnarray}
n_\k = \langle\hat\Psi^\dagger_\k\hat\Psi_\k\rangle,
\label{n_k}
\end{eqnarray}
The field operator in momentum space $\hat\Psi_\k$ is in fact the Fourier transform of $\hat\Psi(\rve)$:
%
\begin{eqnarray}
\left\{
\begin{array}{lll}
\hat\Psi_\k &= &\int e^{-\ii\k\rve} \hat\Psi(\rve) \frac{\dr}{\sqrt{2\pi}^D}\\
\hat\Psi(\rve) &= &\int e^{\ii\k\rve} \hat\Psi_\m \frac{\di \bf{k}}{\sqrt{2\pi}^D}
\end{array}
\right.
\label{Psi relation}
\end{eqnarray}
where $D$ stands for a dimensionality of the system. 
Applying the relation (\ref{Psi relation}) to the equation (\ref{n_k}) one finds the following form of
the momentum distribution
\begin{eqnarray}
n_\k = \frac{1}{2\pi}^D \intint e^{\ii\k\bs{s}} C_1\left(\frac{\bs{s}}{2}+\R,\R-\frac{\bs{s}}{2}\right)\, \di \bs{s} \di \R
\end{eqnarray}

It can be noticed that in a homogeneous bulk system 
the center of mass motion can be integrated separately, as 
the dependence on the momentum is defined by the distance between the particles, not positions themselves. Performing 
this integration for a homogeneous system it yields   
%
\begin{eqnarray}
n_\k = n\int e^{-\ii \k \rve} g_1(\rve)\,\di \rve
\label{nk}
\end{eqnarray}
For example, for a fully Bose-condensed gas, the off-diagonal terms of one-body density matrix
 are constant, $g_1(\rve)=n$, which results in a singular momentum distribution $n_\k=N\delta(\k)$.
 In this case all particles are condensed in $\k=0$ state. 

Another useful quantity is the dynamic structure factor of the system, which characterizes a scattering
process, corresponding to the exchange of energy $\hbar\omega$ and the momentum $\hbar k$ in 
the scattering event. 
The dynamic structure factor $S(k,\omega)$ can be expressed by virtue of the
$k$-component of the density operator at zero temperature 
\begin{eqnarray}
\hat\rho_\k =
\sum_{i=1}^{N}e^{-\ii \k \rve_i}
\label{rhok}
\end{eqnarray}
in the following form
\begin{eqnarray}
S(\k,\omega) = \sum\limits_i |\langle n |\hat\rho^\dagger_\k-
\langle\hat\rho^\dagger_\k\rangle|0\rangle|^2\delta(\hbar\omega-\hbar\omega_i)\,.
\end{eqnarray}
where $\omega_i$ is the frequency of the $i$-th stationary state. 
%
The static structure factor is proportional to the frequency integral of the dynamic structure factor, 
that is it characterizes the overall probability of scattering of a probe particle with the momentum transfer
$\hbar k$. The separate integration over $\omega$ gives
\begin{eqnarray}
S(\k) = \frac{\hbar}{N} \int_0^\infty S(\k,\omega)\,d\omega
=\frac{1}{N}(\langle\hat\rho_\k\hat\rho_{-\k}\rangle - |\langle\hat\rho_\k\rangle|^2)
\label{Skmom}
\end{eqnarray}

The latter expression (\ref{Skmom}) can be sampled directly in Quantum Monte Carlo simulations. The other way to 
write the static structure factor is to relate it to the two-body density matrix by means of the 
equations (\ref{G2}) and (\ref{rhok}), having in mind the commutation properties of the field operator 
$\hat\Psi(r)$
\begin{eqnarray}
S(k) =  1+ \intint \frac{1}{N} e^{\ii (\rve_2-\rve_1) \k}(C_2(\rve_1,\rve_2)-n(r_1)n(r_2))\dr_1 \dr_2
\label{Sknonuni}
\end{eqnarray}

In case of a homogeneous system, the positions of the particles enter the two-body density matrix
only as an interparticle distance $r = |\rve_1-\rve_2|$, thus the static structure factor can be rewritten in terms of the
Fourier transform of the pair distribution (\ref{g2}) of the system: 
%
\begin{eqnarray}
S(k) = 1+n\int e^{\ii r k} (g_2(r)-1)\,\dr
\label{Fourier Sk}
\end{eqnarray}
The static structure factor can yield valuable information on the arrangement and the order of the system, and 
its value can be directly 
accessed in spectroscopy experiments.


\section{Scattering problem\label{secIntrScat}}
\subsection{Introduction}

The construction of a trial wave function for a many-body problem is in most cases a very complex task since
 the exact solution is generally unknown (rare exceptions for analytic solutions are the 1D gas
 of hard rods and hard points (Tonks-Girardeau gas),
 where the analytic solutions 
 are known). A typical approach to develop a trial wave function therefore consists
 in matching long-range behavior with a two-body solution at short distances.     
 In this Section, we will be concerned with the two-body scattering problems in three-dimensional 
 systems, whose solutions are then used in many-body calculations presented in 
 the following chapters.

\subsection{Scattering problem in three-dimensional geometry\label{sec3D}}
\subsubsection{General approach\label{sec3Dsc}}

In this Section, we
formulate a generic two-body scattering problem in a 3D geometry. 
For a three-dimensional system the low-density regime of interparticle interaction
is supposed to be correctly described by two-body collisions.

Consider two particles with coordinates $\rve_1$ and $\rve_2$ and respective masses $m_1$ and $m_2$
staying close enough to see the process as a 
two-particle collision. We suppose that the system is not confined externally, hence the problem may be treated
as translationary invariant, with the center of mass moving with a constant speed.
Our purpose is to obtain the stationary solution $p(\rve_1,\rve_2)$ of the
Schr\"odinger equation
\begin{eqnarray}
\left[-\frac{\hbar^2}{2m_1}\Delta_{\rve_1}-\frac{\hbar^2}{2m_2}\Delta_{\rve_2}
+V(|\rve_1-\rve_2|) \right]p(\rve_1,\rve_2) = \E_{12}\:p(\rve_1,\rve_2)\:\:.
\label{scat0}
\end{eqnarray}
In the center of mass frame, 
the coordinate variables get separated, thus the representation of the Schr\"odinger equation
for the motion of the center of mass $\Rc=(m_1\rve_1+m_2\rve_2)/m_{\rm total}$ gets simple
\begin{eqnarray}
-\frac{\hbar^2}{2m_{\rm total}}\Delta_\Rc f_\Rc(\Rc)= \E_\Rc f_\Rc(\Rc)\:\:,
\label{scatR}
\end{eqnarray}
with $m_{\rm total}=m_1+m_2$ staying for the total mass. The solution 
of Eq.~(\ref{scatR}) is of a free wave form. 
Omitting the normalization, the solution reads 
 $f_\Rc(\Rc) =
\exp[\ii\k_\Rc \Rc]$ with $\k_\Rc$, being the initial center of mass wave number of the
system with the corresponding energy $\E_\Rc=\hbar^2k_\Rc^2/2m_{\rm total}$.

The equation for the position difference $\rve=\rve_1-\rve_2$ contains the pairwise interaction
potential
\begin{eqnarray}
\left(-\frac{\hbar^2}{2\mu}\Delta_{\rve}+V(|\rve|)\right) f(\rve)= \E f(\rve)\:\:,
\label{scatr}
\end{eqnarray}
with
\begin{eqnarray}
\mu = \frac{m_1m_2}{m_1+m_2}
\label{mu}
\end{eqnarray}
denoting the reduced mass. When one finds the solutions of Eqs~(\ref{scatR}-\ref{scatr}), it is possible
to obtain the needed solution of the scattering problem (\ref{scat0}) as 
\begin{eqnarray}
\left\{
\begin{array}{cll}
p(\rve_1,\rve_2) &=&f_\Rc(\Rc) f(\rve)\\
\E_{12}&=&\E + \E_\Rc
\end{array}
\right.
\label{split}
\end{eqnarray}

We will assume that the energy of the incident particle
$E$ is small enough and the solution is spherically symmetric $f(\rve) = f(|\rve|)$.
Under these prescriptions, and using the spherically symmetric representation of
 the Laplacian $\Delta = \frac{\partial^2}{\partial
r^2}+\frac{2}{r}\frac{\partial}{\partial r}$, one can 
conveniently rewrite the equation~(\ref{scatr}), introducing the auxiliary function $w(r)$
\begin{eqnarray}
\label{u(r)}
f(r)&=&\frac{w(r)}{r}
\label{u(0)}
\end{eqnarray}
in a way that its analytic form is similar to a one-dimensional Schr\"odinger equation:
\begin{eqnarray}
-\frac{\hbar^2}{2\mu}w''(r)+V_{\rm int}(r)w(r) = \E w(r)
\label{scatu}
\label{master}
\end{eqnarray}
with the additional requirement of the boundary condition 
\begin{eqnarray}
w(0)&=&0\:\:.
\end{eqnarray}
At distances, large compared to the range of the potential, one can neglect $V_{\rm int}$(r) term
in Eq.~(\ref{scatu}) leaving with a free wave differential equation. 
\begin{eqnarray}
-\frac{\hbar^2}{2\mu}w''(r)= \E w(r)
\label{scatu1}
\end{eqnarray}
The solution for this equation is a plane wave
\begin{eqnarray}
w(r) = A\sin(k_sr+\delta(k_s)),
\label{usol}
\end{eqnarray}
where
\begin{eqnarray}
\hbar k_s=\sqrt{2\mu \E}
\label{ISk}
\end{eqnarray}
stands for the momentum of the incident particle and $\delta(k_s)$ is the scattering phase and $A$ is an arbitrary constant.

Properties of the low-energy scattering depend on a single parameter, namely $s$-wave scattering length $a_{3D}$. Its value 
can be obtained from the phase shift $\delta(k_s)$ as the following limit
\begin{eqnarray}
a_{3D} = -\lim\limits_{k\to 0}\frac{\delta(k_s)}{k_s}
\label{a3D}
\end{eqnarray}

If we consider the asymptotic low-momentum limit $k\to 0$ (slow particles) the scattering solution
(\ref{usol}) may be expanded
\begin{eqnarray}
f(r)\to \textrm{const}\left(1-\frac{a_{3D}}{r}\right)
\label{frlim}
\end{eqnarray}
and it is easy to see, that it has a node at a distance $a_{3D}$. The position of the first node of the
positive energy scattered solution in the limit of low momenta can be seen as an equivalent definition
of the scattering length in a three-dimensional system.

In low-density regime of weekly interacting gas the interparticle distance is large compared to 
the range of the potential. Under such conditions the exact shape of the interaction potential 
is not important and the description in terms of the $s$-wave scattering length is universal.

In the next several sections we will consider the scattering problem for a
hard-sphere potential (\ref{secHS}) as a simplest example, and afterwards the same problem 
for the Yukawa potential (\ref{secSYukawa}) and the common potential between Rydberg atoms (\ref{secSRydberg}) will be solved. We
will obtain explicitly the expressions for the scattered functions, which are of great
importance, since in many cases they can give a deep physical insight into properties of a
many-body system. Indeed, under particular conditions the relation between the correlation 
functions and the scattered wave function $f(r)$ can be found. Another important point to mention 
is that often a many-body trial wave function 
is taken as a product of scattered functions $f(r)$. Hence, these calculations are
of great importance for their further implementation of the Quantum Monte Carlo algorithms.
\subsubsection{Scattering on hard sphere potential\label{secHS}}

As it was mentioned in Sec.~\ref{sec3Dsc}, in the limit of low energy collisions the
 interaction potential is characterized exclusively by one
parameter, the $s$-wave scattering length. It means that when $E\rightarrow 0$ 
the scattering on each potential possessing
equal scattering lengths is the same, and it is said that the scattering becomes {\it universal}.
If the scattering is considered on some repulsive potential, then the easiest choice is the {hard sphere}
(HS) potential:
\begin{equation}
V^{\rm hs}(r)=
\left\{
\begin{array}{cc}
+\infty, & r<a_{3D}\\
0,& r\ge a_{3D}
\end{array}\right.
\label{HS}
\end{equation}

This potential is controlled by a single parameter, which we denote $a_{3D}$ in the definition
(\ref{HS}). It is clear that we can treat this value as a range of the potential.
Simultaneously it has a meaning of the scattering length, as it was 
presented in (\ref{a3D}). It comes out in a natural way from the solution of the
scattering problem, as by definition the $s$-wave scattering length is the position 
of the first node of the analytic continuation of a large-distance free wave solution. 
For the hard sphere potential this free-wave solution is valid for $r>a_{3D}$ with the position
of the node given by $|\rve|=a_{3D}$.

The Schr\"odinger equation (\ref{scatu}) can be rewritten as (the reduced mass is $\mu=m/2$)
\begin{eqnarray}
-\frac{\hbar^2}{m}w''(r)+V^{\rm hs}(r)w(r) = \E w(r)
\end{eqnarray}

A particle is unable to penetrate the potential wall of the hard core 
 thus the solution tend to zero for distances below the size of the hard sphere. Notice that
the energy in this case has only a kinetic component, and the interaction potential does not enter
explicitly. Nevertheless it affects the boundary condition for the solution.
\begin{eqnarray}
\left\{
{\begin{array}{ll}
\displaystyle w(r) = 0,& |r| < a_{3D}\\
\displaystyle w''(r) + \kappa^2 w(r) = 0,&|r| \ge a_{3D}
\end{array}}
\right.
\label{eqHS}
\end{eqnarray}
with a convenient substitution for the frequency $\kappa^2=\frac{m\E}{\hbar^2}$.
The solution of this differential equation (\ref{eqHS}) may be obtained easily.
Joining together with (\ref{u(r)}) one gets:
\begin{eqnarray}
f(r)=
\left\{
{\begin{array}{ll}
\displaystyle 0,& |r| < a_{3D}\\
\displaystyle A\sin(k(r-a_{3D}))\,/r,&|r| \ge a_{3D}
\end{array}}
\right.,
\label{fHS}
\end{eqnarray}
with $A$ for an arbitrary constant. The phase shift
is related linearly to the wave vector of the incident particle $\delta(k) = -\kappa a_{3D}$ 
(\ref{a3D}), showing explicitly that the range of the potential (\ref{HS}) has in fact
the meaning of the three-dimensional scattering length as we said above in the same Section.

\subsubsection{Scattering on Yukawa potential\label{secSYukawa}}
The solution of the two-body scattering problem, as it was mentioned above, is used to construct 
the trial wave functions for subsequent use in the Quantum Monte Carlo simulations. The need of having 
correctly posed short-distance approximation for the trial wave stems from the fact that the majority of 
physically relevant pairwise interactions contain a repulsive core, that can vary in hardness. 
Since the diffusion Monte Carlo 
method is based on a sampling from the particle distribution $\psi_T \psi_0$, an inaccurate choice of  
the trial wave function $\psi_T$ for small $r$ 
 can result in a substantial growth of the overall error of the calculation. This can be manifested by a need of raising
  the number of walkers to reach a convergence as well as by a growth of a common statistical variance. On the other hand, the errors in the long-range part 
 are usually easily ``corrected'' by the DMC algorithm, as this range is statistically well represented, and also the actual discrepancy of the trial wave and the ground state solution is relatively low.    

The two-body scattering problem for the potential of Yukawa is solved in a similar manner, as was used in the solution for a system of hard spheres. 
The solution of the initial Schr\"odinger equation is considered spherically symmetric and therefore rewritten
 as a one-dimensional equation with a single argument, the interparticle distance $r=|\rve_1-\rve_2|$. If we look for the solution in the 
form convenient for subsequent use as a Bijl-Jastrow term in Monte Carlo calculations, then in accordance with Eq.~(\ref{scatu})

\begin{eqnarray}
-\frac{\hbar^2}{m}w''(r)+V_0\frac{\exp (-2 r)}{r}w(r) = \E w(r)\:.
\label{scatuyuk1}
\end{eqnarray}
The solution of the latter equation for short-range distances can be obtained by approximating $\exp(-2r)$ by $1$.
One possibility is to consider the scattering at a finite energy $\E$ and fix the value of $\E$
from continuous matching with the long-range behavior of Bijl-Jastrow term. Equation~(\ref{scatuyuk1}) can be solved 
by means of the hypergeometric functions. A less precise description can be obtained by setting $\E=0$, still 
the obtained Bijl-Jastrow term is well suited for DMC calculation. The $\E=0$ case
\begin{eqnarray}
-\frac{\hbar^2}{m}w''(r)+\frac{\exp (-2 r)}{r}w(r) = 0\, 
\label{scatuyuk2}
\end{eqnarray}
results in a more simple solution which is a linear combination of the modified Bessel functions of the first kind 
and the second kind with particular square root factors,
\begin{eqnarray}
f(r)=C_i \frac{I_1(2\sqrt{r}/\kappa)}{\kappa \sqrt{r}} + C_k \frac{K_1(2\sqrt{r}/\kappa)}{\kappa \sqrt{r}}
\label{fjas}
\end{eqnarray}
with $\kappa = \sqrt{\hbar^2/(mV_0)}$. The first component $I_1$ is finite for $\rve=0$, 
while the second one $K_1$ diverges for $\rve=0$. The arbitrary constant 
is irrelevant for the QMC algorithms that we use, therefore we stick to the following form of the two-body scattering solution
\begin{eqnarray}
f(r)=\frac{I_1(2\sqrt{r}/\kappa)}{\sqrt{r}}=\kappa+\frac{\kappa^3}{2}r+\frac{\kappa^5}{12}r^2 + \mathcal{O}(r^3) \label{yukawajas}
\end{eqnarray}
A more precise analytic solution can be found by using a higher-order expansion of the Yukawa factor $\exp(-2r)$, that is $\exp(-2r)\approx 1-2r$, however the use of a single first term of the expansion is usually enough for practical purposes.

The diffusion Monte Carlo study of Yukawa systems, made by Ceperley \textit{et al.}~\cite{Ceperley2}, was based on the following 
form of the Jastrow term of the trial wave function, typically used in the nuclear matter calculations
\begin{eqnarray}
u(r)=A e^{-Br}(1-e^{-r/D})/r\,.
\label{ceperlyur1}
\end{eqnarray}

It can be shown that the expression $\exp(-u(r))$ can be made coincident in the leading terms with the expansion of the solution (\ref{yukawajas}) of the two-body scattering problem:
\begin{eqnarray}
e^{-u(r)}&=&e^{-A/D}+\frac{A(1+2 B D)e^{-A/D}}{2D^2}r+\\
&\:&+\frac{A(3D-4D+12ABD-12BD^2+12AB^2D^2-12B^2D^3)e^{-A/D}}{24D^2}r^2+\mathcal{O}(r^3)\nonumber
\label{ceperlyur2}
\end{eqnarray}
One can note that Jastrow term (\ref{ceperlyur2}) coincides with the solution of the two-body
scattering problem (\ref{fjas}) for a particular choice of variational parameters $A,B,D$.
 Notice that the parameters $A,B,D$
are subject to a variational optimization within a quantum Monte Carlo framework, although the trial wave function, constructed from a two-body scattering solution, is fixed for any given choice of the mass of the particle and of the interaction strength. 
However in practice it might be convenient to keep the functional form of the solution and to treat its characteristic coefficients as variational parameters.  
A typical Jastrow term of the trial wave function of this symmetrized functional form is given in Fig.~(\ref{yuk_twf}) in 
the Section, devoted to the Monte Carlo methodology.

Worth noticing that in case of the Yukawa potential the pair distribution function 
at zero can be finite for $r=0$, as it happens for the Coulomb potential. 
This can be easily confirmed if one takes a series expansion from Eq. (\ref{ceperlyur1}).
The typical value of the leading component $A/D$ in the conditions of our problem 
is of the order of 10, therefore the zero value of the trial wave $\exp(-A/D)$ is 
very small and practically indistinguishable of zero, 
as one can see from Fig.~(\ref{yuk_twf}).

\subsubsection{Scattering on repulsive van der Waals\label{secSRydberg}}
A similar derivation path can be applied to obtain the solution of the two-body
 scattering for the system of Rydberg atoms, interacting via the simple $1/\rve^6$
interaction potential. The Schr\"odinger equation for the two-body Hamiltonian
 in the reduced units for this problem
 (see Sec.~\ref{secRydberg}) in the system of the center of masses
reads as
\begin{equation}
-w''(r) + \frac{C_6}{r^6} w(r) = E w(r)
\label{eqscatryd}
\end{equation}
where we used a familiar substitution $f(r)=w(r)/r$. The finite low-energy solution of the 
latter equation has the following form (normalization is omitted):
\begin{equation}
f(r)=\frac{1}{\sqrt{r}} K\left(\frac{1}{4},\frac{\sqrt{C_6}}{2 r^2}\right)
\label{ryd_fr}
\end{equation}
where $K(\alpha,x)$ stands for the modified Bessel's function of the second kind.
Its expansion in the series of $r$ is given by

\begin{equation}
\frac{1}{\sqrt{r}} K\left(\frac{1}{4},\frac{\sqrt{C_6}}{r^2}\right)=e^{-\frac{\sqrt{C_6}}{2 r^2}}\left[\frac{\sqrt{2 \pi}}{C_6^{1/4}}\sqrt{r} + \mathcal{O}(r^{5/2})\right]\,.
\end{equation}

\begin{figure}

    \centering

    \subfigure

    {

        \includegraphics[width=0.6\textwidth, angle=270]{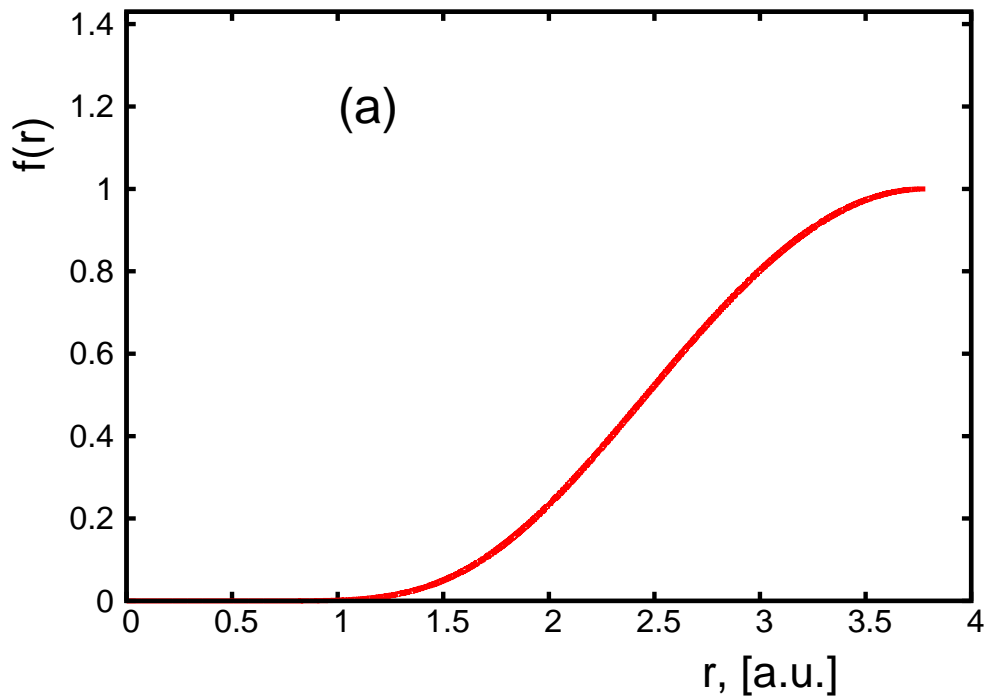}

        \label{fig:first_sub}

    }

    \subfigure

    {

        \includegraphics[width=0.6\textwidth, angle=270]{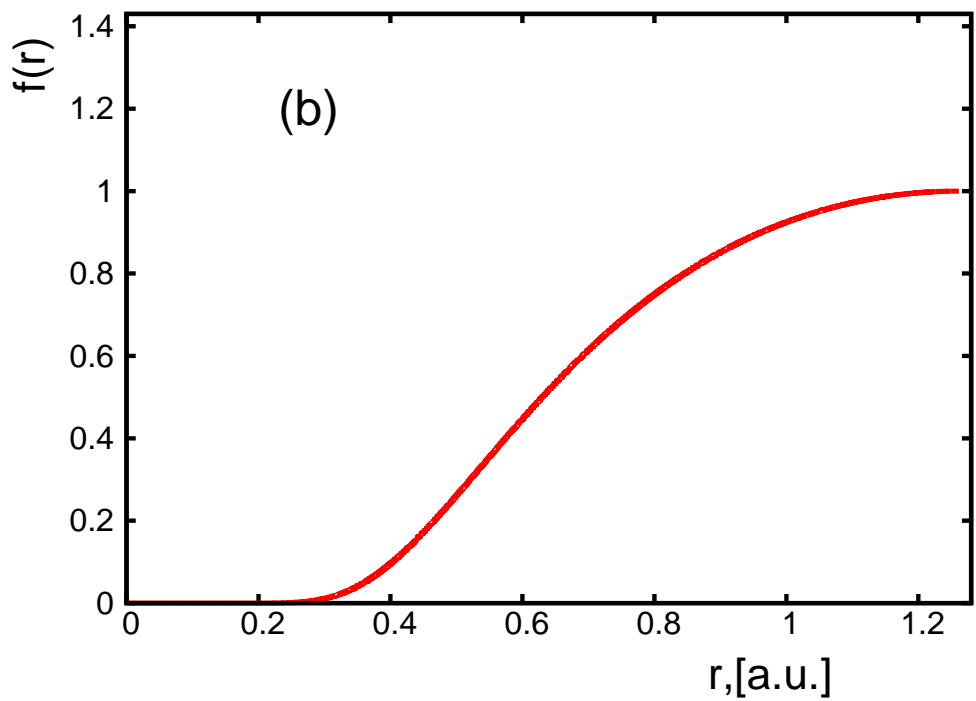}

        \label{fig:second_sub}

    }

    \caption{Typical forms of the trial wave functions for (a) Yukawa potential,  and (b) model potential between Rydberg atoms.}

    \label{typfr}

\end{figure}

The short-range behavior of the function is dominated by the exponential term $e^{-\frac{\sqrt{C_6}}{2 r^2}}$, which smoothly approaches zero as $r\rightarrow 0$. As in the case of the Yukawa potential, 
the interaction constant $C_6$ and the power (-2) can be conveniently treated as variational parameters, while keeping the overall functional form of the trial wave function intact. A 
typical form of a trial wave function and a pair distribution function for the system is presented in Fig.~(\ref{typfr}).

\chapter{Quantum Monte Carlo methods\label{secQMC}}
\section{Introduction}

Quantum Monte Carlo methods (QMC) are very efficient and multi-purpose tools for the investigation of
quantum many-body systems (for a detailed review of the methods see, for instance,
~\cite{Ceperley95},~\cite{Guardiola98}). The use of Quantum Monte Carlo
 techniques provides a deep insight
into the microscopic behavior of quantum states of matter. QMC are essentially {\it ab
initio} methods, relying on a microscopic description of a system and 
gathering valuable information on
its properties of the system through a numerical simulation. In certain cases, 
it turns out that this technique is the only tool for studying complex
problems with reasonable calculation costs. In fact, in order to have a model, accessible for being 
solved analytically in the exact way, a physicist usually faces the necessity 
to make some kinds of approximations,
but this can be avoided to a big extent by virtue of Quantum Monte Carlo simulations. 
As an example, the applicability of  perturbation theory is limited by a small value
of the perturbation parameter, 
while the QMC methods do not present any restrictions of this kind. 
Quantum Monte Carlo technique allow to 
find the ground state solution of the many-body Schr\"odinger equation 
at zero temperature. As follows from the name, the Quantum Monte Carlo methods are based on 
stochastic numerical algorithms of different sorts, that by nature are especially advantageous when 
the system in question possesses multiple degrees of freedom. 
As for any other stochastic procedure, the QMC methods provide results with a certain statistical error
that can be diminished by performing longer measurement series.

We are interested in studying the quantum properties of a given system. The quantum
effects manifest the most when they are not disturbed by the thermal motion, 
that is at the lowest temperatures, when the system maintains in its
ground state. For this case a possible method of choice to address the
problem is the diffusion Monte Carlo (DMC) method. 
For a bosonic system this method allows to obtain the {\it exact} result for the energy of 
the ground state, as well as for any diagonal property of the state. 

In this chapter first of all we discuss the variational Monte Carlo method as the simplest 
one.
Then we will discuss 
the bosonic diffusion Monte Carlo method and give a 
detailed explanation on how trial wave functions are constructed. Finally, we 
will present the main ideas for the implementation of the sampling of the quantities of interest.

\section{Variational Monte Carlo \label{secVMC}}
\subsection{General notes}

The simplest of the Quantum Monte Carlo methods is the {\it variational} method
(VMC). The general idea behind this technique is to find 
an approximate wave function $\psi_T$, called {\it trial wave function} (or, sometimes, {\it
variational wave function}), and then by sampling from the probability distribution
\begin{eqnarray}
p(\rve)~=~|\psi_{\rm trial}(\rve)|^2
\label{p(r)}
\end{eqnarray}
calculate averages of physical quantities. It can be shown that $E_T$, the expectation value of the 
Hamiltonian, is an upper bound to the ground-state energy $E_0$. By expanding the normalized 
trial wave function in the basis of the normalized eigenfunction of the Hamiltonian 
\begin{eqnarray}
\psi_T=\sum_{i=0}^{\infty} c_i \phi_i\\
\sum_{i=0}^{\infty} |c_i|^2=1\,,
\label{psinorm}
\end{eqnarray}
one can rewrite the variational energy as 
\begin{eqnarray}
E_T &=& \frac{\langle\psi_T|\hat H|\psi_T\rangle}{\langle\psi_T|\psi_T\rangle}\\ 
&=&\langle\sum_{i=0}^{\infty} c_i \phi_i|\hat H|\sum_{j=0}^{\infty} c_j \phi_j\rangle=\sum_{i,j=0}^{\infty} c_i^* c_j \langle\phi_i|\hat H| \phi_j\rangle\\
&=&\sum_{i,j=0}^{\infty} c_i^*c_j \delta_{i,j}E_i=\sum_{i=0}^{\infty}E_i |c_i|^2\geq E_0\sum_{i=0}^{\infty}|c_i|^2=E_0 
\label{Et}
\end{eqnarray}
where $E_i$ stands for a corresponding eigenvalue (energy of the $i$-th state).
By minimizing the variational
energy with respect to the parameters entering into it one can optimize the wave function
within the given class of considered wave functions. 
The only situation when the zero variance can be reached is when the wave function is exactly known.
\subsection{Usage of the VMC method}

If the wave function, corresponding to the ground state, is exactly known 
the sampling by means of the Variational Monte Carlo method permits to 
evaluate exactly any static property of the system within some statistical errors. 
Such systems are scarce;
the ground state of a system of hard rods~\cite{Krotscheck99} and the Tonks-Girardeau gas~\cite{Girardeau60}
are among the most famous ones. The Variational Monte Carlo method in this case can provide 
the correlation functions, which could be not accessible directly through the wave function.

%
The Variational Monte Carlo approach provides not only a valuable description of the quantum systems, 
but it also can be used as a first step to deliver a good quality input for the diffusion Monte Carlo method. 
The efficiency and even applicability of this method depends substantially on the optimization of the trial wave function within a chosen class 
of functions.

\subsection{Notes on algorithmic realization}

Let us stick to a coordinate representation in the following description, since 
it is the easiest way to work with external or interparticle potentials. Consider a common $D-$dimensional 
Euclidian space with a system of $N_p$ particles inside. The probability distribution function in this system will be 
a function of $N \cdot D$
variables $p(\R) = p(\rv_1, ..., \rv_{N_p})$. The mean value for an arbitrary operator $\hat Z$
is therefore calculated as an integral of $N\cdot D$ dimensions in the following form:
\begin{eqnarray}
\langle Z\rangle = \frac{\int...\int Z(\rv_1,...,\rv_{N_p})p(\rv_1,...,\rv_{N_p})
d\rve_1...d\rve_{N_p}}{\int...\int p(\rv_1,...,\rv_{N_p})
d\rve_1...d\rve_{N_p}}
\label{A}
\end{eqnarray}

It is clear that the complexity of the estimation of this integral with conventional non-stochastic methods, based 
on a grid calculation,  
grows extremely fast with the number of particles, and already for a few dozens of particles its calculation is unreachable. 
On the other hand, the stochastic procedure on which the Monte Carlo methods are based 
is not affected strongly by the growth of the dimensionality of the problem.
The basic idea of the variational technique is to generate configuration 
states $\R_1,...,\R_m$ with the probability distribution
$p(\R)$ and by means of these states estimate the average value of the operator $\langle Z\rangle$,
\begin{eqnarray}
\langle Z\rangle \approx \frac{1}{m}\sum\limits_{j=1}^m Z(\R_j)\:.
\end{eqnarray}

Every state is obtained only from its preceding configuration, thus 
this set of states is indeed a Markovian chain. The Metropolis algorithm
~\cite{Metropolis53} despite its simplicity is a very efficient tool to produce the chain
with the desired probability $p(\R)$. The transition from the old $\R$
configuration to the new one $\R'$ is accepted with the probability $P(\R
\rightarrow \R')$, expressed by the formula
\begin{eqnarray}
P(\R \rightarrow \R') =
\left\{
\begin{tabular}{ccc}
$p(\R')/p(\R)$, & if $p(\R') < p(\R)$ \\
$1$, & if $p(\R')\ge p(\R)$ \\
\end{tabular}
\right.
\label{Metr}
\end{eqnarray}

In a quantum system, the probability distribution is given by the square of the
wave function. The way we construct the particular trial wave 
function for different quantum systems will be discussed in Sec.~\ref{secWF}. 

There are distinct approaches to perform transitions (or {\it trial moves}) between different particle configurations. 
A straightforward way for creating a new state is to move one particle at once or all the particles
at once $\rve_i' = \rve_i+\bf{\rho}_i,
i=\overline{1,N_p}$, where the displacements $\bf{\rho}_i$ are a set randomly chosen vectors with a 
certain upper bound $|\bf{\rho}_i|<U$. The limiting amplitude $U$ can 
be adjusted in order to have a desired acceptance rate.
It is readily seen that 
for very small values of $U$ the trial moves are accepted almost all the time, but the adjacent states
are strongly correlated, that affects the variance of the overall statistics, 
and in the limit $U \rightarrow 0$ all the states are the same, making the whole calculation pointless. 
On the other hand, the steps of very large amplitude $U$ are accepted only a small fraction of time,  
which also leads to a poor performance. The value of $U$ can be optimized to ensure the maximum total displacement of 
the whole system by means of a set of benchmark calculations, 
but in practice a simple rule of thumb to have the acceptance rate of the trial moves in the range $0.5-0.65$ provides good enough results. 

A generalization of these two strategies of particle displacements, when only a particular fraction of the points 
is shifted together, can bring even faster performance. The group of particles to move in one step can be chosen
randomly. The advantage of this technique lies in the possibility of fine tuning
 the calculation parameters 
in order to achieve an optimal point in terms of the interplay between the correlation of the states and the acceptance rate.

\section{Diffusion Monte Carlo technique\label{secDMC}}

The diffusion Monte Carlo method (DMC) is a stochastic computational technique 
applied to systems at zero temperature, when all of the thermal motion can be neglected. 
The key point of the DMC method is to provide the solution of the time-dependent Schr\"odinger equation in  imaginary time, 
which is known to exponentially decay to the ground state solution in the limit of long times. By means of the diffusion Monte Carlo method 
the equation of state for the system as well as diagonal properties can be calculated exactly with the only cost of controlled statistical noise. 

\subsection{Schr\"odinger equation}

The wave function of a quantum system obeys the Schr\"odinger equation
\begin{eqnarray}
\ii\hbar \frac{\partial}{\partial t} \psi (\R,t) = \hat H \psi (\R,t)\,.
\label{Schrodinger}
\end{eqnarray}
Our aim is to find the ground state properties of the system, rather than its 
actual time evolution. By substituting the time variable by an imaginary one $\tau = -i
t/\hbar$, one arrives to another representation of the Schr\"odinger equation
\begin{eqnarray}
-\frac{\partial}{\partial \tau} \psi (\R,\tau) = (\hat H -E) \psi (\R,\tau),
\label{shifted Schrodinger}
\end{eqnarray}
where $E$ stays for a constant energy shift close to expected ground state energy. 
The latter equation (\ref{shifted Schrodinger}) 
has a formal solution $\psi (\R,\tau) = e^{-(\hat H - E)\tau} \psi (\R,0)$. One can 
expand this solution as the sum over the eigenstates of the Hamiltonian $\hat H\phi_n~=~E_n\phi_n$,
with the eigenstates taken such that the eigenvalues are growing with ascending indexes, that is 
$E_0$ is the lowest among the eigenvalues. Performing the expansion,
\begin{eqnarray}
\psi (\R,\tau) = 
\sum\limits_{i=0}^\infty a_i \phi_i(\R) e^{-(E_i - E)\,\tau}\:.
\label{sum}
\end{eqnarray}
It can be easily seen that the exponents in the sums behave differently (decay or grow) provided the sign of $(E_i-E)$ is positive or negative. For large enough $\tau$ the only component of the sum (\ref{sum}) that survives is the one, corresponding to the ground state. All the other terms decay in time exponentially fast (we suppose the spectrum to be discreet):
\begin{eqnarray}
\psi (\R,\tau) \to a_0~\phi_0(\R)~e^{-\tau (E_0 - E)},\qquad \mbox{if~~} \tau \to \infty
\label{psi limit}
\end{eqnarray}

A general expression for the Hamiltonian of a many-body system of $N_p$ particles, subjected to 
an external force field $V_{\rm ext}(\rve_i)$, depending on a particle's position, and interacting internally through 
a pairwise potential $V_{\rm int}(|\rve_i -\rve_j|)$ can be given as
\begin{eqnarray}
\hat H = -\frac{\hbar^2}{2m} \sum\limits_{i=1}^{N_p} \Delta^2_i
+ \sum\limits_{i=1}^{N_p} V_{\rm ext}(\rve_i)
+ \sum\limits_{i<j}^{N_p} V_{\rm int}(|\rve_i -\rve_j|)
\,.
\label{Hgeneral}
\end{eqnarray}

We take a notation for the constant $D = \hbar^2/2m$, referring to $D$ as 
a ``diffusion'' constant, the sense of which will be clarified later. 
The Schr\"odinger equation (\ref{shifted Schrodinger}) reads as
\begin{eqnarray}
-\frac{\partial}{\partial \tau} \psi (\R,\tau) =-D \Delta_\R \psi (\R,\tau)
+ (V_{\rm total}(\R)  - E)\psi (\R,\tau),
\label{generalShr}
\end{eqnarray}
with the label $\R$ in the Laplacian denoting the differentiation  
with respect to each scalar component of the vector  $\R$.
For shortness, the summation over the internal and the external potentials is denominated by a single
term $V_{\rm total}(\R) =\sum V_{\rm ext}(\rve_i)+ \sum V_{\rm int}(|\rve_{ij}|) $. 

The last term in Eq.~(\ref{generalShr}) $(V_{\rm total}(\R)  - E)\psi (\R,\tau)$ is diagonal, 
and it affects the normalization through a specific step, called {\it branching}. 
As we mentioned above, a key ingredient of the diffusion Monte Carlo technique 
is the use of a trial wave function which allows to radically reduce the 
calculation efforts needed to reach a required accuracy in the result. Hence, one gives an
approximation of the true ground state solution $\psi_0(\R,\tau)$ by a certain trial wave function $\psi_T(\R,\tau)$, which is subject to a correction inside the algorithm by means of the branching. The whole approach, referred to as importance sampling, stems from an analog of the 
Schr\"odinger equation for the product of the true wave function and the trial wave function 
\begin{eqnarray}
f(\R,\tau) = \psi_T (\R) \psi (\R,\tau)\,.
\label{f}
\end{eqnarray}

The expectation value of an operator $Z$ over this product of the wave functions can be thought of as a mixed estimator
$\langle Z\rangle = \int \psi_T Z\,\psi\,\dR/
\int\psi_T\psi\,\dR$. 
 Performing a substitution of $f$ inside the Schr\"odinger equation~(\ref{generalShr}) the 
 following equality can be drawn
\begin{eqnarray}
-\frac{\partial}{\partial \tau} f (\R,\tau) = -D\Delta_\R f (\R,\tau)
+ (E^{\rm local}(\R) - E) f(\R,\tau)+ D\nabla_\R (f(\R,\tau) \F(\R,\tau) )\,.
\end{eqnarray}
where $E^{\rm local}$ is the so-called local energy, that is
 the expectation value of the Hamiltonian with respect to the trial wave function (a broad discussion on the calculation of the
 values is given in Section~\ref{secQuantities}).
\begin{eqnarray}
E^{\rm local}(\R) = \frac{\hat H\psi_T(\R)}{\psi_T(\R)}=(V_{\rm total}(\R)-E)-D\frac{\Delta_\R \psi_T (\R)}{\psi_T (\R)}\,.
\label{local energy}
\label{Eloc}
\end{eqnarray}

The action of the Metropolis algorithm in the variational Monte Carlo method, that is the averaging of an operator over the probability distribution 
$p(\R)=|\psi_T(\R)|^2$, 
is therefore the averaging of the local energy with respect to the set of $N_c$ particle configurations 
\begin{equation}
E=\frac{1}{N_c}\sum_{i=1}^{N_c}E^{\rm local}(\R_i)\,.
\end{equation}
Notice that the local energy is equal to the eigenvalue of the Hamiltonian, when the trial wave function coincides with the eigenfunction. It means that $E^{\rm local}$ is a constant ground state energy with zero variance if $\psi_T(\R,\tau)=\psi_0(\R,\tau)$. The deviations of the trial wave function from 
the exact solution result in a growth of variance in the local energy, which can serve as a quality criterion for the trial wave function. 
The notation $\F$ stands for the {\it drift force}, a vector value, equal to the gradient of the field defined by the trial
wave function, multiplied by a convenient factor,
\begin{eqnarray}
\F = \frac{2\nabla_\R \psi_T(\R)}{\psi_T(\R)}\:.
\label{DriftForce}
\end{eqnarray}
The drift force moves the walkers $\{\R_i\}$ towards the region where $\psi_T$ is large.

Notice that the probability distribution function in a classical system is defined by a Boltzmann factor
$f(\R) \sim \exp(-E_{\rm pot}(\R)/k_B T)$, with $E_{\rm pot}(\R)$ standing
for the potential energy in the system. The force, produced by the field, is equal to its gradient 
with the negative sign, that is $\F(\R) = - \nabla E_{\rm pot} = \nabla \ln p(\R)$. 
Making a formal substitution of the trial wave function instead of the probability density
$p(\R)=\psi_T^2(\R)$, we can recover precisely the same form of the force, as in the definition~(\ref{DriftForce}).  

\subsection{The Green's function\label{secGreen}}

The Schr\"odinger equation~(\ref{generalShr}) can be resolved formally in the following way:
\begin{eqnarray}
\langle\R|f(\tau)\rangle =
\sum\limits_{\R'} \langle\R|e^{-\tau(\hat H-E)}| \R'\rangle
\langle\R'|f(0)\rangle,
\end{eqnarray}
where $f$ stands for time-dependent wave function appearing in the equation.\\
The term $\langle\R|e^{-(\hat H-E)\tau}| \R'\rangle$ is the Green's function $G(\R,\R',\tau)$ of the Hamiltonian, acting as a propagator of the system. The Schr\"odinger equation can be rewritten in the integral form in terms of 
$G(\R,\R',\tau)$ as  
\begin{eqnarray}
f(\R,\tau) = \int G(\R,\R',\tau) f(\R',0)\,{\bf dR'}
\label{Green}
\end{eqnarray}

The last integral equation is an analog of the traditional form of the Schr\"odinger equation (\ref{Schrodinger})
but it allows to obtain the solution $f(\R,\tau)$ by virtue of a multidimensional numerical integration,
where the Monte Carlo methods are applicable. The exact form of the Green's function is not known, however it can be expanded in  power series of $\tau$, therefore with a certain approximation the solution can be reached in a number of integrations, depending on a the time-step size. Hence, the accuracy and the overall computational complexity to obtain the final solution depend on the time step. The solution of Eq.~(\ref{Green}) after a single time step is then given by 
\begin{eqnarray}
f(\R,\tau + \Delta \tau) = \int G(\R,\R',\Delta\tau) f(\R',\tau)\,{\bf dR'}
\label{Green approx}
\end{eqnarray}

As it was stated before, after a large enough number $M$ of time steps the solution of Eq.~(\ref{Green}) decays to the ground state, while all the other components of the solution exponentially disappear. 
\begin{eqnarray}
f(\R,M\Delta\tau)\rightarrow \psi_T(\R)\phi_0(\R),{\rm when}\:\:M\rightarrow\infty
\label{f limit}
\end{eqnarray}

It is natural to separate three different operators inside the Hamiltonian
\begin{eqnarray}
\label{H summands1}
\hat H = (-D \Delta) + D((\nabla_\R \F) + \F \nabla_\R)) + (-E+E^{\rm local}(\R)),
\end{eqnarray}

that is 

\begin{eqnarray}
\label{H summands2}
\hat H = \hat H_1 + \hat H_2 + \hat H_3\, ,
\end{eqnarray}

and write down the Green's function for each of the operators $\hat H_i$,
\begin{eqnarray}
G_i(\R,\R',\tau) = \langle\R|e^{-\tau \hat H_i}|\R'\rangle\:.
\label{GreenCoord}
\end{eqnarray}

If components of an operator do not commute, its Green's function 
(or exponent of the operator) cannot be represented as a product of the Green's functions
of the components. This is evidently the case for the components $\hat H_i$ of the Hamiltonian.
Nonetheless, there is a possibility to use approximations of different orders of $\tau$ for $\exp(-\tau \hat H)$. 
The first-order approximation for the exponential of $\hat H$ corresponds to
\begin{eqnarray}
e^{-\tau\hat H} = e^{-\tau\hat H_1}e^{-\tau\hat H_2}e^{-\tau\hat H_3}+\mathcal{O}(\tau^2)\:.
\label{t2 approximation}
\end{eqnarray}

The integration of this formula, when the term $\mathcal{O}(\tau^2)$ is neglected, 
yields the following expression for the Green's function
\begin{eqnarray}
G(\R,\R',\tau) = \intint
G_1(\R,{\bf R}_1,\tau) G_2({\bf R}_1,{\bf R}_2,\tau) G_3({\bf R}_2,\R',\tau)\,
\di{\bf R}_1 \di {\bf R}_2
\end{eqnarray}

Notice that in accordance with (\ref{GreenCoord}) the Green's function $G$ is a solution of the Bloch
equation
\begin{eqnarray}
\left\{
{\begin{array}{rcll}
\displaystyle -\frac{\partial}{\partial \tau} G(\R,\R',\tau) &=&
\displaystyle \hat H_i\,G(\R,\R',\tau),
&\displaystyle i = 1,2,3\\
\displaystyle G(\R,\R',0)& =&\displaystyle\delta(\R-\R')&\\
\end{array}}
\right.
\label{G}
\end{eqnarray}

The first (kinetic) component $G_1$ satisfies the equation
\begin{eqnarray}
\frac{\partial G_1(\R,\R',\tau)}{\partial \tau} = D\Delta G_1(\R,\R',\tau)
\label{DiffG1}
\end{eqnarray}
which coincides formally with the diffusion equation. The diffusion constant of this equation is equal to $D$, and this is the reasoning of the name that we adopted for the constant. The last equation~(\ref{DiffG1}) can be easily solved in the momentum representation, as the kinetic energy operator is diagonal in this representation. Rewriting the solution again in the space coordinates, we get a well-known formula
\begin{eqnarray}
G_1(\R,\R',\tau) = \frac{1}{(4 \pi \tau D)^{\frac{3N}{2}}}e^{-\frac{(\R-\R')^2}{4\tau D}}
\label{GF1}
\end{eqnarray}

The second term $G_2$, containing the drift force, satisfies the equation 
\begin{eqnarray}
\frac{\partial G_2(\R,\R',\tau)}{\partial \tau} =D \nabla_\R (G_2(\R,\R',\tau) \F )
\end{eqnarray}
is also easily resolved 
\begin{eqnarray}
G_2(\R,\R',\tau) = \delta(\R-\R_c(\tau)),
\label{GF2}
\end{eqnarray}
where $\R_c(\tau)$ stands for the solution of the equation
\begin{eqnarray}
\left\{
{\begin{array}{rcl}
\displaystyle\frac{\di{\bf R}_c(\tau)}{\di\tau}&=&\displaystyle D F(\R_c(\tau)),\\
\displaystyle\R_c(0)&=&\displaystyle\R'
\end{array}}
\right.
\label{importance sampling}
\end{eqnarray}
which defines the motion of the system, subjected to the drift force $F$.

The third equation from the set (\ref{G}) is evident to solve:
\begin{eqnarray}
G_3(\R,\R',\tau) = \exp\{(E-E^{\rm local}(\R))\,\tau\}~\delta(\R-\R')
\label{GF3}
\end{eqnarray}
which is generally referred to as the branching component, as it controls multiplication 
 and annihilation of walkers in the algorithm. 

\subsection{The primitive (first-order) algorithm}

If the wave function of the system $\psi(\R,\tau)$ is real and positive, as it
happens for the ground state of a Bose system, it can be treated as a population
density distribution:
\begin{eqnarray}
\psi(\R,\tau) = C \sum\limits_{i=1}^{N_{\rm walkers}}  \delta ({\bf R - R_i}(\tau)),
\label{Walkers}
\end{eqnarray}
with a constant $C>0$ and ${\bf R_i}(\tau)$ are time-dependent positions of a 
single particle set, referred to as a \textit{walker}. The formula (\ref{Walkers}) should be taken
in a statistical sense, the average of any value $A$ over the left hand side and right 
hand side distributions are equal to each other in the limit when size of the population $N_W$
tends to infinity $\int A(\R) f(\R,\tau)\di{\bf R} =
\lim\limits_{N_w\to\infty} \int A(\R)
\sum\limits_{i=1}^{N_w} C \delta (\R - \R_i(\tau))\di{\R}$. 
The walker resides in the coordinate space of $3N$ dimensions, and the infinitesimal probability $f({\bf
R},\tau)\,{\bf dR}$ is equal to the probability to encounter a walker in
the infinitesimal range ${\bf dR}$ near the point $\R$ in the moment $\tau$.

The algorithmic implementation of the time evolution of the system according to its Hamiltonian
(\ref{H summands1}) is an evolution of a system of walkers upon the 
action of the elementary components of the 
 Green's function, that is the transition matrices (\ref{GF1}), (\ref{GF2}) and (\ref{GF3}).  

The first function $G_1$ determines a diffusion of the whole system of walkers according to 
the equation 
\begin{eqnarray}
{\bf R}_1(\tau+\Delta\tau) = {\bf a} + \R(\tau) ,
\label{R1}
\end{eqnarray}
where ${\bf a}$ stands for a displacement, taken randomly from the three-dimensional normal distribution $\exp\{-{\bf a}^2/(4 \Delta \tau D )\}$.

The second function $G_2$ gives rise to the drift of the set of walkers towards areas in the configuration space where
the trial wave function is large:
\begin{eqnarray}
{\bf R}_2(\tau+\Delta\tau) = \R(\tau) + ( D \Delta\tau ) F(\R)\,.
\label{R2}
\end{eqnarray}

Notice that the Green's functions of the first two kinds (\ref{GF1}),(\ref{GF2}) have a normalization  $\int
G_{1,\,2}(\R,\R',\tau)\,\di\R = \delta(\R')$, that is on each of the steps the number of walkers stays unchanged.

This is not the case for the third propagator, corresponding to the branching term,
\begin{eqnarray}
f^{(3)}(\R, \tau+\Delta\tau) = f(\R,\tau) e^{-(E-E^{\rm local}(\R))\,\Delta\tau}\:,
\label{third step}
\label{branching}
\end{eqnarray}
since the integral over its Green's function $G_3(\R,\R',\tau)$ (\ref{GF3}) is clearly not equal to unity. The most evident interpretation of the action of the third propagator is that each walker has a certain value attached to it, commonly thought as its ``weight''. This value is recalculated on each step for every walker. 
The formula (\ref{branching}) suggests that the walkers of less $E^{\rm local}$ are favored and the contrary are disfavored. A clear disadvantage of this scheme is that one would wish to have a better statistical
representation for the favorable areas of the configuration space, however the generation 
and elimination of walkers is not possible. The alternative and widely applied way to overcome the problem 
is that the value $c=\exp\{-(E-E^{\rm local}(\R))\,\Delta\tau\}$ is treated as a number of exact copies of the walker. 
As one can see, this is a non-integer number that cannot be readily used for it. However, one can apply a randomization 
by throwing a random number from the uniform distribution $[0,\,1)$ and afterwards take $[c]+1$ duplicates of 
the walker, if the number is within $[0,\{c\}]$, and $[c]$ is otherwise (the standard bracket notations $[.]$ and $\{.\}$ 
stand for the integer and the fractional parts, respectively).
The relation $([c]+1)\{c\}+[c](1-\{c\})=c$ ensures the correct mean action of the propagator.    

The equation~(\ref{branching}) also suggests that the branching can be effectively 
controlled by a dynamic choice of the reference energy $E$, that can be essential to avoid 
a collapse of the set of walkers as well as its undesirable expansion, that can make 
the simulation stall.


Looking closely one can notice that the diffusion and the drift steps represent only the 
sampling from the trial wave function. It means that being applied without the branching term, they are equivalent 
to the application of the variational Monte Carlo method (see \cite{LesterBook1994} and Section~\ref{secVMC}). The third
branching step ensures that the system of walkers ``prefers'' the areas of higher 
$\psi_0$ (it is often called a {\it correction} of the trial wave function), and the overall sampling is taken from $\psi_T \psi_0$ rather than $\psi_T^2$. 
As it was commented above, if the trial wave function coincides with the ground-state solution (or, more generally, with an eigenfunction of the Hamiltonian), the local energy 
becomes equal to the eigenvalue, hence the branching factors are the same for all the walkers. The action of the branching step in this case does not affect the final result.
\subsection{Second-order algorithm}

In the previous section, we have described the simplest (first-order in $\tau$) approximation (\ref{t2 approximation}) of the Green's function. The order of the application of the three propagators is clearly irrelevant, since overall results (for instance, the energy of the ground state) will depend linearly on the 
time step $\tau$. To obtain the final result for a quantity of interest one should take a short enough $\tau$ to move the time-related error below the statistical noise or perform a series of simulations with distinct time steps, and then find a linearly extrapolated value. In practice the second approach is much more practical, since the time bias of the result can be very pronounced. However, a clear 
drawback is that the correctness of the linear dependence might be valid for undesirably small times. This is where the higher-order algorithms become useful.

The second-order in $\tau$ expansion of the exponent in the Green's function can be given in the form 
\begin{eqnarray}
e^{-\tau \hat H} =e^{-\tau \hat H_3/2}e^{-\tau \hat H_2/2}e^{-\tau\hat H_1}e^{-\tau \hat H_2 /2}e^{-\tau \hat H_3/2}+\mathcal{O}(\tau^3)
\label{t3 approximation}
\end{eqnarray}
which is not the only possible way of representation, but probably the most efficient for actual application\cite{Chin1990}. The final result for the equation of state of the system, yielded by the formula~(\ref{t3 approximation}), does not have linear in $\tau$ elements, thus for small enough time steps the dependence is quadratic. In this case one can again perform an extrapolation to $\tau \rightarrow 0$ via a series of measurements, or 
alternatively,
find a time step short enough to guarantee the smallness of the time step-related 
error in accordance with a required accuracy level of the simulation. This accuracy is 
generally described by the statistical variance of the result. The choice between 
the two approaches must be taken regarding the interplay between the additional calculation costs, related 
to the extrapolation, and the slower evolution of the system in case of a shorter time step. 

Let us explain the second-order algorithm, that relies on the expansion~(\ref{t3 approximation}).
Each propagation in time corresponds to a constant shift $\Delta\tau/2$ with respect to the current time.
The state of a displaced walker is changed on every step from $\R_{i-1}$ to $\R_{i}$. The effect of the branching propagator is a corresponding 
multiplication or suppression of a walker in question. Since the walker is moved in a loop, the choice of the first step is arbitrary, and the list of the operations can by reordered.\\
The calculation procedure:\\
1) The first propagator, random Gaussian move (\ref{GF1}):\\
$\R = \R_{i-1} + a\,,$ 
$a$ is taken from $\exp(-\bf{x}^2/\Delta\tau)$\\
2) The second propagator, drift with $\F(\R)$ (\ref{GF2}):\\
$\R' = \R + \F(\R)\,\Delta\tau/2$\\
$\R'' = \R + (\F(\R)+\F(\R')\,\Delta\tau/4$\\
$\R''' = \R + \F(\R'')\,\Delta\tau$\\
3) The branching propagator (\ref{GF3}):\\
$\R'''$ unchanged.

\section{Constructing the trial wave functions\label{secWF}}
\subsection{Motivation}

In this Section, the development of trial wave functions for 
bosonic systems is discussed from the technical point of view. The purpose of 
the following discussion is to present the theoretical basis of the construction
of Nosanow-Jastrow trial wave function of a general kind in liquid and solid phases, its relation to 
the actual implementation, the technical issues that one faces, and their solutions.

\subsection{Nosanow-Jastrow trial wave function\label{secNosanow}}


The natural way of constructing a bosonic trial wave function is
to take it in the form of a product of one- and two-body correlation terms: 
%
\begin{eqnarray}
\Psi(\rN) =\prod\limits_{i=1}^{N_p} f_1(\ri)\prod\limits_{j<k}^{N_p} f_2(|\rjk|)\:.
\label{Jastrow}
\end{eqnarray}

This form of the trial wave function is generally referred to as the Jastrow (or Bijl-Jastrow) trial wave function, first proposed by 
Bijl~\cite{Bijl40} and then by Jastrow~\cite{Jastrow55}. 
The pairwise interaction of particles in the system is taken into account by the two-body Jastrow term $f_2(r)$. 
The pair correlation is clearly lost for large enough interparticle distances, that is this term must approach 1 asymptotically. 
The use of the periodic boundary conditions puts additional restrictions to 
$f_2(r)$. Namely, 
in order to avoid additional contributions to the kinetic energy from 
distances $r>d_{\rm plane}$ ($d_{\rm plane}$ stands for 
a distance to the plane, closest to the center of the simulation cell), the Jastrow term 
should be constant in that region. 
For a rectangular simulation box it is one half of a 
minimum box dimension $d_{\rm plane}={\rm min}(L_x,L_y,L_z)/2$.
 For a calculation cell of the shape of truncated octahedron 
(see Appendix B) this distance is equal to 
\begin{equation}
d_{\rm plane}=\frac{3/4}{\sqrt{\frac{1}{1/L_x^2}+\frac{1}{1/L_y^2}+\frac{1}{1/L_z^2}}}\,.
\end{equation}

The value of the Jastrow-Bijl term at distances larger than the cut-off one $r\geq L/2$ is not necessarily
 $f_2(L/2)=1$, but, in principle, can be any arbitrary constant $f_2(L/2)={\rm const},\:r\geq L/2$. Still, 
the unitary value might be convenient to use in the evaluation of the product of Bijl-Jastrow terms 
$\prod\limits_{i<j}f_2(\rve_i-\rve_j)$, which in code is usually implemented as evaluation of 
$\exp\sum\limits_{i<j}\ln f_2(\rve_i-\rve_j)$. With the choice $f_2(L/2)=1,\:r\geq L/2$ the contributions
of pairs with $|\rve_i-\rve_j|\geq L/2$ to the sum is zero. 

The one-body Jastrow term $f_1(\rve)$ is introduced 
to take into account an external potential, present in the system. It can also define symmetry 
properties of the system, for example, the localization of the particles in crystalline nodes. The form 
of this one-body term is typically obtained from a solution of the Schr\"odinger equation for a single 
particle in the chosen potential. 

For a quantum Monte Carlo simulation of a solid phase one might need to induce a corresponding 
 crystalline symmetry to the trial wave function. This is done by ``pegging'' the particles to 
 the nodes of a crystal through a multiplication of the Jastrow term by 
 a particular factor, depending on the coordinates of the particles. 
 A straightforward way to realize such a condition is to consider a one-body term 
 \begin{equation}
 f_1(\rve_i)=g(|\rve_i-\rm \xi_i|) 
 \end{equation} 
 with the configuration $\{\rm \xi_i\}$ standing for a set of crystal nodes' positions and
  $g(\rve)$ is a function, which localizes each particle $\rve_i$ to the site $\rm \xi_i $. This factor is generally referred to as Nosanow 
 term~\footnote[1]{A not so brief explanation of the efficient construction of crystalline guiding wave functions 
 can be found, for instance, in the recent work of Cazorla and collaborators~\cite{Boronat2009NJP}.}, 
 and a corresponding trial wave with a Jastrow two-body term
 \begin{equation}
 \Psi_{\rm NJ}(\rve_1,...,\rve_{N_p})= \prod_{i<j}f_2(\rve_{ij})\prod_{k=1}g(|\rve_i-\rm \xi_i|) 
 \label{NJ_trial_wave_function}
 \end{equation} 
 is called a Nosanow-Jastrow (NJ) trial wave function. 
The localization factor $g(\rve)$ is usually chosen as a 
 Gaussian with a localization parameter $\gamma=1/(2\sigma^2)$:
 \begin{equation}
 g(|\rve_i- \xi_i|)=e^{-\gamma |\rve_i- \xi_i|^2}\,.
 \label{Nosanow_Gaussian}
 \end{equation} 
 since the Gaussian is a quantum mechanical solution for a 3D harmonic oscillator and usually is 
 a good guess for describing a wave function close to a potential minimum. The parameter $\gamma$
 is determined through a variational optimization in VMC calculations.

Quantum Mechanics  
 requires that the bosonic wave function is symmetric with respect to  
 an exchange of equivalent particles, and that the Nosanow term does not satisfy this.
  Evidently, the analytical form of the symmetric trial wave function
  in this case becomes 
 quite cumbersome for implementation and extremely expensive in terms of permanent calculation time, 
 as it should in general contain a sum over all the permutations of the lattice nodes in the system. Nonetheless, 
 the results of the quantum Monte Carlo simulations when the Nosanow-Jastrow term is used are
 for practical purposes indistinguishable from that of the  symmetrized trial wave function \cite{Boronat2008,Boronat2009NJP}, as far as the energetic properties
 of a crystal are concerned, and the exchange energy is usually negligible. In this Thesis we will use this form of the NJ trial wave function throughout all the quantum Monte Carlo
  calculations of solid phases. However, the physical quantities, related to Bose statistics, 
  may not be treated with the NJ trial wave function, since any particle exchange is suppressed.

The quantum Monte Carlo technique requires knowledge of 
not only the trial wave function itself, but also of its
 first two derivatives. 
In the case of the DMC method the second derivative enters in the calculation of energy, which is needed 
for the branching term. Hence the second derivative of the trial wave function is required not also for evaluating 
averages of the observables, but even for the 
time evolution of the system. 

The actual implementation can be substantially improved, if one takes into account that the trial
wave function appears in the implementation of the method in three distinct combinations, which 
can be presented as functions and can be tabulated.

A. The logarithm of the Jastrow term (needed in Metropolis algorithm
in VMC simulations and also for estimations of the non-local quantities, for instance
the one-body density matrix)
\begin{eqnarray}
w(r) = \ln f(r)
\label{u2}
\end{eqnarray}

B. The logarithmic derivative of the trial wave function, which is 
required in the calculation of the drift force (\ref{DriftForce}))
\begin{eqnarray}
{\F}(r) = \frac{\di \ln f(r)}{\di r}=\frac{f'(r)}{f(r)}
\label{F}
\end{eqnarray}

C. The second derivative of the Jastrow term appears only in a linear 
combination with the drift force in the calculation of the kinetic energy. The following representation takes place:
\begin{eqnarray}
E^{\rm local}(r) =-\frac{f''(r)}{f(r)}+\left(\frac{f'(r)}{f(r)}\right)^2
+ \frac{mV_{\rm int}(r)}{\hbar^2}-\frac{d-1}{r}\frac{f'(r)}{f(r)},
\label{e}
\end{eqnarray}
with $d$ stands for for a dimensionality of the problem.

\subsection{Explicit expressions for wave functions\label{secwf3D}}
\subsubsection{Trial wave function for hard sphere potential}

The problem of scattering on a simple hard sphere potential, Yukawa potential and 
common potential between Rydberg atoms was discussed in
Sec.~\ref{secHS}. 
The hard sphere and the considered model potential of Rydberg atoms make the wave function vanish when two particles meet each other, which 
means that three-body collisions are greatly suppressed. The same is true for a Yukawa potential at low density due 
to its similarity to the Coulomb potential. 
If three-body correlations 
are neglected, at small interparticle distances the two-body Jastrow term $f_2(r)$ can be conveniently approximated by the solution $f(r)$
of the two-body scattering problem, that is by the wave function of a system of two particles. At large
distances, the pair wave correlation function asymptotically tends to a constant, as the
particles lose correlation.

Taking these facts into account we introduce the trial function in the following
way~\cite{Giorgini99} (here we use a dimensionless notation in which the
distances $r$ are mesured in units of the hard sphere radius $a_{3D}$ and the energy $E$ is measured in units of
$\hbar^2/(ma^2_{3D})$)
\begin{eqnarray}
f_2(r) =
\left\{
{\begin{array}{ll}
\displaystyle \frac{A\sin(\sqrt{2E}(r-1))}{r},& |r| \le \Rm\\
\displaystyle 1- B\exp\left\{-\frac{r}{\alpha}\,\right\},& |r| > \Rm\\
\end{array}}
\right.
\label{f_PP}
\end{eqnarray}

The Jastrow term has to be smooth at the matching point $\Rm$, that is
\begin{itemize}
\item[A.] the function $f_2(r)$ itself must be continuous:
\begin{eqnarray}
\frac{A\sin(\sqrt{2E}(\Rm-1))}{\Rm} = 1 - B
\exp\left\{-\frac{\Rm}{\alpha}\,\right\}
\end{eqnarray}

\item[B.] derivative $f_2'(r)$ must be continuous
\begin{eqnarray}
\frac{A\sqrt{2E}\cos(\sqrt{2E}(\Rm-1))}{\Rm}
-\frac{A\sin(\sqrt{E}(\Rm-1))}{\Rm^2}
=\frac{B}{\alpha}\exp\left\{-\frac{\Rm}{\alpha}\,\right\}
\end{eqnarray}

\item[C.] the local energy $f_2(r)^{-1}(-\hbar^2\Delta_1/2m-\hbar^2\Delta_1/2m+
V_{\rm int}(\rij))f_2(r)$ must be continuous
\begin{eqnarray}
2E =
\frac{\left(\displaystyle\frac{1}{\alpha^2}-\frac{2}{\Rm\alpha}\right)
B \exp\left(-\displaystyle\frac{\Rm}{\alpha}\,\right)
}{1-B \exp\left(-\displaystyle\frac{\Rm}{\alpha}\,\right)}
\end{eqnarray}
\end{itemize}

The solution of this system is
\begin{eqnarray}
\left\{
{\begin{array}{l}
A =\displaystyle\frac{R}{\sin(u(1-1/R))}\frac{\xi^2-2\xi}{\xi^2-2\xi+u^2},\\
B =\displaystyle\frac{u^2 \exp(\xi)}{\xi^2-2\xi+ u^2},
\end{array}}
\right.
\label{f_PP 1}
\end{eqnarray}
where we used the notation $u = \sqrt{2E}R$ and $\xi = R/\alpha$. The value of
$\xi$ is obtained from the equation
\begin{eqnarray}
1-\frac{1}{R} = \frac{1}{u} \arctg\frac{u(\xi-2)}{u^2+\xi-2}
\label{f_PP 2}
\end{eqnarray}

There are three conditions for the determination of five unknown parameters,
consequently two parameters are left free. The usual way to define them is to minimize
the variational energy in variational Monte Carlo which yields an optimized trial
wave function.

\subsubsection{Trial wave function for Yukawa potential\label{secwfSS}}
The construction of a trial wave function for a Yukawa system can be done in different ways.
The first one, widely employed in our calculations, is a use of the solution of a two-body 
scattering problem, that yields a satisfactory short-range approximation for the trial wave function, 
which is valid for a dilute system.
At large distances
 the trial wave function is intended to arrive smoothly to unity at the half size of the simulation box (see Sec.~(\ref{secIntrScat})).
If the Jastrow term is chosen in a form $f(\rve)=e^{-u(\rve)}$~\cite{Reatto67},
this can be achieved by a symmetrization of the trial wave function with respect to the inversion $r\rightarrow (L-r)$ as 
\begin{equation}
u(r):=u_1(r)+u_1(L-r)-u_1(L/2)
\end{equation}
that brings the logarithm of the Jastrow term and its first derivative in the point $L/2$ to zero. 

According to Eq.~(\ref{yukawajas}), the solution of the two-body scattering problems reads as
\begin{eqnarray}
u_1(r)=\ln\left(\frac{I_1({\rm const}\: \sqrt{r})}{\sqrt{r}}\right)=
\ln \sqrt{A}+\frac{A}{2}r-\frac{A^2}{24}r^2+\mathcal{O}(r^3) \label{yukawajas1}
\end{eqnarray}
where $A$ is a constant, subject to optimization. 
This solution formally coincides with 
the scattering solution provided $A=V_0$ ($V_0$ stands for the interaction strength constant 
of the original two-body scattering problem~(\ref{scatuyuk2})). An optimal value of $A$ should therefore be close to $V_0$.

The most productive way to generate trial wave functions in the case of the Yukawa potential appeared to 
be the hypernetted chain method~\cite{Campbell1979}, based on an iterative solution of a set of 
Euler--Lagrange equations 
\begin{equation}
\frac{\delta}{\delta u_n(\rve_1,...,\rve_n)} \frac{\langle\psi|H|\psi\rangle}{\langle\psi|\psi\rangle}=0,\:i=1,..,N.
\end{equation}
The more detailed explanation of the method is given in Section~\ref{secHNC}.

\begin{figure}[htb]
\centering
\includegraphics[width=0.6\textwidth, angle=270]{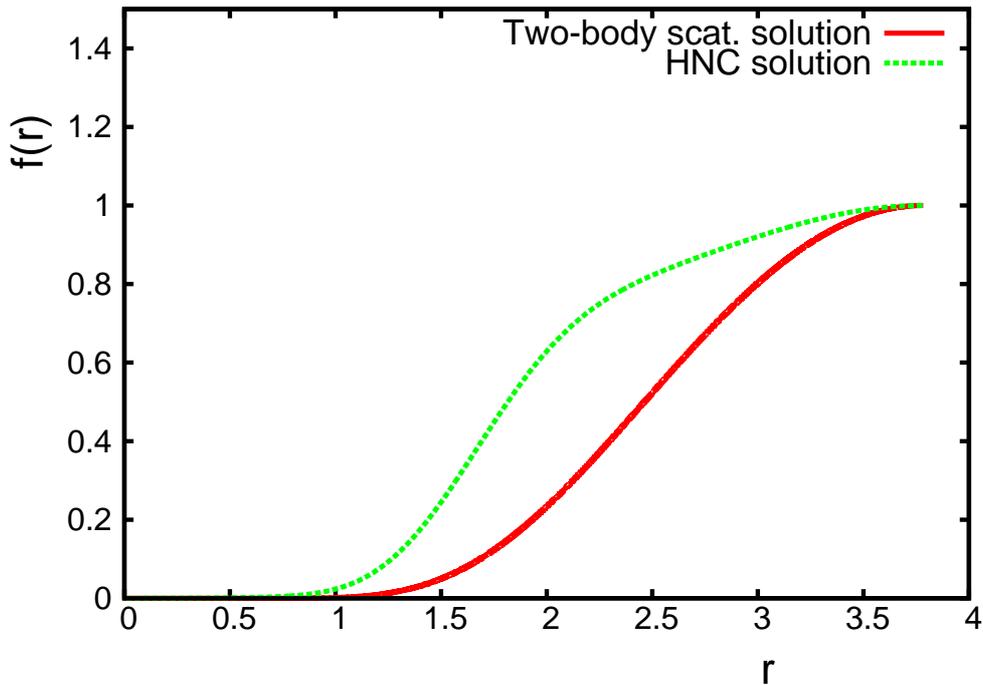}
\caption{Two-body correlation factor for a system, interacting
via a repulsive Yukawa potential in the liquid phase, in the Jastrow parameter-optimized model (red solid line) 
and the HNC one (green dashed line). The particular conditions of the data correspond to 
a DMC calculation with 64 particles in the truncated octahedron cell, $\Lambda=0.46$, $\rho=0.024$. }
\label{yuk_twf}
\end{figure}

A comparison of two-body correlation factors
from Eq. (\ref{yukawajas1}) and the HNC solution is given in Fig.~(\ref{yuk_twf}). 
The particular conditions of the data correspond to 
VMC and DMC calculations with 64 particles in the truncated octahedron cell, $\Lambda=0.46$, $\rho=0.024$ 
(for details on the used model and involved parameters see Section \ref{secYukawa}). 

\begin{table}
\centering

\begin{tabular}{|c|c|c|}
\hline
 & $E/N$  \\
\hline

 VMC Jastrow TWF & 19.862(3)\\
\hline

 VMC HNC TWF & 19.634(3)\\
\hline

 DMC result & 19.595(3)\\

\hline

\end{tabular}    
\caption{The energy per particle without size correction for different t. w. functions compared to the
exact DMC result.}

 \label{yuk_twf_comp}

\end{table}

HNC $f(r)$ leads to better estimates of the energy; a simple comparison 
is shown in Table~(\ref{yuk_twf_comp}) (the conditions of the simulation are the same as in the figures above, no finite size correction added).

\subsubsection{Trial wave function for repulsive van der Waals}
In our simulations of the bulk system with the pairwise van der Waals interaction $1/|\rve|^6$ at zero temperature we use the short-range 
approximation of the two-body scattering problem, as it was solved in Sec.~(\ref{secIntrScat}). The technical procedure to obtain the functional forms of the solution 
follows the derivation, used in the case of the Yukawa potential. The final result for the Jastrow factor reads as a logarithm of the first term in the expansion~(\ref{ryd_fr}):

\begin{eqnarray}
u_1(r)&=&\frac{\sqrt{C_6}}{2r^2}-\ln\frac{\sqrt{2 \pi}}{C_6^{1/4}}-\frac{\ln r}{2}\\
u(r)&=&u_1(r)+u_1(L-r)-2u_1(L/2)\,.	
\end{eqnarray}

\begin{figure}[htb]

\centering
\includegraphics[width=0.6\textwidth, angle=270]{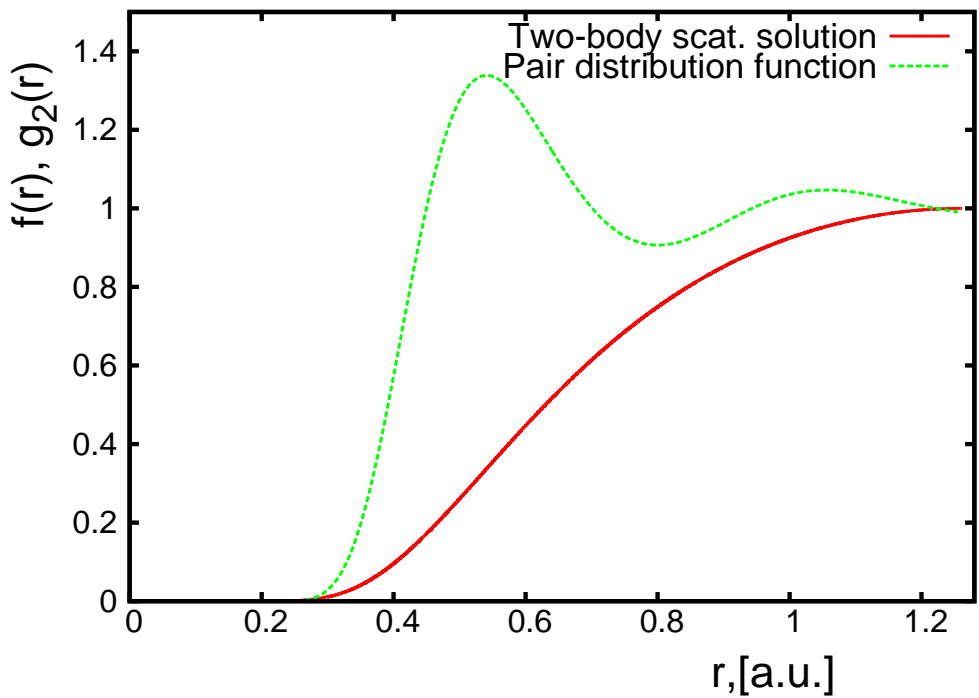}
\caption{A typical two-body correlation factor and pair distribution function for a system, interacting
via a repulsive van der Waals $1/r^6$ potential in the liquid phase. The particular conditions of the data correspond to 
a DMC calculation with 108 particles in the cubic cell, $\rho=6.7$.}
\label{ryd_twf}
\end{figure}

In this equation the factor $C_6$ comes from the interaction strength and is constant. Nonetheless, it can be treated as a parameter and variated in order to optimize the trial wave function by minimizing the VMC energy.
Typical forms of the guiding wave functions and pair distributions are presented in Fig.~\ref{ryd_twf}.

A similar result is provided by the {\it cusp condition} demand, when the leading term 
of the guiding wave function is sought in a suitable exponential form $f(r)=\exp(-a/r^b)$, with 
parameters $a,b$. After a substitution of this functional form into Eq.~(\ref{eqscatryd})) one 
finds that the equation can be satisfied only when $b=2$, while $a$ is still arbitrary. It is 
easily seen, that this procedure yields the leading term of the two-body 
scattering solution.
\subsubsection{Hypernetted chain method\label{secHNC}}
The hypernetted chain method is a technique to solve many-body problems in homogeneous and inhomogeneous media~\cite{HNC},\cite{HNC2}. In this scheme, the static structure factor $S(k)$ that minimizes the variational energy in the subspace of Jastrow wave functions has the form
\begin{equation}
S(k) = { t(k) \over \sqrt{ t^2(k) + 2 t(k) V_{ph}(k)}} \ ,
\label{hncel-b1}
\end{equation}
with $t(k)=\hbar^2k^2/2m$ and $V_{ph}(k)$ the so-called particle-hole interaction. 
Its Fourier transform $FT[V_{ph}(k)] = \tilde V_{ph}(r)$ satisfies the following equation in coordinate space
\begin{equation}
\tilde V_{ph}(r) = g(r) V(r) + {\hbar^2 \over m}\!\mid\!\nabla\!\sqrt{g(r)}\!\mid^2\!
+ (g(r)\!-\!1) \omega_I(r) \ ,
\label{hncel-b2}
\end{equation}
where $V(r)$ and $g(r)$ are the bare two-body potential and the pair distribution function (the Fourier transform of $S(k)$), respectively. Finally, in momentum space the induced interaction $\omega_I(k)$ becomes
\begin{equation}
\omega_I(k) = -{1\over 2}\,t(k) { [ 2 S(k)+1 ][ Sk)-1 ]^2 \over S^2(k) } \ .
\label{hncel-b3}
\end{equation}
In this way, Eqs.~(\ref{hncel-b1}),~(\ref{hncel-b2}) and~(\ref{hncel-b3}) form a set of nonlinear coupled equations that have to be solved   iteratively. The Fourier transform of the resulting $S(k)$ provides $g(r)$ and, in this scheme, the optimal two-body Jastrow factor results from the corresponding HNC/0 equation
\begin{equation}
f_2(r) = \sqrt{g(r)}e^{-N(r)/2}\ ,
\end{equation}
where $N(r)$ is the sum of nodal diagrams, related to $S(k)$ in momentum space by the expression $N(k)=(S(k)-1)^2/S(k)$.


\section{Estimators for physical quantities\label{secQuant}\label{secQuantities}}
\subsection{Local energy\label{secEo}}

The general form of a Hamiltonian of a system of $N$ interacting
bosons in an external field is (\ref{Hgeneral}):
\begin{eqnarray}
\hat H = -\frac{\hbar^2}{2m}\sum\limits_{i=1}^N \Delta_{\ri}
+ \sum\limits_{i=1}^N V_{\rm ext}(\rve) + \sum\limits_{j<k}^N V_{\rm int}(|\rjk|),
\label{Hlocenergy}
\label{H}
\end{eqnarray}
where $m$ is mass of a particle, $V_{\rm ext}(\rve)$ is the external field,
$V_{\rm int}(|\rve|)$ is the two-body particle-particle interaction potential. Given the many-body
wave function $\Psi(\rN)$ the local energy is defined according to (\ref{Eloc}):
\begin{eqnarray}
E^{\rm loc}(\rN) = \frac{\hat H \Psi(\rN)}{\Psi(\rN)}
\end{eqnarray}

The external field and particle-particle interaction are diagonal in
this representation and are calculated trivially as a summation over particles and
pairs of the second and third terms of (\ref{Hlocenergy}). Calculation of the
kinetic energy, first term of (\ref{Hlocenergy}) is more involved, as the Laplacian
operator is not diagonal.

\subsubsection{Local kinetic energy}

In this Section, we will find the expression of the local kinetic energy
\begin{eqnarray}
T^{\rm loc}(\rN) = -\frac{\hbar^2}{2m}\frac{\Delta\Psi(\rN)}{\Psi(\rN)}
\end{eqnarray}

Let us calculate the second derivative in two steps, as the first derivative is
important for the calculation of the drift force. We consider the Jastrow form
(\ref{Jastrow}) of the trial wave function and will express the final results in
terms of one- and two-body terms $f_1$ and $f_2$. The gradient of the
many-body trial wave function is given by
\begin{eqnarray}
{\bf \nabla}_i\Psi(\rN)=\Psi(\rN)\left(\frac{f_1'(\ri)}{f_1(\ri)}\frac{\ri}{r_i}
+\sum\limits_{k\ne i}^N\frac{f_2'(|\rik|)}{f_2(|\rik|)}\frac{\rik}{|\rik|}\right)
\end{eqnarray}

The full expression for the Laplacian is
\begin{eqnarray}
\nonumber
\Delta_{\ri}\Psi(\rN)\qquad=\qquad\Psi(\rN)\left(
\frac{f_1'(\ri)}{f_1(\ri)}\frac{\ri}{r_i}+\sum\limits_{k\ne
i}^N\frac{f_2'(|\rik|)}{f_2(|\rik|)}\frac{\rik}{|\rik|}
\right)^2+\\+\Psi(\rN)\left(
\frac{f_1''(\ri)}{f_1(\ri)}-\left(\frac{f_1'(\ri)}{f_1(\ri)}\right)^2
+\sum\limits_{k\ne i}^N\left[
\frac{f_2''(|\rik|)}{f_2(|\rik|)}-\left(\frac{f_2'(|\rik|)}{f_2(|\rik|)}\right)^2
\right]\right)\,.\nonumber\\
\end{eqnarray}

The kinetic energy can be written in a compact form
\begin{eqnarray}
T^{\rm loc}(\rN) = -\frac{\hbar^2}{2m}\left\{\sum\limits_{i=1}^N\E^{\rm loc}_1(\ri)+
2\sum\limits_{j<k}^N\E^{\rm loc}_2(|\rjk|)-\frac{1}{4}\sum\limits_{i=1}^N |{\F}_i(\rN)|^2\right\}\,,\nonumber\\
\end{eqnarray}
where we introduced notation for the one- and two-body contribution to the local
energy (see, also, (\ref{e}))
\begin{eqnarray}
\E^{\rm loc}_1(\rve)&=& -\frac{f_1''(\rve)}{f_1(\rve)}+\left(\frac{f_1'(\rve)}{f_1(\rve)}\right)^2\\
\E^{\rm loc}_2(r)&=& -\frac{f_2''(r)}{f_2(r)}+\left(\frac{f_2'(r)}{f_2(r)}\right)^2
\end{eqnarray}
and introduced the drift force (see (\ref{F}))
\begin{eqnarray}
\F_i(\rN) =2\left(
\frac{f_1'(\ri)}{f_1(\ri)}\frac{\ri}{r_i}
+\sum\limits_{k\ne
i}^N\frac{f_2'(|\rik|)}{f_2(|\rik|)}\frac{\rik}{|\rik|}\right)
\end{eqnarray}

\subsubsection{Exponentiation}

It is convenient (see Eq.\ref{u2}) to do the exponentiation of the one- and two-
body terms $u_1(\rve) = \ln f_1(\rve)$, $u_2(r) = \ln f_2(r)$. The point is that
numerically a better precision is achieved by working with numbers of the same
order. The formula for the kinetic energy becomes more simple
\begin{eqnarray}
T^{\rm loc}(\rN) = -\frac{\hbar^2}{2m}\left\{
\sum\limits_{i=1}^N u_1''(\ri)+
2\sum\limits_{j<k}^Nu_2''(|\rjk|) +\frac{1}{4}\sum\limits_{i=1}^N |\F_i(\rN)|^2\right\}
\end{eqnarray}
with
\begin{eqnarray}
\F_i(\rN) = 2 u_1'(\ri)\frac{\ri}{r_i} +\sum\limits_{k\ne i}^N u_2'(|\rik|)\frac{\rik}{|\rik|}
\end{eqnarray}

%

\subsection{Static structure factor\label{secSkDMC}}
The static structure factor $S(\k)$ is the correlation 
function of the momentum distribution between
elements $-\k$ and $\k$ (\ref{Skmom}):
\begin{equation}
N S(\k) = \langle\rho_{-\k}\rho_\k\rangle - |\langle\rho_\k\rangle|^2,
\label{Skdef}
\end{equation}

Using the properties of the Fourier component $\rho_{-\k} = (\rho_\k)^*$ it can be
rewritten in a different way
\begin{equation}
N S(\k) = \langle|\rho_\k|^2\rangle - |\langle \rho_\k\rangle|^2\:.
\end{equation}

The density distribution in coordinate space is the sum of $\delta$-functions located 
at the positions of the particles:
%
\begin{equation}
n(\rve) = \frac{1}{V}\sum\limits_{i=1}^N \delta(\rve-\rve_i)\:.
\end{equation}

By means of its momentum space representation (i.e. its Fourier transform)  
\begin{equation}
\rho_\k =
 \sum\limits_{i=1}^N e^{-i\k\rve_i}
=\sum\limits_{i=1}^N \cos\k\rve_i -i \sum\limits_{i=1}^N \sin\k\rve_i
\end{equation}
we obtain a simple expression for the static structure factor
\begin{equation}
N S(\k) =
\left<
\left(\sum\limits_{i=1}^N \cos \k\rve_i\right)^2+
\left(\sum\limits_{i=1}^N \sin \k\rve_i\right)^2
\right>
-\left|\left<\sum\limits_{i=1}^N \cos \k\rve_i\right>\right|^2
-\left|\left<\sum\limits_{i=1}^N \sin \k\rve_i\right>\right|^2\,.
\label{Sk_inhom}
\end{equation}
In a homogeneous system the two last terms in (\ref{Sk}) are vanishing, that is
\begin{equation}
N S(\k) =
\left<
\left(\sum\limits_{i=1}^N \cos \k\rve_i\right)^2+
\left(\sum\limits_{i=1}^N \sin \k\rve_i\right)^2
\right>\,.
\label{Sk}
\end{equation}

If periodic boundary conditions are used, the values
of momenta are quantized and depend on the size of the box
\begin{equation}
k_{n_{x,y,z}} = \frac{2\pi}{L} n_{x,y,z}\:.
\end{equation}

\subsection{Calculation of one-body density matrix}

The one-body density matrix (OBDM) $g_1$ of a homogeneous system described by the
many-body wave function $\psi(\rve_1, ..., \rve_N)$ according to (\ref{g1hom}) is equal
to
\begin{eqnarray}
g_1(|\rve~'-\rve~''|)=N \frac{\int...\int\psi^*(\rve\,', \rve_2, ..., \rve_N)
\psi(\rve\,'', \rve_2, ..., \rve_N)\,d\rve_2 ...d\rve_N}{\int...\int|\psi(\rve_1, ..., \rve_N)|^2\,d\rve_1 ... d\rve_N}\:.
\label{OBDM}
\end{eqnarray}

Since in DMC one does not sample directly the ground-state probability
distribution $\phi_0^2$, but instead the mixed probability $\psi_T\phi_0$ (\ref{f})
one obtains the {\it mixed} one-body density matrix as the output
\begin{eqnarray}
g_1^{mixed}(\rve) =
N \frac{\int...\int \psi^*_T(\rve_1+\rve, \rve_2,..., \rve_N) \phi_0 (\rve_1, \rve_2, ..., \rve_N)\,d\rve_2 ... d\rve_N}
{\int...\int \psi^*_T(\rve_1, ...,\rve_N)\phi_0(\rve_1, ...,\rve_N)\,d\rve_1 ... d\rve_N},
\label{OBDMhom}
\end{eqnarray}

This formula can be rewritten in a way convenient for the Monte Carlo sampling:
\begin{eqnarray}
g_1^{mixed}(r) =
n\frac{\int...\int[\psi^*_T(\rve_1+\rve, \rve_2,..., \rve_N)(\psi^*_T(\rve_1,\rve_2,...,\rve_N))^{-1}]
f(\rve_1, ..., \rve_N)d\rve_1 ... d\rve_N}{\int...\int f(\rve_1, ..., \rve_N)d\rve_1 ... d\rve_N},\nonumber\\
\label{OBDM mixed}
\end{eqnarray}
where we have used the asymptotic formula (\ref{f limit}) and have taken into
account that in a homogeneous system $g_q$ depends only on the module of the
relative distance. If the trial wave function is chosen as a
product of pair functions (\ref{Jastrow}) then using the notation (\ref{u2}) $u(|\rve_i
-\rve_j|) = \ln f_2(|\rve_i -\rve_j|)$) and $f_1 \equiv 1$ one has $\psi_T(\rve_1, ...,\rve_N)
= \prod\limits_{i<j}\exp\{u(|\rve_i-\rve_j|)\}$. Then the ratio of the trial wave function
appearing in (\ref{OBDM mixed}) becomes
\begin{eqnarray}
\frac{\psi_T(\rve_1+\rve, ...,\rve_N)}{\psi_T(\rve_1, ...,\rve_N)}
= \exp\left\{\sum\limits_{j>1} u(|\rve_1+\rve-\rve_j|)-u(|\rve_1-\rve_j|)\right\}
\end{eqnarray}

In order to gain better statistics it is convenient to average over all possible
pairs of particles
\begin{eqnarray}
g_1^{mixed}(r) &=&
\frac{1}{N} \sum\limits_{i=1}^N\frac{\psi_T(\rve_1, ..., \rve_i+\rve, ...,\rve_N)}{\psi_T(\rve_1, ...,\rve_N)}\nonumber\\
&=&\frac{1}{N} \sum\limits_{i=1}^N\exp\left\{\sum\limits_{j \neq i}^N
u(|\rve_i+\rve-\rve_j|)-u(|\rve_i-\rve_j|)\right\}
\end{eqnarray}

The asymptotic limit of the OBDM gives the Bose-Einstein condensate density
\begin{eqnarray}
\lim\limits_{r\to\infty} g_1(r) = \frac{N_0}{V}
\end{eqnarray}
and the condensate fraction is obtained by calculating the asymptotic ratio
\begin{eqnarray}
\lim\limits_{r\to\infty} \frac{g_1(r)}{n} = \frac{N_0}{N}
\end{eqnarray}

\subsection{Two-body density matrix}

The two-body density matrix (TBDM) $\langle\hat\Psi^\dagger(\rve_1')\hat\Psi^\dagger(\rve_2')\hat\Psi(\rve_2)\hat\Psi(\rve_1)\rangle$, depends on 4 vector arguments corresponding to destroying two particles 
at positions $\rve_1$ and $\rve_2$ and inserting them at positions $\rve_1'$ and $\rve_2'$. The diagonal 
element $\rve_1=\rve_1'$ and $\rve_2=\rve_2'$ (see Eq.~(\ref{G2})) is called the pair distribution function. 
Generally, it depends on two vector arguments $\rve_1$ and $\rve_2$. In a translationally invariant system (e.g. in a homogeneous gas)
it is a function of $(\rve_1-\rve_2)$, that is a function of the distance between a pair of particles and two angles. We consider 
spherically averaged pair distribution function
%
\begin{eqnarray}
g_2(|\rve_2-\rve_1|) = \frac{N(N-1)}{n^2}\frac{\int|\psi(\R)|^2\dr_3...\dr_N}{\int|\psi(\R)|^2\,\dR}\:.
\end{eqnarray}

Let us explain now how this formula is implemented in Monte Carlo calculation. We
make summation over all pairs of particles:
\begin{eqnarray}
g_2(r)
=\frac{N(N-1)}{n^2 L^3}
\frac{\int\delta(\rve_1-\rve_2-\rve)|\psi(\R)|^2\,\dR}{\int|\psi(\R)|^2\,\dR}
=\frac{2}{nN}
\frac{\int\sum\limits_{i<j}\delta(r_{ij}-r)|\psi(\R)|^2\,\dR}{\int|\psi(\R)|^2\,\dR}
\end{eqnarray}

If we do a discretization of the coordinate space with spacing $h$ and introduce function
$\vartheta_h(z)$ which is one if $z<h$ and zero otherwise (the distribution is obviously symmetric) one obtains the following 
expressions:
\begin{itemize}
\item In one-dimensional system:
\begin{eqnarray}
g^{1D}_2(z)=\left\langle \frac{2}{2hnN}\sum\limits_{i<j}\vartheta_h(|z_{ij}-z|)|
\right\rangle
\end{eqnarray}

In an uncorrelated system $\vartheta_h(|z|) = 2h/L$ is constant and $g_2(z)=1-1/N$.
The form of the pair distribution function depends on a dimensionality of the problem. 
\item In a three-dimensional system the expression is
\begin{eqnarray}
g^{3D}_2(r)=
\left\langle \frac{2}{4\pi z^2 hnN}\sum\limits_{i<j}\vartheta_h(|r_{ij}-r|)|\right\rangle\,.
\end{eqnarray}
Notice that the distance $r$ enters explicitly in the
expression of the pair distribution function, leading to larger
numerical variance at small distances.
\end{itemize}
\subsection{Pure estimators and extrapolation technique\label{secExtrapolation}}

In a VMC calculation one obtains a {\it variational} esimator for a
given quantity (let it be denoted by an operator $\hat A$), which corresponds to an
average over the trial wave function $\psi_T$:
\begin{eqnarray}
\langle \hat A \rangle_{\rm var} =
\frac{\langle\psi_T|\hat A|\psi_T\rangle}{\langle\psi_T|\psi_T\rangle}
\end{eqnarray}

Instead, the DMC method asymptotically provides a more precise {\it mixed}
estimator given by
\begin{eqnarray}
\langle \hat A \rangle_{\rm mix} =
\frac{\langle\phi_0|\hat A|\psi_T\rangle}{\langle\phi_0|\psi_T\rangle}
\end{eqnarray}

Nonetheless, this type of average can differ from the inbiased (``pure'') ground state average,
which corresponds to the true quantum-mechanical equilibrium value at zero
temperature
\begin{eqnarray}
\langle \hat A \rangle_{\rm pure} = \frac{\langle\phi_0|\hat A|\phi_0\rangle}{\langle\phi_0|\phi_0\rangle}
\end{eqnarray}

The DMC method gives an exact result for the energy, as the mixed average of the
local energy $E_{\rm loc} = \psi_T^{-1}\hat H\psi_T$ coincides with the pure estimator.
This can be easily seen by noticing that when $\langle \phi_0$ acts on $\hat H$, it
gives exactly the ground state energy.

From now on, we will demonstrate that averages of local diagonal 
operators can be calculated in a ``pure'' way.
This means that the pair distribution function, potential energy and static structure
factor can be estimated exactly. Local quantities are diagonal in coordinate space $\langle \R |\hat A
|\R'\rangle = A(\R) \langle \R |\R'\rangle$. The ``pure'' average can be related to
the mixed one in the following way
\begin{eqnarray}
\langle \hat A \rangle_{\rm pure}
=
\frac{\langle\phi_0|A(\R)\frac{\phi_0(\R)}{\psi_T(\R)}|\psi_T\rangle}
{\langle\phi_0|\frac{\phi_0(\R)}{\psi_T(\R)}|\psi_T\rangle}
=
\frac{\langle  A(\R) P(\R) \rangle_{\rm mix}}{\langle P(\R) \rangle_{\rm mix}},
\end{eqnarray}
where $P(\R)$ is defined as
\begin{eqnarray}
P(\R) = \frac{\phi_0(\R)}{\psi_T(\R)} \langle \phi_0|\psi_T\rangle\:.
\label{ndesc}
\end{eqnarray}
The asymptotic number of offsprings of a walker can give $\phi_0(\R)/\psi_T(\R)$ \cite{Liu74}.
By ``tagging'' walkers one can identify, at any time, what is the parent walker. This 
forward walking method \cite{Barnett91,Reynolds86,Runge90,Runge92} permits to sample
pure averages. 

A more simple algorithm was devised by J.~Boronat and J.~Casulleras~\cite{Casulleras95}, 
in which the explicit history should not be recorded and one operates with the 
actual values of an observable. This method is used in our calculations. 

Eq.~(\ref{ndesc}) gives the number of descendants of a walker $\R$ for large times
$\tau\to\infty$. Practically it is enough to wait a sufficiently large, but a finite
time $T$. One makes measurements of a local quantity for all of the walkers, but
calculates the average after the time $T$, so that each walker was replicated
according to the weight $P(\R)$.

An important example of a non-local quantity, for which no ``pure'' is known,  is 
one-body density matrix (see Eq.~(\ref{OBDM mixed})); this quantity deserves a special
attention. We will explain an extrapolation technique, which can be applied for
finding averages of non-local operators. It is also worth noticing that extrapolation
can be used in estimating diagonal quantities, for example, pair correlation function. 

Adopting the notation $\delta \psi$ for the difference between the trial wave function and ground-state wave
function 
\begin{eqnarray}
\phi_0 = \psi_T + \delta \psi\,,
\end{eqnarray}
the ground-state average can be written as
\begin{eqnarray}
\langle\hat A\rangle_{\rm pure} =
\langle\phi_0|\hat A|\phi_0\rangle =
\langle\psi_T|\hat A|\psi_T\rangle + 2\langle\phi_0|\hat A|\delta\psi\rangle
+ \langle\delta\psi|\hat A|\delta\psi\rangle
\end{eqnarray}

If $\delta \psi$ is small the second order term
$\langle\delta\psi|\hat A|\delta\psi\rangle$ can be neglected.
After substitution
$\langle\phi_0|\hat A|\delta\psi\rangle =
\langle\psi_T|\hat A|\phi_0\rangle -
\langle\psi_T|\hat A|\psi_T\rangle$
the extrapolation formula turns into
\begin{eqnarray}
\langle\hat A\rangle_{pure} \approx 2 \langle\hat A\rangle_{\rm mix} - \langle\hat A\rangle_{\rm var}\,.
\label{extrapolation}
\end{eqnarray}

It is possible to write another extrapolation formula of the same order of
accuracy:
\begin{eqnarray}
\langle\hat A\rangle_{\rm pure} \approx \frac{\langle\hat A\rangle^2_{\rm mix}}{\langle\hat
A\rangle_{\rm var}}\,.
\label{extrapolation2}
\end{eqnarray}

An applicability condition for the extrapolation method is that the expressions
(\ref{extrapolation}) and (\ref{extrapolation2}) yield the same final result. Nevertheless, in certain 
situations the use of the second
formula can be advantageous, as it does not change the sign of the function, which is not always true in the case of the first formula, when
the function is very close to 0. This can be useful in an estimation of essentially positive non-local quantities, for instance, the one-body density
matrix.

\chapter{Ewald method for polytropic potentials \label{secEwald}}
\section{Introduction}
The behavior of many-body systems is often governed by the long-range
Coulomb potential between charged particles.  Numerical simulations of such
systems are usually performed by considering a finite number of particles
in a cell with periodic boundary conditions.  The correct estimation of the
potential energy in such systems requires of a summation over all images
created by the periodic boundary conditions. For long-range interaction
potentials such direct summation either converges slowly or it is
conditionally convergent, making its evaluation computationally cumbersome.
Instead, the performance of the calculation can be greatly improved by using
Ewald summation methods~\cite{ewald}. In these methods, the slowly
convergent tail of the sum in the potential energy is represented by a
rapidly convergent sum in momentum space. The method is named after Paul
Peter Ewald who in his pioneering work dated almost a century ago
calculated the electrostatic energy in ionic crystals (a detailed
derivation for the Ewald sums for the Coulomb potential can be found in the
work of  de Leeuw \textit{et al.}~\cite{leeuw}).
An alternative approach to deal with long-range systems
is proposed by Smith~\cite{Smith1994}. In his method, the Hamiltonian and
equations of motion are derived using constraints on the velocities of particles.
Instead, in the following we will stick to a standard model for the Hamiltonian
and will consider ways to improve the convergence in the potential energy.

For a good performance in simulations of large $N$-particle systems, a
number of  modified summation methods has been developed. Historically, the
first efforts to enhance the Ewald method consisted in looking for
appropriate truncation schemes, but all of them were strongly dependent on
the system properties, in particular on the system size. Tabulations of precalculated
terms in both real space and
momentum space sums~\cite{Sangester1976}, as well as polynomial
approximations of the involved
functions~\cite{Lage1947,Brush1966,Hansen1973}, were also proposed to look
for a balance between calculation time and truncation errors. Nevertheless,
these approximate methods suffer from error accumulation in simulations of
large systems, and do not allow for reducing the overall $\mathcal{O}(N^2)$
complexity of the original Ewald summation. The work of Perram \textit{et
al.}~\cite{Perram1988} was the first to give a way to optimize the
splitting of the interparticle potential between the long-range and
short-range parts to yield  a total complexity of $\mathcal{O}(N^{3/2})$.
A special modification of the Ewald method called Wolf
summation~\cite{wolf1999, wolf}, based on a damping of the
Fourier-transformed part of the sum, was posteriorly developed in order to
render the original Ewald summation more efficient for non-periodic systems
and large model sizes. The Ewald technique was also applied to develop the method of
 evaluation of electrostatic potential near the surfaces of ionic crystals~\cite{Parry1975, Parry1976}.

Another way for improving the Ewald method is to perform fast Fourier
transform (FFT) of a reciprocal space sum on a mesh. The oldest algorithm
of this kind is the so-called Particle-Particle Particle-Mesh (P$^3$M)
method, invented and improved to the complexity $\mathcal{O}(N)$ in calculation of forces
 by Hockney and coauthors in the 70's~\cite{Hockney1988, Hockney1973}.
The P$^3$M technique is based on a distribution of
the charge density on a grid using a certain smooth assignment function and
then the discrete Poisson equation is solved using FFT. This algorithm
appeared to be less complex to yield $\mathcal{O}(N \ln N)$ with an
appropriate choice of the free parameters.
The P$^3$M algorithm was recently improved by Ballenegger
\textit{et al.}~\cite{Ballenegger2008} for calculation of energies,
bringing, as claimed, the maximal precision in the energy by an optimization
of the ``influence'' function (a substitution of the potential in the
Fourier-transformed Poisson's equation).
For a comprehensive introduction
to Ewald- and mesh-based techniques  we recommend  to refer to the cited
work of Ballenegger and coauthors  where special attention is paid to
the estimation of both sum truncation-imposed and grid-imposed  errors.
The extension of this method,
called Particle Mesh Ewald~\cite{Darden1993} (PME), makes use of the
analytical form of the sum in the reciprocal space and evaluates potentials
via FFT instead of interpolating them as P$^3$M does. Although PME is
slightly more complex than the P$^3$M algorithm, it is still $\mathcal{O}(N
\ln N)$ and allows to reduce significantly the memory expenses. Later
Particle Mesh Ewald method was reformulated by Essmann \textit{et al.}~\cite{Essmann1995},
making use of cardinal {\it B}-splines to interpolate
structure factors. This approach, called Smooth Particle-Mesh Ewald (SPME)
substantially improved the accuracy of PME with a comparable computational
cost, as it still scales as $\mathcal{O}(N \ln N)$. SPME is also claimed to be
applicable to potentials of the polytropic form $1/|\bs{r}|^k$. In general,
the conventional FFT-based approaches suffer from the severe fallback of
requiring equidistant particle positions. The invention of the variant of
Fourier transform for nonequispaced nodes (NFFT) opened a path to overcome
this shortcoming, while keeping the introduced errors below the specified
target levels. The nonequispaced fast Fourier transform is currently
considered as a promising means to improve the Ewald summation performance,
with open code implementations available~\cite{nfftcode}. The early
variants of the NFFT algorithms are reviewed in the work of
A.~F.~Ware~\cite{Ware1998};  a general approach to the fast summation
methods based on NFFT can be found in the article of
G.~Steidl~\cite{steidl1998}.

The most recent family of algorithms based on the Ewald approach are
 the tree-based algorithms, with the fast multipole method (FMM) being the
 most known and widely used among them.  The algorithm, developed
 primarily by L.~Greengard and V.~Rokhlin~\cite{Greengard1987},
 is based on the idea of
keeping the direct summation of potentials or forces for the nearby atoms
and approximating the interactions of the distant atoms by their multipole
expansions. FMM offers the asymptotically fastest performance among the
Ewald-related algorithms, being linear in $N$ in most cases and not worse
than $\mathcal{O}(N \ln N)$ with explicitly controlled accuracy. The FMM
technique is naturally applicable to inhomogeneous and non-periodic
systems, being also easy to parallelise since it is an entirely real-space
summation.
Since then the algorithm was significantly improved in efficiency,
mostly by introducing new diagonal forms of translation operators~\cite{Greengard1997}.
However, FMM has an intrinsic shortcoming, when applied to molecular dynamics
calculations, as the energy conservation it brings is poor; the method
\textit{per se} is also rather cumbersome in implementation.
Another group of methods, based on the multigrid methods of solving
elliptic (in this particular case -- Poisson's) equations~\cite{Beckers1998},
was developed a decade ago~\cite{Sagui2001}. These methods allow to preserve
the scaling $\mathcal{O}(N)$ and parallelization advantages of
tree-based methods, as well as the applicability in simulations without PBC,
being on the other hand satisfactorily energy-conserving and additionally
accelerated on all length scales. An efficient realization of the multigrid method and its analysis may be found in the work of Sagui \textit{et al.}~\cite{Sagui2004}. An advanced mesh Ewald technique, claimed to reduce significantly the computational costs of charge spreading in multigrid-based methods, was recently proposed by Y.~Shan and coworkers~\cite{Shan2005}.

  A detailed comparison of the optimized $\mathcal{O}(N^{3/2})$
pure Ewald technique, FFT-based summations, and multipole-based methods was
made by H.~G.~Petersen~\cite{Petersen1995} for systems with approximately
uniform charge distributions, taking into account a possible parallel
implementation. According to Petersen, the method of choice with a number
of particles below 10$^4$ is the conventional Ewald summation, PME is
preferable in the range $N\sim 10^4-10^5$, and the fast multipole method
should overperform them with $N >10^5$.
A more recent and ample review of FMM, P$^3$M and pure Ewald methods by
Pollock and Glosli~\cite{Pollock1996a}, based partially on
their own calculations, implies that P$^3$M is faster than the
Ewald summation already for 500 particles, although it is stressed that
the other factors as the ease of the coding, the system geometry, as well as
the code optimization can change the choice. We would also suggest a thorough
survey of different Ewald summation techniques given in the work of Toukmaji
and Board~\cite{Toukmaji1996}.

An approach, alternative to using cubic periodic boundary conditions in a
calculation of long-range interactions, called Isotropic Periodic Sum
(IPS), was recently proposed by Wu and Brooks~\cite{Wu2005}. The main goal
of their approach is to deal with long-range interactions, avoiding
artificial correlations and anisotropy bias
induced by a PBC-based summation in a cubic box.
In this technique, only the
interactions of a particle A with the others within a certain radius
$\bs{r}_c$ are taken into account (as in a plain cut-off scheme), and this
spherical simulation zone is repeated in an infinite number of shifts by
vectors $\bs{r}_{\rm sh}$, such that $|\bs{r}_{\rm sh}|=2N|\bs{r}_c|$.
Therefore, the particle A interacts not only with B (within the sphere
radius), but also with all the images of B, occupying homogeneously the
shells of radii $2N|\bs{r}_c|$, centered in B. The subsequent integration
and summation over the shells allows to obtain explicit expressions of
forces and energies for a number of interactions of most physical interest,
like electrostatic, Lennard--Jones and exponential potentials. The method
is known to yield a performance close to the one shown by the Ewald
summation, but without imposing unwanted symmetry effects.

Since its proposal, the Ewald method has been applied to a large number of
physical problems, although mostly to systems with the Coulomb $1/|\bs{r}|$
interaction potential. In a recent work by R.~E.~Johnson and
S.~Ranganathan~\cite{johnson}, a generalized approach to Ewald summation is
stated to obtain potential energy and forces for systems with a power-law,
Yukawa potential and electronic bilayer systems.
The Ewald method  for
two-dimensional systems with electrostatic interactions was
developed by Parry~\cite{Parry1975}, but his
technique appeared to be computationally inefficient.
Spohr \textit{et al.}.~\cite{Spohr1997}  studied a slab geometry  by treating the
simulation cell as a fully three-dimensional one with the conventional Ewald
summation. Later on, a significant advance was made by Yeh and
Berkowitz~\cite{Yeh1999}, as
the authors managed to obtain the explicit correction term for the
rigorous three-dimensional Ewald summation, that brings the results for
a slab system in a satisfactory agreement with the 2D summation.
The 2D Ewald technique was also applied by Wen Yang \textit{et
al.}~\cite{wenyang}
to calculate the energy of Coulomb particles in a slab system with a uniformly charged
surface. One of the first two-dimensional variants of the Ewald summation was presented in~\cite{Gao1997}, applied to the quasi-2D Stockmayer model with the potential $1/|\bs{r}|^3$. Recent applications to dipolar bosons in a 2D geometry
have been made by  C.~Mora \textit{et al.}~\cite{Mora2007} and Xin Lu
\textit{et al.}~\cite{Xin2008}. On the other hand, the explicit forms of
the Ewald sums for Yukawa interactions have been also reported: in 3D
geometry, with partial periodic boundary conditions~\cite{Salin2000, mazars1}, and in 2D
geometry~\cite{mazars2}. The general approach to the Ewald summation in quasi-two-dimensional systems with power-law potentials and results for several values of power factor are given in~\cite{Mazars2010}.
The Ewald method can also be useful even applied to fast decaying power-law
potentials. For instance, the Ewald formalism was developed in~\cite{Karasawa1989} for the dispersion interaction $1/|\bs{r}|^6$ and later for the Lennard-Jones potential by W.-Z.~Ou-Yang \textit{et al.}~\cite{Ouyang2005}. Also, Shirts \textit{et al.} in their recent work~\cite{Shirts2007}
argue the need for taking into account the effects of cutoffs
in molecular dispersion interactions due to a Lennard-Jones potential,
especially in non-isotropic and inhomogeneous media. The authors developed
two formalisms for the estimation of these cutoff errors in binding free
energy of macromolecular systems, which can in principle be extended to the
other observables. However, it is claimed that the adequate implementation
of the Ewald summation for this kind of systems may render their
corrections unnecessary by mostly eliminating the cutoff-dependent
behavior.

In this Chapter, we report explicit expressions of the Ewald sums for
the general case of particles interacting via a $1/|\bs{r}|^k$ polytropic
potential and in 3D, 2D, and 1D geometries~\cite{EwaldOsychenko}.  The closed derivation of these
sums is given, with special attention being paid to conditionally convergent
potentials.  One of the difficulties of the derivation is that different
terms have to be considered in the cases of short-range, long-range or
``marginal'' potentials. In the case of a short-range interaction, the
original slowly convergent sum is represented as a linear combination of
two rapidly convergent ones. For a long-range interaction, the condition of
charge neutrality in the simulation cell is shown to be necessary to make the energy absolutely convergent within the considered
scheme. The introduction of a uniform neutralizing charged background
(\textit{jellium}), as a particular case of a charge-neutral system, is
also discussed. The explicit forms of the Ewald sums are reported for a
jellium system and for an arbitrary polytropic potential. We explicitly
calculate the expressions for physically relevant interactions as Coulomb,
dipole-dipole, and Lennard-Jones potentials. Finally, we have extended the
Ewald sums to the case of a noncubic simulation cell, that could be
useful in simulations of hexagonal  closed packed (hcp) and two-dimensional
triangular solids. In addition, the general derivation path given in this
work may be used to obtain the forms of Ewald sums for other interaction
potentials.

The computational efficiency is another important issue of the practical
implementation of the method. In fact, one needs to choose correctly a free
parameter, appearing in the integral representation of the sums, and to
decide which number of terms should be kept in spatial and momentum sums in
order to reach the required accuracy. The choice of these three parameters
affects the difference between the calculated result and the exact one as
well as  the calculation complexity. Therefore, a certain optimization of
the parameters is always required. In this Chapter, this optimization
process is formalized and it is shown that following the described
procedure the overall computation time is significantly reduced. The
accuracy of the result is shown to be kept under control, with the only
cost of a preliminary benchmark
calculation.

The rest of the chapter is organized as follows. In
Section~\ref{sec:solforarb}, we formulate the problem, develop the general
Ewald approach and report explicit expressions for the Ewald sums for a
polytropic potential in a three-dimensional cubic simulation cell.
Sections~\ref{sec:ewald2d} and~\ref{sec:ewald1d} contain derivations of the
Ewald sums in two-dimensional and one-dimensional geometries, respectively.
In Section~\ref{sec:ewaldbox}, the case of a simulation cell with different
side lengths is considered for three- and two-dimensional systems. The
final general expressions and their particularization to the  most
physically relevant  cases are presented in Section~\ref{sec:results}. The
practical algorithm for the parameter optimization and an actual
application of the Ewald method is discussed in
Section~\ref{sec:practical}. Summary and conclusions  are drawn in
Section~\ref{sec:conclusions}.

\section{Ewald method in 3D geometry}
\label{sec:solforarb}

\subsection {Basic assumptions and initial sums}
We consider a system of $N$ particles inside a cubic simulation cell of
size $L$ with periodic boundary conditions. Thus,
each particle with coordinates $\bs{r}$ in the initial cell has an
infinite number of images $\bs{r}+\bs{n}L$ in the adjacent cells. The
total potential energy is estimated by
\begin{equation}
U=\frac{1}{2}{\sum_{\bs{n} \in \mz^3}} ^{\prime}
\left[\sum_{i=1}^{N}\sum_{j=1}^{N}\phi(\bs{r}_{ij}+\bs{n}L)\right]
\label{initialhamiltonian1}
\end{equation}
where $\phi(\bs{r})$ is the interparticle potential, $\bs{r}_{ij} \equiv \bs{r}_i -
\bs{r}_j$, and the prime in the first sum means that the summation over
an integer vector $\bs{n}$ must be done omitting the term $\bs{n}=\bs{0}$ when $i=j$.

\subsection{Analytic derivation}
In many physical situations, the interaction potential between two particles
$i$ and $j$ has  the power-law form $q_i q_j/|\bs{r}|^k$ with positive $k$
and $q_i$, $q_j$ being the generalized charges of the particles. This sort of
interaction is generally referred to as \textit{polytropic} potential.

First, let us consider the case of short-range potentials, $k \leq 3$. As we
will see later, the potentials corresponding to $k>3$ give a similar
result. For $k \leq 3$, the right-hand part of
Eq.~(\ref{initialhamiltonian1}) diverges and it can be made
convergent only if the restriction of charge neutrality is required, i.e., when
$\sum_{i=1}^{N}q_i=0$.
It has also been shown~\cite{Fraser1996} that for a pure electrostatic
interaction  the total energy (\ref{initialhamiltonian1}) can be
conditionally convergent even in  a neutral simulation cell because of a
higher multipole contribution. The energy and forces  are therefore
dependent on the order of summation, which can also be implicitly set by a
choice  of a convergence factor. The ambiguity usually appears in a form of
a constant or a position-dependent term, vanishing in the limit
$L\rightarrow \infty$. Hence, the preference in one or another factor should be
dictated either by physical properties of a particular system or by arguments
regarding rates of convergence to the thermodynamic limit. For a general
discussion on the convergence issues appearing in periodic boundary
conditions, see Ref.~\cite{Makov1995}.
The main idea of the Ewald summation technique in the approach proposed by
de Leeuw, Perram, and Smith~\cite{leeuw} is to multiply each component of
the sum by the dimensionless factor $e^{-s n^2}$, with $s>0$ being a
dimensionless regularizing parameter, making the sum absolutely convergent.
Then, the limit $s \rightarrow 0$ is taken, so that the singularity in the
initial sum~(\ref{initialhamiltonian1}) can be explicitly separated into a
term depending only on $s$, that finally can be  cancelled due to the
charge neutrality condition. We take a similar multiplier
$c(\bs{n},\bs{r},s)=e^{-s |\bs{n}+\bs{r}|^2}$ yielding the same rate of
convergence (since $0\leq r\leq 1$ in units of $L$). As the sum, multiplied
by $c$, is invariant to an arbitrary substitution
$\bs{r}\rightarrow\bs{n}+\bs{r}$, the chosen convergence factor allows to
preserve the periodicity of the potential in order to avoid any possible
artefacts in the final results.

For the sake of clearness of the derivation, it is convenient to use
reduced length units, that is to use the size of the box $L$ as unity of
length and substitute $r_{ij}$ by $r_{ij}L$. From now on, and for simplicity,
we use the notation $\bs{r}$ for
$\bs{r}_{ij}$ and, in case of possible ambiguity, we will stick to the
standard notation $\bs{r}_{ij}$.
Also, we rewrite the potential energy by splitting the total sum
(\ref{initialhamiltonian1}) into two terms:
$I_{01}$ (the sum of the interactions between a particle with
all the \textit{other} particles in the box), and $I_{00}$ (the
sum of the interaction of a particle with \textit{its own} images, comprised of the components $i=j$ in Eq. (\ref{initialhamiltonian1})).
Explicitly,
\begin{equation}
U = \frac{1}{L^k}(I_{01} + I_{00}) \ ,
\label{ep1}
\end{equation}
with
\begin{eqnarray}
I_{01} & =& \sum_{\bs{n} \in \mz^3} \left[\sum_{1\leq i<j\leq N}
\frac{q_i q_j e^{-s |\bs{r}_{ij}+\bs{n}|^2}}{|\bs{r}_{ij}+\bs{n}|^k}\right] \label{i01init}\\
I_{00} &= & \frac{1}{2}\sum_{\bs{n} \in \mz^3\backslash \bs{0}}
\frac{e^{-s n^2}}{n^k}\sum_{i=1}^{N}q_i^2 \  ,
\label{i00init}
\end{eqnarray}
where the shorthand notation $n=|\bs{n}|$ is used.

First, let us focus on the $I_{01}$ term, which we rewrite as
\begin{eqnarray}
I_{01}=\sum_{1\leq i<j\leq N}q_i q_j \psi(\bs{r},s) \ ,
\label{tildei10polytropic}
\end{eqnarray}
where we have defined the ``screened'' interaction potential
$\psi(\bs{r},s)=\sum_{\bs{n}}e^{-s |\bs{r}+\bs{n}|^2}/|\bs{r}+\bs{n}|^k$,
extended from a single cell to the whole coordinate space.
Since the total potential energy consists of a sum of pair interaction
components, we may consider a single pair without any loss of generality.

Let us apply the equation
\begin{equation}
x^{-2s}=\frac{1}{\Gamma(s)}\int_0^{\infty}t^{s-1}e^{-tx^2}\di t \ ,
\label{repx1}
\end{equation}
representing the definition of the gamma-function, to the polytropic potential $|\bs{r}+\bs{n}|^{-k}$.
Then the function $\psi$ may be represented in an integral form,
\begin{equation}
\psi(\bs{r},s)=\frac{1}{\Gamma(k/2)}\int_0^{\infty}t^{\frac{k}{2}-1}
\sum_{\bs{n}}e^{-t|\bs{r}+\bs{n}|^2}e^{-s |\bs{r}+\bs{n}|^2}\di t \ .
\label{boro1}
\end{equation}

We expect that the integral (\ref{boro1}) contains a singularity
that will be located in the vicinity of zero. Therefore,
we split this integral into two domains $[0,\alpha^2]$ and
$[\alpha^2,\infty)$,
the corresponding integrals being denoted as $\psi_{{\rm fin}}$ and
$\psi_{{\rm inf}}$, where $\alpha$ is some arbitrary positive
constant,
\begin{equation}
   \psi(\bs{r},s)  = \psi_{{\rm fin}}(\bs{r},s) + \psi_{{\rm inf}}(\bs{r},s) \ .
\label{boro2}
\end{equation}
In the following, we analyze the two terms of the previous sum
(\ref{boro2}).

\begin{enumerate}
\item
The explicit analytical form of the term $\psi_{{\rm inf}}(\bs{r},s)$ can
be found
\begin{equation}
 \psi_{{\rm inf}}(\bs{r},s) =\frac{1}{\Gamma (\frac{k}{2})}
 \sum_{\bs{n}}\int_{\alpha^2}^{\infty}t^{\frac{k}{2}-1}e^{-t|\bs{r}+\bs{n}|^2-s|\bs{r}+\bs{n}|^2}\di t=
\sum_{\bs{n}}\frac{e^{-s |\bs{r}+\bs{n}|^2 }}{|\bs{r}+\bs{n}|^k}
\frac{\Gamma (\frac{k}{2},\alpha ^2|\bs{r}+\bs{n}|^2 )}{\Gamma
(\frac{k}{2})}  \ ,
\end{equation}
where $\Gamma(a,z)$ is the incomplete gamma function. From the large
distance asymptotic expansion of this function, one obtains that the above lattice
sum is absolutely and uniformly convergent if $s \geq 0$ and $\alpha>0$.
Therefore, one may simply take the limit of vanishing
screening $s \rightarrow 0$,
\begin{equation}
\psi_{{\rm inf}}(\bs{r},s) \lims \frac{1}{\Gamma (\frac{k}{2})}\sum_{\bs{n}}
\frac{\Gamma (\frac{k}{2},\alpha ^2|\bs{r}+\bs{n}|^2 )}{|\bs{r}+\bs{n}|^k}
\ .
\label{i01infpolytropic}
\end{equation}

\item
The calculation of  $\psi_{{\rm fin}}(\bs{r},s)$ is done
by making a separate analysis of the $\bs{n}=\bs{0}$ case,
\begin{equation}
\psi_{{\rm fin}}(\bs{r},s) = \psi_{{\rm fin}}^{\bs{m}\neq \bs{0}}(\bs{r},s)
+\psi_{{\rm fin}}^{\bs{m}=\bs{0}}(\bs{r},s) \ . \label{psifinsplit}
\end{equation}
Explicitly,
\begin{eqnarray}
 \psi_{{\rm fin}}^{\bs{m}\neq \bs{0}}(\bs{r},s)  & = &\frac{\pi^{\frac{3}{2}}}{\Gamma(\frac{k}{2})}
\sum_{\bs{m}\neq \bs{0}} \int_0^{\alpha^2}\frac{t^{\frac{k}{2}-1}}{(t+s)^{\frac{3}{2}}}
\exp\left[\frac{-\pi^2m^2}{t+s}+2\pi \ii
\bs{m}\bs{r}\right]\di t \label{psifinnneq0}\\
\psi_{{\rm fin}}^{\bs{m}=\bs{0}}(\bs{r},s)& = &\frac{\pi^{\frac{3}{2}}}{\Gamma(\frac{k}{2})}
\int_0^{\alpha^2}\frac{t^{\frac{k}{2}-1}}{(t+s)^{\frac{3}{2}}}
\di t \ ,
\label{i11tildawodiv2}
\end{eqnarray}
where we have used the Jacobi transformation~\cite{Jacobibook,Whittaker1999}
\begin{equation}
\sum_{\bs{n}} e^{-s|\bs{n}+\bs{r}|^2}=\left(\frac{\pi}{s}\right)^{3/2}
\sum_{\bs{m}}\exp[-\pi^2 m^2/s +2\pi \ii \bs{m}\bs{r}]\;\;{\rm for} \;\;
\bs{n},\bs{m}\in \mz^3  \ ,
\label{3djacobi}
\end{equation}
applied to
\begin{equation}
\exp[-s |\bs{n}+\bs{r}|^2-t|\bs{n}+\bs{r}|^2]=
\exp[-(s+t)|\bs{n}+\bs{r}|^2].
\label{transf1}
\end{equation}

We evaluate the integral $\psi_{{\rm fin}}^{\bs{m}\neq \bs{0}}(\bs{r},s)$
by the following analysis. Consider separately the following factor of the integrated expression from (\ref{psifinnneq0})
\begin{equation}
M=\frac{\exp\left[-\frac{\pi^2n^2}{t+s}\right]}{(t+s)^{\frac{3}{2}}} .
\end{equation}
It is clearly continuous and bounded on $(0,\:+\infty)$ as a function of $(t+s)$,
also notice that $t^{k/2-1}$ is absolutely integrable on $(0,\:\alpha^2)$ for $k>0$.
In accordance with the standard convergence test for improper integrals,
the integral $\psi_{{\rm fin}}^{\bs{m}\neq \bs{0}}(\bs{r},s)$ converges absolutely and uniformly
with $s$ being considered as a parameter. Then, the limit $s\rightarrow 0$ may be carried out
and the integral becomes
\begin{eqnarray}
\psi_{{\rm fin}}^{\bs{m}\neq \bs{0}}(\bs{r},s)=&&\frac{\pi^{\frac{3}{2}}}{\Gamma(\frac{k}{2})}\sum_{\bs{m}\neq \bs{0}}e^{2\pi \ii \bs{m}\bs{r}} \int_{0}^{\alpha^2}t^{\frac{k-5}{2}} \exp\left[-\frac{\pi^2m^2}{t}\right] \di t \label{i11gc1} \\
=&&\sum_{\bs{m}\neq \bs{0}}\frac{\pi^{\frac{3}{2}}\cos(2\pi \bs{m}\bs{r})}{\Gamma(\frac{k}{2})}
\alpha^{k-3} E_{\frac{k-1}{2}}\left(\frac{\pi ^2m^2}{\alpha ^2}\right) \ . \label{i01nnot0fin}
\end{eqnarray}
The function $E_{n}(z)$ is the exponential integral function, and we have cancelled the imaginary
part of the sum (\ref{i11gc1}) by grouping the pairs with $\bs{n}$ and $-\bs{n}$.

Now, we analyze the second term of $\psi_{{\rm fin}}(\bs{r},s) $,
\begin{equation}
 \psi_{{\rm fin}}^{\bs{m} = \bs{0}}(\bs{r},s)  = \frac{\pi^{\frac{3}{2}}}
 {\Gamma(\frac{k}{2})} \int_0^{\alpha^2}\frac{t^{\frac{k}{2}-1}}{(t+s)^{\frac{3}{2}}}
 \di t  \ .
\end{equation}
In terms of a new variable $v=s/(t+s)$,
\begin{equation}
 \psi_{\rm fin}^{\bs{m} = \bs{0}}(\bs{r},s)  = \frac{\pi^{\frac{3}{2}}}
 {\Gamma(\frac{k}{2})} \int_{s/(\alpha^2+s)}^1\frac{(1-v)^{\frac{k}{2}-1}}
 {v^{\frac{k-1}{2}}} s^{(k-3)/2}\di v \ .
 \label{i11n0withexp}
\end{equation}

The integration of $\psi_{{\rm fin}}^{\bs{m} = \bs{0}}(\bs{r},s)$
for a $1/|r|^k$ interaction has to be carefully analyzed as a function of
$k$: $1\leq k<3$, long-range potential; $k=3$, marginal case; and $k>3$,
short-range potential.
\begin{enumerate}
%
%
\item
Suppose $1\leq k<3$. The resulting integral,
\begin{equation}
\psi_{{\rm fin}}^{\bs{m} = \bs{0}}(\bs{r},s) =
 \psi_{\rm fin}^{\bs{m} = \bs{0}}(\bs{r},s)  = \frac{\pi^{\frac{3}{2}}}
 {\Gamma(\frac{k}{2})} \int_{s/(\alpha^2+s)}^1\frac{(1-v)^{\frac{k}{2}-1}}
 {v^{\frac{k-1}{2}}} s^{(k-3)/2}\di v
\label{i11gc2}
\end{equation}
may be given explicitly in terms of incomplete beta- and incomplete
gamma-functions. Expanding the resulting function for small $s$,
\begin{equation}
\psi_{{\rm fin}}^{\bs{m} = \bs{0}}(\bs{r},s) =s^{\frac{k-3}{2}} 2 \pi
\Gamma\left[\frac{3-k}{2}\right]+\frac{2 \pi ^{\frac{3}{2}} \alpha ^{k-3}}{(k-3)
\Gamma\left[\frac{k}{2}\right]} + \mathcal{O}(s)
\label{i01klt3}
\end{equation}
It is easily seen, that the only divergent term in the
expansion~(\ref{i01klt3}) is the first one, which we define as
\begin{equation}
S(s)=s^{\frac{k-3}{2}} 2 \pi  \Gamma\left[\frac{3-k}{2}\right]\ .
\label{Ssklt3}
\end{equation}

We remind that the choice of a convergence factor (that explicitly affects the
summation order) may in principle lead to additional contributions in the total
energy if the convergence of the sum is conditional (like for a
charge-neutral cell of Coulomb particles with non-zero total dipole
moment). In the original derivation of de Leeuw \textit{et al.}~\cite{leeuw}, the
factor $\exp(-sn^2)$ results in an additional dipole-like
component in $\psi_{\rm fin}^{\bs{m} = \bs{0}}$, which breaks the
periodicity of the potential and therefore complicates its use
in simulations with periodic boundary conditions. Moreover, this procedure~\cite{leeuw}
yields a nonvanishing dipole term exclusively for $k=1$ in 3D geometry,
with the rest of the sums remaining unchanged. From our point of view, this
discontinuity points out to an nonphysical character of the dipole term
appearing in the case of the Coulomb potential. Nevertheless, in a number
of studies~\cite{Fraser1996, Makov1995} it is considered as a first order
correction when the convergence to the thermodynamic limit is analyzed. The
mere fact that the results for the two different convergence multipliers
coincide when $k>1$ is a consequence of the absolute convergence of the
higher multipole contributions in this case.

\item Suppose $k=3$. In this marginal case, the expression~(\ref{i11gc2})
may be integrated directly to yield the following logarithmic dependence
\begin{equation}
\psi_{{\rm fin}}^{\bs{m} = \bs{0}}(\bs{r},s)  = \frac{\pi^{\frac{3}{2}}}{\Gamma(\frac{3}{2})}\left(\frac{-2\alpha s-2\alpha^3}{(\alpha^2 +s)^{\frac{3}{2}}}+\ln(s+2\alpha^2+2\alpha\sqrt{s+\alpha^2})-\ln s\right)
\end{equation}
that close to $s=0$ expands as
\begin{equation}
 \psi_{{\rm fin}}^{\bs{m} = \bs{0}}(\bs{r},s) =-2\pi\ln\,s-4\pi+4\pi \ln (2\alpha) + \mathcal{O}(s \ln s)\label{i01keq3}
\end{equation}
with the diverging term
\begin{equation}
S(s)=-2\pi\ln\,s\ . \label{Ssk3}
\end{equation}
%
%
\item
Consider the remaining option $k>3$. In this case,
$(1-v)^{\frac{k}{2}-1}$ is bounded from above  and $(k-1)/2>1$. It means
that the integral converges absolutely and the only finite contribution to the
integral comes from the first (constant) term of the integral expansion
for small $s$,
\begin{equation}
\psi_{{\rm fin}}^{\bs{m} = \bs{0}}(\bs{r},s) = \frac{2 \pi ^{\frac{3}{2}}
\alpha ^{k-3}}{(k-3) \Gamma\left[\frac{k}{2}\right]}
\label{i01kgt3}
\end{equation}

\end{enumerate}
\end{enumerate}

The second term of the total potential energy, $I_{00}$ (\ref{ep1}) can be
derived in a similar form to the first one.
The procedure to find the form of $\psi(\bs{r},s)$ is repeated here with
$\bs{r}_{ij}=0$, hence the results are obtained straightforwardly via
(\ref{i01infpolytropic}), (\ref{i01nnot0fin}), (\ref{i01keq3}) and
(\ref{i01kgt3}),
\begin{eqnarray}
I_{00}=&&\sum_{i=1}^{N}q_i^2\left[\frac{1}{\Gamma (\frac{k}{2})}\sum_{\bs{n}}\frac{\Gamma
(\frac{k}{2},\alpha ^2 n^2 )}{n^k}+\sum_{\bs{m}\neq \bs{0}}\frac{\pi^{\frac{3}{2}}}
{\Gamma(\frac{k}{2})}
\alpha^{k-3} E_{\frac{k-1}{2}}\left(\frac{\pi ^2 m^2}{\alpha ^2}\right)\right. \nonumber \\
&&\left.-\frac{\alpha^k}{\Gamma(\frac{k}{2}+1)}+\psi_{{\rm fin}}^{\bs{m} = \bs{0}}(\bs{r},s)
\right] \ ,
\label{i00poly}
\end{eqnarray}
with the term $\psi_{{\rm fin}}^{\bs{m} = \bs{0}}(\bs{r},s)$ depending on the
potential parameter $k$ via (\ref{i01klt3}), (\ref{i01keq3}) or
(\ref{i01kgt3}).

Putting all together, the potential energy can be written in a more compact form as,
\begin{eqnarray}
U=&&\frac{1}{L^k}(I_{01}+I_{00})\nonumber\\
    =&&\frac{1}{L^k}\sum_{i<j}q_i q_j \psi(r_{ij}/L)+\frac{1}{2L^k}
    \sum_{i=1}^{N} q_i^2 \, \xi+
\frac{1}{L^k}\sum_{i<j}q_i q_j S(s)+\frac{1}{2L^k}\sum_{i}q_i^2 S(s) \ ,
\end{eqnarray}
with the generalized potential,
\begin{equation}
\psi(\bs{r})= \sum_{\bs{n}}R(\bs{n},\bs{r})+\sum_{\bs{m}\neq
0}K(\bs{m},\bs{r})+C_1 \ .
\label{psipoly1}
\end{equation}
A constant shift in the definition of
$\psi$ is introduced to satisfy by the property $\int_{\rm cell}\psi \di \bs{r}=0$,
convenient for a proper treatment of  the background contributions (see
Appendix A). The functions entering in Eq. (\ref{psipoly1}) are defined as
\begin{eqnarray}
R(\bs{n},\bs{r})&=&  \frac{\Gamma (\frac{k}{2},\alpha ^2 |\bs{r} + \bs{n}|^2 )}
{\Gamma (\frac{k}{2})|\bs{r} + \bs{n}|^k}  \label{defr}\\
K(\bs{m},\bs{r})&=&\kappa(\bs{m}) \, \cos(2\pi \bs{m}\bs{r}) \ , \label{defk}\\
\end{eqnarray}
with
\begin{equation}
\kappa(\bs{m})=\frac{\pi^{\frac{3}{2}}\alpha^{k-3}}{\Gamma(\frac{k}{2})}
 E_{\frac{k-1}{2}}\left(\frac{\pi ^2m^2}{\alpha ^2}\right) \ .
 \label{defkappa}
\end{equation}
The explicit form of the function $S(s)$ depends on the $k$ value,
\begin{equation}
S(s)=\left\{
\begin{array}{ll}
s^{\frac{k-3}{2}} 2 \pi  \Gamma\left[\frac{3-k}{2}\right]\, &\textrm{if $k \leq 3$ (singular term)}
\label{singterm}\\
-2\pi \ln s\, &\textrm{if $k=3$ (singular term)} \\
0\, &\textrm{if $k>3$}\end{array}\right. \ ,
\end{equation}
and the term $\xi$ depends only on the choice of $\alpha$,
\begin{equation}
\xi = \sum_{\bs{n}\neq
0}\rho(\bs{n})+\sum_{\bs{m}\neq
0}\kappa(\bs{m})+C_1+C_2 \ ,
\label{xipoly1}
\end{equation}
with
\begin{equation}
\rho(\bs{n})=\frac{\Gamma (\frac{k}{2},\alpha ^2n^2 )}{\Gamma
(\frac{k}{2})n^k} \ ,
\label{defrho}
\end{equation}
and $\kappa(\bs{n})$ defined in Eq. (\ref{defkappa}). The constants $C_1$
and $C_2$ are explicitly,
\begin{eqnarray}
C_{1}&=&\left\{
\begin{array}{ll}
\frac{2 \pi ^{\frac{3}{2}} \alpha ^{k-3}}{(k-3) \Gamma\left[\frac{k}{2}\right]} & \textrm{if $k \neq 3$}\\
-4\pi+4\pi\ln(2\alpha) & \textrm{if $k=3$}\\
\end{array}\right. \label{c1} \\
C_{2}&=&-\frac{\alpha^k}{\Gamma(\frac{k}{2}+1)}\label{c2}\\
\end{eqnarray}

\subsection{Removing singularities for $k\leq 3$}
\label{sec:remsing}

The diverging part $U_s$ (containing a singularity) of the total potential
energy equals to
\begin{equation}
U_s=\frac{1}{L^k}\sum_{i<j}q_i q_j S(s)+\frac{1}{2L^k}\sum_{i}q_i^2 S(s)
=\frac{1}{2L}\left(\sum_{i}q_i\right)^2 S(s)
\label{epotsingular}
\end{equation}
and vanishes, if the charge neutrality condition $\sum_{i}q_i=0$ is taken.

Consider now a charge-neutral system with a neutralizing background consisting
of a large number of identical uniformly distributed
particles of the opposite charge (the ``jellium'' model). We denote the numbers of negatively
charged particles $q_{-}$ and positively charged (background) particles $q_{+}$
as $N_{-}$ and $N_{+}$, respectively. By imposing charge neutrality,
$q_{+}=-[N_{-}/N_{+}]q_{-}$, with $N$ the total number of particles, $N=N_{-}+N_{+}$.

The potential energy for the jellium model can be written as
\begin{equation}
U = \frac{1}{L^k}\sum_{i<j}q_i q_j
\psi(r_{ij}/L)+\frac{N_{-}q_{-}^2+N_{+}q_{+}^2}{2L^k}\xi \ .
\label{ep2}
\end{equation}

The second term in Eq. (\ref{ep2}) has a component proportional to $N_{+}q_{+}^2$.
Note that the negative charges $q_{-}$ and their number $N_{-}$ is defined by the problem and therefore fixed. Hence, in the limit $N_{+} \rightarrow \infty$, this term cancels
$N_{+}q_{+}^2 = (N_{-}^2 q_{-}^2)/N_{+} \rightarrow 0 $, and therefore this background contribution
may be eliminated to yield
\begin{equation}
\frac{N_{-}q_{-}^2+N_{+}q_{+}^2}{2L^k}\xi = \frac{N_{-}q_{-}^2}{2L^k} \xi \ .
\end{equation}
Concerning the first term of Eq.~(\ref{ep2}), let us split it into three
pieces,
\begin{equation}
\frac{1}{L} \sum_{1\leq i<j\leq N}q_i q_j \psi(\bs{r}) =
\frac{1}{L}(S_{--}+2 S_{-+}+S_{++}) \ ,
\label{psibythrees}
\end{equation}
where the first sum corresponds to the interaction between the negative charges
\begin{equation}
S_{--}=\sum_{1\leq i<j\leq N_{-}}q_i q_j \psi(\bs{r}) \ ,
\label{psisum1}
\end{equation}
the second sum is the interaction of the negatively charged particles with the positive charges of the background
\begin{equation}
S_{-+}=\sum_{i=1}^{N_{-}}\sum_{j=1+N_{-}}^{N_{+}+N_{-}}q_i q_j
\psi(\bs{r}) \ ,
\label{psisum2}
\end{equation}
and the third one is the interaction between the background charges
\begin{equation}
S_{++}=\sum_{1+N_{-}\leq i<j\leq N_{+}+N_{-}}q_i q_j \psi(\bs{r}) \ .
\label{psisum3}
\end{equation}
The last two terms $S_{-+}$ and $S_{++}$ are easily shown to be zero in the limit
$N_+\rightarrow \infty$ as a consequence of the zero value of the integral of
$\psi$ over the simulation cell (see Appendix A).

With the above considerations we can finally write the expression for the
potential energy within the jellium model as
\begin{equation}
U^{{\rm jel}} = \frac{q_{-}^2}{L^k}\sum_{i<j}\psi(r_{ij}/L)+\frac{Nq_{-}^2}{2L^k}\xi
\label{epgenfin0}
\end{equation}

In the more general case of different charges in a charge-neutral
simulation cell (with a long-range potential) or a system with an arbitrary
short-range potential the potential energy is given by
\begin{equation}
U^{{\rm gen}} = \frac{1}{L^k}\sum_{i<j}q_i
q_j\psi(r_{ij}/L)+\frac{\sum_{i=1}^{N}q_i^2}{2L^k}\xi \ .
\label{epgenfin}
\end{equation}

A certain analytical conversion of the sum in the reciprocal space is
also possible in order to sum it up faster. Expanding the sum that defines
$K(\bs{m},\bs{r})$ (\ref{defk}), one can simplify it in the following way,
\begin{eqnarray}
\lefteqn{\sum_{i<j}q_i q_j \sum_{\bs{m}\neq 0}K(\bs{m},\bs{r})=\sum_{\bs{m}\neq 0}\kappa(\bs{m})
\sum_{i<j}q_i q_j \cos(2\pi \bs{m}\bs{r})} \nonumber\\
&&=\frac{1}{2}\sum_{\bs{m}\neq 0}\kappa(\bs{m})\sum_{i,j}q_i q_j\left[\cos(2\pi \bs{m}\bs{r}_i)
\cos(2\pi \bs{m}\bs{r}_j)+\sin(2\pi \bs{m}\bs{r}_i)\sin(2\pi
\bs{m}\bs{r}_j)\right] \nonumber \\
&&-\frac{1}{2}\sum_{i}q_i^2\sum_{\bs{m}\neq
0}\kappa(\bs{m}) \nonumber \\
&&=\frac{1}{2}\sum_{\bs{m}\neq 0}\kappa(\bs{m})
\left| \sum_{j}q_j\exp(2\pi \ii \bs{m}\bs{r}_j)\right|^2-\frac{1}{2}\sum_{i}q_i^2\sum_{\bs{m}\neq 0}\kappa(\bs{m})
\label{recipexpansion}
\end{eqnarray}
In this form, the sum over all pairs of particles in the reciprocal space
is represented as a single sum over particles and thus it scales as
$\mathcal{O}(N)$ instead of $\mathcal{O}(N^2)$.
Notice that the number of prefactors $\kappa(\bs{m})$ and exponents in the
sum depends on a chosen cutoff, which in general also might depend on $N$,
making the overall complexity of the $k$-space grow.
Na\"{\i}ve schemes with $\alpha$ and the cutoff not depending on $N$ do not
take into account the interplay between the $r$-space and $k$-space sum
complexities, thus leaving at least $\mathcal{O}(N^2)$ in one of them.
Nevertheless, as we show later, optimization with $\alpha$ and cutoff
depending on $N$ gives a best total complexity of $\mathcal{O}(N^{3/2})$.
An alternative method to sum up the momentum space part is to use Fast
Fourier transform-based techniques (like PME), which is fast as
$\mathcal{O}(N \ln N)$.

The last term in Eq. (\ref{recipexpansion}) cancels  the $\kappa(\bs{m})$
component of $\xi$. Introduce the notation,
\begin{eqnarray}
\tilde{\psi}(\bs{r})&&=\sum_{\bs{n}}R(\bs{n},\bs{r})+C_1\label{tildepsi}\\
\tilde{\xi}&&=\sum_{\bs{n}\neq \bs{0}}\rho(\bs{n})+C_1+C_2\label{tildexi}\\
\tilde{S}_{\rm equal}(\bs{m})&&=q_{-}\sum_{j}\exp(2\pi \ii \bs{m}\bs{r}_j/L)\label{sq}\\
\tilde{S_q}(\bs{m})&&=\sum_{j}q_j\exp(2\pi \ii \bs{m}\bs{r}_j/L) \ ,
\label{se}
\end{eqnarray}
where $\tilde{S}_{\rm equal}$ is used when the system of equally charged particles $q_{-}$ is considered.
 Within this notation the potential energy may be rewritten in the following forms, which are more efficient for
numerical implementation,
\begin{eqnarray}
U^{{\rm jel}} &&= \frac{q_{-}^2}{L^k}\sum_{i<j}\tilde{\psi}(r_{ij}/L)+
\frac{1}{2L^k}\sum_{\bs{m}\neq 0}\kappa(\bs{m})|\tilde{S}_{\rm equal}(\bs{m})|^2+\frac{Nq_{-}^2}{2L^k}\tilde{\xi}\label{epgenfin0eff}\\
U^{{\rm gen}} &&= \frac{1}{L^k}\sum_{i<j}q_i q_j\tilde{\psi}(r_{ij}/L)+
\frac{1}{2L^k}\sum_{\bs{m}\neq
0}\kappa(\bs{m})|\tilde{S_q}(\bs{m})|^2+\frac{\sum_{i=1}^{N}q_i^2}{2L^k}\tilde{\xi} \ ,
\label{epgenfineff}
\end{eqnarray}
with $\bs{r}_i,\:r_{ij}$ in the original length units.

\subsection{Short-range potentials and the marginal case}
In case of a short-range interaction ($k>3$), the potential
energy does not diverge, which is clear from the form of the singular term
$S(s)$(see Eq. \ref{singterm}).
Hence, there is no need to add a neutralizing
background and, even more, the background must be necessarily excluded since it
leads to a divergence in the energy. This is easily seen by considering the
potential energy of the background separately
\begin{equation}
U_{bg}=C \int_{0}^{\rm cell}\frac{\di \bs{r}}{|\bs{r}|^k} \ ,
\end{equation}
that contains a singularity in zero.
The expression for the potential energy is simply equal to
Eq.~(\ref{epgenfin}),
\begin{equation}
U = \frac{1}{L^k}\sum_{i<j}q_i q_j
\psi(r_{ij}/L)+\frac{\sum_{i=1}^{N}q_i^2}{2L^k}\xi \ .
\label{ep22}
\end{equation}

When $k=3$ (marginal case),  both ultraviolet (coming from short-range contributions) and infrared (coming from long-range contributions) divergences arise in zero for the
background as well as in the vicinity of infinity (the logarithmic
divergence in the energy of negative charges). The only coherent model
here is a plain ``quasi-neutral'' gas consisting of a mixture of a finite number of
charges per box with the constraint $\sum q_i = 0$, i.e.,  with the positive
background excluded.

\section{Ewald method in 2D geometry}
\label{sec:ewald2d}

\subsection{General notes for lower dimensions}
The Ewald sums can be extended to two-dimensional (2D) systems interacting
through polytropic potentials.
The difference with the 3D case comes from a different form of the Jacobi imaginary
transformation for the Jacobi $\theta$-functions [its 3D form is given in
Eq.~(\ref{3djacobi})].

The ``third'' Jacobi $\theta$-function $\theta_3(z,\tau)$ is defined as
\begin{equation}
\theta_3(z|\tau)=\sum_{m=-\infty}^{+\infty}e^{\ii\pi \tau m^2} e^{2 m \ii
z} \ ,
\label{theta3}
\end{equation}
and satisfies the Jacobi imaginary transformation,
\begin{equation}
\theta_3(z|\tau)=(-\ii \tau)^{-1/2}e^{\ii \tau'2
z^2/\pi} \, \theta_3(z\tau'|\tau') \ ,
\end{equation}
with $\tau'=-1/\tau$. Under the change of variables, $z=\pi r$ and
$\tau=\ii \pi/s$, the $\theta$-function becomes a Gaussian,
which is the relevant function for performing the Ewald sums,
\begin{equation}
\sum_{n=-\infty}^{+\infty}e^{-s(r+n)^2}=(\pi/s)^{1/2}\sum_{m=-\infty}^{+\infty}
e^{-\pi^2 m^2/s}e^{2\pi \ii m r} \ .
\label{1djacobi}
\end{equation}
This expression will be used later, in the derivation of the Ewald sum in
one-dimensional systems.  Equation (\ref{1djacobi}) may be easily generalized to the
2D geometry,
\begin{equation}
\sum_{\bs{n}}e^{-s|\bs{r}+\bs{n}|^2} = (\pi/s)\sum_{\bs{m}}
e^{-\pi^2 m^2/s}e^{2\pi \ii \bs{m}\bs{r}} \ .
\label{2djacobi}
\end{equation}
Comparing this result for 2D with its 1D (\ref{1djacobi}) and
3D(\ref{3djacobi})  counterparts  one finds that the
dimensionality $D$ affects only the constant multiplier as $(\pi/s)^{D/2}$.

\subsection{Analytic derivation}
The analytical derivation of the Ewald sum in 2D proceeds similarly to
the one already presented for 3D. Equations from  (\ref{ep1}) to (\ref{psifinsplit})
are also valid here because their derivation is done without explicit
reference to the dimensionality of the problem. In particular,
the integral $\psi_{\rm inf}(\bs{r},s)$ converges absolutely and to the same value
\begin{equation}
\psi_{{\rm inf}}(\bs{r},s) \lims \frac{1}{\Gamma (\frac{k}{2})}\sum_{\bs{n}}
\frac{\Gamma (\frac{k}{2},\alpha ^2|\bs{r}+\bs{n}|^2 )}{|\bs{r}+\bs{n}|^k}
\ .
\label{i01infpolytropic2d}
\end{equation}
We make the same decomposition of the integral $\psi_{{\rm
fin}}(\bs{r},s)$ as in 3D,
\begin{equation}
\psi_{{\rm fin}}(\bs{r},s) = \psi_{{\rm fin}}^{\bs{m}\neq \bs{0}}(\bs{r},s)
+\psi_{{\rm fin}}^{\bs{m}=\bs{0}}(\bs{r},s) \ ,
\end{equation}
with
\begin{eqnarray}
 \psi_{{\rm fin}}^{\bs{m}\neq \bs{0}}(\bs{r},s)  & = &\frac{\pi}{\Gamma(\frac{k}{2})}
\sum_{\bs{m}\neq \bs{0}} \int_0^{\alpha^2}\frac{t^{\frac{k}{2}-1}}{(t+s)}
\exp\left[\frac{-\pi^2m^2}{t+s}+2\pi \ii
\bs{m}\bs{r}\right]\di t \label{psifinnneq02d}\\
\psi_{{\rm fin}}^{\bs{m}=\bs{0}}(\bs{r},s)& = &\frac{\pi}{\Gamma(\frac{k}{2})}
\int_0^{\alpha^2}\frac{t^{\frac{k}{2}-1}}{(t+s)}
\di t \ ,
\label{i11tildawodiv22d}
\end{eqnarray}
where the two-dimensional variant of the Jacobi
transformation~(\ref{2djacobi}) is used. The
difference between the pair of equations
(\ref{psifinnneq02d}, \ref{i11tildawodiv22d}) and their
three-dimensional
analogues~(\ref{psifinnneq0}, \ref{i11tildawodiv2}) relies in a
substitution of the 3D factor $(\pi/(t+s))^{3/2}$  by  the 2D one $\pi/(t+s)$.

First, we consider the term $\psi_{{\rm fin}}^{\bs{m}\neq \bs{0}}(\bs{r},s)$.
Following the same analysis as for its 3D counterpart, it can be
shown that this parametric integral also converges absolutely. It yields
\begin{eqnarray}
\psi_{{\rm fin}}^{\bs{m}\neq \bs{0}}(\bs{r},s)=&&\frac{\pi}{\Gamma(\frac{k}{2})}
\sum_{\bs{m}\neq \bs{0}}e^{2\pi \ii \bs{m}\bs{r}} \int_{0}^{\alpha^2}t^{\frac{k}{2}-2}
\exp\left[-\frac{\pi^2m^2}{t}\right] \di t \nonumber \\
=&&\sum_{\bs{m}\neq \bs{0}}\frac{\pi\cos(2\pi \bs{m}\bs{r})}{\Gamma(\frac{k}{2})}
\alpha^{k-2} E_{\frac{k}{2}}\left(\frac{\pi ^2m^2}{\alpha ^2}\right) \ .
\label{i01nnot0fin2d}
\end{eqnarray}

The modification of the integral $\psi_{{\rm fin}}^{\bs{m}=\bs{0}}$ is
less straightforward, since it requires specific integrations and
expansions in the series for small $s$. Namely, we have to evaluate the
integral
\begin{equation}
\psi_{{\rm fin}}^{\bs{m}=\bs{0}} = \frac{\pi}{\Gamma(\frac{k}{2})} \int_{s/(\alpha^2+s)}^{1}
\frac{(1-v)^{\frac{k}{2}-1}}{v^{\frac{k}{2}}}s^{k/2-1}\di v
\label{i11n0withexp2D}
\end{equation}
which is the 2D equivalent of Eq.~(\ref{i11n0withexp}).

In the following, we consider separately the cases of long-range potential ($1\leq k<2$),
marginal interaction ($k=2$) and short-range potential ($k>2$).
\begin{enumerate}
\item
$1\leq k<2$. As in 3D, the integral can be found analytically via the
incomplete beta- and incomplete gamma-function with known series expansions
for small $s$. Omitting these unnecessary intermediate expressions, we
give the final expansion for $\psi_{{\rm fin}}^{\bs{m}=\bs{0}}$,
\begin{equation}
\psi_{{\rm fin}}^{\bs{m}=\bs{0}} =s^{\frac{k-2}{2}} \frac{\pi^2}
{\sin\left(\frac{k\pi}{2}\right) \Gamma\left(\frac{k}{2}\right)}+\frac{2\pi \alpha ^{k-2}}
{(k-2) \Gamma\left[\frac{k}{2}\right]}+ \mathcal{O}(s^{k/2})  \ .
\label{i11k122D}
\end{equation}
The first term of the expansion,
\begin{equation}
S(s)=s^{\frac{k-2}{2}}\frac{\pi^2}{\sin\left(\frac{k\pi}{2}\right)
\Gamma\left(\frac{k}{2}\right)} \ ,
\end{equation}
clearly diverges when $s\rightarrow 0$. Similarly to the 3D case, this term is cancelled
in a charge-neutral cell and  hence,
\begin{equation}
\psi_{{\rm fin}}^{\bs{m}=\bs{0}} =\frac{2\pi \alpha ^{k-2}}{(k-2)
\Gamma\left[\frac{k}{2}\right]}
\label{i11k122Dfin}
\end{equation}

\item
$k=2$. The integration of Eq.~(\ref{i11n0withexp2D}) is performed to yield in the
limit $s\rightarrow 0$ a marginal logarithmic dependence,
\begin{equation}
\psi_{{\rm fin}}^{\bs{m}=\bs{0}}=-\pi \ln s + 2\pi \ln \alpha +
\mathcal{O}(s \ln s) \ .
\label{i11k22D}
\end{equation}
As for the 3D geometry, the jellium model is inapplicable in this
particular case since the energy of the continuous background diverges.
Nonetheless the diverging component
\begin{equation}
S(s)=-\pi \ln s
\end{equation}
can be removed if we consider a charge-neutral system with a finite number
of charges. In this case,
\begin{equation}
\psi_{{\rm fin}}^{\bs{m}=\bs{0}}=2\pi \ln \alpha   \ .
\label{i11k22Dfin}
\end{equation}

\item
$k>2$. The integral (\ref{i11n0withexp2D}) can be evaluated by taking $s=0$, since
its convergence is absolute,
\begin{equation}
\psi_{{\rm fin}}^{\bs{m}=\bs{0}} \lims \frac{2 \pi \alpha ^{k-2}}{(k-2) \Gamma\left[\frac{k}{2}\right]}\label{i11km22Dfin}
\end{equation}
\end{enumerate}

The second potential energy component, $I_{00}$ (\ref{ep1}), is calculated
as in the 3D case. The result for 2D is
\begin{eqnarray}
I_{00}(s) & = & \sum_{i=1}^N q_i^2 (\psi_{{\rm fin}}^{\bs{m}\neq\bs{0}}(\bs{0},s)+
\psi_{\rm inf}(\bs{0},s)-\psi_{{\rm inf}}^{\bs{n}=\bs{0}}(\bs{0},s))
\\
& = & \sum_{i=1}^N q_i^2\left[\sum_{\bs{n}}\frac{\Gamma (\frac{k}{2},\alpha ^2n^2 )}
{\Gamma (\frac{k}{2})n^k}+\sum_{\bs{m}\neq \bs{0}}\frac{\pi \alpha^{k-2}}{\Gamma(\frac{k}{2})}
E_{\frac{k}{2}}\left(\frac{\pi ^2m^2}{\alpha ^2}\right)-
\frac{\alpha^k}{\Gamma(\frac{k}{2}+1)} +\psi_{{\rm fin}}^{\bs{m}=\bs{0}} \right] \ . \nonumber
\label{xi2dfin0}
\end{eqnarray}

\subsection{Final expressions}

With respect to the 3D case, the changes in the 2D Ewald sum
appear in those terms in which the Jacobi transformation is used, that is
in $\kappa(\bs{n})$ and $C_1$,
\begin{eqnarray}
\kappa(\bs{m})& = &\frac{\pi \alpha^{k-2}}{\Gamma(\frac{k}{2})} E_{\frac{k}{2}}
\left(\frac{\pi ^2m^2}{\alpha ^2}\right)  \label{defkappa2d} \\
C_1 & = & \psi_{{\rm fin}}^{\bs{m}=\bs{0}}
\end{eqnarray}
The other terms, namely
$R(\bs{r},\bs{n})$, $\rho(\bs{n})$ and $C_2$, are not affected by
dimensionality and may be taken directly from the previous section.

Within the jellium model for a long-range potential ($k<2$), the Ewald sum
is given by
\begin{equation}
U^{{\rm jel}} =
\frac{q_{-}^2}{L^k}\sum_{i<j}\psi(r_{ij}/L)+\frac{Nq_{-}^2}{2L^k}\xi \ .
\label{epgen2dfin}
\end{equation}
A more general form, applicable to any system with a short-range potential
($k>2$), a charge-neutral system with long-range interaction ($k<2$), or a marginal
($k=2$) potential is expressed as
\begin{equation}
U^{{\rm gen}} = \frac{1}{L^k}\sum_{i<j}q_i q_j\psi(r_{ij}/L)+\frac{\xi}{2L^k}
\sum_{i=1}^N q_i^2 \ .
\label{epgen2dfin2}
\end{equation}

In the same way as for the 3D systems we can modify the sum in the
reciprocal space, and with the same notations
(\ref{tildepsi})~--~(\ref{se}) ($\rho$, $R$ and the constants $C_1$, $C_2$
are the new ones, corresponding to 2D case) the potential energy may be
given by
\begin{eqnarray}
U^{{\rm jel}} &&= \frac{q_{-}^2}{L^k}\sum_{i<j}\tilde{\psi}(r_{ij}/L)+\frac{1}{2L^k}
\sum_{\bs{m}\neq 0}\kappa(\bs{m})|\tilde{S}_{\rm equal}|^2+\frac{Nq_{-}^2}{2L^k}\tilde{\xi}\label{epgen2dfineff}\\
U^{{\rm gen}} &&= \frac{1}{L^k}\sum_{i<j}q_i q_j\tilde{\psi}(r_{ij}/L)+\frac{1}{2L^k}
\sum_{\bs{m}\neq 0}\kappa(\bs{m})|\tilde{S_q}|^2+\frac{\sum q_i^2}{2L^k}\tilde{\xi} \
.
\label{epgen2dfin2eff}
\end{eqnarray}

\section{Ewald method in 1D geometry}
\label{sec:ewald1d}

As it has been commented before for the 2D case, the differences due to
dimensionality are caused by the form of the Jacobi imaginary
transformation. In the derivation for 1D, one needs the following ones
\begin{eqnarray}
x^{-2s} &=&\frac{1}{\Gamma(s)}\int_0^{\infty}t^{s-1}e^{-tx^2}\di t \label{repx1_1d}\\
\sum_{n=-\infty}^{+\infty}e^{-sn^2}&=&(\pi/s)^{1/2}\sum_{m=-\infty}^{+\infty}e^{-\pi^2 m^2/s}
\label{1djacobi2_n}\\
\sum_{n=-\infty}^{+\infty}e^{-s(r+n)^2}&=&(\pi/s)^{1/2}\sum_{m=-\infty}^{+\infty}
e^{-\pi^2 m^2/s}e^{2\pi \ii m r}  \ .
\label{1djacobi2_rn}
\end{eqnarray}
Similarly to what discussed in the previous section, the only
terms to be changed are those where the Jacobi transformation is used, namely
$\psi_{\rm fin}^{m\neq 0}$ (in $I_{01}$ in a radial-dependent form,
in $I_{00}$ for $r=0$). The difference arises from a different power
exponent ($1/2$) in (\ref{1djacobi2_n}) and (\ref{1djacobi2_rn}), that is in
(\ref{i01nnot0fin}) $k$ has to be substituted by $k+2$ (and $\pi^{3/2}$ --
by $\pi^{1/2}$, respectively), yielding
\begin{equation}
\psi_{\rm fin}^{m\neq 0}=\sum_{n\neq 0}\frac{\pi^{1/2} e^{2\pi \ii mr}}{\Gamma(\frac{k}{2})}
\alpha^{k-1} E_{\frac{k+1}{2}}\left(\frac{\pi ^2 m^2}{\alpha ^2}\right) \ .
\label{i11not01D}
\end{equation}

As far as the term $\psi_{\rm fin}^{m=0}$ is concerned, we should perform a
simple integration and do a series expansion for small $s$,
\begin{equation}
\psi_{\rm fin}^{m=0} = \frac{\pi^{1/2}}{\Gamma(\frac{k}{2})}
\int_{s/(\alpha^2+s)}^{1}\frac{(1-v)^{\frac{k-1}{2}}}{v^{\frac{k+1}{2}}}\di v
\label{i11n0withexp1D}
\end{equation}
The estimation of this integral depends on the $k$ value. In the following,
we detail this analysis.

\begin{enumerate}
\item
$k = 1$, the marginal case,
\begin{equation}
\psi_{\rm fin}^{m=0} =  \frac{\pi^{1/2}}{\Gamma(\frac{k}{2})} (-\ln s -2+2 \ln (2\alpha))+
\mathcal{O}(s) \ .
\label{i11n0k11D}
\end{equation}
As before, we keep only the constant term, considering the diverging term
absent due to the charge neutrality condition. Therefore, with
$\Gamma(1/2)=\sqrt{\pi}$ one has
\begin{equation}
\psi_{\rm fin}^{m=0} =  -2+2 \ln (2\alpha)
\label{i11n0k11Dfin}
\end{equation}
\item
$k>1$, the short-range potential,
\begin{equation}
\psi_{\rm fin}^{m=0} =  \frac{\pi^{1/2}}{\Gamma(\frac{k}{2})} \cdot \frac{2\alpha^{k-1}}{k-1}+
\mathcal{O}(s)+\mathcal{O}(s^{(k-1)/2} \ln s) \ .
\label{i11n0km11D}
\end{equation}
In the limit $s\rightarrow 0$, it yields
\begin{equation}
\psi_{\rm fin}^{m=0} =\frac{2\pi^{1/2}\alpha^{k-1}}{(k-1)\Gamma(\frac{k}{2})}
\label{i11n0km11Dfin}
\end{equation}
resembling the 3D result (\ref{i01kgt3}), with the change $k\rightarrow
k+2$ (except in the $\Gamma$ term) and $\pi^{3/2}\rightarrow \pi^{1/2}$.
\end{enumerate}

The final result for the one-dimensional Ewald summation reads
\begin{eqnarray}
\psi(\bs{r}) & =& \sum_{n}R(n,r)+\sum_{m\neq 0}K(m,r))+C_1\\
\xi & = & \sum_{n\neq 0}\rho(n)+\sum_{m\neq 0}\kappa(m)+C_1+C_2 \ ,
\end{eqnarray}
where $C_1=\psi_{fin}^{m=0}$ is taken from the
expressions~(\ref{i11n0k11Dfin}) (if $k=1$) or (\ref{i11n0km11Dfin}) (if
$k>1$).

For $k=1$, the only consistent system is the charge-neutral one with a finite number
of particles. In this case and for a short-range potential ($k>1$) one the
potential energy is given by
\begin{equation}
U^{{\rm gen}} = \frac{1}{L}\sum_{i<j}q_i
q_j\psi(r_{ij}/L)+\frac{\sum_{i=1}^{N} q_i^2}{2L^k}\xi \ .
\label{epgen1dfin}
\end{equation}

Although the Ewald method is applicable to one-dimensional
problems, there is a direct way to calculate the sums for polytropic
potentials
\begin{equation}
U = \frac{1}{L^k}\sum_{n=-\infty}^{n=+\infty}\frac{1}{|r+n|^k} \ .
\label{ep1ddirect}
\end{equation}
For $k>1$, this sum can be represented as
a linear combination of the Hurwitz zeta functions,
\begin{equation}
\frac{1}{L^k}\sum_{n=-\infty}^{+\infty}\frac{1}{|r+n|^k}=\frac{1}{L^k}(H_k(r)+H_k(1-r))
\ .
\label{ep1ddirect2}
\end{equation}
In particular, for $k=2$ the sum converts into a familiar expression used
in the Calogero-Sutherland model~\cite{Sutherland1971,Astra2006},
\begin{equation}
\frac{1}{L^2}\sum_{n=-\infty}^{+\infty}\frac{1}{|r+n|^2}=\frac{\pi^2}{L^2 \sin^2(\pi r) } \ .
\label{ep1ddirect3}
\end{equation}
Notice that the sum (\ref{ep1ddirect2}) may be expressed in terms of
trigonometric functions only for even values of $k$ via $(k-2)$ times
differentiation of Eq. (\ref{ep1ddirect3}). Anyway, the possibility to find
exact expressions for infinite sums in 1D suggests that the use of the
Ewald method might not be needed, but we keep it as a possibly useful mathematical
relation and for completeness.

\section{Generalizations to non-cubic simulation cells}
\label{sec:ewaldbox}

\subsection{3D case}

A special and interesting situation arises if we consider a simulation cell in a more
general way, as a rectangular box with different side lengths
($L_x,\:L_y,\:L_z$ in the corresponding dimensions). The need to deal with
a box of unequal size lengths may occur in the simulation of a solid with
a noncubic lattice (the simplest examples include a hexagonal closed packed
crystal in 3D geometry), since the lattice vectors $\bs{n}$ in the sum over
images on (\ref{initialhamiltonian1}) are no longer orthogonal.
Focusing our analysis to a 3D geometry, the potential energy is now given
by
\begin{equation}
U=\frac{1}{2}{\sum_{\bs{n_a} \in \mz^3}} ^{\prime} \left[\sum_{i=1}^{N}
\sum_{j=1}^{N}\phi(\bs{r}_{ij}+L_0\bs{n}_r)\right] \ ,
\label{initialhamiltonian1nbox}
\end{equation}
with $\bs{n}_r=(\bs{n}_xL_x+\bs{n}_yL_y+\bs{n}_zL_z)/L_0$,
$\bs{n}_{x,y,z}$ being integer vectors along the corresponding
axis $x,\:y,\:z$. We have introduced the geometric average
$L_0=(L_xL_yL_z)^{1/3}$ and we will use reduced $L_0$ units for $r_{ij}$, and hence
$r_{ij}$ will be adimensional.   Repeating the standard procedure,
we multiply the potential energy by a Gaussian term
$\exp(-s|\bs{n}_r+\bs{r}|^2)$ and, at the end, we take the limit $s\rightarrow 0$,
separating the converging part, if present. We group separately the
interaction with images of other particles $I_{01}$ and the interaction of
a particle with its own images $I_{00}$,
\begin{equation}
U = \frac{1}{L_0^k}(I_{01} + I_{00}) \ ,
\label{ep1nbox}
\end{equation}
where
\begin{eqnarray}
I_{01} & = & \sum_{\bs{n} \in \mz^3} \left[\sum_{1\leq i<j\leq N}
\frac{q_i q_j e^{-s |\bs{n}_r+\bs{r}_{ij}|^2}}{|\bs{r}_{ij}+\bs{n}_r|^k}\right]
\label{i01initnbox}\\
I_{00} & =& \frac{1}{2}\sum_{\bs{n} \in \mz^3\backslash \bs{0}}
\frac{e^{-s |\bs{n}_r^2|}}{|\bs{n}_r|^k}\sum_{i=1}^{N}q_i^2 \ .
\label{i00initnbox}
\end{eqnarray}
Comparing the relations (\ref{ep1nbox})~--~(\ref{i00initnbox}) to the cubic
case (\ref{ep1})~--~(\ref{i00init}), one notices that these relations
remain  unchanged if $\bs{n}$ is formally substituted by $\bs{n}_r$, and
the constant coefficient $1/L^k$ is replaced by $1/L_0^k$. Therefore, all
the results found without the Jacobi transformation (\ref{3djacobi}) remain the
same with $\bs{n}_r$ instead of $\bs{n}$. In particular,
Eq.~(\ref{i01infpolytropic}) transforms into the following
\begin{equation}
\psi_{{\rm inf}}= \frac{1}{\Gamma (\frac{k}{2})}\sum_{\bs{n}}
\frac{\Gamma (\frac{k}{2},\alpha ^2|\bs{r}+\bs{n}_r|^2
)}{|\bs{r}+\bs{n}_r|^k} \ .
\label{i01infpolytropicnbox}
\end{equation}
The Jacobi transformation (\ref{3djacobi}) in a noncubic box has the following form
\begin{eqnarray}
\sum_{\bs{n}_r} e^{-s|\bs{n}_r+\bs{r}|^2}=&&\prod_{i=x,y,z}\sum_{n_i} e^{-s(n_iL_i/L_0+r_i)^2}
\nonumber\\
=&& \left[ \prod_{i=x,y,z}\left(\frac{\pi}{s(L_i/L_0)^2}\right)^{1/2}
\right] \prod_{i=x,y,z}\sum_{m_i}\exp\left(-\frac{\pi^2m_i^2}{s(L_i/L_0)^2}\right)
\exp(2\pi \ii m_i r_iL_0/L_i )\nonumber\\
=&&(\pi/s)^{3/2}\sum_{\bs{m}_k}\exp(-\pi^2 |\bs{m}_k|^2/s) \exp(2\pi \ii
\bs{m}_k \bs{r} ) \ ,
\label{3djacobinbox}
\end{eqnarray}
with $\bs{m}_k=\bs{m}_x L_0/L_x+\bs{m}_y L_0/L_y+\bs{m}_z L_0/L_z$
the normalized displacement vector in momentum space.  The last equation is
obtained from the original expression (\ref{3djacobi}) by a formal
substitution of the vector $\bs{m}$ by $\bs{m}_k$.

In order to calculate $\psi_{\rm fin}$ we first modify Eq.~(\ref{transf1}),
\begin{equation}
\exp[-s|\bs{n}_r+\bs{r}|^2-t|\bs{n}_r+\bs{r}|^2]=\exp\left[-(s+t)|\bs{n}_r+\bs{r}|^2\right]
\ ,
\label{transf1nbox}
\end{equation}
then insert it into the relation~(\ref{3djacobinbox}), and finally separate
the summand $\bs{n}=\bs{0}$,
\begin{eqnarray}
\psi_{\rm fin} &= & \frac{\pi^{\frac{3}{2}}}{\Gamma(\frac{k}{2})}\sum_{\bs{m}_k\neq \bs{0}}
\int_0^{\alpha^2}\frac{t^{\frac{k}{2}-1}}{(t+s)^{\frac{3}{2}}}
\exp\left[\frac{-\pi^2\bs{m}_k^2}{t+s}+2\pi \ii \bs{m}_k\bs{r}\right]\di t+
\frac{\pi^{\frac{3}{2}}}{\Gamma(\frac{k}{2})}\int_0^{\alpha^2}\frac{t^{\frac{k}{2}-1}}
{(t+s)^{\frac{3}{2}}} \di t
 \nonumber\\
&= & \psi_{\rm fin}^{\bs{m}\neq \bs{0}}+\psi_{\rm fin}^{\bs{m}=\bs{0}} \ .
\label{i11tildawodiv2nbox}
\end{eqnarray}
The subsequent derivation follows exactly the
derivation for a cubic box, with the change of $\bs{n}$ by
$\bs{n}_r$  and  $\bs{m}$ by $\bs{m}_k$ for  sums
in the real and momentum spaces, respectively. The final
result for a 3D system in a noncubic box can be summarized as follows
\begin{eqnarray}
\psi(\bs{r})=&&\sum_{\bs{n}_r}\frac{\Gamma(k/2,\alpha^2|\bs{n}_r+\bs{r}|^2)}
{\Gamma(k/2)|\bs{n}_r+\bs{r}|^k}
+\sum_{\bs{m}_k\neq \bs{0}}\frac{\pi^{\frac{3}{2}}\alpha^{k-3}\cos(2\pi
\bs{m}_k \bs{r})}{\Gamma(k/2)}E_{\frac{k-1}{2}}\left(\frac{\pi^2|\bs{m}_k|^2}{\alpha^2}\right)
+C_1\nonumber\\
\label{psinbox}\\
\xi=&&\sum_{\bs{n}_r\neq \bs{0}}\frac{\Gamma(k/2,\alpha^2|\bs{n}_r|^2)}
{\Gamma(k/2)|\bs{n}_r|^k}
+\sum_{\bs{m}_k\neq \bs{0}}\frac{\pi^{\frac{3}{2}}\alpha^{k-3}}{\Gamma(k/2)}E_{\frac{k-1}{2}}\left(\frac{\pi^2|\bs{m}_k|^2}
{\alpha^2}\right)
+C_1+C_2\label{xinbox}\\
U = &&\frac{q_{-}^2}{L_0^k}\sum_{i<j}\psi(r_{ij}/L_0)+\frac{Nq_{-}^2}{2L_0^k}\xi\label{epgenfinnbox}
\end{eqnarray}
with the constants $C_1$ and $C_2$ defined in (\ref{c1}) and (\ref{c2}). As
it was done in the cubic box, the potential energy may also be given with the momentum
space sum (linear in $N$). Applying the definitions, similar to
Eqs (\ref{tildepsi})~--~(\ref{se}),
\begin{eqnarray}
\tilde{\psi}(\bs{r})&&=\sum_{\bs{n}_r}R(\bs{n}_r,\bs{r})+C_1\label{tildepsinoncubic}\\
\tilde{\xi}&&=\sum_{\bs{n}_r\neq \bs{0}}\rho(\bs{n}_r)+C_1+C_2\label{tildexinoncubic}\\
\tilde{S}_{\rm equal}(\bs{m}_k)&&=q_{-}\sum_{j}\exp(2\pi \ii \bs{m}_k\bs{r}_j/L)\label{sqnoncubic}\\
\tilde{S_q}(\bs{m}_k)&&=\sum_{j}q_j\exp(2\pi \ii
\bs{m}_k\bs{r}_j/L) \ ,
\label{senoncubic}
\end{eqnarray}
the potential energy for a one-component jellium model converts into
\begin{equation}
U^{\rm jel} = \frac{q_{-}^2}{L_0^k}\sum_{i<j}\tilde{\psi}(r_{ij}/L_0)+\frac{1}{2L_0^k}
\sum_{\bs{m}_k\neq 0}\kappa(\bs{m}_k)|\tilde{S}_{\rm equal}(\bs{m}_k)|^2+\frac{Nq_{-}^2}{2L_0^k}\tilde{\xi} \
,
\label{epjelfinnboxeff}
\end{equation}
with a natural extension to the general case
\begin{equation}
U^{\rm gen} = \frac{1}{L_0^k}\sum_{i<j}q_i q_j\tilde{\psi}(r_{ij}/L_0)+
\frac{1}{2L_0^k}\sum_{\bs{m}_k\neq 0}\kappa(\bs{m}_k)|\tilde{S_q}(\bs{m}_k)|^2+\frac{\sum
q_i^2}{2L_0^k}\tilde{\xi} \ .
\label{epgenfinnboxeff}
\end{equation}
Note that the formulas are derived for an orthogonal basis set. Still many triclinic lattices can be sampled in a similar form. In such cases the crystal is constructed not by translating the smallest-volume unit cell along non-orthogonal vectors but rather by translating a pseudo-unit cell of size $L_x \times L_y \times L_z$ containing more atoms along orthogonal directions. For example hcp crystal can be summed in this way. Nonetheless, as the pseudo-elementary cell technique may be inconvenient in application to triclinic lattices, we would suggest the reader to rely on a reciprocal lattice technique (see~\cite{Ashcroft}).

\subsection{2D case}

The generalization of the formulas found in a square 2D geometry to a rectangular
simulation box comes in a similar manner. It is sufficient to take the
resulting expressions for the two-dimensional problem~(\ref{i01infpolytropic2d})
and~(\ref{i01nnot0fin2d}), and to perform the necessary substitutions
$\bs{n}\rightarrow\bs{n}_r$ and $\bs{n}\rightarrow\bs{m}_k$,
\begin{eqnarray}
\psi(\bs{r})&=&\sum_{\bs{n}_r}\frac{\Gamma(k/2,\alpha^2|\bs{n}_r+\bs{r}|^2)}{\Gamma(k/2)
|\bs{n}_r+\bs{r}|^k}
+\sum_{\bs{m}_k\neq \bs{0}}\frac{\pi\alpha^{k-2}\cos(2\pi  \bs{m}_k \bs{r})}
{\Gamma(k/2)}E_{\frac{k}{2}}\left(\frac{\pi^2|\bs{m}_k|^2}{\alpha^2}\right)
+\psi_{\rm fin}^{\bs{n}=\bs{0}} \nonumber\\
{\: }\\
\xi&=&\sum_{\bs{n}_r\neq \bs{0}}\frac{\Gamma(k/2,\alpha^2|\bs{n}_r|^2)}
{\Gamma(k/2)|\bs{n}_r|^k}
+\sum_{\bs{m}_k\neq \bs{0}}\frac{\pi\alpha^{k-2}}{\Gamma(k/2)}E_{\frac{k}{2}}\left(\frac{\pi^2|\bs{m}_k|^2}
{\alpha^2}\right)
+ \psi_{\rm fin}^{\bs{m}=\bs{0}} - \frac{\alpha^k}{\Gamma(\frac{k}{2}+1)} \:,\nonumber\\
\end{eqnarray}
where $\psi_{\rm fin}^{\bs{m}=\bs{0}}$ is given by the
expressions~(\ref{i11k122Dfin}), (\ref{i11k22Dfin}) or (\ref{i11km22Dfin}).

For a long-range interaction within the jellium model, the potential energy
becomes
\begin{equation}
U^{\rm jel} = \frac{q_{-}^2}{L_0^k}\sum_{i<j}\psi(r_{ij}/L_0)+\frac{Nq_{-}^2}{2L_0^k}\xi
\label{epgen2dfinnbox1}
\end{equation}
with the notation
\begin{eqnarray}
L_0 & =& (L_xL_y)^{1/2}\label{l02d}\\
\bs{n}_r& =& \bs{n}_xL_x/L_0+\bs{n}_yL_y/L_0\label{nl2d}\\
\bs{m}_k &= & \bs{m}_xL_0/L_x+\bs{m}_yL_0/L_y \ .
\label{nr2d}
\end{eqnarray}
For a multicomponent gas (quasi-neutral in case of a long-range potential),
the potential energy is
\begin{equation}
U^{\rm gen} = \frac{1}{L_0^k}\sum_{i<j}q_i q_j\psi(r_{ij}/L_0)+\frac{\sum
q_i^2}{2L_0^k}\xi \ .
\label{epgen2dfinnbox2}
\end{equation}

Finally, the usual modification to calculate the momentum space sum linearly in $N$
is given by
\begin{eqnarray}
U^{\rm jel}&& = \frac{q_{-}^2}{L_0^k}\sum_{i<j}\tilde{\psi}(r_{ij}/L_0)+\frac{1}{2L_0^k}
\sum_{\bs{m}_k\neq 0}\kappa(\bs{m}_k)|\tilde{S}_{\rm equal}(\bs{m}_k)|^2+\frac{Nq_{-}^2}{2L_0^k}\tilde{\xi}\\
U^{\rm gen}&& = \frac{1}{L_0^k}\sum_{i<j}q_i q_j\tilde{\psi}(r_{ij}/L_0)+\frac{1}{2L_0^k}
\sum_{\bs{m}_k\neq 0}\kappa(\bs{m}_k)|\tilde{S_q}(\bs{m}_k)|^2+\frac{\sum
q_i^2}{2L_0^k}\tilde{\xi} \ ,
\end{eqnarray}
with $\tilde{\psi},\:\tilde{\xi},\:\tilde{S}_{\rm equal},\:\tilde{S_q}$ defined by
(\ref{tildepsinoncubic})~--~(\ref{senoncubic}) in their corresponding
two-dimensional variants.

Some of non-orthogonal lattices can be sampled using the concept of pseudo-unit cell. For example a triangular lattice is constructed by translation of a single atom along two vectors with $60^{\circ}$ angle between them. The same filling can be obtained by translation of a  rectangular pseudoelementary cell with two atoms which can be readily calculated with the presented formulas.

\section{Ewald method for Yukawa potential\label{EwaldYukawa}}
As it was mentioned above, the Ewald summation technique can be applied to interaction potentials of a more generic kind, than polytropic ones, for example to the Yukawa class of interaction $E_p(\bs{r}_i-\bs{r}_j=\exp(-a |\bs{r}|)/r$. The interaction potential of this form widely appeared in nuclear physics as a primitive model potential inside nuclei and in simulations of plasmas, where it is used instead of the Coulomb potential to reflect the screening properties of plasma, and in the other applications.

The derivation of Ewald sums in the style of de Leeuw et 
al.~\cite{leeuw}, that we applied in \ref{secEwald}, cannot be used 
directly 
with the Yukawa term $\exp(-a |\bs{r}|)/r$, but its certain modification may actually be used. Let us briefly explain how it can be done, having in mind the general line of derivation, given in the chapters above.

Consider the pairwise potential  $E_p(\bs{r})=\exp(-a |\bs{r}|)/\bs{r}$, with $a$ for a positive constant, defining a screening size. We can repeat all the procedure of obtaining the form of $\rho(\bs{r})$ term (real-term component of the Ewald sum) unchanged for Coulomb potential with the coefficient $\exp(-a |\bs{r}|)$. Therefore this term is obtained straightforwardly as
\begin{equation}
\rho(\bs{n})=\exp(-a |\bs{n}|)\erfc(\alpha |\bs{n}|)/|\bs{n}|\label{rhoyukawa}
\end{equation}
The other component $\xi(\bs{m})$ cannot be found directly in this manner, since the Jacobi transform, relying on a multidimensional variant of the Poisson formula~\cite{Stein1971}
\begin{equation}
\sum_{\bs{n}}f(\bs{r}+\bs{n})=\sum_{\bs{m}}F[f](\bs{m})e^{2 \pi \ii \bs{r}\bs{m}}
\end{equation}
with $f(\bs{u})=\exp(-\bs{u}^2-a|\bs{u}|)$, requires for an analytical form of 3D Fourier transform $F[f]$, which is unknown for $a>0$. Nevertheless, it can be found computationally with an arbitrary precision, thus allowing to discover all the components of $\psi(\bs{r})$ and $\xi$. As the summation in the Fourier space is a summation of cosines with the factors $\kappa(\bs{m})$, depending only on the magnitude of the wave vector $|\bs{m}|$, the latter factors can be precalculated and taken from a one-dimensional array.

The other approach to the derivation of Ewald sums was presented in a number of works~\cite{Salin2000,mazars1,mazars2} and based on a traditional treatment of the Ewald technique for Coulomb systems. The first step of the approach it to consider the charge distribution as $\delta(\bs{r})=(\delta(\bs{r})-d(\bs{r}))+d(\bs{r})=d_1(\bs{r})+d_2(\bs{r})$, where $d(\bs{r})$ stands for a charge density, represented by a function, well localized around the point charge with suitable mathematical properties, normally it is a Gaussian. This background $\rho(\bs{r})$ has to obey the condition, that the integral charge below the Gaussian is equal to the point charge, that is the background is neutralizing. The variance of the Gaussian is taken as an arbitrary parameter and is a subject of optimization in the following treatment. The charge density profiles  are then treated separately, making use of the fact, that the Yukawa potential is a Green's function of a Helmholtz equation of a certain form. The final solution reduces to finding the Green's functions for two Helmholtz equations for charge densities $d_1(\bs{r})$ and $d_2(\bs{r})$, which is a relatively simple analytical problem. The periodic (Ewald) form Yukawa potential is then equal to the sum of these Green's functions.

The final expressions of the Ewald sum for the Yukawa potential in our previous notations $\{\rho(\bs{n}),\kappa(\bs{m}),\xi,C_1,C_2\}$ can be written in the following form:
\begin{eqnarray}
\rho(\bs{n})=&&\frac{\erfc(\alpha|\bs{n}|+a/(2\alpha))e^{a|\bs{n}|}+\erfc(\alpha|\bs{n}|-a/(2\alpha))e^{-a|\bs{n}|}}{2|\bs{n}|}\\
\kappa(\bs{m})=&&\pi\frac{\exp(-\frac{\pi^2\bs{m}^2+a^2/4}{\alpha^2})}{\pi^2\bs{m}^2+a^2/4}\\
C_1=&&4\pi\frac{\exp(-\frac{a^2}{4\alpha^2})}{a^2}\\
C_2=&&-\frac{2\alpha}{\sqrt{\pi}}\exp\left(-\frac{a^2}{4\alpha^2}\right)+a\,\erfc\left(\frac{a}{2\alpha}\right)
\end{eqnarray} 
where $a$ is dimensionless, when $\bs{r}$ is given in the units of $L$. Notice that the limit $a\rightarrow 0$, corresponding to a Coulomb system, yields incorrect results in the constants $C_1$ and $C_2$. This may be seen as a reflection of the fact that the initial sum over $\bs{n}$  for a Yukawa system with any finite parameter $a$ is absolutely convergent, which is not the case for a system of Coulomb charges, thus here the continuous passage to the limit $a\rightarrow 0$ is an incorrect operation.

\section{Summary of the analytic results}
\label{sec:results}
In the previous sections, we have derived general expressions of the Ewald
sums for polytropic $1/|\bs{r}|^k$ potentials in three- two- and
one-dimensional systems. For
integer values of $k$, the polytropic potential reduces to a power-law
interaction, which comprises realizations of high physical relevance.
Integer power-law potentials include
\begin{itemize}
	\item $k=1$ -- Coulomb $1/|\bs{r}|$ interaction;
	\item $k=2$ -- Calogero-Sutherland $1/|\bs{r}|^2$ interaction;
	\item $k=3$ -- isotropic $1/|\bs{r}|^3$ component of dipole-dipole interaction (one-dimensional systems;
two-dimensional system of dipole oriented perpendicularly to the plane);
	\item $k=4,\:5,\:6$ -- interaction between different Rydberg atoms;
	\item $k=6,\:12$ -- van der Waals interaction.
\end{itemize}

The expressions for the potential energy for both the jellium model and
the general case of a charge-neutral simulation cell are the following
\begin{eqnarray}
U^{{\rm gen}} =&& \frac{1}{L_0^k}\sum_{i<j}q_i q_j\psi(r_{ij}/L_0)+\frac{\sum q_i^2}{2L_0^k}\xi\\
U^{{\rm jel}} =&& \frac{q_{-}^2}{L_0^k}\sum_{i<j}\psi(r_{ij}/L_0)+\frac{Nq_{-}^2}{2L_0^k}\xi\\
\psi(\bs{r})=&& \sum_{\bs{n}}R(\bs{n}_r,\bs{r})+\sum_{\bs{m}\neq 0}K(\bs{m}_k,\bs{r})+C_1\\
\xi =&& \sum_{\bs{n}\neq 0}\rho(\bs{n}_r)+\sum_{\bs{m}\neq 0}\kappa(\bs{m}_k)+C_1+C_2\\
R(\bs{n},\bs{r})=&&\rho(\bs{n}+\bs{r})\\
K(\bs{n},\bs{r})=&&\kappa(\bs{m})\cos(2\pi\bs{m}\bs{r})\\
C_{1}^{{\rm 3D}}=&&\left\{
\begin{array}{ll}
\frac{2 \pi ^{\frac{3}{2}} \alpha ^{k-3}}{(k-3) \Gamma\left[\frac{k}{2}\right]} & \textrm{if $k \neq 3$}\\
-4\pi+4\pi\ln(2\alpha) & \textrm{if $k=3$}\\
\end{array}\right.\\
C_{1}^{{\rm 2D}}=&&\left\{
\begin{array}{ll}
\frac{2\pi \alpha ^{k-2}}{(k-2) \Gamma\left[\frac{k}{2}\right]} & \textrm{if $k \neq 2$}\\
2\pi\ln(\alpha) & \textrm{if $k=2$}\\
\end{array}\right.\\
C_{2}=&&-\frac{\alpha^k}{\Gamma(\frac{k}{2}+1)}\\
L_0=&&\left\{
\begin{array}{ll}
(L_xL_yL_z)^{1/3} & \textrm{in 3D}\\
(L_xL_y)^{1/2} & \textrm{in 2D}\\
\end{array}\right.\\
\bs{n}_r=&&(\bs{n}\cdot\bs{L})/L_0,\textrm{ with }\bs{L}=(L_x,L_y,L_z)\\
\bs{m}_k=&&(\bs{m}\cdot\bs{L}')L_0,\textrm{ with
}\bs{L}'=(1/L_x,1/L_y,1/L_z) \ .
\end{eqnarray}
Alternatively, by performing a momentum space sum the above set of
equations become
\begin{eqnarray}
U^{\rm gen} =&& \frac{1}{L_0^k}\sum_{i<j}q_i q_j\tilde{\psi}(r_{ij}/L_0)+\frac{1}{2L_0^k}\sum_{\bs{m}_k\neq 0}\kappa(\bs{m}_k)|\tilde{S_q}(\bs{m}_k)|^2+\frac{\sum q_i^2}{2L_0^k}\tilde{\xi}\label{pigeneff} \\
U^{\rm jel} =&& \frac{q_{-}^2}{L_0^k}\sum_{i<j}\tilde{\psi}(r_{ij}/L_0)+\frac{1}{2L_0^k}\sum_{\bs{m}_k\neq 0}\kappa(\bs{m}_k)|\tilde{S}_{\rm equal}(\bs{m}_k)|^2+\frac{Nq_{-}^2}{2L_0^k}\tilde{\xi}\label{pijeleff}\\
\tilde{\psi}(\bs{r})&&=\sum_{\bs{n}}R(\bs{n}_r,\bs{r})+C_1\\
\tilde{\xi}&&=\sum_{\bs{n}\neq \bs{0}}\rho(\bs{n}_r)+C_1+C_2\\
\tilde{S}_{\rm equal}&&=q_{-}\sum_{j}\exp(2\pi \ii \bs{m}_k\bs{r}_j/L)\\
\tilde{S_q}&&=\sum_{j}q_j\exp(2\pi \ii \bs{m}_k\bs{r}_j/L) \ .
\end{eqnarray}

In accordance with considerations discussed in preceding sections, the simulation cell
has to fulfil the charge neutrality condition ($\sum_{i=1}^N q_i=0$) for
long-range potentials. Also, notice that in the particular case of a cubic
simulation cell, $\bs{n}_r=\bs{n}$, $\bs{m}_k=\bs{m}$.

Explicit expressions of the coefficients $\rho(\bs{n})$ and
$\kappa(\bs{m})$ for the most relevant interactions are summarized
for 3D and 2D systems in
Table~\ref{table3d} and Table~\ref{table2d},
respectively.

\begin{table}[h!b!p!]
\tabnote{Coefficients $\rho(\bs{n})$ and $\kappa(\bs{n})$ taken from Eqs~(\ref{defrho}) and (\ref{i01nnot0fin}) for 3D geometry. LR and SR stand for long range and short range, respectively.}
\begin{tabular*}{1.07\textwidth}{@{\extracolsep{\fill}} | l | l | l |}
\hline
          & $\rho(\bs{n})$ & $\kappa(\bs{m})$  \\ \hline
LR $\frac{1}{|r|}$&$\frac{\erfc (\alpha |\bs{n}|)}{|\bs{n}|}$ &
$\frac{1}{\pi m^2}  e^{-\frac{\pi^2 m^2}{\alpha^2}}   $ \\ \hline
LR $\frac{1}{|r|^2}$&$\frac{e^{-\alpha^2 n^2}}{n^2}$ &
$\frac{\pi}{|\bs{m}|}  \erfc \frac{\pi
|\bs{m}|}{\alpha}  $ \\ \hline
SR $\frac{1}{|r|^4}$&$\frac{\alpha^2 n^2+1}{n^4} e^{-\alpha^2 n^2}$ &
$2\pi\left(\sqrt{\pi}\alpha
e^{-\frac{\pi^2 m^2}{\alpha^2}}-\pi^2|\bs{m}|\erfc \frac{\pi
|\bs{m}|}{\bs{\alpha}}\right)$ \\ \hline
SR $\frac{1}{|r|^5}$&
$\frac{\erfc(\alpha
|\bs{n}|)}{|\bs{n}|^5}+\frac{4e^{-\alpha^2 n^2}}
{3\sqrt{\pi} |\bs{n}|^5 }(\frac{3\alpha |\bs{n}|}{2}+(\alpha |\bs{n}|)^{3})$ &
$\frac{4\pi \alpha^2}{3}\left(e^{-\frac{\pi^2
m^2}{\alpha^2}}-\frac{\pi^2 m^2}{\alpha^2}E_1(\frac{\pi^2
m^2}{\alpha^2})\right)$ \\ \hline
SR
$\frac{1}{|r|^6}$&$(\frac{\alpha^4}{2n^2}+\frac{\alpha^2}{n^4}+\frac{1}
{n^6})e^{-\alpha^2n^2}$ &\small
$\frac{\pi^{3/2}\alpha^3}{3}\left( e^{-\frac{\pi^2 m^2}
{\alpha^2}}(1-\frac{2\pi^2 m^2}{\alpha^2})+\frac{2\pi^{7/2}
|\bs{m}|^3}{\alpha^3}\erfc
\frac{\pi |\bs{m}|}{\alpha}\right) $\normalsize \\ \hline
SR $\frac{1}{|r|^{12}}$ & $\displaystyle\sum\limits_{i=0}^5\frac{(\alpha
n)^{2i}}{i!}\frac{e^{-\alpha^2n^2}}{n^{12}}$ &
$-\displaystyle\sum\limits_{i=0}^4(-2)^{i+1} (7-2i)!!\left(\frac{\pi
m}{\alpha}\right)^{2i}\frac{e^{-\frac{\pi^2 m^2}{\alpha^2}}}{945}$ \\
 & & $-\frac{32\sqrt{\pi}\left(\frac{\pi
m}{\alpha}\right)^{9}}{945}\erfc \frac{\pi m}{\alpha} $ \\ \hline
\end{tabular*}
\label{table3d}
\end{table}

\begin{table}[h!b!p!]
\tabnote{The coefficients $\rho(\bs{n})$ and $\kappa(\bs{m})$ taken from Eqs~(\ref{defrho}) and (\ref{defkappa2d}) for 2D geometry. LR and SR stand for long range and short range, respectively.}
  \begin{tabular*}{1.07\textwidth}{@{\extracolsep{\fill}} | l | l | l |}
\hline
          & $\rho(\bs{n})$ & $\kappa(\bs{m})$  \\ \hline
LR $\frac{1}{|r|}$&  $\frac{\erfc (\alpha |\bs{n}|) }{|\bs{n}|}$ &
$\frac{1}{|\bs{m}|}  \erfc \frac{\pi |\bs{m}|}{\alpha} $ \\ \hline
SR $\frac{1}{|r|^3}$&  $\frac{2\alpha}{\sqrt{\pi}n^2}e^{-\alpha^2n^2}+
\frac{\erfc (\alpha |\bs{n}|) }{|\bs{n}|^3}$ &  $4\left(\sqrt{\pi}\alpha
e^{-\frac{\pi^2 m^2}{\alpha^2}}-\pi^2|\bs{m}|\erfc \frac{\pi
|\bs{m}|}{\bs{\alpha}}\right)$ \\ \hline
SR $\frac{1}{|r|^4}$&  $\frac{\alpha^2
n^2+1}{n^4} e^{-\alpha^2 n^2}$ &
$\pi\alpha^2\left(e^{-\frac{\pi^2 m^2}{\alpha^2}}- \frac{\pi^2
m^2}{\alpha^2} E_1(\frac{\pi^2m^2}{\alpha^2}) \right)$ \\ \hline
SR $\frac{1}{|r|^5}$&  $\frac{\erfc (\alpha |\bs{n}|) }{|\bs{n}|^5}+
\frac{4e^{-\alpha^2n^2}}{3\sqrt{\pi}|\bs{n}|^5}(\frac{3\alpha|\bs{n}|}{2}+(\alpha|\bs{n}|)^{3})$
&
$\frac{8\sqrt{\pi}\alpha^3}{9}\left( e^{-\frac{\pi^2
m^2}{\alpha^2}}(1-\frac{2\pi^2 m^2}
{\alpha^2})+\frac{2\pi^{7/2} |\bs{m}|^3}{\alpha^3}\erfc \frac{\pi
|\bs{m}|}{\alpha}\right)$ \\ \hline
SR $\frac{1}{|r|^6}$&
$(\frac{\alpha^4}{2n^2}+\frac{\alpha^2}{n^4}+\frac{1}
{n^6})e^{-\alpha^2n^2}$ &
$\frac{\pi\alpha^4}{4}\left(e^{-\frac{\pi^2 m^2}{\alpha^2}} (1 -
\frac{\pi^2 m^2}{\alpha^2}) +
  \frac{\pi^4 m^4}{\alpha^4} E_1(\frac{\pi^2 m^2}{\alpha^2})\right)$ \\
  \hline
SR $\frac{1}{|r|^{12}}$&  $\displaystyle\sum\limits_{i=0}^5\frac{(\alpha
n)^{2i}}{i!}\frac{e^{-\alpha^2n^2}}{n^{12}}$ &
$
\displaystyle\sum\limits_{i=0}^4(-1)^i (4-i)!\left(\frac{\pi
m}{\alpha}\right)^{2i}\frac{e^{-\frac{\pi^2 m^2}{\alpha^2}}}{120}$\\
 &    & $-\frac{\left(\frac{\pi
m}{\alpha}\right)^{10}}{120}E_1\left(\frac{\pi^2
m^2}{\alpha^2}\right)$\\  \hline
\end{tabular*}
\label{table2d}
\end{table}

The expressions of the Ewald summation components are consistent in appropriate limits with the earlier published results by M.~Mazars~\cite{Mazars2010, Mazars2011} (inverse power-law potentials in 2D and 3D geometries), by C.~Mora \textit{et al.}~\cite{Mora2007} (2D dipolar bosons), by N.~Karasawa \textit{et al.}~\cite{Karasawa1989} and W.-Z.~Ou-Yang \textit{et al.}~\cite{Ouyang2005} (Lennard--Jones potential).

%
%
%
%
\section{Practical application and optimizations in the Ewald technique}
\label{sec:practical}
\subsection{Optimization scheme with $N^2$ dependence}
\label{subsec:N2}

The basic idea of the Ewald method is to calculate slowly decaying sums in
a rapid manner by means of the Fourier transform of the slowly converging
part. Although conceptually it provides an exact result, the number of terms
which has to be summed in order to reach the needed convergence  is {\it a priori}
unknown.  Once we choose the interaction potential, this fixes the exact
form of the sums to calculate, and the practical remaining question is the
proper choice of the free parameter $\alpha$ and the numbers of terms to
be calculated in the sums, originated from coordinate and momentum spaces.
The total computer time $T$ is obtained from the time needed to evaluate different sums
\begin{equation}
T=(t_r N_r+t_k N_k)N^2/2\label{TofN},
\end{equation}
with the constants $t_r$ and $t_k$ depending on the complexity of the coefficients in the sums and the factor $N^2/2$ approximating the number of pairs. Here $N_r$ and $N_k$ are numbers of terms which are summed for each pair $\bs{r}_{ij}$, in particular the case $N_r=1$ corresponds to the so called minimum image convention.
One can notice that $t_k$ is usually much less then $t_r$, since in the
Jacobi-transformed sum we only calculate cosine functions, which is
generally far less time-consuming than the complicated functions appearing in
$R$. It is clear that the parameter $\alpha$ affects only the resulting error
in the energy. In fact, the value of $\alpha$ being very small or very
large eliminates errors in one of the sums, but amplifies them in the other,
so there is an ``optimal'' point for $\alpha$, yielding a minimum error in
the total energy.

In the following, we discuss a way for error ($\delta E$) minimization assuming the
calculation time $T$ fixed. From our point of view, a useful  approach for
practical implementation is represented by the following scheme
\begin{itemize}
\item
We determine a time law $T=(t_r N_r+t_k N_k)N^2/2$ in a preliminary calculation
and fix the values of $t_r$ and $t_k$.
\item
We take a set of configurations, corresponding to the equilibrated
state using an initial Ewald summation. Then, we calculate the
exact energies $E_{\rm ex}$ (as a converged result of the Ewald
summation) and the energies $E(\alpha,N_r,N_k)$ biased by a choice of $N_r$
and $N_k$.  For each pair $(N_r,\,\,N_k)$, we find an optimal value of
$\alpha=\alpha_{\rm opt}(N_r,N_k)$.
\item
We choose the goal accuracy $\delta E_{\rm acc}$ (normally, well below the statistical
error). We plot the error as a function
of the computer time spent and choose the less time consumption  case among
the points that lie
below $\delta E_{\rm acc}$, therefore obtaining all the parameters required:
$\alpha$, $N_r$ and $N_k$. From now on, these parameters are used in actual
simulations.
\end{itemize}

\subsection{Example of optimization}
\label{subsec:N2example}
Let us illustrate the scheme proposed in the preceding subsection taking as an example the problem of two-dimensional zero-temperature Bose gas of particles, interacting through the pairwise repulsive $C_3/|\bs{r}|^3$ potential. The model corresponds to the dipole-dipole interaction with all dipole moments aligned perpendicularly to the plane of motion. The simulation is performed with $N=108$ particles in a quadratic box ($L_x=L_y=L=1$). The dimensionless Hamiltonian in the present example is taken to be  $\hat{H}=-\sum_{i=1}^N\nabla_i^2/2+\sum_{i<j}1/|\bs{r}_{ij}|^3$ with the unit of energy being $\hbar^2/mL^2$.
To describe  the ground-state properties of the system
we use the variational Monte Carlo (VMC) method and a Jastrow wave function
with a two-body correlation factor which is solution of the
two-body scattering problem~\cite{McMillan65}.

\begin{figure}[tbp]
\begin{center}
\includegraphics*[width=0.85\textwidth]{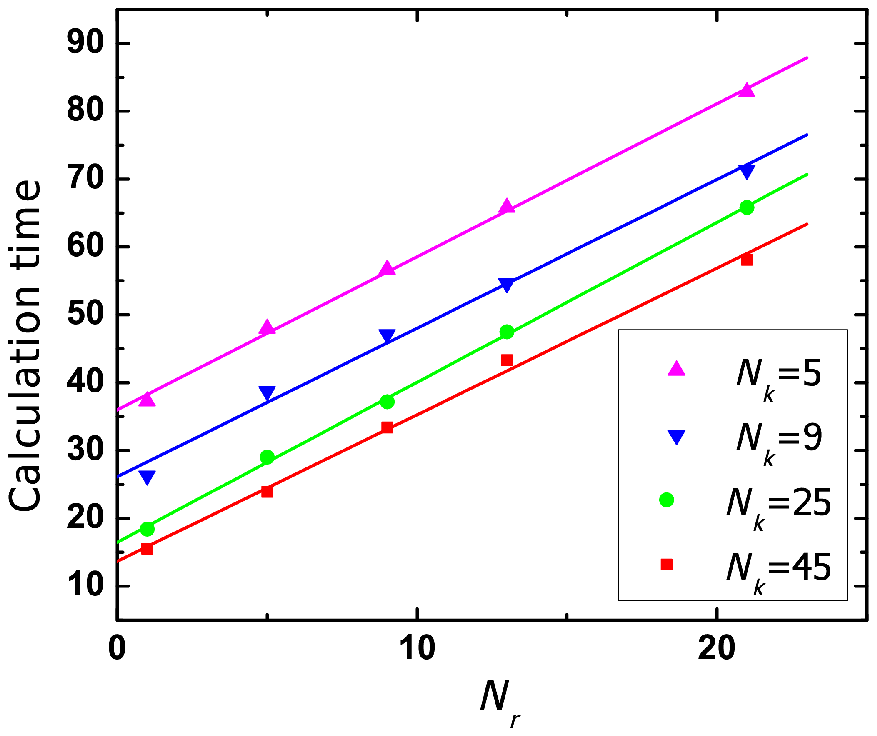}
\end{center}
\caption{Dependence of the calculation time  $T$ on the number of terms $N_r$ in
the coordinate space for fixed numbers of terms in the momentum space $N_k=5,9,25,45$. Symbols are 
calculation times. Lines are linear fits for the data.}
\label{tonnr}
\end{figure}
\begin{figure}[tbp]
\begin{center}
\includegraphics*[width=0.85\textwidth]{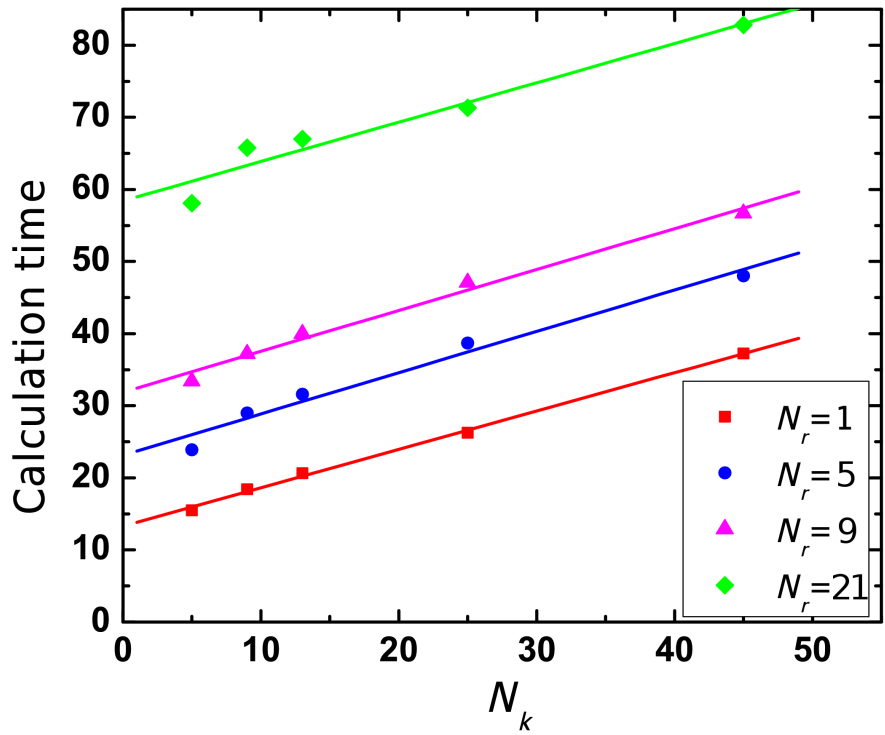}
\end{center}
\caption{Dependence of the calculation time $T$ on the number of terms $N_k$ in
the momentum space for fixed numbers of terms in the coordinate space $N_r=1,5,9,21$. Symbols are 
calculation times. Lines are linear fits for the data.}
\label{tonnk}
\end{figure}

The optimization is done by averaging over $N_{\rm conf}=50$ uncorrelated VMC
configurations, sampled according to the chosen probability distribution.
We define the error $\delta E(\alpha)$ as a sum over $N_{\rm conf}$
configurations of the difference of the Ewald energy $E(i_{\rm
conf},\alpha,N_r,N_k)$, calculated for a given set of parameters
$(\alpha,N_r,N_k)$ and the converged energy $E_{\rm ex}(i_{\rm conf})=\lim_{N_k
\rightarrow \infty}\lim_{N_r \rightarrow \infty}E(i_{\rm
conf},\alpha,N_r,N_k)$. The dependence of the computer time $T$, needed for the
evaluation of Ewald sums, on the parameter set is shown in Figs~\ref{tonnr} and
\ref{tonnk}. In
Fig.~\ref{tonnr}, we show the dependence of $T$ on the number of terms
$N_r$ in real space for different fixed numbers of terms
$N_k$ in the momentum space. The computation time is proportional to the
number of terms and the resulting dependence is linear in $N_r$. A fixed
number of terms $N_k$ requires a certain amount of calculations which
results in a constant shift. Similarly, keeping $N_r$ fixed and varying
$N_k$ produces a linear dependence in $N_k$ with a constant shift which
depends on $N_r$, as shown in Fig.~\ref{tonnk}.

As one sees in Figs.\ref{tonnr} and \ref{tonnk}, the time dependence is linear both on
$N_k$ and $N_r$,
although the point corresponding to (0,0) in $(N_r,\,N_k)$ does not
necessarily gives $T=0$, since the reported time also contains some
initializing calculations.
The total error in the potential, as it is defined above, is given by
\begin{equation}
\delta E(\alpha)=\sqrt{\sum_{i_{\rm{conf}}=1}^{N_{\rm{conf}}}
\frac{\left(E(\alpha,i_{\rm{conf}})-E_{\rm{ex}}(i_{\rm{conf}})\right)^2}{N_{\rm{conf}}}}
\ .
\label{dalpha1}
\end{equation}
According to our previous considerations, in the
case of very small or very large values of $\alpha$ the error coming from one of
two sums, that is in the real or momentum space, grows and dominates over
the error coming from the other sum; for a certain ``optimal'' range of
$\alpha$ these two errors are of the same order. Notice that for each
particular configuration, and each pair $(N_r,\,N_k)$, it is possible to find
$\alpha_{\rm opt}(i_{conf})$, such that
$E(\alpha_{\rm opt}(i_{conf}),i_{conf})-E_{\rm ex}(i_{conf})=0$. Instead, our task
is to obtain a ``universal'' parameter $\alpha_0$, minimizing the total
error~(\ref{dalpha1}). The mean over the configuration set of the biased
energies $\bar{E}(\alpha_0, i_{\rm{conf}})$ is used as an estimation for
the mean of the exact energies $\bar{E}_{ex}$, introducing an inevitable
systematic error.  As it appears in typical calculations, this error is at
least one order of magnitude smaller than the statistical error~(\ref{dalpha1}) given by
the minimization of $\delta E(\alpha)$.
 In our benchmark calculations we also checked the dependence of the total energy on the value $\alpha$ for different pairs $(N_r,N_k)$, which revealed characteristic plateaus for certain ranges of $\alpha$ (of the order of 1). It means that the Ewald summation indeed converges fast to a universal result. Nevertheless, the optimum value of the parameter $\alpha$, minimizing the cumulative error, depends on the cut-off numbers in the sums in the range from $\alpha=1.45$ ($N_r=21$, $N_k=5$) to $\alpha=5.25$ ($N_r=1$, $N_k=45$).

\begin{figure}[tbp]
\begin{center}
\includegraphics*[width=0.85\textwidth]{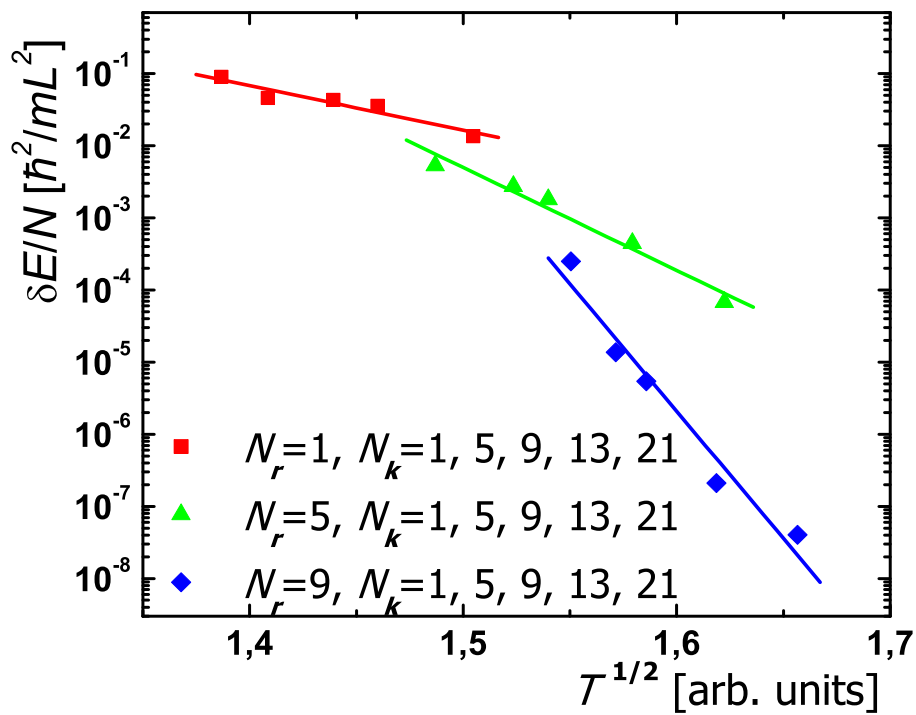}
\end{center}
\caption{Resulting error of the energy per particle as a function of the computer time for different parameter sets. Symbols are 
the errors, related to truncation in the Ewald method. Lines are exponential fits for the data.}
\label{erront4}
\end{figure}

A second step is the study of the dependence of the error and time on different pairs
$(N_r,\,N_k)$. The calculation time can be split as the sum of times for
summing up in coordinate and momentum spaces,
$
T=(N_r t_r+N_k t_k)N^2/2
$
as in Eq.~(\ref{TofN}),
with $N_r,\;N_k$ being the numbers of terms in each sum for a single pair. Every
one of these sums converges when $N_r,\;N_k\rightarrow \infty$ to a certain
value, depending on $\alpha$, while the sum of the limiting values is a
constant. We can take into account the errors, corresponding to each of the
sums separately. For $\alpha\rightarrow 0$ the error for the real space term
is zero and the other one tends to infinity (and vice versa as
$\alpha\rightarrow \infty$). The minimum total error should therefore
correspond to the value of $\alpha$, satisfying the relation
$\di (\delta E_r + \delta E_k)/\di \alpha = 0$.

Focusing on the 2D system of our example, we note
that the long-range expansions of the terms in (\ref{defrho}) and
(\ref{defkappa2d})  are
similar, in a sense that the leading terms in both expressions are
Gaussians,
\begin{eqnarray}
\frac{\Gamma(k/2,\alpha^2n^2)}{\Gamma(k/2)n^k} & = & \exp(-\alpha^2n^2)\left[\frac{C_r}{n^2}+
O\left(\frac{1}{|\bs{n}|^3}\right)\right] \label{3drealtermexpand}\\
\frac{\pi^{\frac{3}{2}}\alpha^{k-3}}{\Gamma(k/2)}E_{\frac{k-1}{2}}\left(\frac{\pi^2m^2}{\alpha^2}
\right) & = & \exp\left(-\frac{\pi^2m^2}{\alpha^2}\right)\left[\frac{C_k}{m^2}+
O\left(\frac{1}{|\bs{m}|^3}\right)\right] \ .
\label{3dmomtermexpand}
\end{eqnarray}
The power-law terms in $\bs{n},\:\bs{m}$ and the constants $C_r,\:C_k$ may be
neglected since the leading behavior is driven by the Gaussian. The cut-off
errors due to finite numbers of elements in the sums can be evaluated by
ignoring the discrete structure of the images and approximating the sums by
uniform integrals,
\begin{eqnarray}
\delta E& =& \frac{N^2}{2}\int_R^{\infty}\exp(-\alpha^2 r^2) 2\pi r \di r +
\frac{N^2}{2}\int_K^{\infty}\exp\left(-\frac{\pi^2k^2}{\alpha^2}\right) 2\pi k \di k \nonumber \\
& = & \frac{\pi N^2}{2}\left[\frac{\exp(-\alpha^2 N_r/\pi)}{\alpha^2}+
\frac{\alpha^2 \exp\left(- \pi N_k/\alpha^2\right)}  {\pi^2}\right]
\label{dE1}
\end{eqnarray}
with $R\simeq \sqrt{N_r/\pi}$ and $K\simeq \sqrt{N_k/\pi}$ the
approximate cut-off lengths in  real and momentum spaces, respectively.
The optimal value for $\alpha$ can be obtained by solving the equation $\di \delta E_r/\di \alpha = -\di \delta E_r/\di \alpha $. The first-order approximation of this equation is found by taking logarithms of both sides and omitting constants and terms, depending on $\alpha$ logarithmically, that is
\begin{equation}
A_k/\alpha^2-A_r \alpha^2= 0
\label{dE2}
\end{equation}
with $A_k=\pi N_k$ and $A_r=N_r/\pi$, which yields
\begin{equation}
\alpha=(A_k/A_r)^{1/4}=\left(\pi^2 N_k/N_r\right)^{1/4} \ .
\end{equation}
Then, at lowest order one finds~(\ref{dE1}),
\begin{equation}
\delta E\sim  \frac{N^2}{2}\exp(-\alpha^2 N_r/\pi)=\frac{N^2}{2}\exp(-\sqrt{N_k N_r}) \ .
\end{equation}
Since the calculation time is linear with the numbers of elements $N_r$ and
$N_k$, we may conclude that with $N_k$ fixed and comparatively large $N_r$,
$\;\ln(\delta E) \sim \sqrt{N_r} \sim \sqrt{T}$ and vice versa, with
$N_r$ fixed and large $N_k$, $\ln(\delta E) \sim \sqrt{N_k} \sim \sqrt{T}$.
This power law may be easily checked in our calculations, as it is shown in
Fig.~(\ref{erront4}). Note that for the obtained value of $\alpha$ the errors of the real- and momentum-space cutoffs are of the same order of magnitude, that is
$\delta E_r \approx \delta E_k$, which may serve as a rough criterion to optimize the parameter $\alpha$.

\subsection{Optimization scheme with $N^{3/2}$ dependence}
\label{subsec:N32}
A more advanced procedure for optimization of the parameters, proposed by Perram \textit{et al.}~\cite{Perram1988}, yields an asymptotic scaling $N^{3/2}$, with $N$ the number of particles. It is based on the form of Ewald summation with the rearranged momentum space
sum, linear in $N$ (\ref{pigeneff}). Note that here the linear $N$ dependence is obtained after a summation over particle coordinates, while in Sections~(\ref{subsec:N2})~and~(\ref{subsec:N2example}) the momentum space sum was evaluated over the pairs of the particles and had $N^2$ dependence.
 Suppose the values of the calculation
time $t_r$, $t_k$ to perform unit computations in both sums are known and
the target error level $\exp(-p)$ is fixed. Then, the total execution time
in the real and momentum spaces is
\begin{equation}
T=T_r+T_k=N^2\pi R^2 t_r+N \pi K^2 t_k
\end{equation}
with $p=\alpha^2 R^2=\pi^2 K^2/\alpha^2$. Expressing $K$ as $K=p/(\pi R)$
we can see that the minimum of the total time T corresponds to
\begin{eqnarray}
R_{\rm opt}=&&\left(\frac{p}{\pi}\right)^{1/2}\left(\frac{t_k}{t_r}\right)^{1/4}N^{-1/4}\\
K_{\rm opt}=&&\left(\frac{p}{\pi}\right)^{1/2}\left(\frac{t_k}{t_r}\right)^{-1/4}N^{1/4}\\
\alpha_{\rm opt}=&&\sqrt{\pi}\left(\frac{t_k}{t_r}\right)^{-1/4}N^{1/4} \ .
\end{eqnarray}
The computation time is equally divided between the real and
momentum space parts (this was also stated in our simple optimization
scheme), with a scaling of the whole summation given by
\begin{equation}
T=2 N^2 \pi R^2 t_r= 2 p\sqrt{t_r t_k}N^{3/2}
\end{equation}
Notice that the values of the free parameters change very slowly when
the simulation cell is enlarged, and in particular $\alpha$ is not affected
by the choice of the precision. Similar formulas for the optimized
parameters in three-dimensional systems, with a discussion of different
techniques to improve performance of the Ewald summation, are given by
Fincham~\cite{Fincham1994}. A more precise and detailed analytic study
of the cut-off errors with verifications of the analytic results in actual
calculations can be found in the work of Kolafa and Perram~\cite{Kolafa1992}.
An optimized method for treating the truncation error in Ewald sums with
generic potentials was proposed by  Natoli and
Ceperley~\cite{Natoli1995}. While the needed CPU time scales as $\mathcal{O}
(N \ln N)^{3/2}$, it was shown that in the example of the Coulomb potential the
method resulted in greatly improved accuracy compared to that of standard
Ewald technique for a comparable computational effort.
This method is based on an expansion of the real
space function in an arbitrary radial basis with a parametric set of
numbers in place of the $k$-dependent prefactors of $\exp(2\pi\ii
\bs{m}\bs{r})$. The subsequent minimization of $\chi^2$ with respect to the
whole set of parameters yields a final optimal solution, that is the real
space expansion coefficients and the $k$-space factors.  This technique
was also applied to derive the optimized summation formulas for the
two-dimensional Coulomb system~\cite{Holtzmann2005}.

In general, the unit computation time in momentum space is 2--4 times faster
than the one in real space. Taking the following reasonable assumptions
$p=4\pi$, $t_k/t_r=3$, we find $R_{\rm opt}\approx 2.6/N^{1/4}$ (the box size is taken to be 1). We want
$R$ to be below 0.5, since in this case the summation in the real space
reduces to the accumulation of the single component $\bs{n}=0$ (``minimum image convention''). This
condition $R_{\rm opt}=0.5$, with our previous assumptions, corresponds to
\begin{equation}
N_{\rm opt}= 770,\:\:K_{\rm opt}=8.0,\:\:\alpha_{\rm opt}=7.1
\end{equation}
In smaller systems, the other components of the real sum, starting from
$|\bs{n}|=1$, should be considered.

It is worth pointing out that if the interaction is very
strong at short distances (as for the Lennard-Jones potential), then
in principle the real-space cut-off $R$ can be chosen below the ``hard core
radius'' with a large enough value of $\alpha$. This leads to the possibility
of dropping completely the real-space part of the total sum  and treat the
$k$-space only. This can be advantageous in different aspects, especially
with the current progress in the development of efficient FFT-based
methods. Nonetheless, we are not aware of any present
application of a similar technique.

\section{Conclusions}
\label{sec:conclusions}

In the present Chapter, we have applied the Ewald summation method to
$1/|\bs{r}|^k$ polytropic potentials in three-, two- and
one-dimensional geometries in a simulation box with periodic boundary
conditions. We have found the explicit functional forms for all the
components of the sums in both real and momentum spaces, with
special attention being paid to the cases of long-range interactions, that
is conditionally convergent or divergent potentials (corresponding to
$k<D$, with $D$ standing for the dimensionality), ``marginal''
interactions ($k=D$), and short-range interactions (with $k>D$). For the
latter case of short-range interaction potentials, where in principle a
straightforward summation of the initial sum (\ref{initialhamiltonian1}) is
possible, the Ewald method is shown to be useful, as it yields the faster
(Gaussian) convergence rate. A condition of charge neutrality of the simulation cell is
stated to be necessary for conditionally convergent and divergent
potentials; a homogeneous positive charge background (``jellium'' model) is
introduced as the most relevant and frequently used kind of neutralization.
The conditionality of the convergence for a charge-neutral system, governed
by the Coulomb interaction, is discussed with a justification of the use of
a specific periodicity-preserving convergence factor. The derivation
technique, presented in our work, is consistent with the arguments of
de Leeuw \textit{et al.}~\cite{leeuw}.

The results are first presented for the case of a 3D system
in a cubic simulation box in order to explain the
general mathematical procedure, which for the specific case of the Coulomb
potential recovers well-known results~\cite{allen}. Later on, the same
mathematical technique is applied to 2D and 1D geometries.
For the one-dimensional case the
initial sum for the potential energy  is explicitly evaluated (\ref{ep1ddirect2}),
nonetheless the Ewald summation is developed for this
case too and may be used as a mathematical equality. The special
representations of the reciprocal space sums, linear in the number of
particles $N$ and hence more efficient in actual modeling, are presented
for 3D and 2D systems. The explicit expressions for the terms of the Ewald
sums are given in a tabular form for physically relevant potentials
with small integer power indexes $k$, as dipole-dipole interaction
potential, Lennard-Jones potential and others in both three- and
two-dimensional geometries
(see Tables~1 and 2).

When the simulation box cannot be chosen cubic, for example in a modeling
of a three-dimensional hcp crystal structure, the Ewald method can also be
applicable after a certain modifications. Formally, it consists in the choice
of an appropriate rectangular simulation box and a substitution of the
vector $\bs{n}$ by $\bs{n}_r=(\bs{n}_xL_x+\bs{n}_yL_y+\bs{n}_zL_z)/L_0$ and
$\bs{m}_k=(\bs{m}_x /L_x+\bs{m}_y /L_y+\bs{m}_z /L_z)L_0$ in the real
and momentum space sums, respectively [see
(\ref{i01infpolytropicnbox}) and (\ref{3djacobinbox})].

The optimization of the involved parameters, that is cut-off numbers in
both sums and the integration parameter $\alpha$, is a necessary operation
in order to improve the convergence rates and avoid excessive calculations.
The main idea of the optimization, proposed in the present work, is to
perform a benchmark calculation, minimizing the variance of the result. A
particular example of the application of the technique is presented for a
calculation of the potential energy of a two-dimensional gas of dipoles, aligned
perpendicular to the plane of motion. This practical optimization technique
is thought to be efficient for stationary and nearly uniform systems that
appear, for instance, in Monte Carlo simulations. In spite of being very
simple, it allows to find rather quickly adequate parameter ranges. The
analytical estimations of the parameters are given as well and are proven
to be consistent with the results, obtained in our method. A more
sophisticated method to optimize the calculation parameters, taking
advantage of the $\mathcal{O}(N)$ representation of the Fourier transform
sum, is also presented with explicit estimations of the parameters for a
typical system simulated by Quantum Monte Carlo methods.

\chapter{Phase diagram of a Yukawa system\label{secYukawa}}
\section{Introduction}
Recent advances in trapping and controlling ultracold dilute gases have permitted to realize highly tunable and extremely pure Fermi systems~\cite{RPMcoldgases1,RPMcoldgases2}. This has provided new insight in the study of fundamental problems in condensed matter physics. For example, the original BCS theory~\cite{BCS} was developed to explain superconductivity in metals, where the control over interactions and densities is very limited. However, in recent experiments with ultracold Fermi gases in the BCS-BEC crossover the strength of the interactions is controlled by external magnetic fields in the vicinity of a Feshbach resonance, while the geometry is tuned by means of magnetic or optical confinement. This has allowed, for instance, to measure the equation of state in the BCS-BEC crossover in high precision experiments~\cite{EOS:experiment1,EOS:experiment2}. Numerically, the best calculation of the zero-temperature equation of state is obtained in quantum Monte Carlo simulations
~\cite{EOS:BCSBEC3D1,EOS:BCSBEC3D2,EOS:BCSBEC3D3,EOS:BCSBEC3D4,EOS:BCSBEC3D5}

After the big success achieved with single species there is nowadays a growing interest in fermionic mixtures. Quite recently, fermionic mixtures consisting of atoms with different masses have been realized experimentally~\cite{exp:LiK1,exp:LiYb1} 
\y{and studied theoretically~\cite{th:LiK1,th:LiK2,th:LiK3}}. Novel physical phenomena like Efimov states~\cite{theory:Efimov1,theory:Efimov2,theory:Efimov3,theory:Efimov4,theory:Efimov5}, trimer and cluster formation might be observed~\cite{exp:Efimov1,exp:Efimov2,exp:Efimov3,exp:Efimov4,exp:Efimov5,exp:Efimov6,exp:Efimov7,exp:Efimov8} in these systems. The case of large mass imbalance is especially interesting, and mixtures of $^6$Li and $^{40}$K are being investigated experimentally~\cite{exp:LiK1,exp:LiK2,exp:LiK3,exp:LiK4,exp:LiK5,exp:LiK6,exp:LiK7,exp:LiK8}. Even larger mass ratios are reached in mixtures of $^6$Li and $^{173}$Yb~\cite{exp:LiYb1,exp:LiYb2}. In this chapter we present results for the phase diagram of Fermi mixtures as a function of the mass ratio using quantum Monte Carlo methods and determine how crystallization of this system can be realized.

From the theoretical point of view, it was proposed in Ref.~\cite{YukawaPot2D} that an effective Yukawa interaction, induced between heavy-light pairs of fermions, might lead to crystallization in quasi-two-dimensional systems. In this work we extend that discussion and analyze the possibility of realizing a gas-crystal phase transition at zero temperature in three-dimensional systems. We obtain the phase diagram and discuss how large mass ratios have to be for reaching crystallization.

The interest in the phase diagram of quantum Yukawa particles is rather old as the Yukawa potential has long been used, for instance, as a model for neutron matter~\cite{Baym1975,neutron2}.
The Yukawa potential also describes interactions in dusty plasmas where charged dust particles are surrounded by plasma which introduces screening~\cite{dusty_plasma1,dusty_plasma2,dusty_plasma3}. The Yukawa potential is often used as well as a model for suspensions of charged colloidal particles~\cite{yukawa:colloids1,yukawa:colloids2,yukawa:colloids3,yukawa:colloids4,yukawa:colloids5,yukawa:colloids6}. The classical finite temperature phase diagram has been extensively studied 
~\cite{dusty_plasma1,dusty_plasma2,dusty_plasma3,yukawa:colloids1,yukawa:colloids2,yukawa:colloids3,yukawa:colloids4,yukawa:colloids5,yukawa:colloids6} while much less is known about the full quantum phase diagram.

In the 70's, Ceperley and collaborators~\cite{Ceperley1,Ceperley2} used the diffusion Monte Carlo algorithm to estimate the zero-temperature phase diagram of the Yukawa Bose fluid. In their work the phase diagram was built assuming that the Lindemann ratio remains constant along the solid-gas coexistence curve, with the explicit value being evaluated only at a single point. In the present Chapter we carry out a full study of the transition curve and present the phase diagram in terms of experimentally relevant densities and mass ratios of heavy to light fermions. The Lindemann criterion prediction has turned out to be quite precise apart from the region of high densities.

\section{Model Hamiltonian}
Mixtures of fermions with different masses have been realized recently in a new generation of experiments~\cite{exp:LiK1,exp:LiK2,exp:LiK3,exp:LiK4,exp:LiK5,exp:LiK6,exp:LiK7,exp:LiK8,exp:LiYb1,exp:LiYb2}. The interactions can be tuned to allow the formation of two-component molecules. The $s$-wave interactions within a single component are prohibited due to Pauli principle. Yet, an effective interaction between same-spin fermions can be induced by the presence of the other component. The limit of large mass ratio has been analytically addressed in Ref.~\cite{YukawaPot2D}. The effective interaction between heavy particles, which was obtained in the limit of large distances within first Born approximation, has the form of a screened Coulomb (Yukawa) potential. This leads to a description of the system in terms of a composite (molecular) bosonic gas interacting with an effective potential.

We study a system of heavy fermions of mass $M$ interacting among themselves and moving on a background of light fermions of mass $m$. The net effect induced by the movement of the light fermions can be characterized by a Yukawa potential, leading to the following effective  Hamiltonian~\cite{YukawaPot2D} describing the interaction between composite bosons formed by pairs of heavy and light atoms
\begin{equation}
\hat H=-\frac{\hbar^2}{2M}\sum_i\Delta_{i} + \sum_{i<j}\frac{2\hbar^2}{m}\frac{\exp(-2|{\rve}_i-{\rve}_j|/a)}
{a|{\rve}_i-{\rve}_j|},
\label{HYukawa}
\end{equation}
where $a$ is the atom-atom $s$-wave scattering length between two atomic species and ${\rve}_i$ are positions of heavy atoms while the positions of light atoms have been integrated out. The ground-state properties of the system are then governed by two dimensionless parameters, namely the gas parameter $na^3$ and the mass ratio $M/m$. Equivalently, Hamiltonian (\ref{HYukawa}) describes a bosonic system interacting via the screened Coulomb potential $V_{\rm int}(r) = q \exp(-\lambda r)/r$ by mapping the charge to $q = 2\hbar^2/ma$ and the screening length to $\lambda^{-1} = a/2$.

We calculate the ground-state properties corresponding to the Hamiltonian~(\ref{HYukawa}) by means of the diffusion Monte Carlo (DMC) algorithm~\cite{Boronat94}. This method solves stochastically the Schr\"odinger equation in imaginary time providing the exact energy within controllable statistical errors. The coexistence curves can then be traced by direct comparison of the energies of the solid and gas phases. The efficiency of the DMC method is greatly enhanced when importance sampling is used. This is done by multiplying the (unknown) ground-state wave function $\psi({\rve}_1, \ldots, {\rve}_N)$ by a guiding wave function $\psi_T({\rve}_1, \ldots, {\rve}_N)$ and solving the equivalent Schr\"odinger equation for the product. As a result, the points in phase space where the guiding function is large get sampled more frequently and this improves convergence to the ground state (see Section~\ref{secQMC} of this Thesis).

The properties of the gas phase are studied by constructing the guiding function in a Jastrow two-body product form $\psi_T({\bf   r}_1, \ldots, {\rve}_N) = \prod_{i<j}f_2(|{\rve}_i-{\rve}_j|)$. We determine the optimal two-body Jastrow term $f_2(r)$ by solving the corresponding Euler--Lagrange hypernetted-chain equations~\cite{HNC} (HNC/EL), discussed previously in Section~\ref{secHNC}, discarding the contribution of the elementary diagrams. The resulting wave function captures basic ingredients coming both from the two- and many-body physics of the problem. On the other hand, the energy of the solid phase is obtained by using a Nosanow-Jastrow guiding wave function $\psi_T({\rve}_1, \ldots, {\rve}_N) = \prod_{i=1}^N f_1({\rve}_i-{\rve}_i^{\text{latt.}})\prod_{i<j}f_2(|{\rve}_i-{\rve}_j|)$ with Gaussian one-body terms $f_1({\rve}_i-{\rve}_i^{\text{latt.}}) = \exp(-\alpha({\rve}-{\rve}_i^{\text{latt.}})^2)$ describing the localization of particles close to the lattice sites ${\rve}_i^{\text{latt.}}$. The parameter $\alpha$ controls the localization strength and is optimized by minimizing the variational energy.

In order to find the energy in the thermodynamic limit, we carry out simulations of a system of $N$ particles in a box with periodic boundary conditions, and take the limit $N\to\infty$ while keeping the density fixed. In the simulation of the crystal the number of particles should be commensurate with the box which restricts the allowed number of particles. For fcc packing the simulation box supports
$N=4i^3=4, 32, 108, 256, \ldots$. In order to add more values we also use periodic boundary conditions on a truncated octahedron (see Appendix~B), which allows simulations with $N=2i^3 = 2, 16, 54, 128, 250, 432, \ldots$ particles with a larger effective volume of the simulation box and reduced anisotropy effects. Finally, the convergence is further improved by the Ewald summation technique~\cite{ewald,EwaldOsychenko} in the cubic box, which we use in the calculations at large densities.

\begin{figure}
\centering
\includegraphics[width=0.6\textwidth, angle=-90]{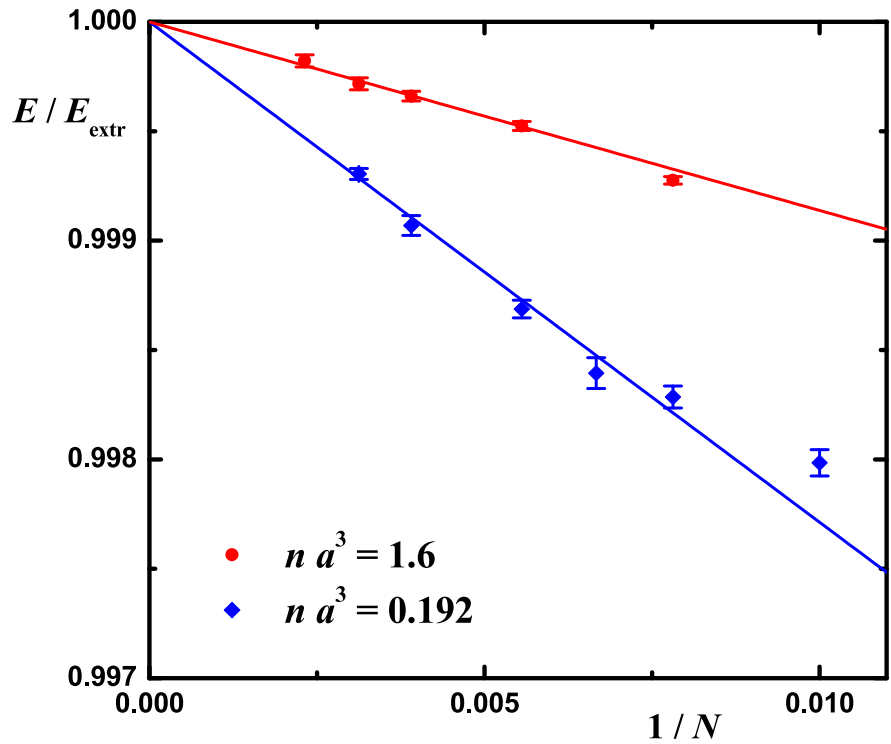}
\caption{An example of finite-size dependence of
the energy in the gas with periodic boundary conditions
in truncated octahedron for $M/m=187$ at two different
densities $na^3 = 1.6$ (upper set of data points) and
$na^3 = 0.192$ (lower set of data points). Symbols, DMC energy; lines, linear fit to energy for large system sizes.
Energies are scaled with the thermodynamic value $E_{extr}$, obtained in $1/N\to 0$ extrapolation.}
\label{Fig:finite size}.
\end{figure}

In Fig.~\ref{Fig:finite size} we show two characteristic examples of the finite-size
dependence of the energy at two different densities.
For large enough system sizes, the energy is well fitted by a linear
dependence in $1/N$. For small number of particles the
behavior is no longer linear, especially at large densities
 due to strong interparticle correlations. We find that system
sizes of $N>100$ have to be used in order to ensure the linear regime at
considered densities. The thermodynamic energy is then obtained as a
result of a linear extrapolation $1/N \to 0$.

\section{Phase diagram}
An intrinsic property of Coulomb particles is to self-assemble into a Wigner crystal at low densities and to remain in a gaseous phase in the opposite limit, due to the long-range character of the interaction~\cite{Wigner}. The Yukawa potential is similar to the Coulomb one at densities large enough for the interparticle distance to be much smaller than the screening length, which is fixed by the $s$-wave scattering length $a$ between two different species of atoms. One then concludes that the Yukawa system stays in a gaseous phase at large densities. In the opposite regime of small densities, $na^3\ll 1$, the interaction potential decays exponentially fast showing a short-range behavior that leads the system to a gaseous phase. For example, the fcc crystal of hard-sphere bosons of diameter $a_s$ melts at density $n a_s^3 \approx 0.24$~\cite{HS3D1,HS3D2,HS3D3}. The intermediate regime $na^3\approx 1$ is the most interesting one, as crystallization may or may not take place depending on the strength of the interaction, which in the current case of the Hamiltonian in Eq.~(\ref{HYukawa}) is governed by the mass ratio $M/m$. A relevant question then is what is the minimal mass ratio at which crystallization can be observed.

\begin{figure}[htb]
\centering
\includegraphics[width=0.6\textwidth, angle=270]{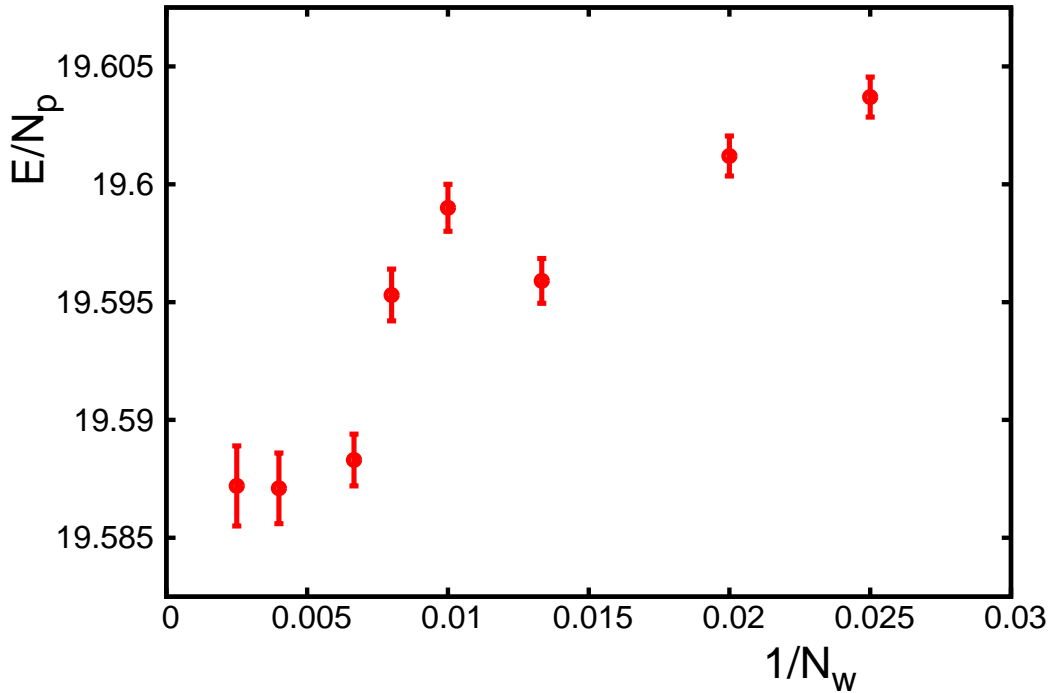}
\caption{Dependence of the energy per particle $E/N_p$ on the inverse number of walkers $1/N_w$ for the Yukawa system in a liquid phase, 64 particles.}
\label{yuk_nw}
\end{figure}

In order to obtain an accurate description of the phase diagram, we study the finite size dependence and extrapolate the energy to the thermodynamic limit. As we mentioned in the sections, devoted to the methodology, the diffusion Monte Carlo method is also generally biased by the average size of the population of walkers $N_w$ (sets of $N_p$ particle configurations), that can be controlled in the simulation. In practice the results of different calculations converge to a stable value, when $N_w$ is large enough, although the convergence can be achieved with much less $N_w$ with a better trial wave. Usually the population of walkers, when the convergence is reached, is between 250 and 500 and is not affected much by the size of a system, therefore a preliminary simulation can be carried out fast with a small number of particles. Figure~(\ref{yuk_nw}) demonstrates the analysis of $N_w$ convergence for the Yukawa system of 64 particles in a liquid phase (no size correction added).  The convergence is seen to 
be reached for $N_w \approx 200$.
 
\begin{figure}[htb]
\centering
\includegraphics[width=0.6\textwidth, angle=270]{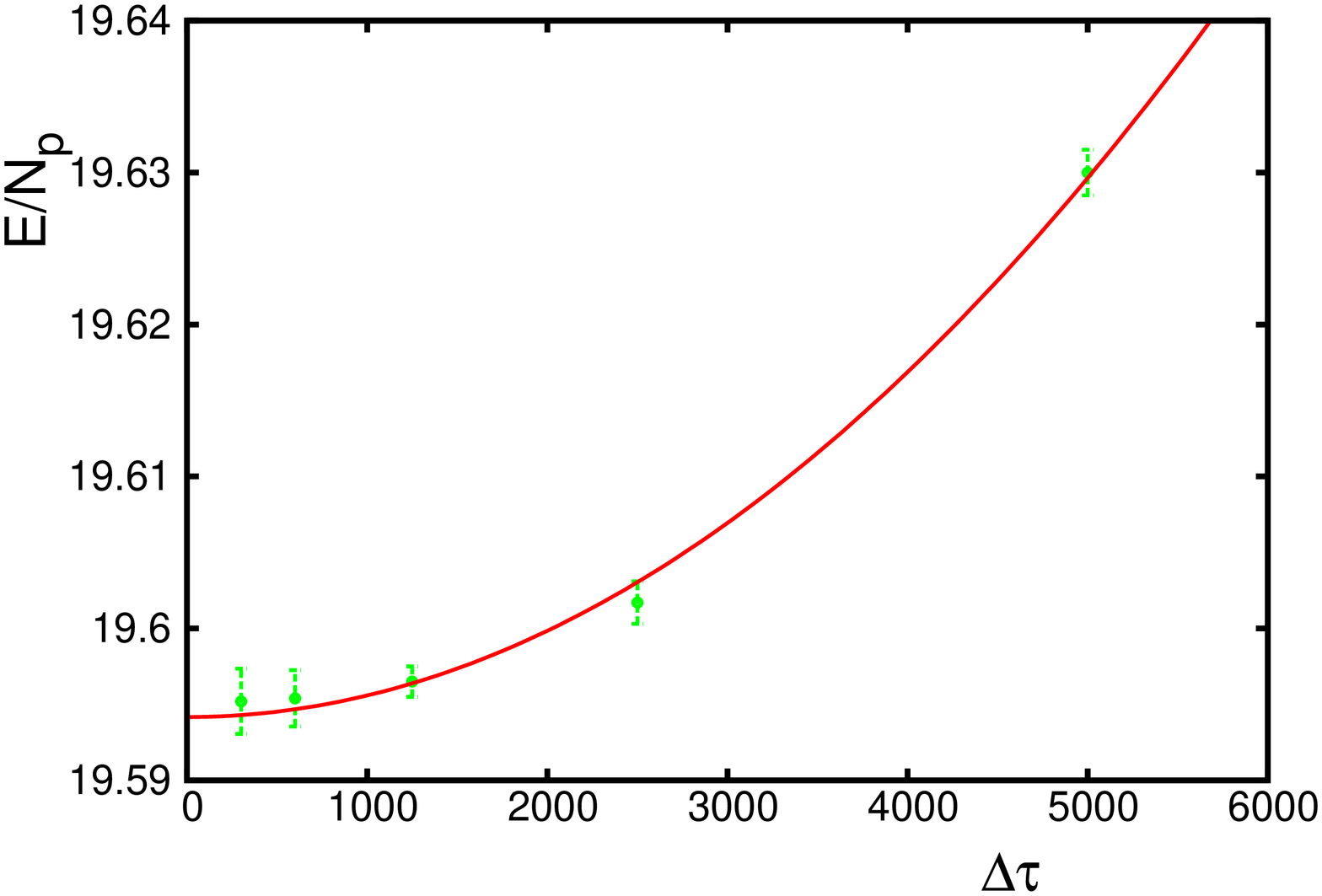}
\caption{Energy per particle $E/N_p$ (green points with errorbars) versus the time step $\Delta\tau$ for the Yukawa system in a liquid phase, 64 particles; red solid line for a quadratic fit. Energy 
units are $E_1=\hbar^2/ma^2$, time units are $\hbar/E_1 \cdot 10^{-6}$.}
\label{yuk_dt}
\end{figure}

Throughout all the DMC simulations that we run we use the second-order in time step approximation of the Green's function of the Hamiltonian, which means that the time step bias in the results is also $\mathcal{O}(\tau^2)$. This dependence can be observed by performing a set of trial calculations in order to find a 
value of $\Delta\tau$, when the time-related error complies with our accuracy goal. In Figure~(\ref{yuk_dt}) we plotted the energy per particle versus 
time step and extrapolated the data to 0 by a parabola. Here we can suggest that the acceptable time step is approximately 1000 (the required accuracy level is around the statistical noise for each single calculation). 

\begin{figure}
\centering
\includegraphics[width=0.75\textwidth, angle=-90]{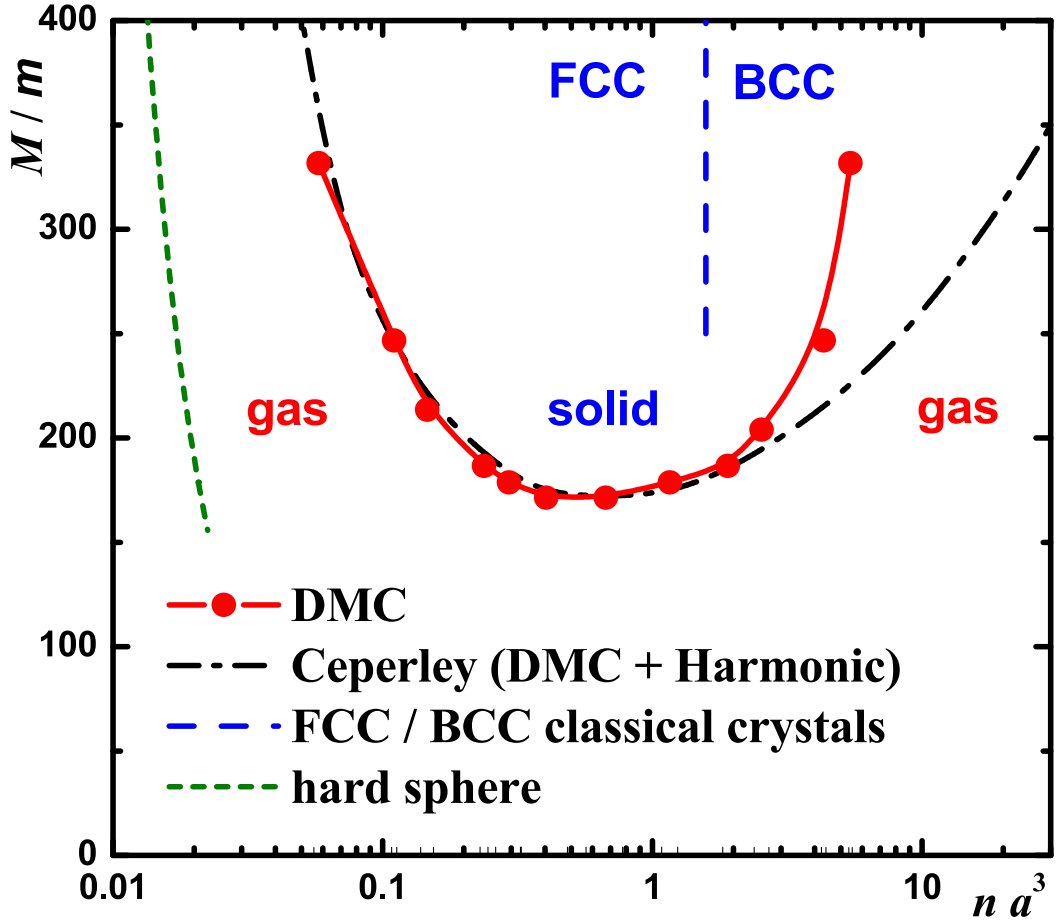}
\caption{Zero-temperature phase diagram of the Yukawa potential corresponding to the Hamiltonian in Eq.~(\ref{HYukawa}) in terms of the gas parameter $na^3$ and the mass ratio $M/m$. Red symbols: transition point as obtained from the double-tangent Maxwell construction applied to the Monte Carlo data energies extrapolated to the thermodynamic limit; dashed line: critical density $n a^3 = 1.58\ldots$ at which the energy of perfect fcc and bcc packings are equal; dash-dotted line: prediction of Ceperley \textit{et al.}~\cite{Ceperley1,Ceperley2} obtained by imposing a constant Lindemann ratio; short-dashed line: $na_s^3=0.24$~\cite{HS3D1,HS3D2,HS3D3}.}
\label{Fig:phase diagram}.
\end{figure}

 The resulting energies of the gas and solid phases are then analyzed using the double-tangent Maxwell construction which provides the melting and freezing densities. The zero-temperature phase diagram parameterized in terms of the dimensionless density $na^3$ and the mass ratio, is shown in Fig.~\ref{Fig:phase diagram}. We find that for mass ratios smaller than the critical value $M/m\approx 180$ the gas phase is energetically preferable at any density. On the other hand, for larger mass ratios there is always a gas-solid transition at low densities and a solid-gas transition at large ones. Energetically, both the fcc and bcc lattices are possible in the solid phase. It is very difficult to discern numerically which packing is preferred as the energies in different crystalline phases are extremely close. Still, in the large potential energy limit, corresponding to a mass ratio $M/m\gg 1$, it is enough to compare the potential energy of the classical crystals with different packings. A simple, geometrical construction assuming that particles are tightly tied to their equilibrium positions leads to a transition density $n a^3 \approx 1.58$. This prediction is depicted as a blue dashed line in Fig.~\ref{Fig:phase diagram}. In the low-density limit we numerically find the value of the $s$-wave scattering length $a_s$ of the Yukawa potential~(\ref{HYukawa}) and fit it as $a_s/a = 0.436 \ln(M/m)$ with accuracy below 1\% in the region of interest. Note that $a$ is the $s$-wave scattering length of fermionic particles which lead to the effective bosonic Hamiltonian~(\ref{HYukawa}) while $a_s$ is the $s$-wave scattering length between bosonic Yukawa particles. For the sake of comparison we also plot in Fig.~\ref{Fig:phase diagram} the gas-solid transition line of hard spheres of size $a_s$ given by $M/m = \exp(1.424/(n^{1/3}a))$.

The figure also shows the results of Ceperley \textit{et al.}~\cite{Ceperley1,Ceperley2} which were obtained by doing DMC calculations for three characteristic points in the phase diagram close to the solid-gas transition line. Overall, the agreement between that prediction and our results is good, the main differences affecting the region of large density where Coulomb effects are strong. To our best knowledge this is the first time that the high-density quantum solid-gas phase transition is observed in a simulation of Yukawa systems.

In the case of the fermionic molecules, the resulting critical mass ratio is much larger than $M/m\approx 13.6$ for which the system is unstable due to formation of Efimov states~\cite{theory:Efimov1,theory:Efimov2,theory:Efimov3,theory:Efimov4,theory:Efimov5}. The obtained phase diagram describes properties of metastable fermionic molecules while the true ground state corresponds to a many-body bound state. The stronger the effective interaction is (that is, the larger the mass ratio), the more distant are heavy fermions and the smaller the overlap with localized Efimov states is.

\section{Large mass ratios}
According to our results, the minimal mass ratio for which the crystalline phase can exist is $M/m\approx 180$ and it is achieved at the somewhat large value of the gas parameter $na^3\approx 0.3$. At these densities the fermionic nature of the molecules becomes important as the Hamiltonian~(\ref{H}) is derived under the assumption that $na^3\lesssim 1/8$~\cite{YukawaPot2D}. Our bosonic model is expected to be reliable at smaller densities where the critical mass ratio is further increased.

The mixtures of different fermionic atoms have already been successfully realized in experiments~\cite{exp:LiYb1,exp:LiYb2} but at significantly smaller mass ratios. Probably, the largest directly achievable mass ratio currently is that of Yb and Li atoms, $M/m = 29$, which is still much smaller than the critical mass ratio needed to observe the formation of an ultracold crystal.

An alternative way to realize a fermionic mixture with a large and variable mass ratio is to confine one of the components to an optical lattice. At low filling fraction the distances between atoms are large compared with the lattice spacing, and the separation of length scales allows the description of the movement of a particle in the lattice as that of a quasiparticle with an effective mass moving in a medium where the lattice is absent. In a deep lattice interactions between particles are much weaker than the confining energy and so, to a first approximation, one can consider that as the problem of a single particle diffusing in the lattice.

An optical lattice created by counter-propagating laser beams imposes an external potential  $V_{\text{latt.}}(x,y,z)=V_{0}\left(\sin^2{kx}+\sin^2{ky}+\sin^2{kz}\right)$ on every particle. The diffusion of a particle over a large distance is then governed by the tunneling rate between neighboring sites. The diffusion is largely suppressed (and the effective mass greatly increased) when the amplitude of the optical lattice is large, i.e. when $V_0\gg E_r$ with $E_r=\hbar^2k^2/2m$ the recoil energy. The excitation spectrum in the lowest band can be described by Bloch waves of quasi-momentum $\mathbf{q}$ and energy $\varepsilon_0(\mathbf{q})=\frac{3}{2}\hbar\omega_0 - 2J\left(\cos{q_xd}+\cos{q_yd} +\cos{q_zd}\right) +\ldots$ with $d=\pi/k$ the lattice constant~\cite{revmodBloch}. At small momenta the spectrum is quadratic in $q$ and can be interpreted as the spectrum $\varepsilon_0(q) = E_0 + \hbar^2q^2/2m^*$ of a free quasiparticle with an effective mass $m^*$. Within the lowest band approximation the effective mass is inversely proportional to the hopping parameter $J$,
\begin{equation}
\frac{m^*}{m} = \frac{1}{\pi^2}\frac{E_r}{J} \ .
\label{m:band}
\end{equation}
The tunneling is greatly suppressed in the deep optical lattice limit $V_0\gg E_r$. To better understand the contribution of the tunneling term in the present case, a semiclassical treatment within the Wentzel-Kramers-Brillouin(WKB) approximation can be used to calculate the tunneling probability $p$. One finds that it is proportional to 
\begin{equation}
p\sim J^2 \propto \exp\{-2\int_{x_1}^{x_2}\;dx\,\sqrt{2m(V(x)-E)}/\hbar\}\:,
\end{equation}
where $x_1$ and $x_2$ are the classical turning points. In the deep optical lattice limit 
one can assume $V(r) - E \approx V(r)$, with $x_1$ and $x_2$ corresponding to the positions of two neighboring minima. The resulting integral can be easily evaluated and predicts an exponential form $J \propto \exp(-\sqrt{V_0/E_r})$. A more precise expression can be obtained from the width of the lowest band in the 1D Mathieu-equation~\cite{revmodBloch}, yielding
\begin{equation}
J=\frac{4}{\sqrt\pi}E_r\left(\frac{V_0}{E_r}\right)^{3/4}
\exp\left\{-2\left(\frac{V_0}{E_r}\right)^{1/2}\right\} \, .
\label{J:band}
\end{equation}
This expression, together with Eq.~(\ref{m:band}), provides an analytic approximation for the effective mass $m^*$.

\begin{figure}
\begin{center}
\includegraphics[width=0.6\textwidth, angle=-90]{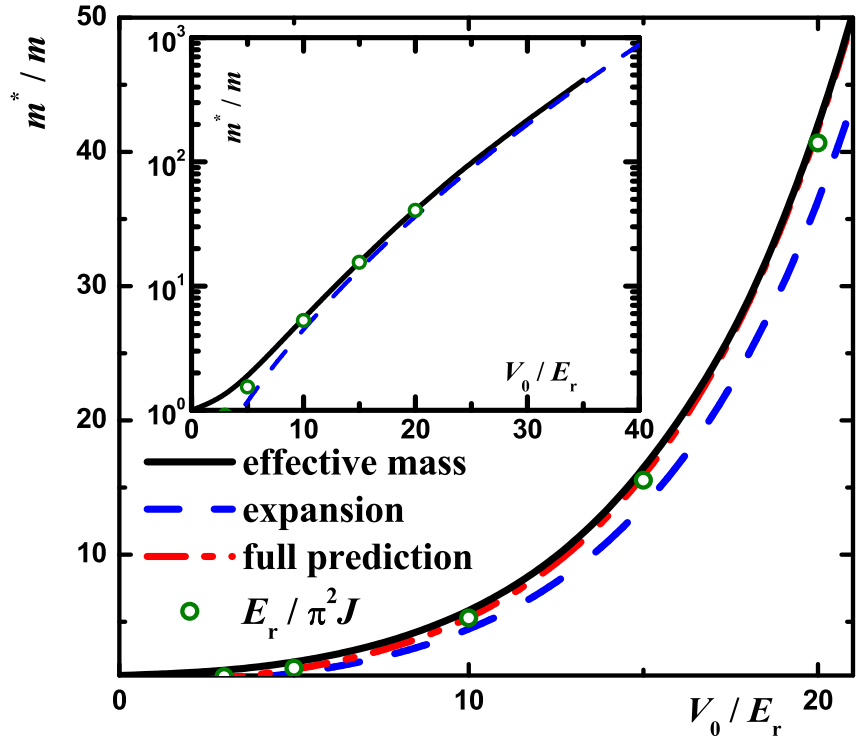}
\caption{ Effective mass as a function of the lattice amplitude $V_0$ in units of the recoil energy $E_r$. Solid line: results obtained from the diffusion constant evaluated by propagation in imaginary-time; circles: lowest band approximation of Eq.~(\ref{m:band}) with values of $J$ taken from~\cite{revmodBloch}; dashed line: same results with $J$ from the expansion in Eq.~(\ref{J:band}); dash-dotted line, same expansion with $V_0$ shifted by $-3/4 E_r$. Inset: same results on a semi-logarithmic scale.}
\label{Fig:meff}
\end{center}
\end{figure}

In order to determine the dependence of $m^*$ on the lattice parameters in a non-perturbative way we evaluate the diffusion constant $D$ of a real particle moving on the lattice and compare it to the diffusion constant $D_0 = \hbar^2/2m^*$ of a free quasiparticle of effective mass $m^*$. The diffusion constant is obtained by means of DMC propagation in imaginary time by measuring the mean-square displacement
$\langle({\rve}(\tau) - {\rve}(0))^2 \rangle = \langle(x(\tau) - x(0))^2 \rangle+\langle(y(\tau)-y(0))^2 \rangle+\langle(z(\tau)-z(0))^2 \rangle$ where ${\rve}=(x,y,z)$ denote particle coordinates. The diffusion constant is then extracted as $D = \lim\limits_{\tau \to \infty} \hbar \langle({\rve}(\tau) - {\rve}(0))^2 \rangle/(2\tau d)$, where $d=3$ is the system dimensionality. The resulting dependence of $m^*$ on the lattice amplitude is shown in Fig.~\ref{Fig:meff}. The figure shows the Monte Carlo prediction (solid line) compared with the approximation of Eq.~(\ref{m:band}) with $J$ taken from  Ref.~\cite{revmodBloch} (circles) and from Eq.~(\ref{J:band}) (dashed line). As it can be seen, there is an almost constant shift between $m^*$ obtained in the Monte Carlo simulation and Ref.~\cite{revmodBloch} compared to Eqs.~(\ref{m:band}-\ref{J:band}). We have found that the description in the relevant region of interest is very much improved by subtracting a constant shift $E^{(1)}=-3/4 E_r$ from $V_0$ in the argument of Eq.~(\ref{J:band}). This last prediction is shown by a thin line in Fig.~\ref{Fig:meff} and provides a good approximation for $V_0\gtrsim 10 E_r$.

One can understand these results in the following way: in the absence of the optical lattice the effective mass and the bare mass coincide, so $m^* = m$. As the amplitude $V_0$ of the lattice is increased, the particle movement is slowed down and the effective mass increases. In the deep optical lattice limit the effective mass grows as $m^*/m\propto\exp(\sqrt{V_0/E_r})$ and so the ratio can be made arbitrarily large by increasing the amplitude $V_0$ (for instance $m^*/m \sim 1000$ at $V_0/E_r = 40$; see the inset in Fig.~\ref{Fig:meff}). This mechanism allows for increasing the mass of one of the two components while keeping the other one unaltered, so that the ratio $M/m$ of the fermionic mixture can be made as large as desired when the mass of the heavy component is identified with the effective mass $m^*$. Consequently, and according to the phase diagram shown in Fig.~\ref{Fig:phase diagram}, there is a wide range of densities where one could find the system in the crystalline superlattice phase. Heights of optical lattices as large as $(35-60) E_r$ are readily achieved in current experiments~\cite{large_recoil1,large_recoil2} and correspond to sufficiently large effective mass ratios for the crystallization to be realized.

Both small density and large density transition lines are accessible for Yukawa interaction caused by screening in dusty plasma, colloids and neutron matter. On the contrary, in two-component Fermi gas only the left part of the phase diagram can be realized since the effective Yukawa interaction is valid only at low densities. In fact, the validity criterion for the interaction potential in Eq.~(\ref{H}) was studied in Ref.~\cite{YukawaPot2D} and was found to be well satisfied for distances larger then $r \approx 2a$ which leads to the condition $\rho a^3 \lesssim 1/8$ when $r$ is identified with the mean interparticle distance. In this way, for example, for $\rho a^3 = 0.1$ and mass ratio $M/m = 300$ the system is expected to be in a crystalline form. Much larger effective mass ratios can be achieved for realistic~\cite{large_recoil1,large_recoil2} lattice heights of $(35-60) E_r$. We thus conclude that by using an optical lattice, a fermionic mixture of very different mass components can be used to test the phase diagram of the equivalent Yukawa model.

\section{Conclusions}
To summarize, in this Chapter we have obtained the zero-temperature phase diagram of bosons interacting through Yukawa forces. We have used a diffusion Monte Carlo simulation starting from a very good approximation to the optimal variational ground-state wave function obtained by solving the corresponding Euler--Lagrange hypernetted chain equations. The resulting phase diagram is very similar to the one originally obtained by Ceperley and collaborators~\cite{Ceperley1,Ceperley2}, although significant differences arise at large densities. The phase diagram shows that any fermionic mixture of pure elements will always be seen in gaseous form, as the mass ratios required for crystallization of weakly bound fermionic molecules are far beyond the ones that can be achieved in nature. Finally, we investigate an alternative mechanism based on the confinement of one of the species to a deep optical lattice which exponentially increases its effective mass as a function of the confining amplitude. The  resulting mass ratio of the mixture created in this way can then be tuned at will and could be used to check experimentally the predicted phase diagram both in the gas and crystal (superlattice) phases.

\chapter{Phase diagram of Rydberg atoms \label{secRydberg}}
\section{Introduction}
Rydberg atoms have one electron excited to a high energy level.
Such atoms exhibit strong and highly tunable interactions which may
have an extraordinarily long range.
Optically excited from suspended clouds of cold atoms,
Rydberg atoms 
interact both between themselves and with the surrounding unexcited atoms, 
resulting in a rich behavior of the Rydberg systems.

Due to the strong interactions, a Rydberg atom shifts
the levels of nearby atoms sufficiently to prevent their subsequent excitation.
A large number of studies
deal with a \emph{local} blockade regime.
In such a regime a Rydberg atom blocks excitations in its vicinity, 
and the atomic clouds may be injected with well over $10^3$ Rydberg excitations
before the existing excitations block any further ones %
~\cite{Phau2007-EvidenceForCoherentCollectiveRydbergExcitationInTheStrongBlockadeRegime,%
Gould2004-LocalBlockadeOfRydbergExcitationInAnUltracoldGas,%
Weidemuller2004-SuppressionOfExcitationAndSpectralBroadeningInducedByInteractionsInAColdGasOfRydbergAtoms}.
Unfortunately, the arrangement of the excited atoms in such experiments is not directly accessible
and has been a subject of intense investigation.
Understanding the ordering of Rydberg atoms may be important 
for interpretation of the experimental results, for example
for the antiblockade effect predicted in~\cite{Pohl2007-ManyBodyTheoryOfExcitationDynamicsInAnUltracoldRydbergGas}.
It was also suggested that a spatially
ordered state may allow for a better control over quantum states in such experiments %
~\cite{Pohl2010-DynamicalCrystallizationInTheDipoleBlockadeOfUltracoldAtoms}.
Finally, there is an exciting possibility of observing phase transitions in these 
versatile systems, 
especially to states with long-range ordering~%
~\cite{Phau2008-QuantumCriticalBehaviorInStronglyInteractingRydbergGases,%
Pfau2009-UniversalScalingInAStronglyInteractingRydbergGas}.

Quantum many-body treatments attempting modelling of realistic Rydberg systems 
have been developed in the past %
~\cite{%
Gould2004-LocalBlockadeOfRydbergExcitationInAnUltracoldGas,%
Robicheaux2005-ManyBodyWaveFunctionInADipoleBlockadeConfiguration,%
Pohl2007-ManyBodyTheoryOfExcitationDynamicsInAnUltracoldRydbergGas,%
Cote2010-ManyBodyDynamicsOfRydbergExcitationUsingTheOmegaexpansion,%
Pohl2010-DynamicalCrystallizationInTheDipoleBlockadeOfUltracoldAtoms,%
Pohl2009-MesoscopicRydbergEnsemblesBeyondThePairwiseInteractionApproximation%
}, and were successful in reproducing a number of important experimental features 
~\cite{%
Gould2004-LocalBlockadeOfRydbergExcitationInAnUltracoldGas,%
Pohl2009-MesoscopicRydbergEnsemblesBeyondThePairwiseInteractionApproximation,%
Pfau2009-UniversalScalingInAStronglyInteractingRydbergGas,%
Weidemuller2010-EvidenceOfAntiblockadeInAnUltracoldRydbergGas,
Weidemuller2010-CoherentPopulationTrappingWithControlledInterparticleInteractions%
}.
Due to complexity, it is often difficult to consider long-range order with such calculations.
Nonetheless, strong short-range spatial correlations between Rydberg atoms 
were obtained in the calculations of Refs.~\cite{%
Robicheaux2005-ManyBodyWaveFunctionInADipoleBlockadeConfiguration,%
Weidemuller2010-EvidenceOfAntiblockadeInAnUltracoldRydbergGas,%
Cote2010-ManyBodyDynamicsOfRydbergExcitationUsingTheOmegaexpansion%
}, 
as the atoms avoid each other due to the blockade. 
Successful observation of the antiblockade effect
was also a demonstration  of a creation of the strong short-range correlations %
~\cite{Weidemuller2010-EvidenceOfAntiblockadeInAnUltracoldRydbergGas}.
Possibility of long-range ordering (crystallization) of Rydberg atoms was recently predicted 
for systems coupled to specially selected chirped laser pulses
~\cite{Pohl2010-DynamicalCrystallizationInTheDipoleBlockadeOfUltracoldAtoms,%
Kokkelmans2011-AdiabaticFormationOfRydbergCrystalsWithChirpedLaserPulses}.
Ordering was also considered, and crystalline phase found, in theoretical calculations of both 
one and two-dimensional optical lattices~%
~\cite{%
Lesanovsky2010-DynamicalCrystalCreationWithPolarMoleculesOrRydbergAtomsInOpticalLattices,%
Weimer2010-TwoStageMeltingInSystemsOfStronglyInteractingRydbergAtoms,%
Lesanovsky2011-TwoDimensionalRydbergGasesAndTheQuantumHardSquaresModel,%
Markus2011-DislocationMediatedMeltingOfOneDimensionalRydbergCrystals%
}.
Remarkable 
non-commensurate crystalline phases in optical lattices emerged in~Ref.~%
~\cite{%
Weimer2010-TwoStageMeltingInSystemsOfStronglyInteractingRydbergAtoms}.

Given the complex nature of the interactions in the Rydberg systems,
it is important to know how much of the behavior of
large assemblies of Rydberg atoms stems directly from the 
pair potential of the interaction between the atoms.
For this reason we aim to study ordering in the simplest model of the Rydberg systems.
Because of the large number of Rydberg-excited atoms in the experiments, we consider the thermodynamic limit. 
While the results are established in the thermodynamic equilibrium,
many present experiments with Rydberg atoms are too short to reach equilibrium. 
Thus comparison in such cases must be made cautiously.

\section{Model and methods}
The dominant interactions in the Rydberg systems are usually the
F\"orster-resonant dipole-dipole interactions between the excited atoms. 
It was shown by Walker and Saffman %
~\cite{Saffman2005-ZerosOfRydbergRydbergFosterInteractions, %
Saffman2008-ConsequencesOfZeemanDegeneracyForTheVanDerWaalsBlockadeBetweenRydbergAtoms} %
that, given a pair of Rydberg atoms in the same state, 
the interaction will not have zeroes as
a result of the hyperfine structure or alignment of the atoms
only if the resonant coupling is from the $s$ to $p$ states.
Furthermore, interactions in the $s+s\rightarrow p+p$
channels depend only weakly on the hyperfine structure of the $p$ states,
resulting in a nearly isotropic interaction, to within $10^{-2}$.
This perhaps in part motivates the use of the $ns$ Rydberg states in current experiments %
~\cite{Phau2007-EvidenceForCoherentCollectiveRydbergExcitationInTheStrongBlockadeRegime%
,Phau2008-RydbergExcitationOfBoseEinsteinCondensates%
,Pfau2009-UniversalScalingInAStronglyInteractingRydbergGas%
,Weidemuller2010-CoherentPopulationTrappingWithControlledInterparticleInteractions}.
Neglecting the hyperfine structure, the interaction for this resonance is isotropic and 
its matrix element is given  in terms of the F\"orster defect $\delta$ as
~\cite{Saffman2005-ZerosOfRydbergRydbergFosterInteractions}  
\begin{equation}
V(r)=\frac{\delta}{2}-\operatorname{sign}(\delta)\sqrt{\left(\frac{C_3}{r^3}\right)^2+\frac{\delta^2}{4}},
\end{equation} 
which changes from $V = C_3/r^{3}$ 
to van der Waals' $V = C_6/r^{6}$ (with $C_6=-C_3^2/\delta$) for distances much 
larger than the crossover $R_{3\rightarrow 6}\sim(-C_6/\delta)^{1/6}$.
In the case of a strong local blockade, the blockade radius is often larger than the crossover distance.
In such a case, the excited atoms are more likely to be found at 
distances where the interaction is already of the van der Waals type.

The above arguments motivate the repulsive van der Waals model
for the Rydberg atoms in the local blockade regime.
We disregard any energy transfer or interactions with the underlying gas of the ground-state atoms,
and particles are treated as spinless bosons in three-dimensional space with the many-body Hamiltonian
\begin{equation}
\mathcal{H}=-\frac{\hbar^2}{2m}\sum_i\nabla_i^2+\sum_{i<j}\frac{C_6}{{\left|\bm{r}_i-\bm{r}_j\right|}^6} .
\end{equation}
Defining the reduced units of length and energy as
\begin{equation}
r_0=\left(\frac{mC_6}{\hbar^2}\right)^\frac{1}{4} 
\; , \;
E_0=\frac{\hbar^3}{m^{3/2}C_6^{1/2}}
\label{eq:reducedunits}
\end{equation}
allows to describe the properties of this model universally in terms of just two parameters,
the dimensionless density $\rho r_0^3$ and temperature $\kBoltzmann T/E_0$.
The units are selected to satisfy $E_0=\hbar^2/mr_0^2=C_6/r_0^6$. 
The mass $m$ in Eq.~(\ref{eq:reducedunits}) is the mass of the atom. 

It is important to establish the applicability of the bulk phase diagram to finite systems. 
For a cloud of size $R$ and
number density $\rho$, the tail potential energy per particle can be estimated as $\rho C_6/R^3$.
In order for the phase transition to occur at the same parameters in the limited system 
as in a bulk one, it is sufficient that the missing potential energy is
much smaller than the kinetic energy.
In the case $T \kBoltzmann\gg E_0$, this reduces to $R\gg \sqrt[3]{\rho C_6 /T \kBoltzmann}$.
When $T \kBoltzmann\ll E_0$, kinetic energy is estimated as $\hbar^2\rho^{2/3}/m$  and thus $R\gg r_0 (\rho r_0^3)^{1/9}$.

\section{Results for the phase diagram}
\begin{figure}
\centering
\includegraphics[angle=270,width=0.9\textwidth]{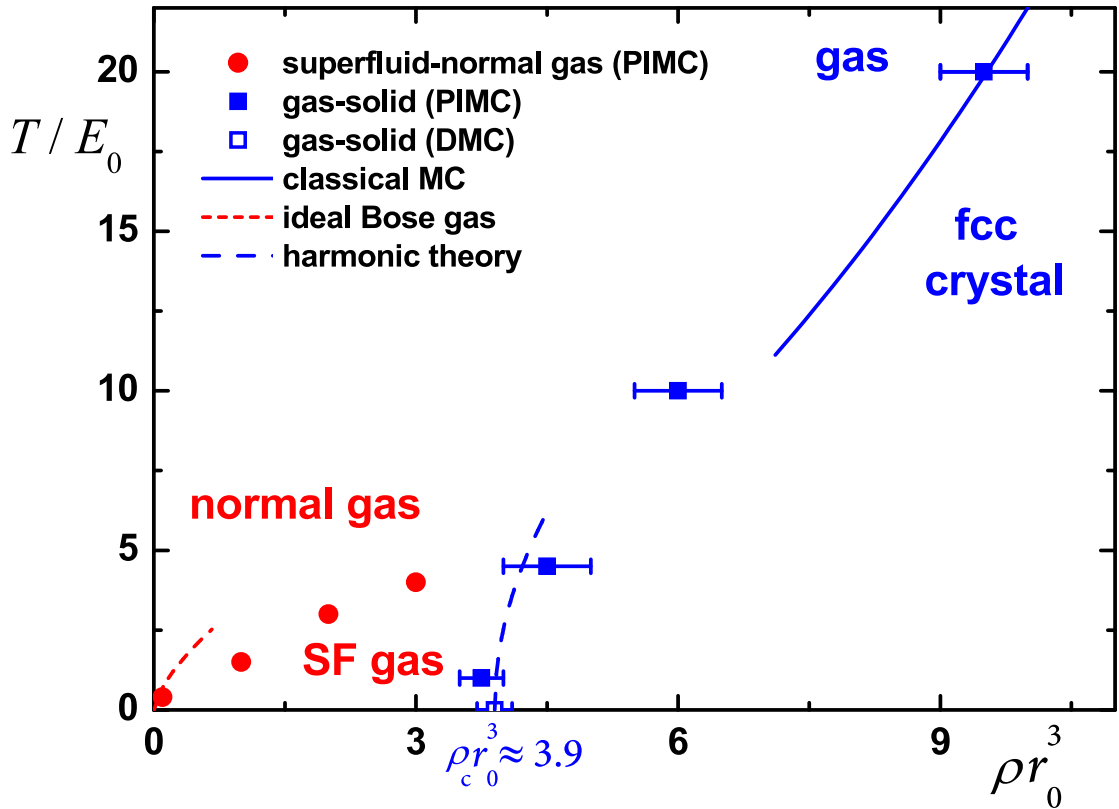}
\caption{\label{fig:phasediagram}  
Phase diagram of the repulsive $C_6/r^6$ interaction, scaled to units given in Eq.~(\ref{eq:reducedunits}).
Location of the gas-to-solid transition at zero temperature, determined with \abbrev{dmc}, 
is shown with an open blue square on the $T=0$ axis. 
Dashed blue line shows gas-to-solid transition as 
found with harmonic theory [Eq.~(\ref{eq:Tcharmonic})].
Solid blue line shows classical gas-to-solid transition [Eq.~(\ref{eq:Tclassical})].
Solid blue squares show location of the gas-to-solid transition as determined with \abbrev{pimc}.
Red short-dashed line marks the Bose--Einstein condensation of the ideal gas [Eq.~(\ref{eq:TcBECideal})].
Filled red bullets show Bose--Einstein condensation temperatures found with \abbrev{pimc}.
}
\end{figure}

The phase diagram of the model includes a solid 
at high densities and, at lower densities, a gas phase 
that Bose-condenses at sufficiently low temperatures~\cite{RydbergOsychenko}. 
To locate these phase regions, we employed a number of 
methods, each suitable in a certain area of the phase diagram.
At zero temperature, the model was treated with the diffusion Monte Carlo (\abbrev{dmc}), 
a projector method which provides an exact ground-state energy for bosonic systems (see Section~\ref{secQMC} of this Thesis).
\abbrev{Dmc} has been used successfully in the past to calculate the equations of state and 
locate quantum phase transitions for a variety of systems. 
Transitions at non-zero temperature were studied with path integral Monte Carlo (\abbrev{pimc})~\cite{Gillan90,Chakravarty97,Ceperley95,Sakkos09}, 
a first principles method which allows to compute the averages 
of quantum operators by summing over the quantum partition function of the system.
Both \abbrev{dmc} and \abbrev{pimc} methods allow to treat systems with several hundred 
particles under periodic boundary conditions, with thermodynamic limits obtained 
by a suitable extrapolation.
Additionally, classical limits were established with classical Monte Carlo calculations.
In two regimes the location of phase transitions 
could be expressed in a semi-analytical form. 
In the first case, the transition between superfluid and normal gas 
was expressed in terms of the scattering length of the potential by means of a known relationship.
In the second, the solid-to-gas transition was located at low temperatures with the harmonic theory.
The results are summarized in the phase diagram shown in Fig.~\ref{fig:phasediagram}.

\begin{figure}[htb]
\centering
\includegraphics[width=0.9\textwidth, angle=0]{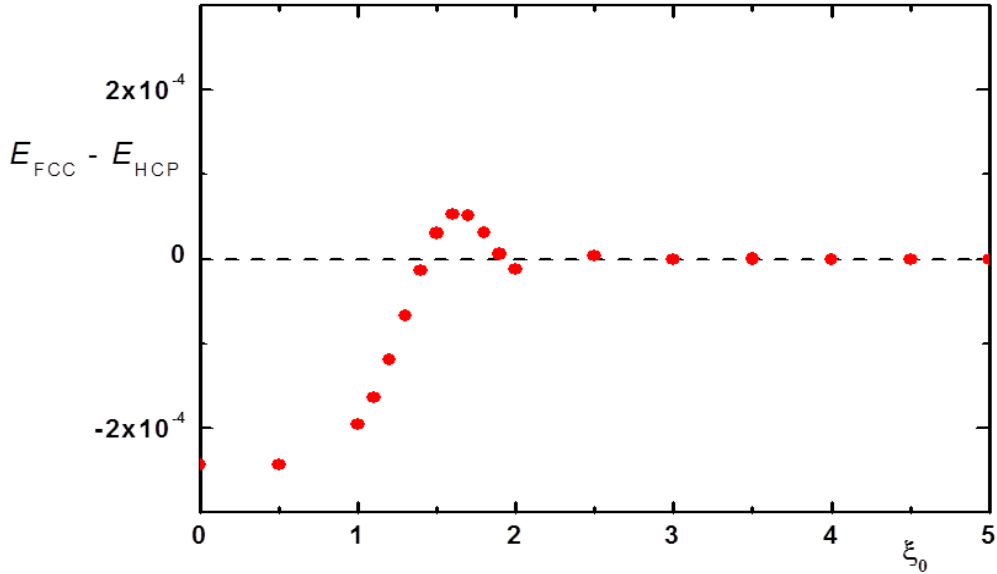}
\caption{Dependence of the difference of the Madelung energies of ideal fcc and hcp crystals in the units of $E_0$ on blockade radius $\xi_0$.}
\label{fccvshcp}
\end{figure}

At sufficiently high density, the atoms are expected to form a crystalline solid.
Summing the potential energy of the \emph{perfect} lattice structures, 
we conclude that the preferred symmetry is fcc.
While other structures may be excluded on the energetic grounds, 
the energy of the hcp structure is very close to that of the fcc.
The difference between the perfect crystal energies,
$E_\text{hcp}-E_\text{fcc}=2 \times 10^{-4} (\rho r_0^3)^2 E_0$,
is small enough to be comparable to or even swamped by the temperature effects 
in present experiments (for example, in works %
~\cite{Gould2004-LocalBlockadeOfRydbergExcitationInAnUltracoldGas%
,Phau2008-RydbergExcitationOfBoseEinsteinCondensates%
,Weidemuller2010-CoherentPopulationTrappingWithControlledInterparticleInteractions}).
The hcp phase is anticipated to be metastable with respect to the transition to the fcc phase.
Zero-point motion and temperature effects are expected to keep the fcc symmetry preferred to hcp.
If the dressed interaction~\cite{Li2012, Pohl2010-ThreeDimensionalRotonExcitationsAndSupersolidFormationInRydbergExcitedBoseEinsteinCondensates} in the form $V_{g-g}=1/(\xi_0^2+r^6)$ between the ground-state atoms is considered 
($\xi$ stands for a blockade radius), the preferable crystal packing can change, as shown in Fig.~(\ref{fccvshcp}).

In the rest, we assume the system crystallizes in the fcc structure.

Investigation on the zero-temperature line was done with the \abbrev{dmc} method~\cite{LesterBook1994,Boronat1994}. 
For importance sampling in the gas phase we used a Jastrow form 
\begin{equation}
f_{ij}(r)=\prod_{i<j} \{\exp\left[ -1/2 (b/r_{ij})^2 \right]+\exp\left[ -1/2 (b/(L-r_{ij}))^2 \right]\}\:,
\end{equation} 
$r_{ij}<L/2$, for a periodic box of size $L$.
The second power in $1/r$ arises from the cusp condition 
of the scattering problem with the repulsive $1/r^6$ potential
and is also compatible with the presence 
of long-wavelength phonons~\cite{Reatto67}.
The parameter $b$ was variationally optimized beforehand.
The Nosanow--Jastrow wave function was used for importance sampling 
in the solid phase~\cite{Nosanow1964,Boronat2008}.
It consists of the product of the above Jastrow term
and a site-localizing Nosanow term $\prod_i \exp\left[-(\bm{r}_i-\bm{l}_i)^2/2\gamma \right]$,
where $\bm{r}_i$ and $\bm{l}_i$ denote correspondingly the coordinates of the
atoms and lattice sites, and $\gamma$ is  
the second optimized parameter (for the detailed discussion see Section~\ref{secNosanow}). 
The breaking of exchange symmetry between particles in the solid affects the energy only negligibly~\cite{Boronat2009NJP}.
Within the statistical errors of the \abbrev{dmc}, results for the energies 
of the fcc and hcp lattices are indistinguishable and both are lower than the energies 
derived using bcc configuration.

\begin{figure}
  \begin{center}
    \begin{tabular}{c}
      \resizebox{120mm}{!}{\includegraphics[angle=270]{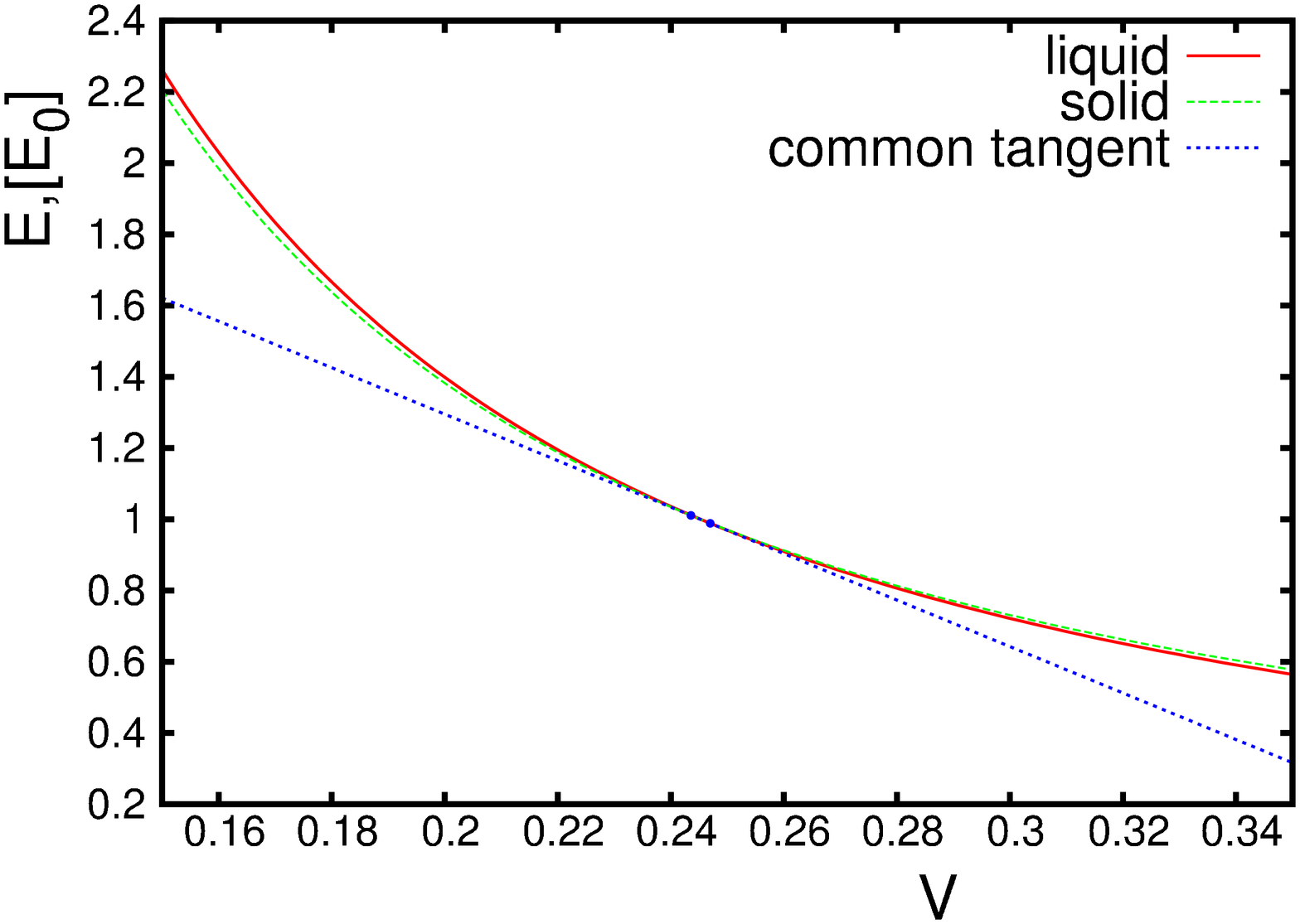}} \\
      \resizebox{120mm}{!}{\includegraphics[angle=270]{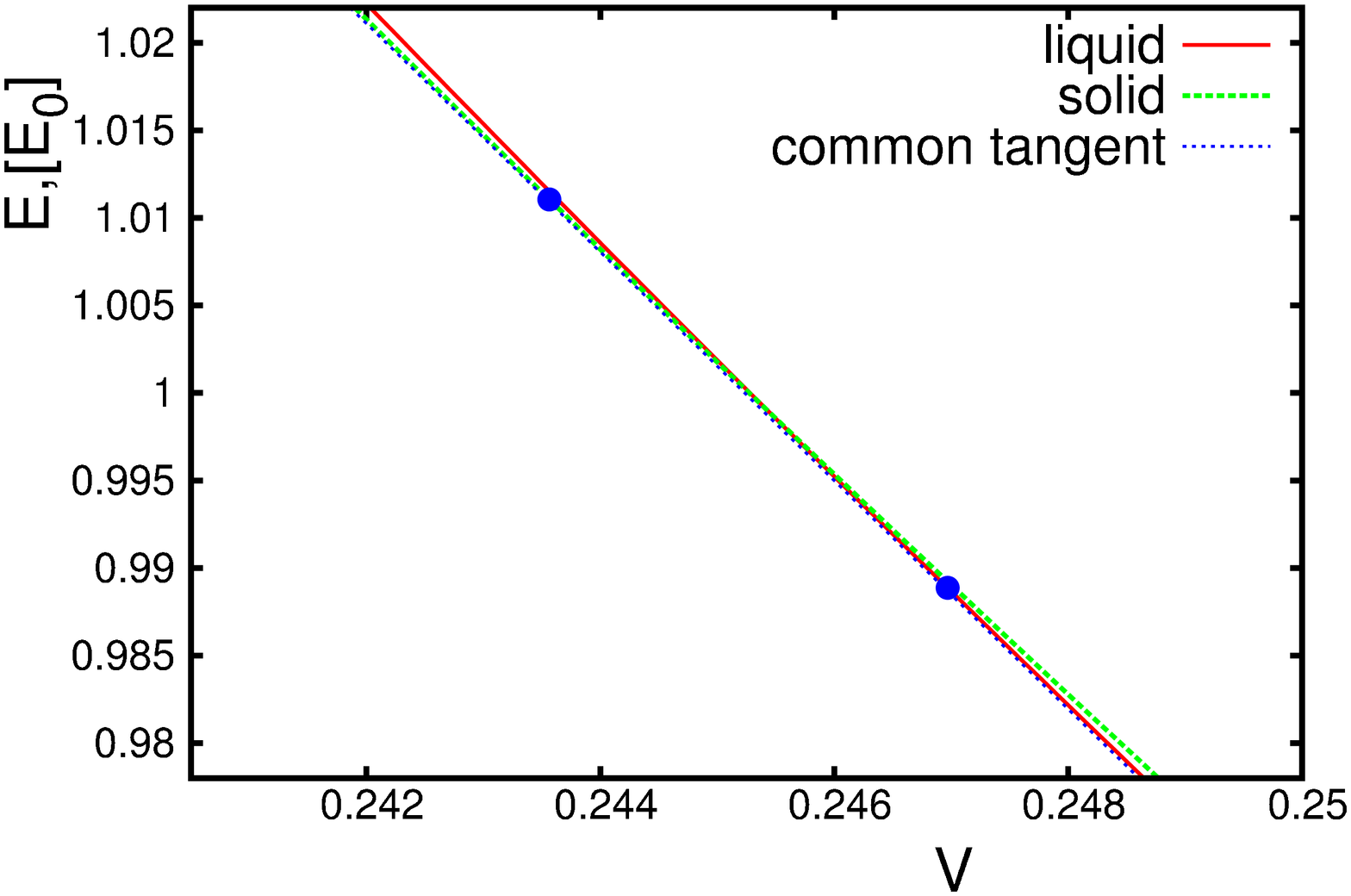}} \\
    \end{tabular}
    \caption{Study of double-tangent Maxwell construction. Equations of states $E(V)$ (the volume $V=1/\rho$, dimensionless) calculated with DMC for the system of 256 particles for liquid (red solid line) and solid (green dashed line) phases
    of the Rydberg system, normalized by the intersection energy $E_0$. The common tangent is a blue dotted line, blue circles stand for two tangent points. The upper and lower plots demonstrate the same data with diffenerent ranges of the inverse density. }
    \label{ry_mc}
  \end{center}
\end{figure}

While the phase transitions are conventionally reported as a function of pressure rather
than density, density of the Rydberg atoms is more accessible and controllable experimentally.
We therefore choose to express the transition locations in terms of density, 
even for the first-order solidification transition (in this case one needs to specify the coexistence region).
We find that the equations of state for the fcc solid and gas phases cross at the \emph{transition density}
\begin{equation}
\rho_c \, r_0^3=3.9 \pm 0.2,
\label{eq:TcDMC}
\end{equation}
expressed in the reduced units with the help of Eq.~(\ref{eq:reducedunits}).
The coexistence region of the solid and gas phase at zero temperature, 
determined using the double-tangent Maxwell construction, is narrow 
and is in fact smaller than the above error for the transition density (which arises mostly 
from the extrapolation to the thermodynamic limit; calculations were performed with up to 256 particles).
The double-tangent Maxwell construction\footnote[1]{For a detailed discussion on the topic see, for instance,
~\cite{Huang2005}} is a standard procedure to describe the phase transitions. It relies on equality of chemical 
potentials of two phases at the same pressure in the coexistence region. 
Alternatively it can be presented as
a linear dependence of
the Helmholtz free energy $A$ on the volume $V=1/\rho$ in the coexistence region:
\begin{equation}
\frac{\partial A_1}{\partial V}(V_1)=\frac{A_2-A_1}{V_2- V_1}
\end{equation} 
and a demand of a constant pressure
\begin{equation}
\frac{\partial A_1}{\partial V}(V_1)=\frac{\partial A_2}{\partial V}(V_2)
\end{equation} 
that is equivalent to finding of a common tangent of the equations of two states (at $T=0$ free energy 
is equal to the total energy of the system). The Maxwell construction at zero temperature is given in Fig.~(\ref{ry_mc}). 
The lines cubic polynomial fits to the Monte Carlo data.
The

Although the coexistence region is relatively narrow, 
its width can be obtained with a satisfactory accuracy. Let us explain how it is
 done in our case. First of all we need to evaluate the effect of the statistical uncertainty on 
 these results. The Monte Carlo data for the equation of state of the liquid and
 crystalline phases, with the statistic errors for each point 
(it is convenient to work with volume $V=1/\rho$ as an argument), is approximated with a cubic polynomial. 
On the first step, the value of energy at each point is moved by a random Gaussian shift
  with the variation, equal to the corresponding statistical error. Then, the coefficients of the optimal  
 polynomial approximations for $E_{\rm liquid}(V)$ and $E_{\rm solid}(V)$ are recalculated, and therefore
 one finds new positions of the melting and crystallization points, the intersection point and the 
 width of the coexistence zone. If this procedure is carried out many times, one can obtain the variance
 of each of the values. In this case the error is approximately 1.5\%. A separate source of the error in the estimation of the 
 width of the coexistence zone is the finite-size dependence of the results, that may be estimated 
 by extrapolating the data to $N_p\rightarrow \infty$. 
The relative error of this estimation is of the same order, as for the phase transition density, given above, that is $\sim 5\%$. 
The overall result for the width of the coexistence region can be evaluated as 
\begin{equation}
 \Delta \rho = 0.056\pm 0.003\,.
\end{equation} 

The transition line between solid and gas phases at small temperatures
can be determined with the harmonic theory~\cite{Ashcroft}, 
assuming the Lindemann ratio remains unchanged on the transition line. 
The value of the Lindemann parameter at melting 
may be extracted from the \abbrev{dmc} calculations of the transition density at zero temperature. 
The resulting low-temperature dependence of the gas-to-solid transition density is given by
\begin{equation}
T^{\text{harmonic}}_c=C\sqrt{\left(\rho-\rho_c\right)r_0^3}\frac{E_0}{\kBoltzmann},
\label{eq:Tcharmonic}
\end{equation}
where $\rho_c$ is the transition density at zero temperature, Eq.~(\ref{eq:TcDMC}), and
the constant $C=8.0$ is determined numerically from the dispersion 
curves of the solid and depends on the interactions and geometry of the fcc lattice.

A quantum solid melts at lower temperatures 
than the classical one due to the zero-point motion of the atoms.
The classical transition was located in the canonical ensemble 
by Metropolis sampling of the Boltzmann factor. %
As the potential energy $C_6/r^{6}$ is exactly proportional 
to the square of the density, the transition temperature 
for the classical system also scales exactly as $T\propto\rho^2$. 
We find that 
\begin{equation}
T_c^\text{classical}=0.22\left(\rho r_0^3\right)^2 \frac{E_0}{\kBoltzmann}.
\label{eq:Tclassical}
\end{equation}
As expected, such scaling removes the Planck constant from the classical transition temperature,
which in fact simplifies to $T_c^\text{classical} \kBoltzmann=0.22 \rho^2 C_6$.

To fully account for quantum effects, the gas-to-solid transition at $T\ne 0$
was also located with \abbrev{pimc} calculations. 
We used a decomposition of the action operator that is accurate 
beyond the fourth order
~\cite{Chin2004-QuantumStatisticalCalculationsAndSymplecticCorrectorAlgorithms}.
For details of the method and implementation, 
see Ref.~\cite{Sakkos09}. 
The transition was located  by observing 
melting or solidification while working in the canonical ensemble, 
beginning with configurations of atoms placed on a randomly distorted lattice.
Used in this way, the calculations determine a range in which the transition density is located.
\abbrev{Pimc} results confirm the validity of the harmonic approximation at low temperatures.
At higher temperatures the transition density follows the classical melting curve (\ref{eq:Tclassical}).

The above results establish the solidification transition of the repulsive van der Waals model.
Additionally, the dynamic nature of the Rydberg gas
raises a possibility for the spatial ordering to
be induced kinetically, as the combination 
of decay and strong blockade
will favor supplanting excitations to be equidistant
from their immediate neighbors. 
We modelled such a process and observed that 
replacement of decaying excitations in local blockade regime
indeed creates a short-distance order,
but not a true long-distance crystalline ordering.
These finding are consistent with much more elaborate dynamic models of 
Refs.~\cite{%
Robicheaux2005-ManyBodyWaveFunctionInADipoleBlockadeConfiguration,%
Weidemuller2010-EvidenceOfAntiblockadeInAnUltracoldRydbergGas,%
Cote2010-ManyBodyDynamicsOfRydbergExcitationUsingTheOmegaexpansion%
}.

At low temperature, the gas phase of the model is expected to form a Bose--Einstein condensate (\abbrev{bec}). 
Transition between the \abbrev{bec} and normal gas phases
at low densities lies slightly above  the ideal Bose gas condensation temperature,
\begin{equation}
T_\text{\abbrev{bec}}^\text{ideal}=2\pi\left(\frac{\rho r_0^3}{2.612\dots}\right)^{2/3}\frac{E_0}{\kBoltzmann},
\label{eq:TcBECideal}
\end{equation} 
due to the repulsive interaction 
between particles~\cite{Giorgini2008-CriticalTemperatureOfInteractingBoseGasesInTwoAndThreeDimensions}. 
The correction is governed by the scattering 
length of the potential $a_s$, which can be found to be equal to
$a_s=2\, \Gamma\left(3/4\right)/ \, \Gamma\left(1/4\right) r_0=0.676\dots r_0$.
The transition temperature is then given by 
$
T_\text{\abbrev{bec}}=T_\text{\abbrev{bec}}^\text{ideal}\left(1+c a_s \rho^{1/3} \right)
$,
where $c$ is a positive constant of the order of unity 
(for details, see Ref.~\cite{Giorgini2008-CriticalTemperatureOfInteractingBoseGasesInTwoAndThreeDimensions} and references therein). 
In the present case this expression is only valid at
very low densities 
(one needs to satisfy at least $\rho r_0^3 < 5\times 10^{-2}$ to make the description 
in terms of the zero-momentum scattering length meaningful), 
where the magnitude of the correction is not significant.

At higher densities the \abbrev{bec}-to-normal gas transition is no longer universal 
and depends on the form of the potential. 
We determined the location of this second-order transition with the \abbrev{pimc} method 
by calculating the superfluid transition from the winding number estimator~\cite{Ceperley1987-PathIntegralComputationOfSuperfluidDensities}.
The \abbrev{pimc} calculations show that at higher densities 
the interactions deplete the condensate 
and the transition temperature is lower than for the ideal Bose gas.
Combining the \abbrev{pimc} results, the region
in which the triple point is located was determined 
as 
$4.5 < T/(E_0/\kBoltzmann)   < 6.5$ and $4 < \rho r_0^3 < 5$, 
which we consider sufficiently narrow for practical considerations.

\section{Comparison with experimental conditions}

Because the interaction constant $C_6$ enters 
the reduced units [Eq.~(\ref{eq:reducedunits})], the effective
temperature and density can be varied over many orders of magnitude.
Most of the present experiments are deeply in the ``classical''
region of the phase diagram (Fig.~\ref{fig:phasediagram}).
As an example, we consider the conditions of the experiments presented 
in Ref.~\cite{Phau2008-RydbergExcitationOfBoseEinsteinCondensates}.
For the excitation with 170~ns laser pulses, the system parameters 
at $4~\mu\text{K}$ are $T/(E_0/\kBoltzmann)\approx33\times10^{5}$ 
and $\rho r_0^3\approx1.9\times10^3$, which in fact correspond to 
the gas phase of the equilibrium phase diagram.
For 320~ns excitation pulses and $T=1~\mu\text{K}$, 
$T/(E_0/\kBoltzmann)\approx8.2\times10^{5}$ and $\rho r_0^3\approx7.4\times10^3$, well below 
the gas-to-solid transition.
Therefore, the achievable temperature and density 
are already in the range suitable for investigating the equilibrium phase diagram.
Increasing the excitation number increases the interaction 
constant $C_6$ and moves the system deeper 
into the classical regime where the gas and solid phases are
separated by the simple condition of Eq.~(\ref{eq:Tclassical}).
The quantum regime of the phase diagram may be accessed by decreasing 
the excitation numbers or increasing the F\"orster defect $\delta$.
 
Whether Rydberg atoms in actual experiments will reach or even approach
an equilibrium phase depends on their lifetime, 
the experiment's duration and availability of a relaxation mechanisms.
Because of the short lifetimes of the Rydberg states, 
most current experiments are performed on such short 
timescales as to make the thermal motion negligible. 
It is therefore said that the experiments are performed with Rydberg excitations of a frozen gas.
If the experiments are extended closer to the currently achievable 
lifetimes of the Rydberg states, which can be as large as 100~$\mu$s %
~\cite{Robicheaux2008-SpatiallyResolvedObservationOfDipoleDipoleInteractionBetweenRydbergAtoms,%
Pfau2009-UniversalScalingInAStronglyInteractingRydbergGas},
some degree of thermal equilibration will already be achieved.
Besides the thermal motion there are, however, at least two other kinds of motion that may 
need to be considered.
The first one is the motion of the excited atoms due to the strong forces between them. 
The characteristic timescale associated with such a motion is the time that it 
takes for a Rydberg atom to travel the mean distance 
between Rydberg atoms. Given the mean distance $\xi\approx\rho^{-1/3}$  and the imbalance force of the order of $C_6/\xi^6$, this time is given by
\begin{equation}
t_\text{ballistic}\sim\sqrt{\frac{m}{C_6 \rho^{8/3}}}\:.
\label{eq:ballistictime}
\end{equation}
For a small fixed number of Rydberg excitations time (\ref{eq:ballistictime}) decreases rapidly 
with the excitation number $n$ as $n^{-11/2}$; 
as the local blockade is reached, $\rho \propto C_6^{-1/2}$, and $t_\text{ballistic}$ instead grows as $n^{11/6}$. 
For example, Rydberg systems created by the $1.970~\mu\text{s}$ pulses from $1\mu K$ gas
in the experiment of Ref.~\cite{Phau2008-RydbergExcitationOfBoseEinsteinCondensates} have 
$t_\text{ballistic}\approx12~\mu\text{s}$.
For the setup of Ref.~\cite{Pfau2009-UniversalScalingInAStronglyInteractingRydbergGas},
$t_\text{ballistic}\approx60~\mu\text{s}$ while the clouds 
could be successfully studied for as long as $20~\mu\text{s}$.
Collisional ionization and heating could potentially hamper such relaxation~\cite{Weidemuller2007-ModelingManyParticleMechanicalEffectsOfAnInteractingRydbergGas}.

\section{Excitation spectrum of a classical crystal}
In the previous parts of this Section the properties of a quantum system comprised of Rydberg atoms at zero and finite temperature were investigated, and the quantum phase diagram was presented.

\begin{figure}[h!]
  \centering
\includegraphics[scale=0.75]{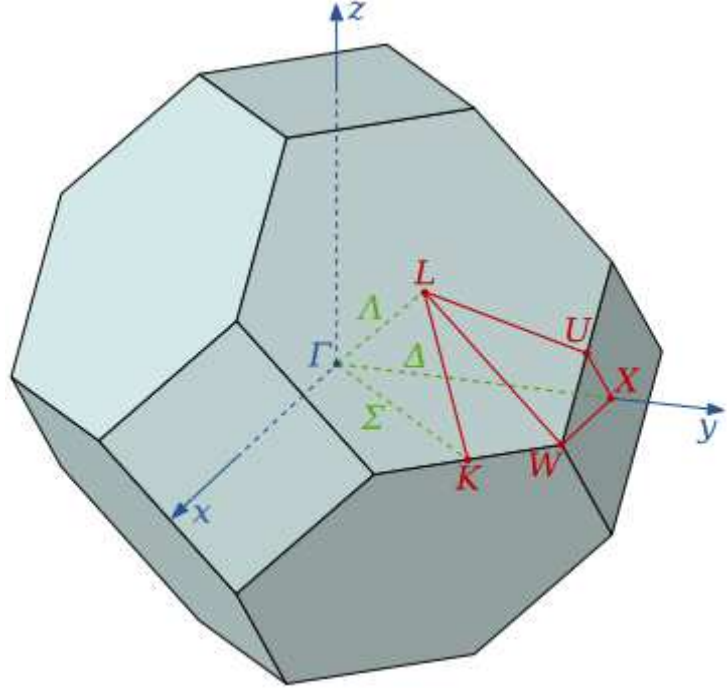}
  \caption{Brillouin zone of fcc crystal packing. Figure is taken from Wikipedia webpage 
{\it http://en.wikipedia.org/wiki/Brillouin\_zone}.}
\label{fccbrill}
\end{figure}

In the present subsection, we consider a classical system, governed by the van der Waals interaction $1/r^6$ and look for its excitation spectrum in the low temperature regime (or equivalently in the regime of very strong interactions).
In accordance with the comparison of classical potential energies of different kinds of crystal lattice, for $T=0$ we chose the face-centered cubic (fcc) lattice as preferable, with the Madelung energy of the hexagonal close pack (hcp) being only slightly higher. In practice the fcc packing is also advantageous, since its elementary cell is ``cubic'', and its first Brillouin zone has central, plane and axis symmetry.\footnote[1]{The first Brillouin zone is a primitive cell, not reducible to a smaller one by translations of the lattice in reciprocal space} The first Brillouin zone of fcc formation is known to be a truncated octahedron with the critical points K,L,U,W,X (see Fig.~\ref{fccbrill}).

The Hamiltonian of the system takes the form
\begin{equation}
H=-\sum_{i=1}^N \nabla^2 + \sum_{i>j}\frac{C_6}{|\bs{r}_i-\bs{r}_j|^6}\,.\label{rydbhamspec}
\end{equation}
We consider the case of low kinetic energy, that is when the interaction strength constant $C_6$ is large enough, and the system may be treated in a perturbative manner within a harmonic approximation.
The positions of the crystal nodes can be fixed at $\bs{R}_i$. In order to find the excitation spectrum of this system we can follow the procedure of Refs.~\cite{Bonsall59, Mora2007}. 
First of all, we give to the particles small arbitrary displacements $\bs{u}_i$ from the nodes. The potential energy is then represented by
\begin{equation}
E_p=\frac{C_6}{2}\sum_{i\neq j}\frac{1}{|\bs{r}_{ij}+\bs{u}_{ij}|^k}\label{epotharm1}
\end{equation}
with the notations $\bs{r}_{ij}=\bs{r}_{i}-\bs{r}_{j}$, $\bs{u}_{ij}=\bs{u}_{i}-\bs{u}_{j}$ and $k=6$.

Then, we perform an expansion of the potential energy $E_p$ in powers of the displacements $u_i^{\alpha}$, 
where $\alpha$ labels the Cartesian coordinates $\alpha=x,y,z$ and $U$ stands for the Madelung energy of a chosen crystal packing.
\begin{equation}
E_p=U+\frac{1}{2}\sum_{i\alpha}\sum_{j \beta}\Phi_{\alpha \beta}(\bs{r}_{ij})u_i^{\alpha}u_j^{\beta}+...\label{epotharm2}
\end{equation}
The first-order term here vanishes due to the fact that $\{\bs{R}_i\}$ is a minimum-energy configuration. The Hessian matrix $\Phi_{\alpha \beta}(\bs{r}_{ij})$ is obtained through a simple double differentiation:
\begin{eqnarray}
\Phi_{\alpha \beta}(\bs{r}_{ij})&=&-\frac{C_k}{2}\left(\frac{k(k+2)r_{ij}^{\alpha}r_{ij}^{\beta}}{r_{ij}^{k+4}}-
\frac{k \delta_{\alpha \beta}}{r_{ij}^{k+2}}\right), \:\:{\rm if }\:\:\:\: i\neq j,\label{phiharm1}\\
\Phi_{\alpha \beta}(\bs{r}_{ii})&=&\frac{C_k}{2}\sum_{j\neq i}\left(\frac{k(k+2)r_{ij}^{\alpha}r_{ij}^{\beta}}{r_{ij}^{k+4}}-
\frac{k \delta_{\alpha \beta}}{r_{ij}^{k+2}}\right)\,.\label{phiharm2}
\end{eqnarray}

The Fourier transformed dynamical matrix $C_{\alpha \beta}(\bs{q})$ 
is related to $\Phi_{\alpha \beta}(\bs{r}_{ij})$ by the expression

\begin{equation}
C_{\alpha \beta}(\bs{q})=\frac{1}{m}\sum_{i,j}\Phi_{\alpha \beta}(\bs{r}_{ij})\exp[-\ii \bs{q} \bs{r}_{ij}] \label{dynmatrix1}
\end{equation}
where $m$ stands for the mass of a particle in the kinetic term of the Hamiltonian.
In order to simplify the treatment and get rid of the conditional definitions for $\Phi_{\alpha \beta}$,
we introduce the matrix $S_{\alpha \beta}(\bs{q})$ such that
\begin{eqnarray}
S_{\alpha \beta}(\bs{q})=\sum_{i \neq j}\left(\frac{k(k+2)r_{ij}^{\alpha}r_{ij}^{\beta}}{r_{ij}^{k+4}}-
\frac{k \delta_{\alpha \beta}}{r_{ij}^{k+2}}\right)\exp[-\ii \bs{q} \bs{r}_{ij}] \label{dynmatrix2}
\end{eqnarray}
then it can be seen that
\begin{equation}
C_{\alpha \beta}(\bs{q})= -\frac{C_k}{2m}(S_{\alpha \beta}(\bs{q})-S_{\alpha \beta}(0))\label{dynmatrix3}
\end{equation}
The dynamical matrix can be used to obtain the normal mode frequencies by solving the eigenvalue problem
\begin{equation}
C_{\alpha \beta}(\bs{q})\bs{e}(\bs{q})=\omega^2\bs{e}(\bs{q}).
\end{equation}
Since $C_{\alpha \beta}$ is a real symmetric 3x3 matrix, 
there exists a complete and orthonormal set of three eigenvectors, 
written in which the Hamiltonian is diagonal. The wave vectors $\bs{q}$
are taken from the irreducible Brillouin zone.

For our case of the chosen model potential between Rydberg atoms $k=6$, one has to evaluate the following sums
\begin{eqnarray}
S_{\alpha \beta}(\bs{q})&=&
\sum_{i \neq j}
\left(\frac{48 r_{ij}^{\alpha}r_{ij}^{\beta}}{r_{ij}^{10}}-
\frac{6 \delta_{\alpha \beta}}{r_{ij}^8}\right)\exp[-\ii \bs{q} \bs{r}_{ij}] 
\label{dynmatrix4}\\
S_{\alpha \beta}(0)&=&
\sum_{i \neq j}\left(\frac{48 r_{ij}^{\alpha}r_{ij}^{\beta}}{r_{ij}^{10}}-
\frac{6 \delta_{\alpha \beta}}{r_{ij}^8}\right)
\label{dynmatrix40}
\end{eqnarray}
The Hamiltonian is rewritten in dimensionless form as
\begin{equation}
H=-\frac{1}{2 r_s}\sum\limits_{i=1}^N \Delta_i^2 + \sum\limits_{i<j} \frac{1}{|\rve_i-\rve_j|^6}
\label{hrs}
\end{equation} 
where distance and energy units are 
\begin{eqnarray}
r_0=1/\rho^{1/3}\label{r0rydspec}\\
E_0=C_6/r_0^6\:, \label{e0rydspec}
\end{eqnarray} 
 and the following form for the dimensionless parameter $r_s$
\begin{equation}
\frac{1}{r_s}=\frac{\hbar^2}{m r_0^2 E_0}=\frac{\hbar^2 }{ m C_6 \rho^{4/3}}\label{1divrs}
\end{equation} 
which characterizes the ratio between the kinetic and potential energies. 
It can be noticed that the parameter $r_s$ plays the role 
of the mass $m$ in Eq.~(\ref{dynmatrix1}). 
The units of frequency of the normal modes are taken as 
\begin{equation}
\omega_0=\frac{1}{\sqrt{r_s}}=\frac{\hbar}{ (m C_6)^{1/2} \rho^{2/3}}\,.\label{omega0}
\end{equation}
In reduced units the problem can be rewritten as 
\begin{eqnarray}
\omega^2(\bs{q})\bs{u}=\Phi_{\alpha\,\beta}(\bs{q})\bs{u}	\label{exdimless}
\end{eqnarray}
with
\begin{eqnarray}
\Phi_{\alpha\,\beta}(\bs{q})=\frac{\partial^2}{\partial u_{\alpha}\partial u_{\alpha}}
\left(\sum\limits_{j\neq 0}\frac{1-\exp(\ii \bs{q}\rve_j)}{|\bs{r}_j+\bs{u}|^6}\right)\left|_{\bs{u}\rightarrow 0}\right.\,.\label{phiab}
\end{eqnarray}
The kernel $\Phi_{\alpha\,\beta}(\bs{q})$ can be calculated by means of the Ewald summation technique. 

\begin{figure}[htb]
\centering
\includegraphics[width=0.85\textwidth, angle=0]{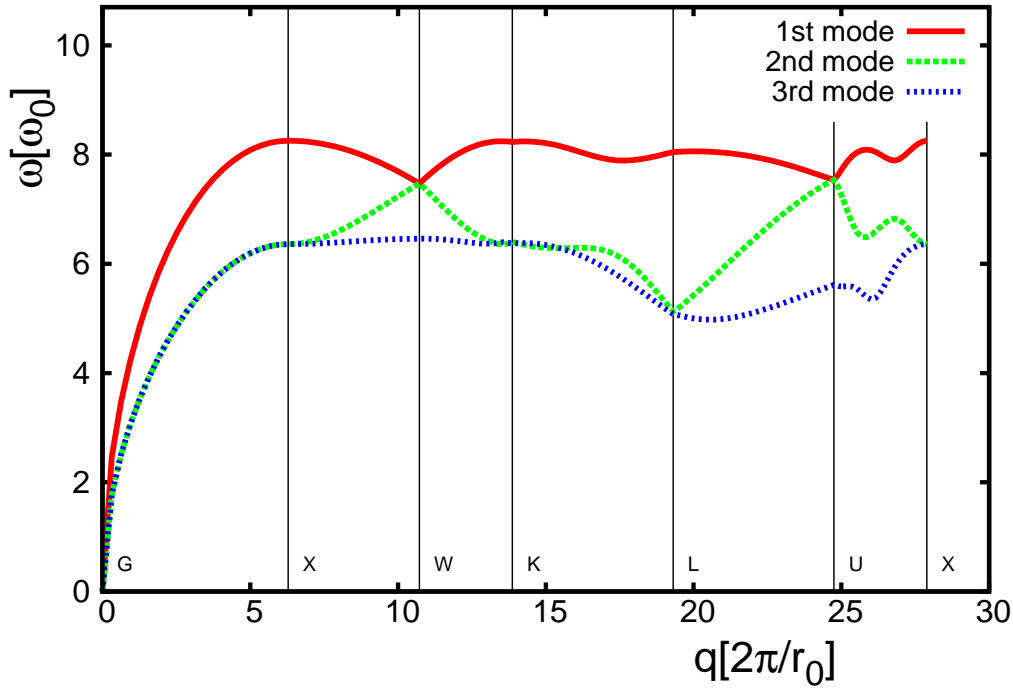}
\caption{Excitation spectrum of the van der Waals $1/r^6$ interaction. The wave vector $\bs{q}$ follows the path G,X,K,L,U,W,X in the first Brillouin zone with the metrics of the vector preserved. The units for frequency and wave vector are $\omega_0$ and $2\pi/r_0$ respectively, see Eqs.~(\ref{r0rydspec})~and~(\ref{omega0}).}
\label{r6spec}
\end{figure}

The excitation spectrum in this units is shown in Fig.~\ref{r6spec}.
In our description the three distinct branches of the spectrum can be degenerate (the second and the third coincide on the path $GX$). 
For small values of momentum $\bs{q}$ the excitations are in phononic (linear) regime
modes with the sound velocity $c=\omega/|\bs{q}|$, which is dependent on a direction in a crystal. 
For $GX$ direction in the crystal two sound velocities
(of two coinciding transverse modes and one longitudinal mode) are equal to 
\begin{eqnarray}
c_{1}&=&19.41\nonumber\\
c_{2,3}&=&10.04\nonumber
\end{eqnarray} 
in the units of $\omega_0 r_0$.
It can be noticed that the longitudinal mode is also phononic unlike the case of the 
three-dimensional Wigner crystal,
where a typical long-wavelength behavior of the frequency is $\omega_l(q)\sim \sqrt{q}$.
The spectrum is periodic as one reaches a border of the first Brillouin zone,
the values of the frequencies are also continuous, as seen in Fig.~\ref{r6spec}, when a momentum vector approaches 
the point $X$ from different directions. 
At finite temperature the spectrum is expected to be smeared.

\section{Discussion and conclusions}
A very interesting and open question is the possibility of 
supersolidity in Rydberg systems.
Ground state atoms dressed in Rydberg states exhibit weak van der Waals interactions 
at large distances, as described in~\cite{Pohl2010-ThreeDimensionalRotonExcitationsAndSupersolidFormationInRydbergExcitedBoseEinsteinCondensates}.
The question of supersolidity of such atoms was addressed in Refs.~\cite{Boninsegni2010-SupersolidDropletCrystalInADipoleBlockadedGas%
,Pohl2010-ThreeDimensionalRotonExcitationsAndSupersolidFormationInRydbergExcitedBoseEinsteinCondensates}.
Here, we consider an alternative scenario in which the gas is additionally allowed to have a lattice 
of Rydberg excitations.
Such a lattice would in turn impose weak but long-ranged spatial correlations onto the ground state atoms.
At the same time, the ground-state atoms may be Bose condensed~\cite{Phau2008-RydbergExcitationOfBoseEinsteinCondensates}. 
However, it may be impossible to identify which of the atoms was excited within a certain proximity, 
as was demonstrated, for example, by the superatom analysis of the experimental 
results in Refs.~\cite{Phau2007-EvidenceForCoherentCollectiveRydbergExcitationInTheStrongBlockadeRegime,%
Gould2004-LocalBlockadeOfRydbergExcitationInAnUltracoldGas}. 
However, motion will lead to dephasing of this state 
~\cite{Pfau2011-ArtificialAtomsCanDoMoreThanAtomsDeterministicSinglePhotonSubtractionFromArbitraryLightFields}.
If the atoms are indeed prepared 
in such a mixed state, combining the ground $|g\rangle$ and excited $|e\rangle$ states as $|g\rangle+\alpha|e\rangle$, 
$\bar{N}_g^{-1}\ll\alpha\ll 1$,
then both the lattice-forming and the \abbrev{bec} components are indistinguishable and may be said to be formed by the same atoms.
Therefore, such a system would consist of particles which
would simultaneously break translational symmetry and possess off-diagonal long-range order, 
which is a realization of supersolid. 
While our model does not include the light field, the conditions for the phases 
of both excited and ground state atoms may be immediately extracted from Fig.~\ref{fig:phasediagram}, 
just with different reduced units for the two species.

In conclusion, it is possible to 
parameterize a model with isotropic van der Waals interactions 
into a universal phase diagram.
We have characterized the phase diagram 
of Rydberg atoms by considering a model of bosons with
repulsive van der Waals interaction, and determined
solidification and Bose--Einstein condensation conditions.
Relaxation mechanisms other than thermal motion should be
considered if one considers Rydberg systems on timescales of several tenth of microseconds. We have also studied the excitation spectrum 
within the approximation of a classical harmonic crystal.
Finally, it is worth mentioning that 
interactions between Rydberg excitations 
open a possibility of
new supersolid scenarios.

\chapter{Para-hydrogen at low temperature \label{secHydrogen}}
\section{Introduction \label{secHintro}}
\hyphenation{me-tas-ta-ble}
Superfluidity and Bose-Einstein condensation (BEC) have been stunningly shown in
metastable dilute alkali gases, magnetically confined at ultralow
temperatures.~\cite{Pitaevskii03} The extreme diluteness of these gases 
allows for the achievement of BEC with an almost full occupation of the zero-momentum
state that has been possible to observe and measure quite easily. This
contrasts with the difficulties encountered in the measure of the
condensate fraction in liquid $^4$He, which amounts only 8\% at the
equilibrium density.~\cite{glyden0} However, liquid $^4$He is a stable superfluid below the
lambda transition $T_\lambda=2.17$~K and therefore a system more easily
accessible. Before the blowup produced in the field of quantum fluids by the first 
experimental realization of BEC gases, liquid helium was the only paradigm
of a superfluid. From long time ago, there has been great interest in the
search of superfluid condensed phases other than liquid helium.
Spin-polarized atomic deuterium and tritium are predicted to be fermionic 
and bosonic liquids, respectively, in the limit of zero
temperature.~\cite{panoff,ivana}
However, its experimental study has proven to be very elusive due to its
high recombination rate, and only the case of atomic hydrogen, whose 
ground state is a gas, has been experimentally driven to its BEC
state.~\cite{klepner}    
The next candidate for superfluidity is molecular hydrogen, which has been studied for a long
time.~\cite{silvera} This seems a priori an optimal system due to its very light mass but
it crystallizes at relatively high temperature as a
consequence of the intensity of its intermolecular attraction, without
exhibiting any superfluid transition in the liquid phase. In the
present work, we study the properties of metastable liquid or glass
molecular hydrogen at very low temperatures using quantum Monte Carlo
methods.

In 1972, Ginzburg and Sobyanin~\cite{Ginsburg72} proposed that any Bose liquid
should be superfluid below a certain temperature $T_{\lambda}$,
unless it solidifies at temperature $T_f$ higher than
$T_{\lambda}$. To give a first estimation of $T_{\lambda}$, 
they used the ideal Bose gas theory, obtaining
\begin{equation}
T_{\lambda} = 3.31 \, \frac{\hbar^2}{g^{2/3} m k_B} \rho^{2/3} \ ,
\label{eq:Tlambdaideal}
\end{equation}
where $m$ is the atomic mass, $g$ is the spin degeneracy, $k_B$ is
the Boltzmann constant and $\rho$ is the density of the system.
Ginzburg and Sobyanin proposed molecular \textit{para}-hydrogen (\ph) as a plausible 
candidate for superfluidity: being a
spinless boson ($g=1$) with a small mass, \ph \  should
undergo a superfluid transition at a relatively high
temperature (according to Eq. (\ref{eq:Tlambdaideal}), $T_{\lambda}
\simeq 6$ K).

The estimation of $T_{\lambda}$, given by Eq. (\ref{eq:Tlambdaideal}), 
is clearly inaccurate in the case of dense liquids because it cannot 
account for the
observed dependence of $T_{\lambda}$ with the density. In fact, 
$T_{\lambda}$ slightly decreases in liquid $^4$He when $\rho$ increases, a
manifestly opposite behavior to the increase with $\rho^{2/3}$ given by the
ideal gas formula (\ref{eq:Tlambdaideal}). In order to
provide a more reasonable estimation of $T_{\lambda}$, Apenko~\cite{Apenko99}
proposed a phenomenological prescription for the superfluid
transition, similar to the Lindemann criterion for classical crystal melting.
In this way, he was able to take into
account quantum decoherence effects due to the strong
interatomic potential and to relate the critical temperature for
superfluidity with the mean kinetic energy per particle above the
transition. For \ph, he concluded that $T_{\lambda}$
should vary between $1.1$ K and $2.1$ K, depending on the density of
the system.

Superfluid \ph \ is not observed in a stable form because it
crystallizes at temperature $T_f = 13.8$ K, which is significantly higher
than the expected $T_{\lambda}$.
Several studies about crystal nucleation in \ph \  have been performed 
in order to understand if the liquid can
enter a supercooled phase, i.e., a metastable phase in which the
liquid is cooled below its freezing temperature without forming a
crystal. Maris \textit{et al.}~\cite{Maris83}
calculated the rate $\Gamma(T)$ of homogeneous nucleation of the
solid phase from the liquid as a function of the temperature $T$,
showing a maximum of $\Gamma$ around $T = 7$ K and a rapid
decrease at lower temperature. This suggests that, if it would be
possible to supercool the liquid through the range where $\Gamma$
is large, one might be able to reach a
low-temperature region where the liquid is essentially stable.
However, recent experiments have indicated that, even at $T \sim
9$ K, the rate of crystal growth is so high that the liquid phase
freezes quickly into a metastable polymorph crystal.~\cite{kuhnel11}

Even though several supercooling techniques have been proposed 
to create a metastable liquid phase in bulk 
\ph,~\cite{Maris87,Vorobev00,Grisenti06} none of them has proven so
far to be
successful and no direct evidence of superfluidity has been
detected. However, there are evidences of superfluidity in several  
spectroscopic studies of  small doped \ph \ clusters.  
In 2000, Grebenev \textit{et al.}~\cite{Grebenev00} analyzed the rotational
spectra of a linear carbonyl sulfide (OCS) molecule surrounded by 14 to 16
\ph \  molecules absorbed in a larger helium  droplet, which
fixes the temperature of the cluster. 
When \ph \ is immersed in a $^4$He droplet ($T= 0.38$ K), 
the measured spectra shows a peak indicating the
excitation of angular momentum around the OCS axis. On the other hand, 
if the small
\ph \  cluster is put inside a colder $^4$He-$^3$He droplet
($T = 0.15$ K), the peak disappears: the OCS
molecule is then able to rotate freely inside the hydrogen cluster,
pointing to the superfluidity of the surrounding \ph \ molecules. 
These results have been confirmed in a later experiment on small
\ph \  clusters doped with carbon dioxide.~\cite{Li10} From a
precise analysis of the rotational spectra, it has been possible
to measure the effective momentum of inertia of these small
systems, and thus of their superfluid fraction $\rho_s$, providing
a clear evidence of superfluidity in clusters made up of $N \le
18$ \ph \  molecules. These clusters are too small for extracting reliable
predictions of a metastable liquid phase and larger clusters would be
desirable. To this end, Kuyanov-Prozument and Vilesov~\cite{Prozument08}  have been
able to stabilize liquid clusters with an average size of
$N \approx 10^4$ \ph \  molecules down to temperature $T= 2$ K, 
but they do not see any evidence of superfluidity. Other attempts of
producing liquid \ph \ well below $T_f$  ($T=1.3$ K) are based on the generation
of continuous hydrogen filaments of macroscopic
dimensions.~\cite{Grisenti06}

The search for a superfluid \ph \ phase has been intense also from the
theoretical point of view. The rather simple radial form of the \ph-\ph
interaction and the microscopic accuracy achieved by quantum Monte Carlo
methods have stimulated a long-standing effort for devising possible scenarios
where supercooled \ph \ could be studied. In practically all the cases, the
search is focused on systems of reduced dimensionality or in finite
systems. PIMC simulations of \ph \ films adsorbed on a surface with
impurities observed superfluidity for some arrangements of these
impurities,~\cite{Gordillo97} but these results were posteriorly questioned by
other PIMC studies.~\cite{Boninsegni05H2} In a one-dimensional channel, like
the one provided experimentally by narrow carbon nanotubes, it has been
predicted a stable liquid phase in the limit of zero temperature.~\cite{h21d}
The largest number of theoretical works are devoted to the study of small
clusters, both pure~\cite{Sindzingre91,Mezzacapo06,Mezzacapo07,
Khairallah07,Mezzacapo08,ester,guardiola} and doped with impurities.~\cite{Kwon02,
Paesani05,Kwon05} All these simulations show that \ph \  
becomes superfluid below a certain temperature $T=1$-$2$ K and that the 
superfluid fraction depends on the number of molecules
of the cluster. When the cluster becomes larger than a certain molecular number 
($N > 18$-$25$), solid-like structures are observed and the superfluidity 
vanishes.   

We address the
calculation of the equation
of state of the metastable liquid \ph \ phase in the limit of zero
temperature using the diffusion Monte Carlo (DMC) method. The simulation of
the liquid phase  in this limit is easier than at finite temperature and 
therefore it is able to provide accurate information on its main energetic
and structure properties.    

The rest of the chapter is organized as follows. In Sec. ~\ref{secHmeth}, we introduce
the quantum Monte Carlo methods used in the study 
and report specific details on how the simulations are carried out. Sec.
~\ref{secHres} contains the results of the equation of state, structure properties,
and condensate fraction of metastable liquid \ph \ at zero temperature.
and finally
the main conclusions of the present work are discussed in Sec.~\ref{secHconcl}.

\section{Methodology and construction of trial wave functions \label{secHmeth}}
The H$_2$ molecule, which  is composed of two hydrogen
atoms linked by a covalent bond, is spherically symmetric in the
\textit{para}-hydrogen state (total angular momentum zero).  
The energy scale involved in electronic excitations   
($\sim 10^{5}$ K) is orders of magnitude larger than the intermolecular
one  ($\sim 10^{1}$ K), thus to model the \ph-\ph interaction by means of a 
radial  pair-potential and to
consider the molecules as point-like turns out to be justified upon the
condition  of low or moderate pressures.  In this work, we have chosen the
well-known and commonly used semiempirical Silvera-Goldman pair 
potential.~\cite{SilveraGoldman} 
This potential has proved to be accurate at low  temperature and
at the pressure regimes in which we are interested.

The study in the limit of zero temperature has been  performed
with  the DMC method. DMC is a first-principles method which can access 
exactly the ground state of bosonic systems.
It is a form of Green's Function Monte Carlo
which samples the projection of the ground state from the initial configuration
with the operator $\exp{\left[-(\mathcal{H}-E_0)\tau\right]}$. Here, $\mathcal{H}$
is the Hamiltonian
\begin{equation}
\mathcal{H} = -\frac{\hbar^2}{2m} \sum_{i=1}^{N} {\bm \nabla}_i^2 +
\sum_{1=i<j}^{N} V(r_{ij}) \ ,
\label{hamiltonian}
\end{equation}
$E_0$ is a norm-preserving adjustable constant and $\tau$
is the imaginary time. The simulation is 
performed by advancing in $\tau$ via a combination of diffusion, drift and branching
steps on walkers $\bm{R}$ (sets of $3N$ coordinates) representing the wavefunction of 
the system.~\cite{Boronat94}  The imaginary-time evolution of the walkers is
``guided'' during the drift stage by a guiding wavefunction $\phi_G$, which
is usually a good guess for the  wavefunction of the system. This function
contains basic ingredients of the system as its symmetry, phase and
expected behaviors at short and long distances according to its
Hamiltonian. Technically,  $\phi_G$ allows  importance sampling and thus
reduces the variance of the ground-state estimations. It is straightforward 
to show that for the Hamiltonian  $\mathcal{H}$ and any operator commuting with 
it, the expectation value  is computed exactly within statistical error.
Other diagonal operators which do not fulfill this condition require of a special
treatment, known as pure estimation,~\cite{pures} which leads also for this case to
unbiased results (for details see Section~\ref{secQuant} of this Thesis).

The phase of the system is imposed within the typical imaginary-time length
by the guiding wave function. This property of the DMC method is here a key
point if we are pursuing a prospection on the properties of 
the metastable liquid \ph \ phase. Then, for the liquid phase $\phi_G$ is
taken in a Jastrow form
\begin{equation}
\phi_G (\bm{R}) = \prod_{1=i<j}^{N} f(r_{ij}) \ ,
\label{jastrow}
\end{equation}
with a two-body correlation function~\cite{f2reatto}
\begin{equation}
f(r) = \exp \left[ - \frac{1}{2} \, \left( \frac{b}{r} \right)^5 
-\frac{L}{2} \, \exp \left[ - \left( \frac{r - \lambda}{\Lambda} \right)^2
\right] \right] \ .
\label{f2}
\end{equation}
In order to compare the results obtained for the liquid phase with the ones
corresponding to the stable hcp solid we have carried out some simulations
with a guiding wave function of Nosanow-Jastrow type
\begin{equation}
\phi_G^s (\bm{R}) = \prod_{1=i<j}^{N} f(r_{ij}) \, \prod_{i=1}^{N}
g(r_{iI}) \ ,
\label{nosanow}
\end{equation}
the set $\{ \bm{r}_I \}$ being the lattice points of a perfect hcp lattice.
Optimal values for the parameters entering Eq. (\ref{f2}) are  $b=3.68 {\angstrom}$,
$L=0.2$, $\lambda=5.24 {\angstrom}$, and $\Lambda=0.89 {\angstrom}$ \ for the liquid phase,
and $b=3.45 {\angstrom}$, $L=0.2$, $\lambda=5.49 {\angstrom}$, and $\Lambda=2.81 {\angstrom}$ \ for
the solid one. The Nosanow term is chosen in Gaussian form, $g(r)=\exp
(-\gamma r^2)$.  The density dependence of 
the parameters in the Jastrow term is small, and 
neglected in practice when used in DMC, whereas  the Nosanow term 
parameter $\gamma$ is optimized for the whole range of densities. 
We have used 256 and 180 particles per
simulation box for the liquid and hcp solid phases, respectively. The number of
walkers and time-step have been adjusted to reduce any bias coming from
them to the level of the statistical noise.

\section{Results at zero temperature \label{secHres}}
We have calculated the main properties of the metastable liquid and stable
hcp solid phases of \ph. Our main goal has been to know the properties of
a hypothetical bulk liquid phase and compare them with the ones of the
stable solid. In order to achieve reliable estimations of liquid \ph \ it
is crucial to work with a guiding wave function of liquid type, as we have
discussed in the preceding Section. Within the typical imaginary-time
length of our simulations we have not seen the formation of any crystal
structure, i.e.,  no signatures of
Bragg peaks in the structure function $S(k)$ have been registered so far.

\begin{figure}
\centerline{
\includegraphics[width=0.87\textwidth]{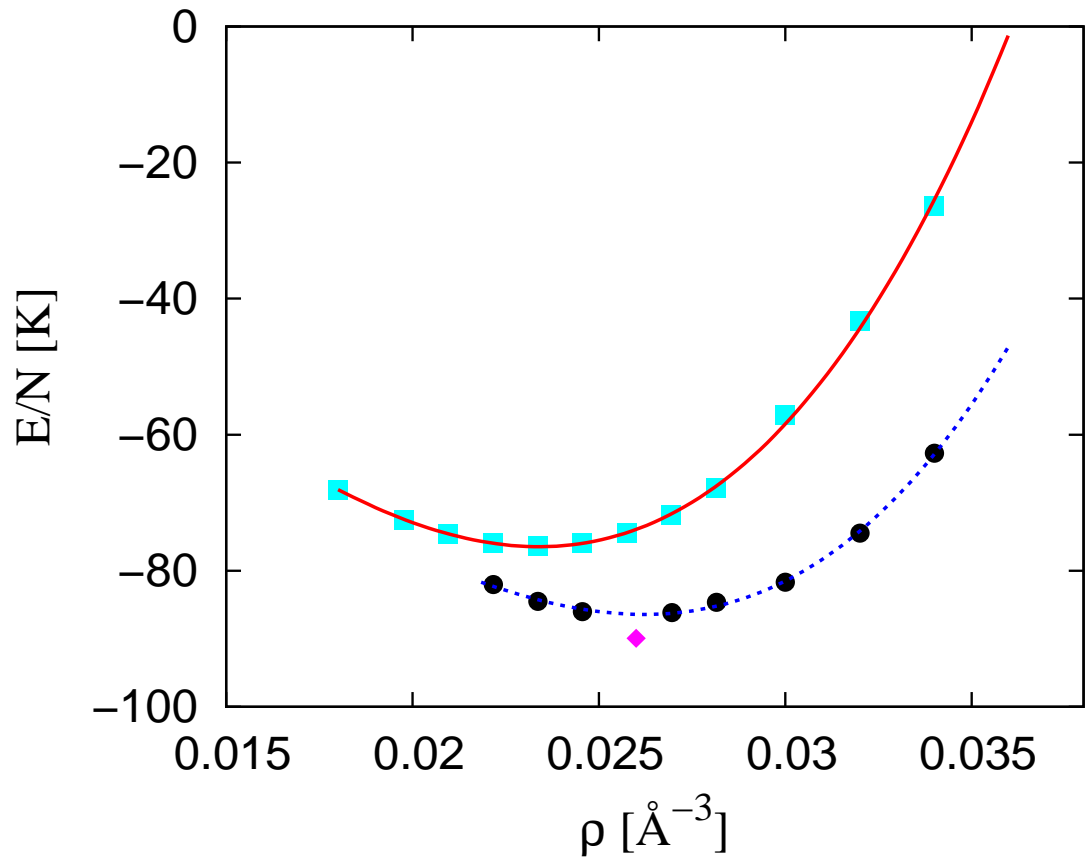}}%
\caption{DMC energies per particle as a function of the density. Squares
and circles correspond to the liquid and solid phases, respectively. Solid
and dashed lines are the polynomial fits to the DMC energies for the liquid
and solid, respectively. The diamond is the experimental energy of hcp molecular
hydrogen from Ref.~\cite{phexperimental}}  
\label{fig:ener0t}
\end{figure}

\begin{figure}[b]
\centerline{
\includegraphics[width=0.87\textwidth]{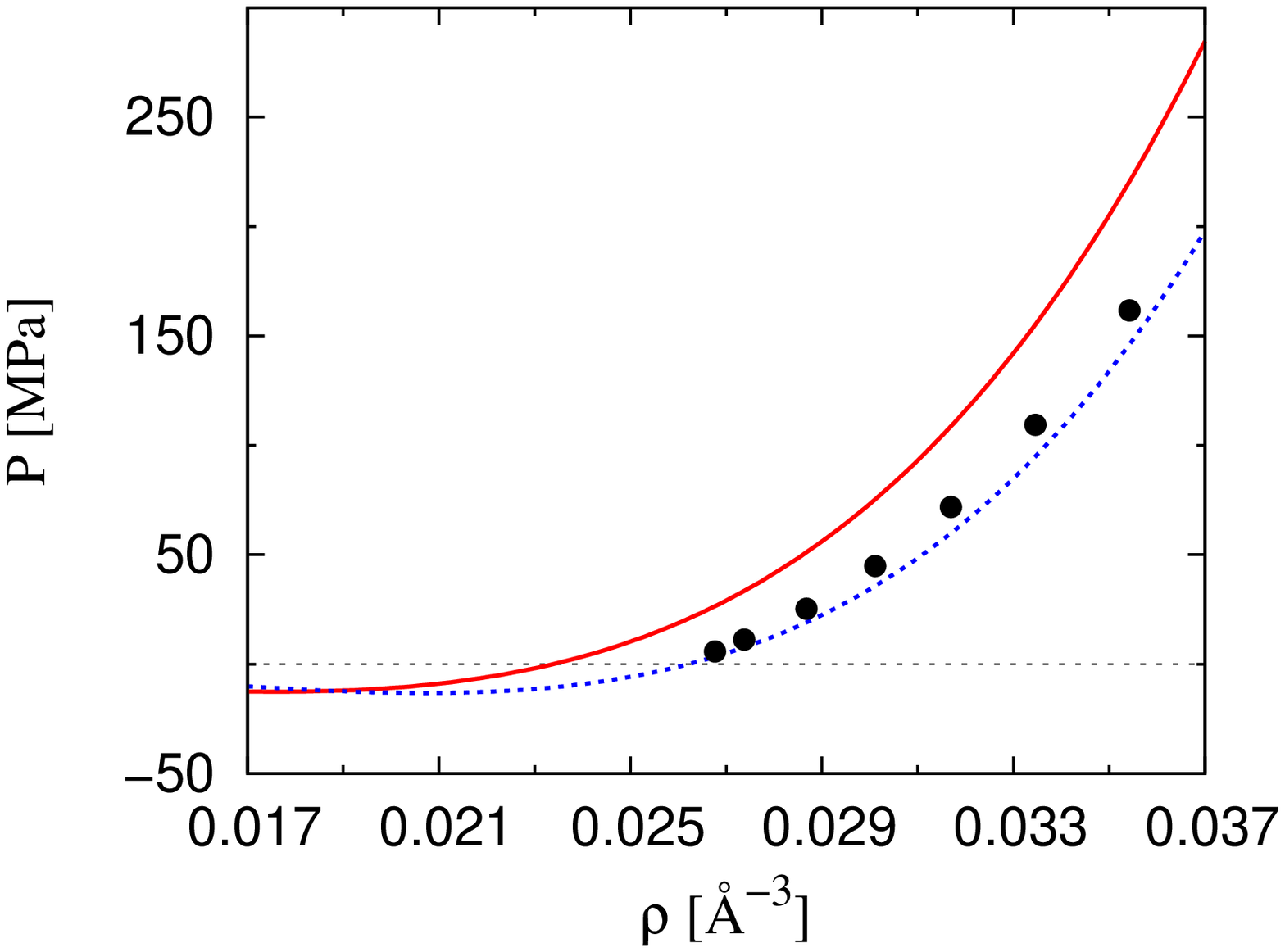}}
\caption{Pressure of the liquid (solid line) and solid (dashed line) \ph \ phases as a 
function of the density. Experimental points for the solid phase~\cite{phexperimental2} 
are show as solid circles. 
}  
\label{fig:pres0t}
\end{figure}

In Fig. \ref{fig:ener0t}, we plot the DMC energies per particle of metastable
liquid \ph \ as a function of the density. For comparison, we also report
the results obtained for the hcp crystal phase. Our hcp energies are in
close agreement with the ones reported in Ref.~\cite{pederiva} using
the same Silvera-Goldman potential. In the figure, we also show the 
experimental estimation at $T=0$ K from Ref.~\cite{phexperimental},
$E/N=-89.9$ K, that lies a bit below of our results. This is again in
agreement with previous DMC results~\cite{pederiva} which show that the experimental
energy is, in absolute value, underestimated and overestimated by the
Silvera-Goldman and Buck potential,~\cite{buck} respectively. Our results for both
phases are well reproduced by the polynomial law
\begin{equation}
\frac{E}{N} = \left( \frac{E}{N} \right)_0 + A \left( \frac{\rho-\rho_0}{\rho_0}
\right)^2 + B \left( \frac{\rho-\rho_0}{\rho_0} \right)^3 \ ,
\label{eqstate}
\end{equation}  
$(E/N)_0$ and $\rho_0$ being the equilibrium energy per particle and
equilibrium density, respectively. These equations of state are shown in
Fig. \ref{fig:ener0t} with lines. The optimal parameters of the fits are:
$\rho_0=0.026041(20) {\angstrom}^{-3}$, $(E/N)_0=-86.990(37)$ K, $A=232(2)$ K,
$B=156(11)$ K for the solid, and $\rho_0=0.023386(40) {\angstrom}^{-3}$, 
$(E/N)_0=-76.465(51)$ K, $A=188(1)$ K, $B=131(10)$ K for the liquid. As
expected, our DMC results shows that the solid phase is the stable one with
a difference in energy per particle at the respective equilibrium points of
$\sim 10$ K, the equilibrium density of the liquid being $\sim 10$ \%
smaller than the solid one. The same trend was observed in a DMC simulation
of two-dimensional \ph, but there the differences were significantly
smaller.~\cite{claudi2d} It is worth noticing that about one half of the 
energy difference in the bulk systems
comes from the decrease of the kinetic energy per particle going from the
liquid to the solid: at density $\rho=0.03\, {\angstrom}^{-3}$, it amounts $93.3(1)$
and $89.1(1)$ K for the liquid and solid, respectively.

\begin{figure}[t]
\centerline{
\includegraphics[width=0.87\textwidth]{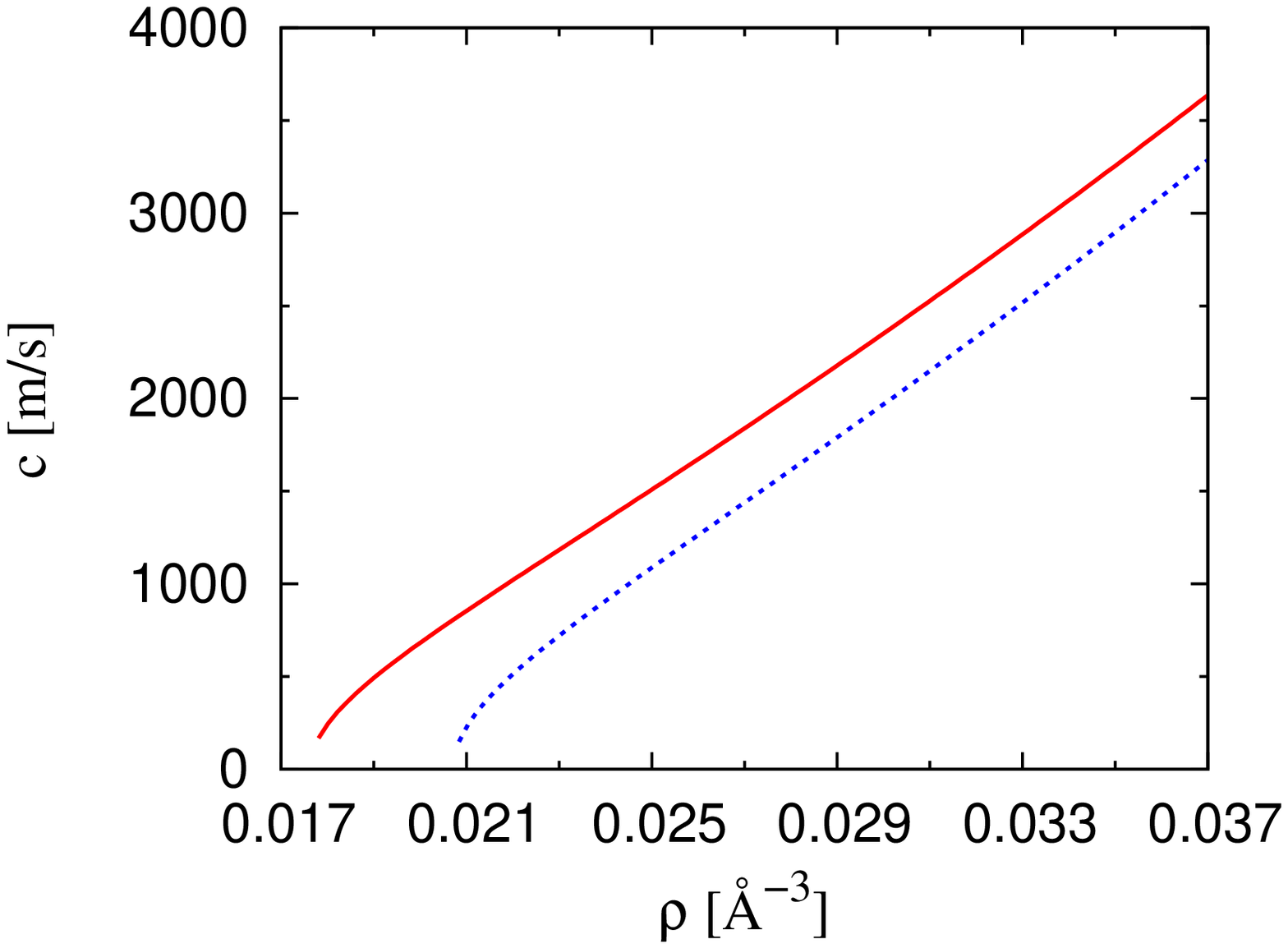}}%
\caption{Speed of sound of the liquid (solid line) and solid (dashed line) \ph \ phases as a 
function of the density.  
}  
\label{fig:velocitat0}
\end{figure}

From the equations of state (\ref{eqstate}), it is easy to know the
pressure of the system at any density using the relation $P(\rho)= \rho^2 [
d (E/N) / d \rho]$. The results obtained for metastable liquid and stable
solid phases are shown in Fig. \ref{fig:pres0t}. As one can see, at a given
density the pressure of the liquid is larger than the one of the solid
mainly due to the different location of the equilibrium densities ($P=0$).
The results for the solid are compared with experimental data from Ref.
~\cite{phexperimental2} showing a good agreement especially for not
very large pressures. The density at which the function $P(\rho)$ has a
zero slope defines the spinodal point; beyond this limit the system is no
more thermodynamically stable as a homogeneous phase. At this point, the
speed of sound $c(\rho)= [ m^{-1} (d P / d \rho)]^{1/2}$ becomes zero.
Results for $c(\rho)$ are shown for both phases in Fig.
\ref{fig:velocitat0}. The speed of sound decreases when the density is
reduced and drops to zero at the spinodal point: ($\rho_c=0.0176(1)\,
{\angstrom}^{-3}$,
$P_c=-12.6(5)$ MPa) and   ($\rho_c=0.0195(1)\, {\angstrom}^{-3}$, $P_c=-17.5(5)$ MPa) for
liquid and solid, respectively.

\begin{figure}[]
\begin{center}
\includegraphics[width=0.87\textwidth]{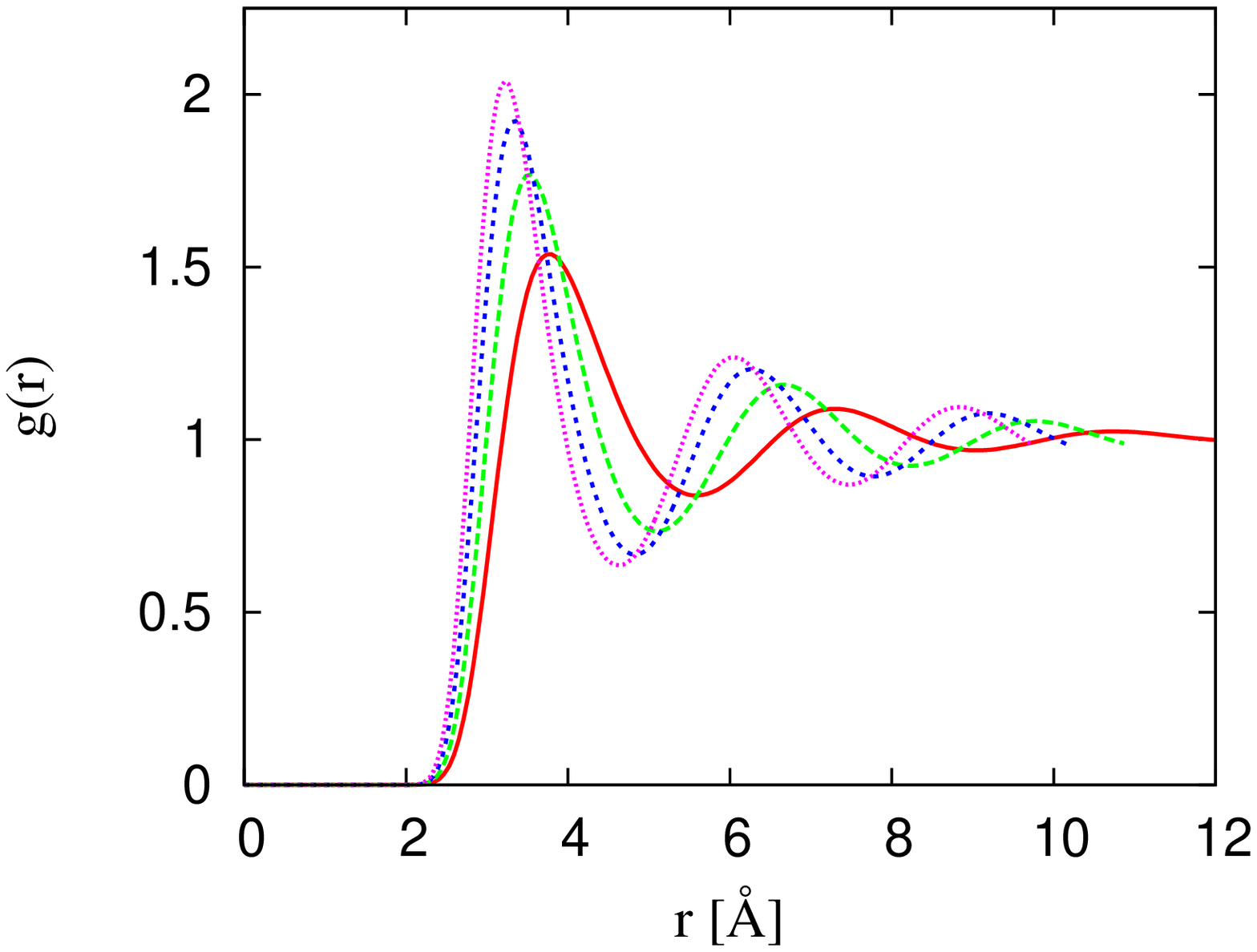} 
\\
\includegraphics[width=0.87\textwidth]{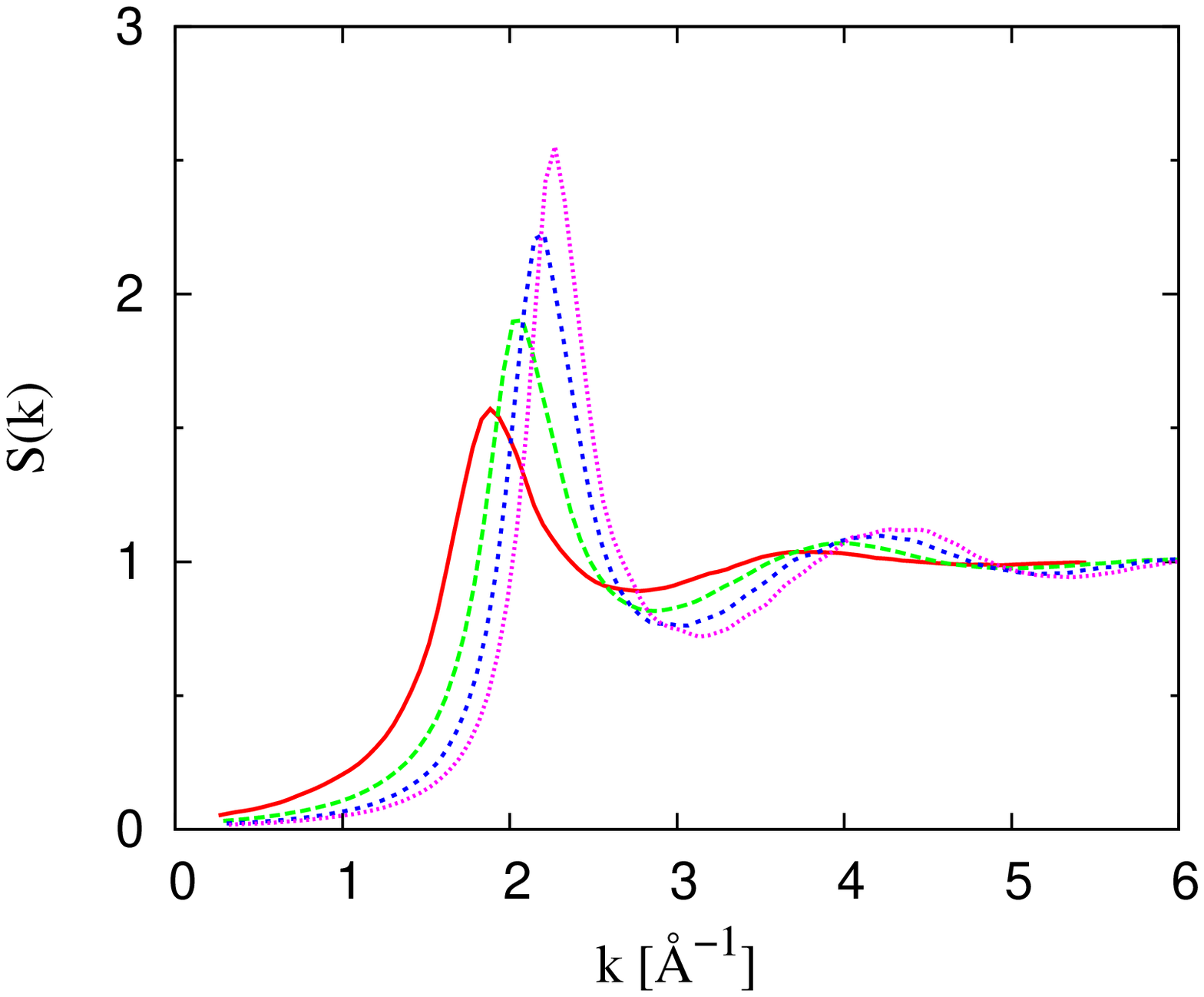}
\end{center}
\caption{\textit{Top panel:} Two-body radial distribution function of the
liquid \ph \ phase at different densities: solid, long-dashed, 
short-dashed, and dotted lines stand for densities 
$\rho=0.0180$, $0.0245$, $0.0300$, and $0.0340\, {\angstrom}^{-3}$, respectively.
\textit{Bottom panel:} Static structure factor of the liquid phase. Same
densities and notation than in the top panel.  
}  
\label{fig:grskdens}
\end{figure}

\begin{figure}[b]
\centerline{
\includegraphics[width=0.87\textwidth]{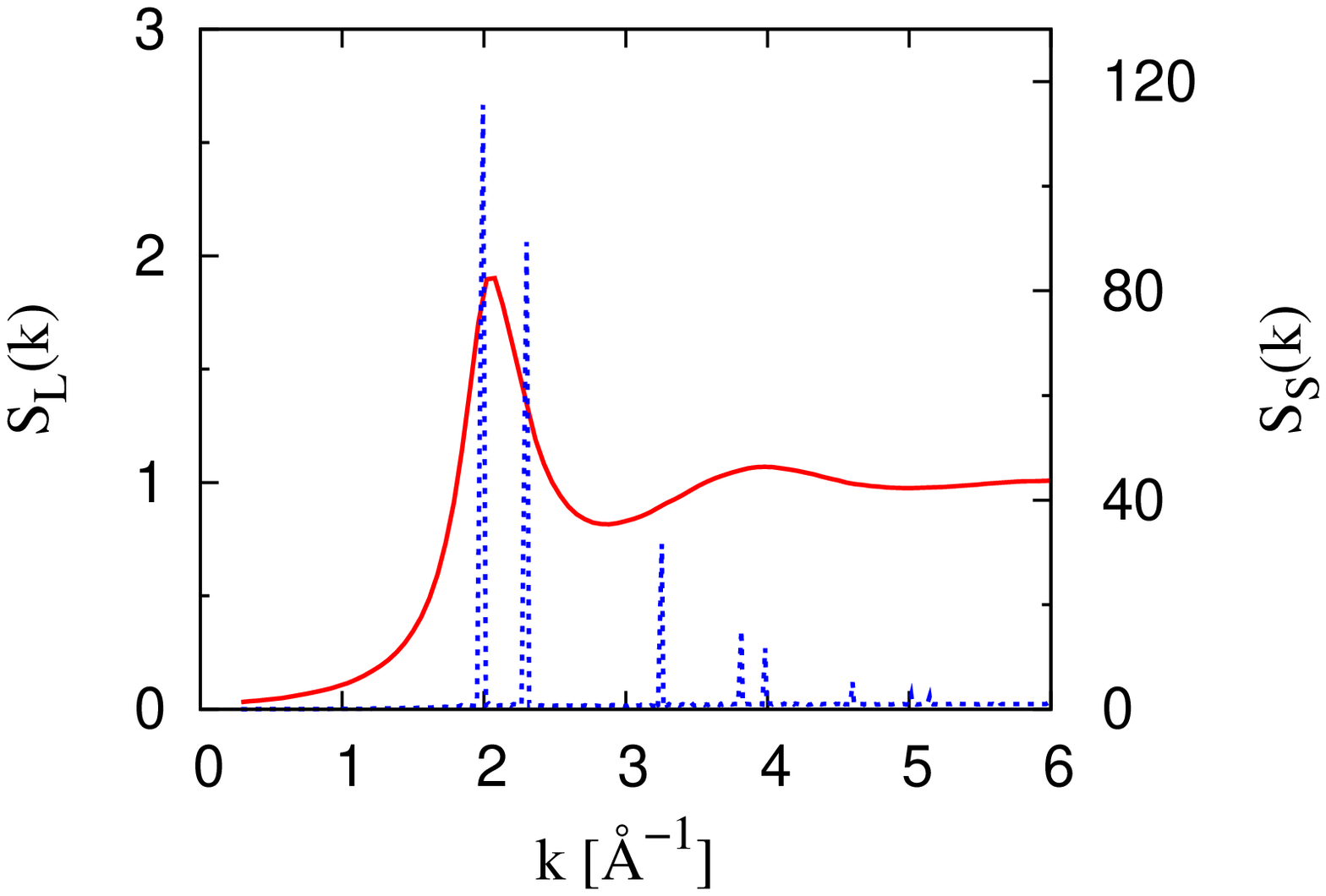}}%
\caption{Static structure function of liquid and solid \ph \ at  
density $\rho=0.0245\, {\angstrom}^{-3}$. The result for the liquid $S_L(k)$ (left
scale) is shown with a solid line; the one for the hcp solid $S_S(k)$
(right scale) with a dashed line.
}  
\label{fig:skcomp}
\end{figure}

DMC produces also accurate results for the structure of the bulk system. In
Fig. \ref{fig:grskdens}, we show results for the two-body radial distribution
function $g(r)$ of the liquid \ph \ phase for a set of densities. This
function is proportional to the probability of finding two molecules
separated by a distance $r$. Increasing the density, the main peak becomes
higher and moves to shorter interparticle distances; at least three peaks
are observed. All these features point to the picture of a very dense quantum
liquid, with much more structure than in stable liquid $^4$He. In the same
Fig. \ref{fig:grskdens}, we show results for the static structure factor
$S(k)$, related to $g(r)$ by a Fourier transform. As one can see, the main
peak increases quite fast with the density suggesting a highly structured
metastable liquid. Nevertheless, we have not observed within the scale of
the simulations the emergence of any Bragg peak which would point to
formation of crystallites in the simulation box. In Fig. \ref{fig:skcomp},
we illustrate the comparison between $S(k)$ for the liquid and solid
systems at a density $\rho=0.0245$~\rve{A}$^{-3}$, close to the equilibrium
density of the liquid. The difference is the one expected between a liquid
and a solid: oscillating function towards one at large $k$ for the liquid
and a sequence of Bragg peaks, corresponding to the hcp lattice, for the
solid.

\begin{figure}[t]
\centerline{
\includegraphics[width=0.87\textwidth]{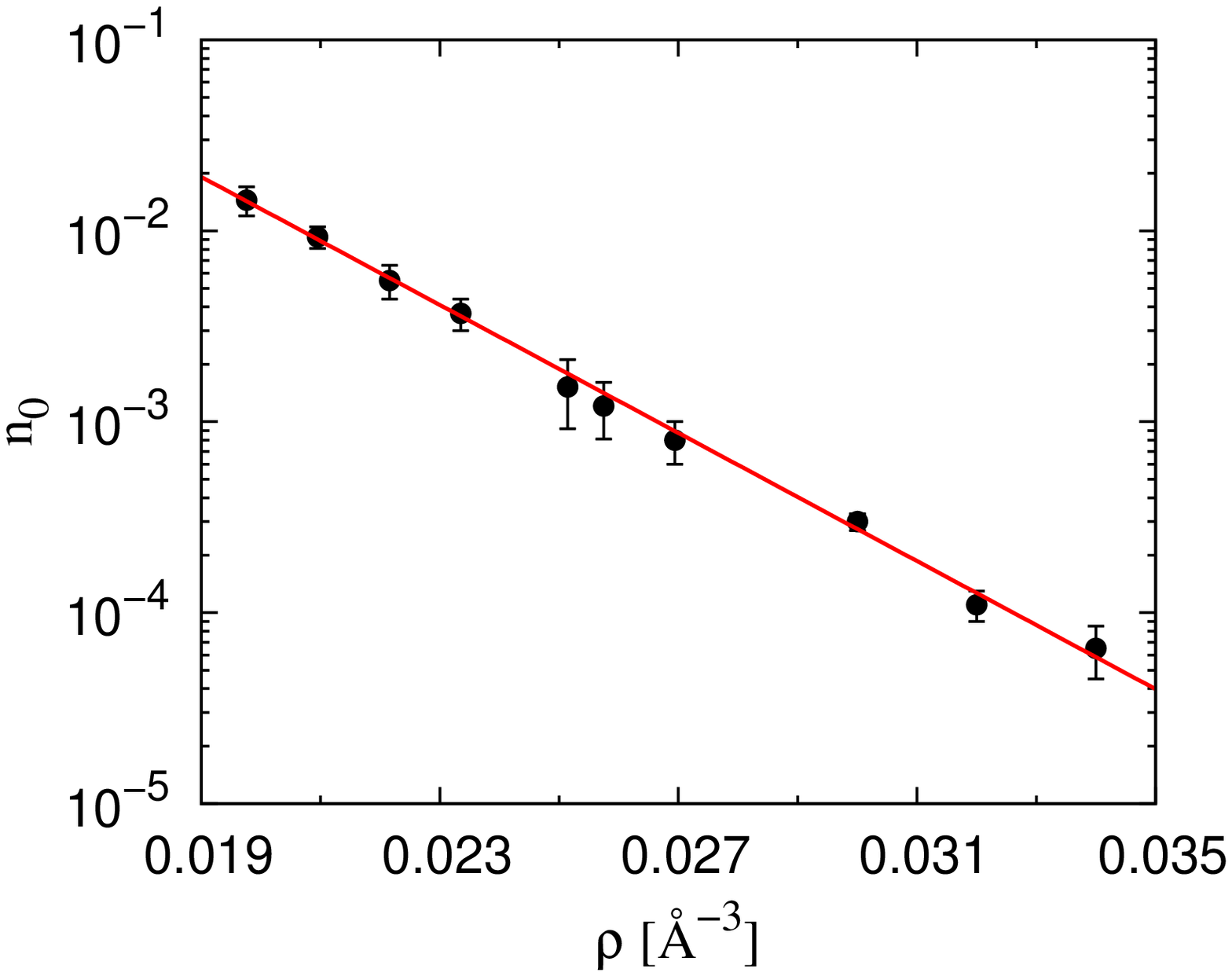}}%
\caption{Condensate fraction of metastable liquid \ph \ as a function of
the density. The points are the DMC results and the line is an exponential
fit to them.}  
\label{fig:condensate}
\end{figure}

One of the most relevant properties of a superfluid is the mean occupation
of the zero-momentum state, i.e., the condensate fraction $n_0$. As it is
well known, $n_0$ can be obtained from the asymptotic behavior of the
one-body density matrix $\rho(r)$,
\begin{equation}
n_0= \lim_{r \rightarrow \infty}  \rho(r)   \ , 
\label{condensate}
\end{equation}
with $\rho(r)$ being obtained as the expectation value of the operator
\begin{equation}
\left \langle \frac{ \Phi({\rve}_1,\ldots,{\rve}_i+{\rve},\ldots,{\bf
r}_N )}{\Phi( {\rve}_1,\ldots,{\rve}_N ) }  \right \rangle \ .
\label{obdm2}
\end{equation}
DMC results for the condensate fraction of liquid \ph  \ as a function of
the density,
obtained using the extrapolated estimator (there are no reliable pure estimators
for non-diagonal operators), are shown in Fig. \ref{fig:condensate}. The
decrease of $n_0$ with the density is well described by an exponential
decay (line in the figure). The strong interactions induced by the deep
attractive potential well produce a big depletion of the condensate state.
At the equilibrium density, our estimation for the condensate fraction is
$n_0=0.0037(7)$. This value is more than one order of magnitude smaller
than the measured condensate fraction~\cite{glyden0} of liquid $^4$He at equilibrium
($0.08$).

\section{Conclusions \label{secHconcl}}
A possible observation of superfluidity in bulk or clustered systems of para-hydrogen atoms \ph
was a subject of an extensive study in the last decades (~\cite{Ginsburg72},~\cite{Sindzingre91})
 due to a variety of advantageous properties of such atoms. Nevertheless the experimental observation has not been performed yet,
mostly because of a relatively high solidification temperature of \ph, that can bring a perspective superfluid transition deeply 
into the metastable phase. 

Our Group performed a multifold study of the system of para-hydrogen atoms \ph
 at low temperatures deeply below the point of crystallization by means of Quantum Monte Carlo 
methods. The zero-temperature simulation was performed in order to 
investigate the properties of a metastable liquid phase and to find the fraction of 
the Bose-Einstein condensate in the relevant range of densities. The methods of choice for the zero-temperature
simulations of the \ph system were the Variational Monte Carlo and the diffusion Monte Carlo techniques.
The latter is an efficient and versatile instrument to calculate the quantum properties of the system, including 
the non-local ones. The results of the zero-temperature simulations suggest that the metastable liquid para-hydrogen
is a strongly correlated liquid, that again might be a sign of instability of this hypothetical 
system. The calculation of the Bose-Einstein condensate show that the condensate fraction is substantially 
lower than in the liquid helium $^4$He.

\chapter{Conclusions and list of publications\label{secConclusions}\label{secPapers}}
Here we summarize the results, presented in the Thesis.

In Chapter~\ref{secEwald}, devoted to the generalization of the Ewald summation technique, we derived 
explicit expressions for the Ewald sums in three-dimensional systems, governed by a generic $1/|\bs{r}|^k$ power-law 
interaction, with periodic boundary conditions applied. We also extended the 
derivation to the cases of two-dimensional and one-dimensional geometry. The importance of 
these generalizations can be seen, as they apply to many physically important interaction potentials 
as the dipole-dipole interaction, van der Waals interaction etc. 
In this Thesis we give the functional forms 
for the terms of the Ewald sums (both in momentum and coordinate space), ready for implementation 
in actual calculations. The derivation and the functional form of the results differs in
  the cases of short-ranged ($k>D$), long-ranged ($k<D$) and 
``marginal'' ($k=D$) forces, where $D$ is the system dimensionality. The cases of long-range forces
require separate calculations because of the 
divergence of the potential energy. This unphysical divergence disappears when demanding 
a charge-neutrality of the system, which can be restored within the ``jellium'' model. 
The resulting expressions in the jellium model are explicitly given. It is argued 
that in the case of some short-range potentials the Ewald method can be advantageous with respect to 
a direct summation
due to a faster convergence rate, typically, of a Gaussian kind versus a certain 
power-law rate. We also give a discussion of the convergence properties of a quasi-neutral Coulomb system.

The results are first presented for the case of a 3D system
in a cubic simulation box in order to explain the
general mathematical procedure, which for the specific case of the Coulomb
potential recovers well-known results~\cite{allen}. Later on, the same
mathematical technique is applied to 2D and 1D geometries.
For the one-dimensional case the
initial sum for the potential energy  is explicitly evaluated in Eq.~(\ref{ep1ddirect2}),
nonetheless the Ewald summation is developed as well for this
case and may be used as a mathematical equality. The more efficient
representations of the reciprocal space sums, which have a complexity $\mathcal{O}(N)$ instead of $\mathcal{O}(N^2)$
 with the number of
particles $N$, are presented
for 3D and 2D systems. The explicit expressions for the terms of the Ewald
sums are given in a tabular form for physically relevant potentials
with small integer power indices $k$, as dipole-dipole interaction
potential, Lennard-Jones potential and others in both three- and
two-dimensional geometries
(refer to Tables~\ref{table3d} and \ref{table2d}).

When the simulation box cannot be chosen cubic, for example in a modelling
of a three-dimensional hcp crystal structure, the Ewald method can also be
applied after certain modifications. Formally, it consists in the choice
of an appropriate rectangular simulation box and a substitution of the
vector $\bs{n}$ by $\bs{n}_r=(\bs{n}_xL_x+\bs{n}_yL_y+\bs{n}_zL_z)/L_0$ and
$\bs{m}_k=(\bs{m}_x /L_x+\bs{m}_y /L_y+\bs{m}_z /L_z)L_0$ in the real
and momentum space sums, respectively [see
Eqs.~(\ref{i01infpolytropicnbox}) and (\ref{3djacobinbox})].

The optimization of the involved parameters, that is the integration parameter $\alpha$ 
and the cut-off lengths in
coordinate and momentum space, is a necessary operation
in order to improve the convergence rates and to avoid excessive calculations.
The main idea of the optimization, that we propose in the this Thesis, is to
perform a benchmark calculation, minimizing the variance in the energy. 
We show how this optimization technique works on the example of 
a two-dimensional gas of dipoles, aligned
perpendicularly to the plane of motion. This proposed optimization technique
is mostly efficient in simulations of gases and crystals. In spite of being very
simple, it allows to find rather quickly adequate parameter ranges. The
analytical estimations of the optimal parameters are given as well and are proven
to be consistent with the results, obtained in an explicit optimization. A more
sophisticated method to optimize the calculation parameters, taking
advantage of the $\mathcal{O}(N)$ representation of the Fourier transform
sum, is also presented with explicit estimations of the parameters for a
typical system simulated by Quantum Monte Carlo methods.

Another problem studied is the phase diagram of Yukawa systems. The Coulomb 
potential is probably one of the most basic and widespread interactions in the nature 
as it describes forces between charges. In a charge neutral system, the presence of 
a second component introduces screening. 
The Yukawa potential as well appears in the problem of mass-imbalanced two-component Fermi gas. 
It is known that in the case of a large mass imbalance, the light fermions introduce an 
effective Yukawa potential between the heavy fermions. It was predicted in~\cite{YukawaPot2D}
that this effective interaction might lead to crystallization in two-dimensional systems, 
although no estimations were done in three-dimensional case. The study of Ceperley \textit{et al.}~\cite{Ceperley1} 
partially addresses the problem of the zero-temperature phase diagram. In their work the transition
line was drawn by using approximate Lindemann melting criterion. For the first time a fully 
quantum mechanical calculation is done in order to find the zero-temperature phase diagram 
of Yukawa particles. The diffusion Monte Carlo method is expected to give a ground state energy 
exactly, so the phase diagram should be exact. The melting and solidification points are found 
by using the double-tangent Maxwell construction. The Lindemann criterion indeed applies in a 
large part of the phase diagram apart from the high density regime where notable differences compared 
to the prediction of Ceperley \textit{et al.} are visible. 

We discovered that an improvement to the efficiency of the diffusion Monte Carlo method can be obtained by using 
the hypernetted chain (HNC) method (based on a solution of the corresponding Euler-Lagrange hypernetted chain equations) in a construction of Jastrow guiding wave functions. 
The HNC method allows to find a very good numerical approximation for the two-body Jastrow terms provides a result, which is 
already optimized in a many-body sense. The HNC solutions are also advantageous, since they do not require cumbersome multiparametric optimizations, and are quite precise 
and very fast.

The phase diagram of Yukawa particles has a peculiar shape. That is, a double transition is possible, when 
for a fixed interaction strength, a change in the density might cause reentrant gas-solid-gas transition. This unusual behavior is caused 
by a competition between long-range Coulomb part of the Yukawa potential (melts at large densities) and exponential screening (melts at small densities).
To the best of our knowledge this is the first time that the high-density melting was observed in this system in a quantum simulation. 

 Based on our calculations we find that the mass ratios in 
 achievable fermionic mixtures of pure elements are too small to undergo a transition to a crystal phase regardless of the density of the bulk. Nonetheless, larger effective mass ratios can be reached 
 if the heavier component is confined by an optical lattice, that enhances strongly the strength of 
 the lattice. This alternative procedure of obtaining large effective mass ratios is discussed in the Thesis, and  based on the available data on the current and perspective experimental set-ups the principal feasibility to produce experimental conditions for observation of the crystallization is argued. The heights of optical lattices are typically tunable by the intensity of laser beams, so that the correctness of our theoretical predictions for the phase transition can be checked.

In order to localize more accurately the phase transition point in the high density regime we applied the Ewald summation technique, that yields 
the potential energy of the simulation cell, replicated infinitely in the space. This method proved to be efficient in enhancing the rate of 
the convergence of the DMC energy with the number of particles in spite of introducing additional calculation costs. The specific Ewald summation 
for the case of a Yukawa system was implemented in the simulation code, used in our research Group. The idea to apply the HNC method to produce 
highly optimized Jastrow terms of trial wave functions can be used in future Monte Carlo simulations of other physical systems.

In the next part of the work, given in the Thesis, we presented a
quantum Monte Carlo study of a bulk system of bosonic Rydberg atoms, that is 
alkali atoms with a single electron residing on a very high orbital,
interacting through van der Waals interaction 
potential $1/|\rve|^6$. A perspective utilization of
Rydberg atomic clouds as quantum gates or for an observation of 
Bose--Einstein condensation, requires an utmost stable and predictable 
system, that often implies that the interaction is repulsive and independent on 
spacial directions, hence our simple model should be physically relevant. 
The asymptotic van der Waals $C_6/|\rve|^6$ is a common leading 
term of the interaction of Rydberg atoms, derived by virtue of the perturbation theory. 
One of the aims that we pursue in the study is to understand how much of the 
behavior of a real system comes from this leading term, and to which extent 
the $1/|\rve|^6$ interaction of the excited atoms in the cloud 
can describe crucial properties of experimentally relevant systems like 
mixtures of excited and ground state atoms.  

The zero-temperature properties of a system of Rydberg atoms, as presented in Section~\ref{secRydberg}, 
are controlled by a unified dimensionless parameter $\rho r_0^3$, which depends on the density $\rho$ and the interaction strength $r_0=(mC_6/\hbar^2)^{1/4}$. The behaviors of distinct 
systems with different parameters like particle mass and interaction strength constant, are therefore identical, if they possess the 
same value of $\rho r_0^3$, for instance, the systems solidify and melt at the same dimensionless density. Making use of the diffusion Monte Carlo method, we found that for 
a system, governed by a model Hamiltonian
\begin{equation*}
\mathcal{H}=-\frac{\hbar^2}{2m}\sum_i\nabla_i^2+\sum_{i<j}\frac{C_6}{{\left|\bm{r}_i-\bm{r}_j\right|}^6} .
\end{equation*} 
rewritten in a proper dimensionless form, the zero-temperature transition point happens in the density range
\begin{equation*}
\rho_c \, r_0^3=3.9 \pm 0.2\,.
\end{equation*}
The position of the phase transition is obtained by applying the double-tangent Maxwell construction technique implying that the pressure and the free energy stay unchanged along the melting curve. Possible types of the crystal packing, preferred by the system in its solid state, have been discussed. The calculations of the Madelung energy for different lattices suggest that face-centered cubic (fcc) and hexagonal close-packed (hcp) are the preferable packings with a slight advantage of fcc, although the introduction of a quantum defect correction $\delta$ in the interaction potential can give the advantage to hcp. We stick to the fcc lattice, which was used throughout our quantum Monte Carlo simulations as a preferable crystal formation in the solid phase. 

The finite temperature properties of the system were studied both with classical and quantum Monte Carlo methods. First of all we applied the classical Monte Carlo technique, based on the evolution of the system 
in accordance with its classical partition function. The estimates for the liquid-solid phase transition curve in case of high temperatures, where the classical approach is valid, have been found. The region of low temperatures, where the quantum description is required, was studied with the path integral Monte Carlo (PIMC) method. This technique allowed to find the region of 
the solid-gas phase transition as well as to localize the transition between normal fluid and superfluid phases. The results of the PIMC simulations completed the phase diagram of the system as a function of dimensionless density and dimensionless temperature, on the other hand confirming the correctness of the DMC and classical calculations, as the PIMC data came as a smooth transition between the two.

We also present a discussion of a possible treatment of the crystallized Rydberg atom clouds as a model for perspective research of the supersolid phase.

In Section~\ref{secHydrogen} we performed extensive quantum Monte Carlo calculations of atomic 
para-hydrogen \ph \ at
zero temperature below its solidification curve in a range of densities. 
In this study our principal motivation was 
to understand better the properties of the metastable liquid/glass phase at
low temperatures. In the limit of zero temperature we have used the DMC
method, which is a very efficient tool to sample this metastable
phase through the use of a trial wave function with a corresponding symmetry. 
The data provided by diffusion Monte Carlo method suggests that the Bose-Einstein 
condensate fraction is subject to a large depletion compared to that of the stable liquid $^4$He.
%
%

In conclusion in this Thesis we successfully applied quantum Monte Carlo techniques 
for studying the quantum phase diagrams in a number of physical systems which are relevant to current 
and future experiments. Our research Group 
\begin{itemize}
\item learned to carry out the Ewald summation for different physical systems and to use it for solving
the finite size correction problem; 
\item studied and implemented in the code the modified 
periodic boundary conditions (truncated tetrahedron), which might be used to enhance the 
efficiency of our Monte Carlo calculations;
\item realized the usefullness the hypernetted chain (HNC) method for constructing optimized Jastrow terms 
of trial wave functions;
\item improved the understanding of the ways to study quantum phase transitions by virtue of quantum and classic 
Monte Carlo method.

\end{itemize}
As a consequence of this work, and as a result of this Thesis, we published the following articles:

\begin{itemize}
\item[1.] O. N. Osychenko, G. E. Astrakharchik, Y. Lutsyshyn, Yu. E. Lozovik, and J. Boronat: ``{\it Phase
diagram of Rydberg atoms with repulsive van der Waals interaction}'', Phys. Rev. A {\bf 84}, 063621 (2011).
\item[2.] O. N. Osychenko, G. E. Astrakharchik, and J. Boronat: ``{\it Ewald method for polytropic potentials in
arbitrary dimensionality}'', Mol. Phys. {\bf 110}, 4, 227-247 (2012).
\item[3.] O. N. Osychenko, G. E. Astrakharchik, F. Mazzanti, and J. Boronat: ``{\it Zero-temperature phase
diagram of Yukawa bosons}'', Phys. Rev. A {\bf 85}, 063604 (2012).
\item[4.] O. N. Osychenko, R. Rota, and J. Boronat: ``{\it Superfluidity of metastable glassy bulk para-hydrogen
at low temperature}'', Phys. Rev. B {\bf 85}, 224513 (2012).
\end{itemize}

\addcontentsline{toc}{chapter}{Appendix:}
\begin{appendix}

\chapter{Ewald method for polytropic potentials}
We prove that the sums $S_{-+}$ and $S_{++}$
(\ref{psisum2}--\ref{psisum3}) vanish on average, allowing to calculate the
potential energy over the negatively charged particles' positions only.
\begin{itemize}
\item
First, let us show that the integral of $\psi$ over the cell is zero.
Since the distances are in the units of $L$, consider the cubic cell
$\Omega = (x,y,z) \in [-1/2,1/2]^3$, that yields
\begin{equation}
\int_{\Omega}\psi(\bs{r}) \di \bs{r} = J_1 + J_2 + C_1
\label{intpsi1}
\end{equation}
where
\begin{eqnarray}
J_1 &&= \int_{\Omega}\di \bs{r} \sum_{\bs{n}}R(\bs{n},\bs{r})\label{j11} \\
J_2 &&= \int_{\Omega}\di \bs{r} \sum_{\bs{m}\neq \bs{0}}K(\bs{m},\bs{r}) \
.
\label{j21}
\end{eqnarray}
It can be easily seen, that the second integral $J_2$ is zero,
\begin{eqnarray}
J_2   &&= \sum_{\bs{m}\neq
\bs{0}}\kappa(\bs{m},\bs{r})\int_{\Omega}\cos(2 \pi \bs{m}
\bs{r})\di \bs{r}\nonumber \\
&&= \sum_{\bs{m}\neq \bs{0}}\kappa(\bs{m}) \frac{\sin(2 \pi m_x+m_y+m_z))}
{(2 \pi)^3 m_x m_y m_z} = 0
\end{eqnarray}

As far as the integral $J_1$ is concerned, we can notice that the
regions $\Omega'(\bs{n})=\bs{r}+\bs{n}$, where $\bs{r} \in
\Omega,\;\bs{n} \in \mz^3$
are the same cubic unit cells, displaced by an integer vector, thus
covering all
the coordinate space with only zero-measure intersections.
It means that the summation of the integrals in (\ref{j11}) over the
cell $\Omega$
can be substituted by the integration over the whole coordinate space,
\begin{eqnarray}
J_1 &&= \sum_{\bs{n}} \int_{\Omega'(\bs{n})}R(\bs{n},\bs{r})\di \bs{r}=
\sum_{\bs{n}} \int_{\Omega'(\bs{n})}\frac{\Gamma (\frac{k}{2},\alpha
^2|\bs{r}+\bs{n}|^2 )}
{\Gamma (\frac{k}{2})|\bs{r}+\bs{n}|^k}\nonumber \\
&&= \alpha^{k-3} \int_{\mr^3 \setminus 0}\frac{\Gamma
(\frac{k}{2},\bs{\rho}^2)}
{\Gamma (\frac{k}{2})\bs{\rho}^k}\di \bs{\rho} = -\frac{2 \pi
^{\frac{3}{2}}
\alpha ^{k-3}}{(k-3) \Gamma\left[\frac{k}{2}\right]}=-C_1 \ ,
\end{eqnarray}
and thus the whole integral (\ref{intpsi1}) is equal to zero.

\item
Consider two species of the particles: negative charges $q_i$ on
positions $\bs{r}_i$ and a positively charged and uniformly distributed background
with a total charge $q_{+}N_{+}=-q_i N_i$, ensuring the neutrality of the cell.
 Let us demonstrate that $S_{-+}$ is equal to zero, when the number of background charges tends to infinity.
In this case the sum (\ref{psisum2}) for $S_{-+}$ may be rewritten as an integral over
 the background charges' positions
\begin{equation}
S_{-+}=\sum_i q_i \int_{\Omega}\psi(\bs{r}_p-\bs{r}_i)\sigma \di
\bs{r}_p = \sum_i q_i \int_{\Omega_i}\psi(\bs{r})\sigma \di  \bs{r} \ ,
\label{Sepapp1}
\end{equation}
where we did the change of variables $\bs{r}=\bs{r}_p-\bs{r}_i$. The
regions $\Omega$ and $\Omega_i$ refer to the original simulation cell
and the same cell, moved by the vector $\bs{r}_i$,
 and $\sigma$ stands for the background charge density $\sigma=-q_i N_i/V(\Omega)$. It is clear that
every vector $\bs{r}=(x,y,z)\in \Omega_i$ can be displaced into the cell
$\Omega$ by the corresponding shift
$\tilde{\bs{r}}=(\tilde{x},\tilde{y},\tilde{z})=(x-aL,y-bL,z-cL)\in
\Omega$ with integers $a,\:b,\:c$. The Jacobian $J$ of the change of
variables $\bs{r}\rightarrow\tilde{\bs{r}}$ is obviously 1. On the other
hand, due to the periodicity of $\psi$,
\begin{equation}
\psi(\bs{r}) = \psi(\tilde{\bs{r}})\ ,
\end{equation}
and $\tilde{\bs{r}}$ runs over the whole region $\Omega$ due to the
conservation of the volume with $J=1$. Finally, Eq.~(\ref{Sepapp1}) can
be written as
\begin{equation}
S_{-+}=\sum_i q_i \int_{\Omega}\psi(\tilde{\bs{r}}) \sigma \di
\tilde{\bs{r}} = 0\ .
\end{equation}

In the similar manner,
 the interaction between the charges of the background $S_{++}$ in the limit $N_{+}\rightarrow\infty$ is
given by the double integral
\begin{equation}
S_{++}=\frac{1}{2}\int_{\Omega}\di \bs{r}_1\int_{\Omega} \di \bs{r}_2
\psi(\bs{r}_1-\bs{r}_2) \sigma^2  =0 \ ,
\label{Sppapp1}
\end{equation}
since $\int_{\Omega}\psi(\bs{r}_1-\bs{r}_2) \di \bs{r}_2=0$, following
the same arguments as for the case of $S_{-+}$.
\end{itemize}

\chapter{Truncated octahedron boundary conditions\label{secTO}}

Apart from the standard cubic periodic boundary conditions we expored a possibility to apply the  truncated octahedron periodic boundary conditions. This kind of periodic conditions has an advantage to reduce effects of anisotropy as well as enhance the overall efficiency of the simulation.

The truncated octahedron simulation cell, represented by a cube with removed angles, possesses a periodicity with the steps $\{\bs{L}\}_p=(\pm L/2,\pm L/2, \pm L/2)$, which means that only a crystal formation of a cubic class (simple cubic, bcc, fcc) with a composition, corresponding to even values of $L/a$ ($a=(N_{cell}/\rho)^{1/3}$ stands for a size of the elementary cell) is commensurate with the cell. It means that the simulation in periodic boundary conditions is only possible for a certain reduced set of particles in the box. In practice it is equivalent to the exclusion of the crystal configurations with odd values of $L/a$; for instance, in case of fcc lattice one has to omit $N_p=2(2i+1)^3=2,\:64,\:250,\:686...$.

The situation regarding hcp formation is similar, with the only difference that as it always happens in the case of hcp the simulation should be performed in a cell with unequal size lengths $L_x,L_y,L_z$. If the truncation of the interaction (minimum image convention) is to be applied, one takes the distance from the center to the nearest plane, that is the plane with the equation $X/L_x+Y/L_y+Z/L_z=3/4$. The spherical cut-off radius in this case is equal to
\begin{equation}
r_{\rm cut}=\frac{3/4}{\sqrt{1/L_x^2+1/L_y^2+1/L_z^2}}\, .
\end{equation}
The values of $L_x$, $L_y$, $L_z$ are chosen such that the cell is commensurate with the lattice. We also require that an arbitrary periodic translation $\bs{r}\rightarrow \bs{r}+\{\bs{L}\}$ brings a lattice site again to a site. This condition results in a demand of having all numbers $L_x/a_x$, $L_y/a_y$, $L_z/a_z$ even, that again reduces the variety of options for a number of particles in a cell to $N_p=32,256,864,...$. 

The number of particles $N_p$ in a simulation in a truncated octahedron cell corresponds to a cut-off radius 
\begin{equation}
r_c=\frac{\sqrt{3}}{4}(2 N_p/\rho)^{1/3}
\end{equation}
which is a distance to the nearest plane of the cell. The same distance in the cubic p.b.c. is obtained by $3\sqrt{3}/4 N_p\approx 1.3$ or, as diffusion Monte Carlo technique is $\mathcal{O}(N_p^2)$, by about 68\% of additional calculation time. This effect is hindered by a more complex implementation of a particle motion, but in practice the overall efficiency gap is never below 30\%. 

As it was mentioned above, a truncated octahedron cell replicated periodically with the displacements from the set $\{\bs{L}\}$ fills entirely the coordinate space. This makes the application of the Ewald technique theoretically possible, once the original potential energy sum is rewritten in a suitable form, compatible with the standard $x,y,z$-axis periodicity, required by the Ewald method. Indeed, as one can easily notice, the total set of cell images can be classified into two groups: the images, produced by an even number steps with a resulting cell displacement $(iL,jL,kL)$  ($i$, $j$, $k$ are arbitrary integers) and by an odd number of step with a displacement $(iL,jL,kL)+(L/2,L/2,L/2)$.
If one thinks of a simulation cell as a compound of the original cell $\Omega$ and the displaced cell $\Omega_d=(L/2,L/2,L/2)+\Omega$, the total space will be covered again by its replications along the axis $x,y,z$ with the period $L$. This allows to apply the Ewald summation technique directly to this complex cell with the number of particles equal to $2N_p$, but due to the similarity of $\Omega$ and $\Omega_d$, the sums may be simplified in the following way:

\begin{eqnarray}
\frac{1}{2}\sum_{i\neq j}^{2N_p}\sum_{\bs{n}}R(\bs{r}_{ij})&=&\frac{1}{2}\sum_{\bs{n}}\left[\sum_{i,j=1; i\neq j}^{N_p}
+\sum_{i,j=N_p+1; i\neq j}^{2N_p}\right.+\left.\sum_{i=1}^{N_p}\sum_{j=N+1}^{2N_p}+\sum_{i=N_p+1}^{2N_p}\sum_{j=1}^{N_p}\right]R(\bs{r}_{ij})\nonumber\\
&=&\sum_{\bs{n}}\left[\sum_{i,j=1; i\neq j}^{N_p}+\sum_{i=1}^{N_p}\sum_{j=N+1}^{2N_p}\right]R(\bs{r}_{ij})\nonumber\\
&=&\sum_{\bs{n}}\left[\sum_{i,j=1; i\neq j}^{N_p}R(\bs{r}_{ij})+\sum_{i,j=1}^{N_p}R(\bs{r}_{ij}+\bs{L}/2)\right]\nonumber\\
&=&\sum_{\bs{n}}\left[\sum_{i,j=1; i\neq j}^{N_p}(R(\bs{r}_{ij})+R(\bs{r}_{ij}+\bs{L}/2))
+\sum_{i=1}^{N_p}R(\bs{L}/2)\right]\, .
\end{eqnarray}
where we place the points $1,...,N_p$ to the cell $\Omega$ and their corresponding images $1+Np,...,2Np$ to $\Omega_d$. In a similar line of thought we can represent the Fourier-transformed part of the Ewald sum:
\begin{equation}
\frac{1}{2}\sum_{i\neq j}^{2N_p}\sum_{\bs{m}\neq 0}K(\bs{r}_{ij})=
\sum_{\bs{m}\neq 0}(\sum_{i,j=1; i\neq j}^{N_p}(K(\bs{r}_{ij})+K(\bs{r}_{ij}+\bs{L}/2)))
+\sum_{i=1}^{N_p}K(\bs{L}/2))\, .
\end{equation}

The constants $\xi$, $C_1$, $C_2$ (refer to Eqs.~(\ref{xipoly1},\ref{c1},\ref{c2})) stay clearly unchanged, as they characterize the self-image interactions, not biased by a specific cell geometry. Notice that the number of particles, appearing as a factor of $\xi$ in the expression for the total energy is now $2N_p$.

\end{appendix}
\bibliography{ref2}
\bibliographystyle{alpha}

\chapter*{Acknowledgements}
\addcontentsline{toc}{chapter}{Acknowledgements}
I acknowledge the financial support of DGI and Generalitat
de Catalunya (Spain) through the grants, that allowed to bring this work to life.
I would like to express my deep gratitude to my advisers, Jordi and 
Grigory, and all my other collaborators, for I learned a lot from them both 
professionally and personally. I dare to hope that working at the University I
found not only colleagues, but also good friends. I would like to thank 
Yaroslav Lutsyshyn for his careful reading of the manuscript. Finally, I can't forget to 
mention my family, that despite being far from me, was always
very caring and supportive.

\end{document}